\documentclass[traditabstract,longauth]{aa}

\usepackage{txfonts}
\usepackage{graphicx}
\usepackage{longtable}
\usepackage{multirow}
\usepackage{dcolumn}
\usepackage{lscape}
\usepackage{natbib}
\usepackage{multibib}
\usepackage{hyperref}

\bibpunct{(}{)}{;}{a}{}{,} 

\def\setsymbol#1#2{\expandafter\def\csname #1\endcsname{#2}}
\def\getsymbol#1{\csname #1\endcsname}

\def\Planck{{\it Planck\/}}

\newbox\tablebox    \newdimen\tablewidth
\def\leaderfil{\leaders\hbox to 5pt{\hss.\hss}\hfil}
\def\tablenote#1 #2\par{\begingroup \parindent=0.8em
    \abovedisplayshortskip=0pt\belowdisplayshortskip=0pt
    \noindent
    $$\hss\vbox{\hsize\tablewidth \hangindent=\parindent \hangafter=1 \noindent
    \hbox to \parindent{\sup{\rm #1}\hss}\strut#2\strut\par}\hss$$
    \endgroup}

\def\L2{\ifmmode L_2\else $L_2$\fi}

\def\DeltaT{\ifmmode \Delta T\else $\Delta T$\fi}
\def\deltat{\ifmmode \Delta t\else $\Delta t$\fi}
\def\fknee{\ifmmode f_{\rm knee}\else $f_{\rm knee}$\fi}
\def\Fmax{\ifmmode F_{\rm max}\else $F_{\rm max}$\fi}
\def\solar{\ifmmode{\rm M}_{\mathord\odot}\else${\rm M}_{\mathord\odot}$\fi}

\def\inv{\ifmmode^{-1}\else$^{-1}$\fi}
\def\mo{\ifmmode^{-1}\else$^{-1}$\fi}
\def\sup#1{\ifmmode ^{\rm #1}\else $^{\rm #1}$\fi}
\def\expo#1{\ifmmode \times 10^{#1}\else $\times 10^{#1}$\fi}
\def\,{\thinspace}
\def\lsim{\mathrel{\raise .4ex\hbox{\rlap{$<$}\lower 1.2ex\hbox{$\sim$}}}}
\def\gsim{\mathrel{\raise .4ex\hbox{\rlap{$>$}\lower 1.2ex\hbox{$\sim$}}}}

\def\simprop{\mathrel{\raise .4ex\hbox{\rlap{$\propto$}\lower 1.2ex\hbox{$\sim$}}}}
\def\deg{\ifmmode^\circ\else$^\circ$\fi}
\def\pdeg{\ifmmode $\setbox0=\hbox{$^{\circ}$}\rlap{\hskip.11\wd0 .}$^{\circ}
          \else \setbox0=\hbox{$^{\circ}$}\rlap{\hskip.11\wd0 .}$^{\circ}$\fi}
\def\arcs{\ifmmode {^{\scriptstyle\prime\prime}}
          \else $^{\scriptstyle\prime\prime}$\fi}
\def\arcm{\ifmmode {^{\scriptstyle\prime}}
          \else $^{\scriptstyle\prime}$\fi}
\newdimen\sa  \newdimen\sb
\def\parcs{\sa=.07em \sb=.03em
     \ifmmode \hbox{\rlap{.}}^{\scriptstyle\prime\kern -\sb\prime}\hbox{\kern -\sa}
     \else \rlap{.}$^{\scriptstyle\prime\kern -\sb\prime}$\kern -\sa\fi}
\def\parcm{\sa=.08em \sb=.03em
     \ifmmode \hbox{\rlap{.}\kern\sa}^{\scriptstyle\prime}\hbox{\kern-\sb}
     \else \rlap{.}\kern\sa$^{\scriptstyle\prime}$\kern-\sb\fi}
\def\ra[#1 #2 #3.#4]{#1\sup{h}#2\sup{m}#3\sup{s}\llap.#4}
\def\dec[#1 #2 #3.#4]{#1\deg#2\arcm#3\arcs\llap.#4}
\def\deco[#1 #2 #3]{#1\deg#2\arcm#3\arcs}
\def\rra[#1 #2]{#1\sup{h}#2\sup{m}}

\def\dots{\relax\ifmmode \ldots\else $\ldots$\fi}
\def\WHzsr{\ifmmode $W\,Hz\mo\,sr\mo$\else W\,Hz\mo\,sr\mo\fi}
\def\mHz{\ifmmode $\,mHz$\else \,mHz\fi}
\def\GHz{\ifmmode $\,GHz$\else \,GHz\fi}
\def\mKs{\ifmmode $\,mK\,s$^{1/2}\else \,mK\,s$^{1/2}$\fi}
\def\muKs{\ifmmode \,\mu$K\,s$^{1/2}\else \,$\mu$K\,s$^{1/2}$\fi}
\def\muKRJs{\ifmmode \,\mu$K$_{\rm RJ}$\,s$^{1/2}\else \,$\mu$K$_{\rm RJ}$\,s$^{1/2}$\fi}
\def\muKHz{\ifmmode \,\mu$K\,Hz$^{-1/2}\else \,$\mu$K\,Hz$^{-1/2}$\fi}
\def\MJysr{\ifmmode \,$MJy\,sr\mo$\else \,MJy\,sr\mo\fi}
\def\MJysrmK{\ifmmode \,$MJy\,sr\mo$\,mK$_{\rm CMB}\mo\else \,MJy\,sr\mo\,mK$_{\rm CMB}\mo$\fi}
\def\microns{\ifmmode \,\mu$m$\else \,$\mu$m\fi}
\def\micron{\microns}
\def\muK{\ifmmode \,\mu$K$\else \,$\mu$\hbox{K}\fi}
\def\microK{\ifmmode \,\mu$K$\else \,$\mu$\hbox{K}\fi}
\def\muW{\ifmmode \,\mu$W$\else \,$\mu$\hbox{W}\fi}
\def\kms{\ifmmode $\,km\,s$^{-1}\else \,km\,s$^{-1}$\fi}
\def\kmsMpc{\ifmmode $\,\kms\,Mpc\mo$\else \,\kms\,Mpc\mo\fi}
\setsymbol{LFI:center:frequency:70GHz:units}{70.3\,GHz}
\setsymbol{LFI:center:frequency:44GHz:units}{44.1\,GHz}
\setsymbol{LFI:center:frequency:30GHz:units}{28.5\,GHz}

\setsymbol{LFI:center:frequency:70GHz}{70.3}
\setsymbol{LFI:center:frequency:44GHz}{44.1}
\setsymbol{LFI:center:frequency:30GHz}{28.5}

\setsymbol{LFI:center:frequency:LFI18:Rad:M:units}{71.7\GHz}
\setsymbol{LFI:center:frequency:LFI19:Rad:M:units}{67.5\GHz}
\setsymbol{LFI:center:frequency:LFI20:Rad:M:units}{69.2\GHz}
\setsymbol{LFI:center:frequency:LFI21:Rad:M:units}{70.4\GHz}
\setsymbol{LFI:center:frequency:LFI22:Rad:M:units}{71.5\GHz}
\setsymbol{LFI:center:frequency:LFI23:Rad:M:units}{70.8\GHz}
\setsymbol{LFI:center:frequency:LFI24:Rad:M:units}{44.4\GHz}
\setsymbol{LFI:center:frequency:LFI25:Rad:M:units}{44.0\GHz}
\setsymbol{LFI:center:frequency:LFI26:Rad:M:units}{43.9\GHz}
\setsymbol{LFI:center:frequency:LFI27:Rad:M:units}{28.3\GHz}
\setsymbol{LFI:center:frequency:LFI28:Rad:M:units}{28.8\GHz}
\setsymbol{LFI:center:frequency:LFI18:Rad:S:units}{70.1\GHz}
\setsymbol{LFI:center:frequency:LFI19:Rad:S:units}{69.6\GHz}
\setsymbol{LFI:center:frequency:LFI20:Rad:S:units}{69.5\GHz}
\setsymbol{LFI:center:frequency:LFI21:Rad:S:units}{69.5\GHz}
\setsymbol{LFI:center:frequency:LFI22:Rad:S:units}{72.8\GHz}
\setsymbol{LFI:center:frequency:LFI23:Rad:S:units}{71.3\GHz}
\setsymbol{LFI:center:frequency:LFI24:Rad:S:units}{44.1\GHz}
\setsymbol{LFI:center:frequency:LFI25:Rad:S:units}{44.1\GHz}
\setsymbol{LFI:center:frequency:LFI26:Rad:S:units}{44.1\GHz}
\setsymbol{LFI:center:frequency:LFI27:Rad:S:units}{28.5\GHz}
\setsymbol{LFI:center:frequency:LFI28:Rad:S:units}{28.2\GHz}

\setsymbol{LFI:center:frequency:LFI18:Rad:M}{71.7}
\setsymbol{LFI:center:frequency:LFI19:Rad:M}{67.5}
\setsymbol{LFI:center:frequency:LFI20:Rad:M}{69.2}
\setsymbol{LFI:center:frequency:LFI21:Rad:M}{70.4}
\setsymbol{LFI:center:frequency:LFI22:Rad:M}{71.5}
\setsymbol{LFI:center:frequency:LFI23:Rad:M}{70.8}
\setsymbol{LFI:center:frequency:LFI24:Rad:M}{44.4}
\setsymbol{LFI:center:frequency:LFI25:Rad:M}{44.0}
\setsymbol{LFI:center:frequency:LFI26:Rad:M}{43.9}
\setsymbol{LFI:center:frequency:LFI27:Rad:M}{28.3}
\setsymbol{LFI:center:frequency:LFI28:Rad:M}{28.8}
\setsymbol{LFI:center:frequency:LFI18:Rad:S}{70.1}
\setsymbol{LFI:center:frequency:LFI19:Rad:S}{69.6}
\setsymbol{LFI:center:frequency:LFI20:Rad:S}{69.5}
\setsymbol{LFI:center:frequency:LFI21:Rad:S}{69.5}
\setsymbol{LFI:center:frequency:LFI22:Rad:S}{72.8}
\setsymbol{LFI:center:frequency:LFI23:Rad:S}{71.3}
\setsymbol{LFI:center:frequency:LFI24:Rad:S}{44.1}
\setsymbol{LFI:center:frequency:LFI25:Rad:S}{44.1}
\setsymbol{LFI:center:frequency:LFI26:Rad:S}{44.1}
\setsymbol{LFI:center:frequency:LFI27:Rad:S}{28.5}
\setsymbol{LFI:center:frequency:LFI28:Rad:S}{28.2}

\setsymbol{LFI:white:noise:sensitivity:70GHz:units}{134.7\muKs}
\setsymbol{LFI:white:noise:sensitivity:44GHz:units}{164.7\muKs}
\setsymbol{LFI:white:noise:sensitivity:30GHz:units}{143.4\muKs}

\setsymbol{LFI:white:noise:sensitivity:70GHz}{134.7}
\setsymbol{LFI:white:noise:sensitivity:44GHz}{164.7}
\setsymbol{LFI:white:noise:sensitivity:30GHz}{143.4}

\setsymbol{LFI:white:noise:sensitivity:LFI18:Rad:M:units}{512.0\muKs}
\setsymbol{LFI:white:noise:sensitivity:LFI19:Rad:M:units}{581.4\muKs}
\setsymbol{LFI:white:noise:sensitivity:LFI20:Rad:M:units}{590.8\muKs}
\setsymbol{LFI:white:noise:sensitivity:LFI21:Rad:M:units}{455.2\muKs}
\setsymbol{LFI:white:noise:sensitivity:LFI22:Rad:M:units}{492.0\muKs}
\setsymbol{LFI:white:noise:sensitivity:LFI23:Rad:M:units}{507.7\muKs}
\setsymbol{LFI:white:noise:sensitivity:LFI24:Rad:M:units}{462.2\muKs}
\setsymbol{LFI:white:noise:sensitivity:LFI25:Rad:M:units}{413.6\muKs}
\setsymbol{LFI:white:noise:sensitivity:LFI26:Rad:M:units}{478.6\muKs}
\setsymbol{LFI:white:noise:sensitivity:LFI27:Rad:M:units}{277.7\muKs}
\setsymbol{LFI:white:noise:sensitivity:LFI28:Rad:M:units}{312.3\muKs}
\setsymbol{LFI:white:noise:sensitivity:LFI18:Rad:S:units}{465.7\muKs}
\setsymbol{LFI:white:noise:sensitivity:LFI19:Rad:S:units}{555.6\muKs}
\setsymbol{LFI:white:noise:sensitivity:LFI20:Rad:S:units}{623.2\muKs}
\setsymbol{LFI:white:noise:sensitivity:LFI21:Rad:S:units}{564.1\muKs}
\setsymbol{LFI:white:noise:sensitivity:LFI22:Rad:S:units}{534.4\muKs}
\setsymbol{LFI:white:noise:sensitivity:LFI23:Rad:S:units}{542.4\muKs}
\setsymbol{LFI:white:noise:sensitivity:LFI24:Rad:S:units}{399.2\muKs}
\setsymbol{LFI:white:noise:sensitivity:LFI25:Rad:S:units}{392.6\muKs}
\setsymbol{LFI:white:noise:sensitivity:LFI26:Rad:S:units}{418.6\muKs}
\setsymbol{LFI:white:noise:sensitivity:LFI27:Rad:S:units}{302.9\muKs}
\setsymbol{LFI:white:noise:sensitivity:LFI28:Rad:S:units}{285.3\muKs}

\setsymbol{LFI:white:noise:sensitivity:LFI18:Rad:M}{512.0}
\setsymbol{LFI:white:noise:sensitivity:LFI19:Rad:M}{581.4}
\setsymbol{LFI:white:noise:sensitivity:LFI20:Rad:M}{590.8}
\setsymbol{LFI:white:noise:sensitivity:LFI21:Rad:M}{455.2}
\setsymbol{LFI:white:noise:sensitivity:LFI22:Rad:M}{492.0}
\setsymbol{LFI:white:noise:sensitivity:LFI23:Rad:M}{507.7}
\setsymbol{LFI:white:noise:sensitivity:LFI24:Rad:M}{462.2}
\setsymbol{LFI:white:noise:sensitivity:LFI25:Rad:M}{413.6}
\setsymbol{LFI:white:noise:sensitivity:LFI26:Rad:M}{478.6}
\setsymbol{LFI:white:noise:sensitivity:LFI27:Rad:M}{277.7}
\setsymbol{LFI:white:noise:sensitivity:LFI28:Rad:M}{312.3}
\setsymbol{LFI:white:noise:sensitivity:LFI18:Rad:S}{465.7}
\setsymbol{LFI:white:noise:sensitivity:LFI19:Rad:S}{555.6}
\setsymbol{LFI:white:noise:sensitivity:LFI20:Rad:S}{623.2}
\setsymbol{LFI:white:noise:sensitivity:LFI21:Rad:S}{564.1}
\setsymbol{LFI:white:noise:sensitivity:LFI22:Rad:S}{534.4}
\setsymbol{LFI:white:noise:sensitivity:LFI23:Rad:S}{542.4}
\setsymbol{LFI:white:noise:sensitivity:LFI24:Rad:S}{399.2}
\setsymbol{LFI:white:noise:sensitivity:LFI25:Rad:S}{392.6}
\setsymbol{LFI:white:noise:sensitivity:LFI26:Rad:S}{418.6}
\setsymbol{LFI:white:noise:sensitivity:LFI27:Rad:S}{302.9}
\setsymbol{LFI:white:noise:sensitivity:LFI28:Rad:S}{285.3}

\setsymbol{LFI:knee:frequency:70GHz:units}{29.5\mHz}
\setsymbol{LFI:knee:frequency:44GHz:units}{56.2\mHz}
\setsymbol{LFI:knee:frequency:30GHz:units}{113.7\mHz}

\setsymbol{LFI:knee:frequency:70GHz}{29.5}
\setsymbol{LFI:knee:frequency:44GHz}{56.2}
\setsymbol{LFI:knee:frequency:30GHz}{113.7}

\setsymbol{LFI:knee:frequency:LFI18:Rad:M:units}{16.3\mHz}
\setsymbol{LFI:knee:frequency:LFI19:Rad:M:units}{15.1\mHz}
\setsymbol{LFI:knee:frequency:LFI20:Rad:M:units}{18.7\mHz}
\setsymbol{LFI:knee:frequency:LFI21:Rad:M:units}{37.2\mHz}
\setsymbol{LFI:knee:frequency:LFI22:Rad:M:units}{12.7\mHz}
\setsymbol{LFI:knee:frequency:LFI23:Rad:M:units}{34.6\mHz}
\setsymbol{LFI:knee:frequency:LFI24:Rad:M:units}{46.2\mHz}
\setsymbol{LFI:knee:frequency:LFI25:Rad:M:units}{24.9\mHz}
\setsymbol{LFI:knee:frequency:LFI26:Rad:M:units}{67.6\mHz}
\setsymbol{LFI:knee:frequency:LFI27:Rad:M:units}{187.4\mHz}
\setsymbol{LFI:knee:frequency:LFI28:Rad:M:units}{122.2\mHz}
\setsymbol{LFI:knee:frequency:LFI18:Rad:S:units}{17.7\mHz}
\setsymbol{LFI:knee:frequency:LFI19:Rad:S:units}{22.0\mHz}
\setsymbol{LFI:knee:frequency:LFI20:Rad:S:units}{8.7\mHz}
\setsymbol{LFI:knee:frequency:LFI21:Rad:S:units}{25.9\mHz}
\setsymbol{LFI:knee:frequency:LFI22:Rad:S:units}{15.8\mHz}
\setsymbol{LFI:knee:frequency:LFI23:Rad:S:units}{129.8\mHz}
\setsymbol{LFI:knee:frequency:LFI24:Rad:S:units}{100.9\mHz}
\setsymbol{LFI:knee:frequency:LFI25:Rad:S:units}{38.9\mHz}
\setsymbol{LFI:knee:frequency:LFI26:Rad:S:units}{58.9\mHz}
\setsymbol{LFI:knee:frequency:LFI27:Rad:S:units}{104.4\mHz}
\setsymbol{LFI:knee:frequency:LFI28:Rad:S:units}{40.7\mHz}

\setsymbol{LFI:knee:frequency:LFI18:Rad:M}{16.3}
\setsymbol{LFI:knee:frequency:LFI19:Rad:M}{15.1}
\setsymbol{LFI:knee:frequency:LFI20:Rad:M}{18.7}
\setsymbol{LFI:knee:frequency:LFI21:Rad:M}{37.2}
\setsymbol{LFI:knee:frequency:LFI22:Rad:M}{12.7}
\setsymbol{LFI:knee:frequency:LFI23:Rad:M}{34.6}
\setsymbol{LFI:knee:frequency:LFI24:Rad:M}{46.2}
\setsymbol{LFI:knee:frequency:LFI25:Rad:M}{24.9}
\setsymbol{LFI:knee:frequency:LFI26:Rad:M}{67.6}
\setsymbol{LFI:knee:frequency:LFI27:Rad:M}{187.4}
\setsymbol{LFI:knee:frequency:LFI28:Rad:M}{122.2}
\setsymbol{LFI:knee:frequency:LFI18:Rad:S}{17.7}
\setsymbol{LFI:knee:frequency:LFI19:Rad:S}{22.0}
\setsymbol{LFI:knee:frequency:LFI20:Rad:S}{8.7}
\setsymbol{LFI:knee:frequency:LFI21:Rad:S}{25.9}
\setsymbol{LFI:knee:frequency:LFI22:Rad:S}{15.8}
\setsymbol{LFI:knee:frequency:LFI23:Rad:S}{129.8}
\setsymbol{LFI:knee:frequency:LFI24:Rad:S}{100.9}
\setsymbol{LFI:knee:frequency:LFI25:Rad:S}{38.9}
\setsymbol{LFI:knee:frequency:LFI26:Rad:S}{58.9}
\setsymbol{LFI:knee:frequency:LFI27:Rad:S}{104.4}
\setsymbol{LFI:knee:frequency:LFI28:Rad:S}{40.7}

\setsymbol{LFI:slope:70GHz:units}{$-1.03$\mHz}
\setsymbol{LFI:slope:44GHz:units}{$-0.89$\mHz}
\setsymbol{LFI:slope:30GHz:units}{$-0.87$\mHz}

\setsymbol{LFI:slope:70GHz}{$-1.03$}
\setsymbol{LFI:slope:44GHz}{$-0.89$}
\setsymbol{LFI:slope:30GHz}{$-0.87$}

\setsymbol{LFI:slope:LFI18:Rad:M:units}{$-1.04$\mHz}
\setsymbol{LFI:slope:LFI19:Rad:M:units}{$-1.09$\mHz}
\setsymbol{LFI:slope:LFI20:Rad:M:units}{$-0.69$\mHz}
\setsymbol{LFI:slope:LFI21:Rad:M:units}{$-1.56$\mHz}
\setsymbol{LFI:slope:LFI22:Rad:M:units}{$-1.01$\mHz}
\setsymbol{LFI:slope:LFI23:Rad:M:units}{$-0.96$\mHz}
\setsymbol{LFI:slope:LFI24:Rad:M:units}{$-0.83$\mHz}
\setsymbol{LFI:slope:LFI25:Rad:M:units}{$-0.91$\mHz}
\setsymbol{LFI:slope:LFI26:Rad:M:units}{$-0.95$\mHz}
\setsymbol{LFI:slope:LFI27:Rad:M:units}{$-0.87$\mHz}
\setsymbol{LFI:slope:LFI28:Rad:M:units}{$-0.88$\mHz}
\setsymbol{LFI:slope:LFI18:Rad:S:units}{$-1.15$\mHz}
\setsymbol{LFI:slope:LFI19:Rad:S:units}{$-1.00$\mHz}
\setsymbol{LFI:slope:LFI20:Rad:S:units}{$-0.95$\mHz}
\setsymbol{LFI:slope:LFI21:Rad:S:units}{$-0.92$\mHz}
\setsymbol{LFI:slope:LFI22:Rad:S:units}{$-1.01$\mHz}
\setsymbol{LFI:slope:LFI23:Rad:S:units}{$-0.95$\mHz}
\setsymbol{LFI:slope:LFI24:Rad:S:units}{$-0.73$\mHz}
\setsymbol{LFI:slope:LFI25:Rad:S:units}{$-1.16$\mHz}
\setsymbol{LFI:slope:LFI26:Rad:S:units}{$-0.79$\mHz}
\setsymbol{LFI:slope:LFI27:Rad:S:units}{$-0.82$\mHz}
\setsymbol{LFI:slope:LFI28:Rad:S:units}{$-0.91$\mHz}

\setsymbol{LFI:slope:LFI18:Rad:M}{$-1.04$}
\setsymbol{LFI:slope:LFI19:Rad:M}{$-1.09$}
\setsymbol{LFI:slope:LFI20:Rad:M}{$-0.69$}
\setsymbol{LFI:slope:LFI21:Rad:M}{$-1.56$}
\setsymbol{LFI:slope:LFI22:Rad:M}{$-1.01$}
\setsymbol{LFI:slope:LFI23:Rad:M}{$-0.96$}
\setsymbol{LFI:slope:LFI24:Rad:M}{$-0.83$}
\setsymbol{LFI:slope:LFI25:Rad:M}{$-0.91$}
\setsymbol{LFI:slope:LFI26:Rad:M}{$-0.95$}
\setsymbol{LFI:slope:LFI27:Rad:M}{$-0.87$}
\setsymbol{LFI:slope:LFI28:Rad:M}{$-0.88$}
\setsymbol{LFI:slope:LFI18:Rad:S}{$-1.15$}
\setsymbol{LFI:slope:LFI19:Rad:S}{$-1.00$}
\setsymbol{LFI:slope:LFI20:Rad:S}{$-0.95$}
\setsymbol{LFI:slope:LFI21:Rad:S}{$-0.92$}
\setsymbol{LFI:slope:LFI22:Rad:S}{$-1.01$}
\setsymbol{LFI:slope:LFI23:Rad:S}{$-0.95$}
\setsymbol{LFI:slope:LFI24:Rad:S}{$-0.73$}
\setsymbol{LFI:slope:LFI25:Rad:S}{$-1.16$}
\setsymbol{LFI:slope:LFI26:Rad:S}{$-0.79$}
\setsymbol{LFI:slope:LFI27:Rad:S}{$-0.82$}
\setsymbol{LFI:slope:LFI28:Rad:S}{$-0.91$}

\setsymbol{LFI:FWHM:70GHz:units}{13\parcm01}
\setsymbol{LFI:FWHM:44GHz:units}{27\parcm92}
\setsymbol{LFI:FWHM:30GHz:units}{32\parcm65}

\setsymbol{LFI:FWHM:70GHz}{13.01}
\setsymbol{LFI:FWHM:44GHz}{27.92}
\setsymbol{LFI:FWHM:30GHz}{32.65}

\setsymbol{LFI:FWHM:LFI18:units}{13\parcm39}
\setsymbol{LFI:FWHM:LFI19:units}{13\parcm01}
\setsymbol{LFI:FWHM:LFI20:units}{12\parcm75}
\setsymbol{LFI:FWHM:LFI21:units}{12\parcm74}
\setsymbol{LFI:FWHM:LFI22:units}{12\parcm87}
\setsymbol{LFI:FWHM:LFI23:units}{13\parcm27}
\setsymbol{LFI:FWHM:LFI24:units}{22\parcm98}
\setsymbol{LFI:FWHM:LFI25:units}{30\parcm46}
\setsymbol{LFI:FWHM:LFI26:units}{30\parcm31}
\setsymbol{LFI:FWHM:LFI27:units}{32\parcm65}
\setsymbol{LFI:FWHM:LFI28:units}{32\parcm66}

\setsymbol{LFI:FWHM:LFI18}{13.39}
\setsymbol{LFI:FWHM:LFI19}{13.01}
\setsymbol{LFI:FWHM:LFI20}{12.75}
\setsymbol{LFI:FWHM:LFI21}{12.74}
\setsymbol{LFI:FWHM:LFI22}{12.87}
\setsymbol{LFI:FWHM:LFI23}{13.27}
\setsymbol{LFI:FWHM:LFI24}{22.98}
\setsymbol{LFI:FWHM:LFI25}{30.46}
\setsymbol{LFI:FWHM:LFI26}{30.31}
\setsymbol{LFI:FWHM:LFI27}{32.65}
\setsymbol{LFI:FWHM:LFI28}{32.66}

\setsymbol{LFI:FWHM:uncertainty:LFI18:units}{0.170\arcm}
\setsymbol{LFI:FWHM:uncertainty:LFI19:units}{0.174\arcm}
\setsymbol{LFI:FWHM:uncertainty:LFI20:units}{0.170\arcm}
\setsymbol{LFI:FWHM:uncertainty:LFI21:units}{0.156\arcm}
\setsymbol{LFI:FWHM:uncertainty:LFI22:units}{0.164\arcm}
\setsymbol{LFI:FWHM:uncertainty:LFI23:units}{0.171\arcm}
\setsymbol{LFI:FWHM:uncertainty:LFI24:units}{0.652\arcm}
\setsymbol{LFI:FWHM:uncertainty:LFI25:units}{1.075\arcm}
\setsymbol{LFI:FWHM:uncertainty:LFI26:units}{1.131\arcm}
\setsymbol{LFI:FWHM:uncertainty:LFI27:units}{1.266\arcm}
\setsymbol{LFI:FWHM:uncertainty:LFI28:units}{1.287\arcm}

\setsymbol{LFI:FWHM:uncertainty:LFI18}{0.170}
\setsymbol{LFI:FWHM:uncertainty:LFI19}{0.174}
\setsymbol{LFI:FWHM:uncertainty:LFI20}{0.170}
\setsymbol{LFI:FWHM:uncertainty:LFI21}{0.156}
\setsymbol{LFI:FWHM:uncertainty:LFI22}{0.164}
\setsymbol{LFI:FWHM:uncertainty:LFI23}{0.171}
\setsymbol{LFI:FWHM:uncertainty:LFI24}{0.652}
\setsymbol{LFI:FWHM:uncertainty:LFI25}{1.075}
\setsymbol{LFI:FWHM:uncertainty:LFI26}{1.131}
\setsymbol{LFI:FWHM:uncertainty:LFI27}{1.266}
\setsymbol{LFI:FWHM:uncertainty:LFI28}{1.287}

\setsymbol{HFI:center:frequency:100GHz:units}{100\,GHz}
\setsymbol{HFI:center:frequency:143GHz:units}{143\,GHz}
\setsymbol{HFI:center:frequency:217GHz:units}{217\,GHz}
\setsymbol{HFI:center:frequency:353GHz:units}{353\,GHz}
\setsymbol{HFI:center:frequency:545GHz:units}{545\,GHz}
\setsymbol{HFI:center:frequency:857GHz:units}{857\,GHz}

\setsymbol{HFI:center:frequency:100GHz}{100}
\setsymbol{HFI:center:frequency:143GHz}{143}
\setsymbol{HFI:center:frequency:217GHz}{217}
\setsymbol{HFI:center:frequency:353GHz}{353}
\setsymbol{HFI:center:frequency:545GHz}{545}
\setsymbol{HFI:center:frequency:857GHz}{857}

\setsymbol{HFI:Ndetectors:100GHz}{8}
\setsymbol{HFI:Ndetectors:143GHz}{11}
\setsymbol{HFI:Ndetectors:217GHz}{12}
\setsymbol{HFI:Ndetectors:353GHz}{12}
\setsymbol{HFI:Ndetectors:545GHz}{3}
\setsymbol{HFI:Ndetectors:857GHz}{4}

\setsymbol{HFI:FWHM:Maps:100GHz:units}{9\parcm88}
\setsymbol{HFI:FWHM:Maps:143GHz:units}{7\parcm18}
\setsymbol{HFI:FWHM:Maps:217GHz:units}{4\parcm87}
\setsymbol{HFI:FWHM:Maps:353GHz:units}{4\parcm65}
\setsymbol{HFI:FWHM:Maps:545GHz:units}{4\parcm72}
\setsymbol{HFI:FWHM:Maps:857GHz:units}{4\parcm39}
\setsymbol{HFI:FWHM:Maps:100GHz}{9.88}
\setsymbol{HFI:FWHM:Maps:143GHz}{7.18}
\setsymbol{HFI:FWHM:Maps:217GHz}{4.87}
\setsymbol{HFI:FWHM:Maps:353GHz}{4.65}
\setsymbol{HFI:FWHM:Maps:545GHz}{4.72}
\setsymbol{HFI:FWHM:Maps:857GHz}{4.39}

\setsymbol{HFI:beam:ellipticity:Maps:100GHz}{1.15}
\setsymbol{HFI:beam:ellipticity:Maps:143GHz}{1.01}
\setsymbol{HFI:beam:ellipticity:Maps:217GHz}{1.06}
\setsymbol{HFI:beam:ellipticity:Maps:353GHz}{1.05}
\setsymbol{HFI:beam:ellipticity:Maps:545GHz}{1.14}
\setsymbol{HFI:beam:ellipticity:Maps:857GHz}{1.19}

\setsymbol{HFI:FWHM:Mars:100GHz:units}{9\parcm37}
\setsymbol{HFI:FWHM:Mars:143GHz:units}{7\parcm04}
\setsymbol{HFI:FWHM:Mars:217GHz:units}{4\parcm68}
\setsymbol{HFI:FWHM:Mars:353GHz:units}{4\parcm43}
\setsymbol{HFI:FWHM:Mars:545GHz:units}{3\parcm80}
\setsymbol{HFI:FWHM:Mars:857GHz:units}{3\parcm67}

\setsymbol{HFI:FWHM:Mars:100GHz}{9.37}
\setsymbol{HFI:FWHM:Mars:143GHz}{7.04}
\setsymbol{HFI:FWHM:Mars:217GHz}{4.68}
\setsymbol{HFI:FWHM:Mars:353GHz}{4.43}
\setsymbol{HFI:FWHM:Mars:545GHz}{3.80}
\setsymbol{HFI:FWHM:Mars:857GHz}{3.67}

\setsymbol{HFI:beam:ellipticity:Mars:100GHz}{1.18}
\setsymbol{HFI:beam:ellipticity:Mars:143GHz}{1.03}
\setsymbol{HFI:beam:ellipticity:Mars:217GHz}{1.14}
\setsymbol{HFI:beam:ellipticity:Mars:353GHz}{1.09}
\setsymbol{HFI:beam:ellipticity:Mars:545GHz}{1.25}
\setsymbol{HFI:beam:ellipticity:Mars:857GHz}{1.03}

\setsymbol{HFI:CMB:relative:calibration:100GHz}{$\lsim 1\%$}
\setsymbol{HFI:CMB:relative:calibration:143GHz}{$\lsim 1\%$}
\setsymbol{HFI:CMB:relative:calibration:217GHz}{$\lsim 1\%$}
\setsymbol{HFI:CMB:relative:calibration:353GHz}{$\lsim 1\%$}
\setsymbol{HFI:CMB:relative:calibration:545GHz}{}
\setsymbol{HFI:CMB:relative:calibration:857GHz}{}

\setsymbol{HFI:CMB:absolute:calibration:100GHz}{$\lsim 2\%$}
\setsymbol{HFI:CMB:absolute:calibration:143GHz}{$\lsim 2\%$}
\setsymbol{HFI:CMB:absolute:calibration:217GHz}{$\lsim 2\%$}
\setsymbol{HFI:CMB:absolute:calibration:353GHz}{$\lsim 2\%$}
\setsymbol{HFI:CMB:absolute:calibration:545GHz}{}
\setsymbol{HFI:CMB:absolute:calibration:857GHz}{}

\setsymbol{HFI:FIRAS:gain:calibration:accuracy:statistical:100GHz}{}
\setsymbol{HFI:FIRAS:gain:calibration:accuracy:statistical:143GHz}{}
\setsymbol{HFI:FIRAS:gain:calibration:accuracy:statistical:217GHz}{}
\setsymbol{HFI:FIRAS:gain:calibration:accuracy:statistical:353GHz}{2.5\%}
\setsymbol{HFI:FIRAS:gain:calibration:accuracy:statistical:545GHz}{1\%}
\setsymbol{HFI:FIRAS:gain:calibration:accuracy:statistical:857GHz}{0.5\%}

\setsymbol{HFI:FIRAS:gain:calibration:accuracy:systematic:100GHz}{}
\setsymbol{HFI:FIRAS:gain:calibration:accuracy:systematic:143GHz}{}
\setsymbol{HFI:FIRAS:gain:calibration:accuracy:systematic:217GHz}{}
\setsymbol{HFI:FIRAS:gain:calibration:accuracy:systematic:353GHz}{}
\setsymbol{HFI:FIRAS:gain:calibration:accuracy:systematic:545GHz}{7\%}
\setsymbol{HFI:FIRAS:gain:calibration:accuracy:systematic:857GHz}{7\%}

\setsymbol{HFI:FIRAS:zero:point:accuracy:100GHz:units}{0.8\MJysr}
\setsymbol{HFI:FIRAS:zero:point:accuracy:143GHz:units}{}
\setsymbol{HFI:FIRAS:zero:point:accuracy:217GHz:units}{}
\setsymbol{HFI:FIRAS:zero:point:accuracy:353GHz:units}{1.4\MJysr}
\setsymbol{HFI:FIRAS:zero:point:accuracy:545GHz:units}{2.2\MJysr}
\setsymbol{HFI:FIRAS:zero:point:accuracy:857GHz:units}{1.7\MJysr}

\setsymbol{HFI:FIRAS:zero:point:accuracy:100GHz}{0.8}
\setsymbol{HFI:FIRAS:zero:point:accuracy:143GHz}{}
\setsymbol{HFI:FIRAS:zero:point:accuracy:217GHz}{}
\setsymbol{HFI:FIRAS:zero:point:accuracy:353GHz}{1.4}
\setsymbol{HFI:FIRAS:zero:point:accuracy:545GHz}{2.2}
\setsymbol{HFI:FIRAS:zero:point:accuracy:857GHz}{1.7}

\setsymbol{HFI:unit:conversion:100GHz:units}{0.2415\MJysrmK}
\setsymbol{HFI:unit:conversion:143GHz:units}{0.3694\MJysrmK}
\setsymbol{HFI:unit:conversion:217GHz:units}{0.4811\MJysrmK}
\setsymbol{HFI:unit:conversion:353GHz:units}{0.2883\MJysrmK}
\setsymbol{HFI:unit:conversion:545GHz:units}{0.05826\MJysrmK}
\setsymbol{HFI:unit:conversion:857GHz:units}{0.002238\MJysrmK}

\setsymbol{HFI:unit:conversion:100GHz}{0.2415}
\setsymbol{HFI:unit:conversion:143GHz}{0.3694}
\setsymbol{HFI:unit:conversion:217GHz}{0.4811}
\setsymbol{HFI:unit:conversion:353GHz}{0.2883}
\setsymbol{HFI:unit:conversion:545GHz}{0.05826}
\setsymbol{HFI:unit:conversion:857GHz}{0.002238}

\setsymbol{HFI:colour:correction:alpha=-2:V1.01:100GHz}{0.9893}
\setsymbol{HFI:colour:correction:alpha=-2:V1.01:143GHz}{0.9759}
\setsymbol{HFI:colour:correction:alpha=-2:V1.01:217GHz}{1.0007}
\setsymbol{HFI:colour:correction:alpha=-2:V1.01:353GHz}{1.0028}
\setsymbol{HFI:colour:correction:alpha=-2:V1.01:545GHz}{1.0019}
\setsymbol{HFI:colour:correction:alpha=-2:V1.01:857GHz}{0.9889}

\setsymbol{HFI:colour:correction:alpha=0:V1.01:100GHz}{1.0008}
\setsymbol{HFI:colour:correction:alpha=0:V1.01:143GHz}{1.0148}
\setsymbol{HFI:colour:correction:alpha=0:V1.01:217GHz}{0.9909}
\setsymbol{HFI:colour:correction:alpha=0:V1.01:353GHz}{0.9888}
\setsymbol{HFI:colour:correction:alpha=0:V1.01:545GHz}{0.9878}
\setsymbol{HFI:colour:correction:alpha=0:V1.01:857GHz}{1.0014}

\def\arbb{{$\alpha_{\rm R-O(Thermal)}$}}
\def\aoxbb{$\alpha_{\rm UV-X(radio-quietQSO)}$}

\def\gr{{$\gamma$-ray}}
\def\grs{{$\gamma$-rays}}
\def\mw{{microwave}}

\def\nup{$\nu_{\rm peak}$}
\def\nupS{$\nu_{\rm peak}^{\rm S}$}
\def\nupIC{$\nu_{\rm peak}^{\rm IC}$} 
\newcommand{\etal}{et al.}
\newcommand{\planck}{\textit{Planck}}
\newcommand{\ROSAT}{\textit{ROSAT}}
\newcommand{\swift}{\textit{Swift}}
\newcommand{\fermi}{\textit{Fermi}}
\newcommand{\wmap}{\textit{WMAP}}
\begin{document}

\title{Simultaneous \planck, \swift, and \fermi\ observations
of X-ray and $\gamma$-ray selected blazars}

\author{\small
P.~Giommi\inst{2,3}
\and
G.~Polenta\inst{2, 23}
\and
A.~L\"{a}hteenm\"{a}ki\inst{1, 19}
\and
D.~J.~Thompson\inst{5}
\and
M.~Capalbi\inst{2}
\and
S.~Cutini\inst{2}
\and
D.~Gasparrini\inst{2}
\and
J.~Gonz\'{a}lez-Nuevo\inst{43}
\and
J.~Le\'{o}n-Tavares\inst{1}
\and
M.~L\'{o}pez-Caniego\inst{32}
\and
M.~N.~Mazziotta\inst{33}
\and
C.~Monte\inst{14, 33}
\and
M.~Perri\inst{2}
\and
S.~Rain\`{o}\inst{14, 33}
\and
G.~Tosti\inst{35, 15}
\and
A.~Tramacere\inst{28}
\and
F.~Verrecchia\inst{2}
\and
H.~D.~Aller\inst{4}
\and
M.~F.~Aller\inst{4}
\and
E.~Angelakis\inst{41}
\and
D.~Bastieri\inst{13, 34}
\and
A.~Berdyugin\inst{45}
\and
A.~Bonaldi\inst{37}
\and
L.~Bonavera\inst{43, 7}
\and
C.~Burigana\inst{26}
\and
D.~N.~Burrows\inst{10}
\and
S.~Buson\inst{34}
\and
E.~Cavazzuti\inst{2}
\and
G.~Chincarini\inst{46}
\and
S.~Colafrancesco\inst{23}
\and
L.~Costamante\inst{47}
\and
F.~Cuttaia\inst{26}
\and
F.~D'Ammando\inst{27}
\and
G.~de Zotti\inst{22, 43}
\and
M.~Frailis\inst{24}
\and
L.~Fuhrmann\inst{41}
\and
S.~Galeotta\inst{24}
\and
F.~Gargano\inst{33}
\and
N.~Gehrels\inst{5}
\and
N.~Giglietto\inst{14, 33}
\and
F.~Giordano\inst{14}
\and
M.~Giroletti\inst{25}
\and
E.~Keih\"{a}nen\inst{12}
\and
O.~King\inst{42}
\and
T.~P.~Krichbaum\inst{41}
\and
A.~Lasenby\inst{6, 38}
\and
N.~Lavonen\inst{1}
\and
C.~R.~Lawrence\inst{36}
\and
C.~Leto\inst{2}
\and
E.~Lindfors\inst{45}
\and
N.~Mandolesi\inst{26}
\and
M.~Massardi\inst{22}
\and
W.~Max-Moerbeck\inst{42}
\and
P.~F.~Michelson\inst{47}
\and
M.~Mingaliev\inst{44}
\and
P.~Natoli\inst{16, 2, 26}
\and
I.~Nestoras\inst{41}
\and
E.~Nieppola\inst{1, 17}
\and
K.~Nilsson\inst{17}
\and
B.~Partridge\inst{18}
\and
V.~Pavlidou\inst{42}
\and
T.~J.~Pearson\inst{8, 29}
\and
P.~Procopio\inst{26}
\and
J.~P.~Rachen\inst{40}
\and
A.~Readhead\inst{42}
\and
R.~Reeves\inst{42}
\and
A.~Reimer\inst{31, 47}
\and
R.~Reinthal\inst{45}
\and
S.~Ricciardi\inst{26}
\and
J.~Richards\inst{42}
\and
D.~Riquelme\inst{30}
\and
J.~Saarinen\inst{45}
\and
A.~Sajina\inst{11}
\and
M.~Sandri\inst{26}
\and
P.~Savolainen\inst{1}
\and
A.~Sievers\inst{30}
\and
A.~Sillanp\"{a}\"{a}\inst{45}
\and
Y.~Sotnikova\inst{44}
\and
M.~Stevenson\inst{42}
\and
G.~Tagliaferri\inst{21}
\and
L.~Takalo\inst{45}
\and
J.~Tammi\inst{1}
\and
D.~Tavagnacco\inst{24}
\and
L.~Terenzi\inst{26}
\and
L.~Toffolatti\inst{9}
\and
M.~Tornikoski\inst{1}
\and
C.~Trigilio\inst{20}
\and
M.~Turunen\inst{1}
\and
G.~Umana\inst{20}
\and
H.~Ungerechts\inst{30}
\and
F.~Villa\inst{26}
\and
J.~Wu\inst{39}
\and
A.~Zacchei\inst{24}
\and
J.~A.~Zensus\inst{41}
\and
X.~Zhou\inst{39}
}
\institute{\small
Aalto University Mets\"{a}hovi Radio Observatory, Mets\"{a}hovintie 114, FIN-02540 Kylm\"{a}l\"{a}, Finland\\
\and
Agenzia Spaziale Italiana Science Data Center, c/o ESRIN, via Galileo Galilei, Frascati, Italy\\
\and
Agenzia Spaziale Italiana, Viale Liegi 26, Roma, Italy\\
\and
Astronomy Department, University of Michigan, 830 Dennison Building, 500 Church Street, Ann Arbor, Michigan 48109-1042, U.S.A.\\
\and
Astroparticle Physics Laboratory, NASA/Goddard Space Flight Center, Greenbelt, MD 20771, U.S.A.\\
\and
Astrophysics Group, Cavendish Laboratory, University of Cambridge, J J Thomson Avenue, Cambridge CB3 0HE, U.K.\\
\and
Australia Telescope National Facility, CSIRO, P.O. Box 76, Epping, NSW 1710, Australia\\
\and
California Institute of Technology, Pasadena, California, U.S.A.\\
\and
Departamento de F\'{\i}sica, Universidad de Oviedo, Avda. Calvo Sotelo s/n, Oviedo, Spain\\
\and
Department of Astronomy and Astrophysics, Pennsylvania State University, 525 Davey Lab, University Park, PA 16802, U.S.A.\\
\and
Department of Physics and Astronomy, Tufts University, Medford, MA 02155, U.S.A.\\
\and
Department of Physics, Gustaf H\"{a}llstr\"{o}min katu 2a, University of Helsinki, Helsinki, Finland\\
\and
Dipartimento di Fisica G. Galilei, Universit\`{a} degli Studi di Padova, via Marzolo 8, 35131 Padova, Italy\\
\and
Dipartimento di Fisica M. Merlin dell'Universit\`{a} e del Politecnico di Bari, 70126 Bari, Italy\\
\and
Dipartimento di Fisica, Universit\`a degli Studi di Perugia, 06123 Perugia, Italy\\
\and
Dipartimento di Fisica, Universit\`{a} di Ferrara, Via Saragat 1, 44122 Ferrara, Italy\\
\and
Finnish Centre for Astronomy with ESO (FINCA), University of Turku, V\"{a}is\"{a}l\"{a}ntie 20, FIN-21500, Piikki\"{o}, Finland\\
\and
Haverford College Astronomy Department, 370 Lancaster Avenue, Haverford, Pennsylvania, U.S.A.\\
\and
Helsinki Institute of Physics, Gustaf H\"{a}llstr\"{o}min katu 2, University of Helsinki, Helsinki, Finland\\
\and
INAF - Osservatorio Astrofisico di Catania, Via S. Sofia 78, Catania, Italy\\
\and
INAF - Osservatorio Astronomico di Brera, via E. Bianchi 46, 23807, Merate (LC), Italy\\
\and
INAF - Osservatorio Astronomico di Padova, Vicolo dell'Osservatorio 5, Padova, Italy\\
\and
INAF - Osservatorio Astronomico di Roma, via di Frascati 33, Monte Porzio Catone, Italy\\
\and
INAF - Osservatorio Astronomico di Trieste, Via G.B. Tiepolo 11, Trieste, Italy\\
\and
INAF Istituto di Radioastronomia, Via P. Gobetti 101, 40129 Bologna, Italy\\
\and
INAF/IASF Bologna, Via Gobetti 101, Bologna, Italy\\
\and
INAF/IASF, Sezione di Palermo, via Ugo La Malfa 153, I-90146 Palermo, Italy\\
\and
ISDC Data Centre for Astrophysics, University of Geneva, ch. d'Ecogia 16, Versoix, Switzerland\\
\and
Infrared Processing and Analysis Center, California Institute of Technology, Pasadena, CA 91125, U.S.A.\\
\and
Institut de Radioastronomie Millim\'{e}trique (IRAM), Avenida Divina Pastora 7, Local 20, 18012 Granada, Spain\\
\and
Institute f\"{u}r Astro- und Teilchenphysik and Institut f\"{u}r Theoretische Physik, Leopold-Franzens-Universit\"{a}t Innsbruck, A-6020 Innsbruck, Austria\\
\and
Instituto de F\'{\i}sica de Cantabria (CSIC-Universidad de Cantabria), Avda. de los Castros s/n, Santander, Spain\\
\and
Istituto Nazionale di Fisica Nucleare, Sezione di Bari, 70126 Bari, Italy\\
\and
Istituto Nazionale di Fisica Nucleare, Sezione di Padova, I-35131 Padova, Italy\\
\and
Istituto Nazionale di Fisica Nucleare, Sezione di Perugia, 06123 Perugia, Italy\\
\and
Jet Propulsion Laboratory, California Institute of Technology, 4800 Oak Grove Drive, Pasadena, California, U.S.A.\\
\and
Jodrell Bank Centre for Astrophysics, Alan Turing Building, School of Physics and Astronomy, The University of Manchester, Oxford Road, Manchester, M13 9PL, U.K.\\
\and
Kavli Institute for Cosmology Cambridge, Madingley Road, Cambridge, CB3 0HA, U.K.\\
\and
Key Laboratory of Optical Astronomy, National Astronomical Observatories, Chinese Academy of Sciences, 20A Datun Road, Chaoyang District, Beijing 100012, China\\
\and
Max-Planck-Institut f\"{u}r Astrophysik, Karl-Schwarzschild-Str. 1, 85741 Garching, Germany\\
\and
Max-Planck-Institut f\"{u}r Radioastronomie, Auf dem H\"{u}gel 69, 53121 Bonn, Germany\\
\and
Owens Valley Radio Observatory, Mail code 249-17, California Institute of Technology, 1200 E California Blvd, Pasadena, CA 91125, U.S.A.\\
\and
SISSA, Astrophysics Sector, via Bonomea 265, 34136, Trieste, Italy\\
\and
Special Astrophysical Observatory, Russian Academy of Sciences, Nizhnij Arkhyz, Zelenchukskiy region, Karachai-Cherkessian Republic, 369167, Russia\\
\and
Tuorla Observatory, Department of Physics and Astronomy, University of Turku, V\"ais\"al\"antie 20, FIN-21500, Piikki\"o, Finland\\
\and
Universit\`{a} degli studi di Milano-Bicocca, Dipartimento di Fisica, Piazza delle Scienze 3, 20126 Milano, Italy\\
\and
W. W. Hansen Experimental Physics Laboratory, Kavli Institute for Particle Astrophysics and Cosmology, Department of Physics and SLAC National Accelerator Laboratory, Stanford University, Stanford, CA 94305, U.S.A.\\
}

\titlerunning{Simultaneous \planck, \swift, and \fermi\ observations of blazars}
\authorrunning{P.~Giommi \etal}
\date{Received 4 August 2011 / Accepted 31 January 2012}

\abstract{
We present  simultaneous \planck, \swift, \fermi, and ground-based data for 105 blazars belonging to three  samples with flux limits in the 
soft X-ray, hard X-ray, and \gr~bands, with additional 5\,GHz flux-density limits to ensure a good probability of a \planck\ detection. 
We compare our results to those of a companion paper presenting simultaneous \planck\ and multi-frequency observations 
of 104 radio-loud northern active galactic nuclei selected at radio frequencies. 
While we confirm several previous results, our unique data set allows us to  demonstrate that the selection method strongly influences the results, producing biases that cannot be ignored.
Almost all the BL Lac objects have been detected by the \fermi\ Large Area Telescope (LAT), whereas 30\% to 40\% of the flat-spectrum radio quasars (FSRQs) in the radio, soft X-ray, and hard X-ray selected samples are still below the \gr~detection limit even after integrating  27 months of \fermi-LAT data. 
The radio to sub-millimetre spectral slope of blazars is quite flat, with $\langle\alpha\rangle \sim 0$ up to about 70\,GHz, above which it 
steepens to  $\langle\alpha\rangle \sim -0.65$. The BL Lacs have significantly flatter spectra than FSRQs at higher frequencies. 
The distribution of the rest-frame synchrotron peak frequency (\nupS) in the spectral energy distribution (SED) of FSRQs is the same in {\it all} the blazar samples with  $\langle$~\nupS$\rangle = 10^{13.1 \pm 0.1}$ Hz, while the mean inverse Compton peak frequency, $\langle$\nupIC$\rangle$, ranges from $10^{21}$ to $10^{22}$ Hz.
The distributions of \nupS\ and \nupIC\ of  BL Lacs are much broader and are shifted to higher energies than those of FSRQs; their shapes strongly depend on the selection method.
The Compton dominance of blazars, defined as the ratio of the inverse Compton to synchrotron peak luminosities, ranges from less than 0.2 to nearly 100, with only FSRQs reaching values larger than about 3. Its distribution is broad and depends strongly 
on the selection method, with \gr~selected blazars peaking at $\sim 7$ or more, and radio-selected blazars at values close to 1, thus implying that the  
common assumption that the blazar power budget is largely dominated by high-energy emission is a selection effect. 
A comparison of our multi-frequency data with theoretical predictions shows that simple homogeneous SSC models cannot explain the simultaneous SEDs of most of the \gr~detected blazars in all samples. The SED of the blazars that were not detected by \fermi-LAT may instead be consistent with SSC emission. Our data challenge the correlation between bolometric luminosity and \nupS\ predicted by the blazar sequence.
}

\keywords{}

\maketitle


\section{Introduction}

Blazars are jet-dominated extragalactic objects characterized by the emission of strongly variable and 
polarized non-thermal radiation across the entire electromagnetic spectrum, from radio waves to \gr s 
(e.g., \citealt{UP95}). 
As  the extreme properties of these sources are due to the relativistic amplification of radiation 
emitted along a jet pointing very close to the line of sight (e.g., \citealt{bla78,UP95}), they are
rare compared to both objects pointing their jets at random angles and radio quiet QSOs where the
emitted radiation is due to thermal or reflection mechanisms ultimately powered by accretion onto a supermassive black hole \citep[e.g.,][]{abdoseds}.  
Despite that, the strong emission of blazars at all wavelengths makes them the dominant type of extragalactic sources
in the radio, \mw, \gr, and TeV bands where accretion and other thermal emission processes do not produce significant amounts of radiation \citep{toffolatti1998,giommi2004,hartman99,abdoseds,costaghis02,colaf06,Weekes2008}. 
For these reasons, blazars are hard to distinguish from other sources at optical and X-ray frequencies, while they dominate the  \mw~and $\gamma$-ray 
sky at high Galactic latitudes. The advent of the \fermi\  \citep{atwood2009} and \planck\ \footnote{\Planck\
      \emph{(http://www.esa.int/\Planck)} is a project of the European
Space Agency -- ESA --
      with instruments provided by two scientific
consortia funded by ESA
      member states (in particular the lead
countries: France and
      Italy) with contributions from NASA (USA), and
telescope reflectors
      provided in a collaboration between ESA and a
scientific consortium led
      and funded by Denmark.} \citep{tauber2010a,planck2011-1.1}
satellites, which are surveying these 
two observing windows, combined with the versatility of the \swift\ observatory \citep{gehrels04}, is giving us the unprecedented
opportunity to collect multi-frequency data for very large samples of blazars and open the way 
to a potentially much deeper understanding of the physics and demographics of these still puzzling objects.

Blazars can be categorized by their optical properties and the shape of their broad-band spectral energy distributions (SEDs). Blazar SEDs
always show two broad bumps in the $\log \nu$ -- $\log \nu F_\nu$ space; the lower energy one is usually attributed to synchrotron radiation while the more energetic one is attributed to inverse Compton scattering.
Blazars displaying strong and broad optical emission lines are usually called flat-spectrum radio quasars (FSRQs), while objects with no 
broad emission lines (i.e., rest-frame equivalent width, EW, $< 5$\,\AA) are called BL Lac objects. 
\cite{pagio95}  introduced the terms LBL and HBL to distinguish between 
BL Lacs with low and high values of the peak frequency of the synchrotron bump (\nupS).  \cite{abdoseds} extended this definition to all types of blazars and defined the terms LSP, ISP, and HSP (corresponding to low, intermediate, and 
high synchrotron peaked blazars)  for the cases where  \nupS $< 10^{14}$\,Hz,  $10^{14}$\,Hz $<$ \nupS $< 10^{15}$\,Hz, and  \nupS $> 10^{15}$\,Hz, respectively. 
In the rest of this paper, we use the LSP/ISP/HSP nomenclature.

It is widely recognized that one of the most effective ways of studying the physics of blazars is through the use of multi-frequency data that is ideally 
simultaneous. There are several examples of studies following this approach (e.g., \citealt{gioansmic,vonmontigny95, sambruna96,fossati98,giommisax,nieppola06,pad06}), but in most cases the samples are heterogeneous and the data are sparse and non-simultaneous.

The compilation of simultaneous and well-sampled SEDs requires the organization of complex multifrequency observation campaigns, 
involving the coordination of observations from several observatories.
Such large efforts have been carried out  only rarely, and almost
exclusively on the occasion of large flaring events in a few bright and well-known blazars, e.g., 3C\,454.3
\citep{giommi454.3,Abdo454.3,vercellone09}, Mkn\,421,
\citep{donnarumma09,abdomkn421}, and PKS\,2155$-$304 \citep{Aharonian09}.

Significant progress has been made with the publication of a compilation of quasi-simultaneous 
SEDs of a large sample of \gr~bright blazars \citep{abdoseds}. This is an important step forward from previous compilations, as the sample presented is statistically representative of the population of bright \gr~selected blazars, and the data were quasi-simultaneous, that is collected within three months of the \fermi-LAT observations.

With \planck,  \swift, and \fermi-LAT simultaneously in orbit, complemented by
other space and ground-based observatories, it is now possible to assemble high-quality multi-frequency datasets that allow us to build simultaneous and well-sampled broad-band spectra of large and statistically well-defined samples of active galactic nuclei (AGNs).

In this and a companion paper \citep{planck2011-6.3a}, we present the first results  of a large cooperative program 
between the  \planck, \fermi-LAT, and \swift~satellites and a number of ground-based observatories, carried out to collect multi-frequency 
data on large samples of blazars selected using different criteria and observed when the sources lie in the field of view of \planck. 

In this paper, we concentrate on blazars selected in the soft X-ray, hard X-ray, and \gr~bands. We  
present the simultaneous data, test for flux correlations, and estimate some key parameters characterizing the SEDs. We then
compare the results obtained for the different samples. Detailed fits to models, variability studies, and more complete theoretical interpretations will be presented elsewhere.

Throughout this paper, we define the radio-to-submillimetre spectral index  $\alpha$ by $S(\nu)\propto \nu^{\alpha}$,
and we adopt a $\Lambda$CDM cosmology with $H_0 = 70$\,km\,s$^{-1}$\,Mpc$^{-1}$, $\Omega_m = 0.27$, and $\Omega_\Lambda = 0.73$  \citep{Komatsu2011}.

\section{Sample selection}

To explore the blazar parameter space from different viewpoints, we used several different criteria to select the blazars to be observed simultaneously by \planck, \swift, and \fermi.  
In this paper, we considered three samples of blazars that are flux-limited in the high-energy part of the electromagnetic spectrum:
soft X-ray (0.1--2.4\,keV) sources from the \ROSAT~All-Sky Survey Bright Source Catalog (1RXS, \citealt{rass}, hereafter referred to as the RASS
sample), hard X-ray (15--150\,keV) sources from the \swift-BAT 54-month source catalog (\citealt{batpa}, hereafter referred to as the
BAT sample), and \gr\ sources from the \fermi-LAT 3-month Bright AGN Source List (\citealt{abdoAGNpaper}, hereafter referred to as the
\fermi-LAT~sample).

These high-energy-selected samples were complemented by a radio flux-limited sample of northern sources (hereafter referred to as the radio sample), 
which is presented in the companion paper \citep{planck2011-6.3a}. 
We used these four samples, defined in widely different  parts of the electromagnetic spectrum, to try to disentangle the intrinsic properties of blazars from the 
heavy selection effects that often afflict blazar samples. In total we considered 175 sources. 

We based our classification of different blazar types on the Roma-BZCAT catalog  \citep{bzcat10}, which is a compilation of known blazars that were carefully checked 
to determine their blazar type in a uniform and reliable way.  \citet{bzcat10} divided blazars into three categories: BZQ/FSRQ, in which the optical spectrum
has broad emission lines; BZB/BL Lac objects, in which the optical spectrum is featureless or contains only absorption lines from the host galaxy; BZU/uncertain type, comprising objects for which
the authors could not find sufficient data to safely determine the source classification, and objects that have peculiar characteristics \citep[see][for details]{bzcat10}.
According to this classification, 96 of our objects are of the FSRQ type, 40 are BL Lacs, and the rest are of uncertain type.  About 160 were 
observed by \swift\ simultaneously with \planck, mostly by means of dedicated target of opportunity (ToO) pointings. In the following we describe the selection criteria for each high-energy selected sample. Details of the radio-selected sample are given in  \cite{planck2011-6.3a}.

\subsection{The issue of blazar classification}

The classification of blazars as either featureless (BL Lacs) or broad-lined objects  (FSRQs), although very simple in principle, is neither unambiguous, nor robust. The borderline between the two blazar subclasses, namely 5\,\AA\ in the source rest-frame for the EW of any emission line, was originally defined as a result of the optical identification campaigns of the sources discovered in the first well-defined and complete samples of (bright) radio and X-ray selected objects \citep{stickel91,stocke91}. 
However,  we now know that well-known BL Lac objects such as OJ\,287  -- and BL\,Lac itself -- exhibit emission lines  with EWs well above the 5\,\AA\ limit on some occasions \citep{vermeulen95,corbett96}. Several other BL Lac objects  have strong emission lines with EWs just below, and sometimes above the 5\,\AA\ threshold, depending on the variable continuum level  \citep[see e.g.][and references therein]{lawrence96,ghisellini11}. Well-known FSRQs such as 3C\,279 also appear nearly featureless during bright states \citep{pian99}. The detection of broad Lyman-$\alpha$ emission in the UV spectrum of classical BL Lacs such as Mkn\,421 and Mkn\,501 \citep{stocke11} contributes to the blurring of the distinction between the two types of blazars. 

It is difficult to differentiate between BL Lac objects and radio galaxies as  BL Lacs have been defined as sources for which the 4000\,\AA\ Ca H\&K break (a stellar absorption feature in the host galaxy) is diluted by non-thermal radiation more than a certain amount that was first quantified by  \cite{stocke91} and then revised by   \cite{marcha96}, \cite{landt02}, and  \cite{landt04}. The level of non-thermal blazar light around 4000\,\AA\ reflects the intrinsic radio power of the jet; it can be highly variable and depends strongly on the position of the peak of the synchrotron emission, thus ensuring that the border between BL Lacs and radio galaxies is quite uncertain.
\cite{giommisimplified} tackled the problem of blazar classification using extensive Monte Carlo simulations and showed that the observed differences can be interpreted within a simple scenario where FSRQs and BL Lacs share the basic non-thermal emission properties.

In the present study, we relied on the blazar classification given in the Roma-BZCAT catalog \citep{bzcat10}, which re-assessed the blazar subclass of each object after a critical review of the optical data available in the literature and in large public databases such as the SDSS  \citep{york00}. 
Despite that, some uncertainties remain, which may in turn influence our conclusions about the differences between BL Lacs and FSRQs. However, the large size of our samples ensures that a few misclassifications should not significantly affect our results. To assess the impact of both blazar misclassification and transitional objects in a quantitative way, it is necessary to perform detailed simulations.

\subsection{The \fermi-LAT (\texorpdfstring{\gr}{gamma-ray}~flux-limited) sample}

Our \gr~flux-limited sample was created from the \fermi~LAT Bright Source List\footnote{\url{http://www.asdc.asi.it/fermibsl/}} \citep{abdoAGNpaper}. We selected all the high Galactic latitude ($|b|>10\degr$) blazars detected with high significance ($\mathrm{TS}>100$)\footnote{The test statistic (TS) is defined as $\mathrm{TS} = -2 \ln(L_0/L_1)$ with $L_0$ the likelihood of the null-hypothesis model and  $L_1$ the likelihood of a competitive model \citep[see e.g.][]{Abdo1FGL}.}.
To reduce the size of the sample and ensure that all the sources are well above the \planck\ sensitivity limit for one survey, 
we considered only the sources with radio flux density (taken from BZCAT) $S_{5\,\rm GHz} > 1$\,Jy. We realized that this is a double cut, with a TS limit at \gr~energies and a flux-density limit in the radio band. 
A TS limit translates into different \gr~flux limits depending on the \gr~spectral slope, with higher sensitivity to flat-spectrum sources \citep[see fig. 7 of][]{abdoAGNpaper}. Hence,  
for our statistical considerations we also considered the subsample with a flux cut of $F( E>100~{\rm MeV})> 8\times10^{-8}$ ph\,cm$^{-2}$\,s$^{-1}$, which removed this dependence on the spectral slope.

The sample so defined includes 50 sources, 40 of which are brighter than the \gr~flux limit of $8\times10^{-8}$ ph\,cm$^{-2}$\,s$^{-1}$.  Relaxing the radio flux density cut 
would have provided a purely \gr~flux-limited sample and increased the number of sources to $\approx70$, but with about 40--50\% of the objects with $S_{5\,\rm GHz} < 1$\,Jy being undetected by \planck. 

\begin{table*}[t]
\caption{The \fermi-LAT (\gr~TS/flux-limited) sample.}
\begin{center}
\scriptsize
\begin{tabular}{llcrcD{.}{.}{2.2}D{.}{.}{2.2}D{.}{}{5.0}ccl}
\hline
\hline
      &                    &         &         &   &      & \multicolumn{1}{c}{X-ray Flux}  & \multicolumn{1}{c}{Flux Density} & 
\multicolumn{1}{c}{\fermi~Flux} & & \\
      &                    & R.A.    & Dec.~~~~     &   &      & \multicolumn{1}{c}{0.1--2.4\,keV} & \multicolumn{1}{c}{1.4\,GHz \tablefootmark{a}} & 
\multicolumn{1}{c}{1--100\,GeV} & & \\
Source name & \fermi-LAT~name (1FGL)  & (J2000) & (J2000)~~ & $z$ &  \multicolumn{1}{c}{Rmag} & \multicolumn{1}{c} {\tablefootmark{b}} & \multicolumn{1}{c}{mJy}
 & \multicolumn{1}{c}{\tablefootmark{c}} & Swift obs. date & Blazar type\\
 \hline
1Jy\,0118$-$272  & 1FGLJ\,0120.5$-$2700 &  01 20 31.6 & $-$27 01 24 &  0.557 &  15.5 &    0.72&   934. &  3.7 $\pm$  0.4  &          \tablefootmark{d}  &  BL\,Lac - LSP\\
S4\,0133+47    & 1FGLJ\,0137.0+4751 &  01 36 58.5 &  47 51 29 &  0.859 &  17.6 &    1.04 &  1138. &  9.6 $\pm$  0.6  & 2010-02-05 &FSRQ -LSP \\
PKS\,0202$-$17  & 1FGLJ\,0205.0$-$1702 &  02 04 57.6 & $-$17 01 18 &  1.740 &  17.3 &    0.57 &  1220. &  1.5 $\pm$  0.3  & 2010-01-08 &FSRQ - LSP \\
PKS\,0208$-$512  & 1FGLJ\,0210.6$-$5101 &  02 10 46.2 & $-$51 01 01 &  1.003 &  14.8 &    0.75 &  3493. &  7.1 $\pm$  0.6  & 2009-11-26 &Uncertain - LSP\\
PKS\,0215+015 & 1FGLJ\,0217.9+0144 &  02 17 48.9 &  01 44 49 &  1.715 &  18.7 &    2.56 &   751. &  6.0 $\pm$  0.5  &       \tablefootmark{d} &FSRQ - LSP\\
1Jy\,0218+357  & 1FGLJ\,0221.0+3555 &  02 21 05.5 &  35 56 14 &  0.944 &  20.0 &    0.85 &  1707. &  6.4 $\pm$  0.5  &  2010-08-19 &Uncertain - LSP \\
4C\,28.07      & 1FGLJ\,0237.9+2848 &  02 37 52.4 &  28 48 09 &  1.213 &  18.8 &    0.58 &  2197. &  3.7 $\pm$  0.4  &  2010-02-05 &FSRQ - LSP\\
PKS\,0235+164  & 1FGLJ\,0238.6+1637 &  02 38 38.9 &  16 36 59 &  0.940 &  18.5 &    1.24 &  1941. & 32.7 $\pm$  1.1  &  2010-01-30& BL\,Lac - LSP \\
PKS\,0332$-$403  & 1FGLJ\,0334.2$-$4010 &  03 34 13.6 & $-$40 08 25 &   \tablefootmark{e} &  17.5 &    0.73 &  1042. &  3.8 $\pm$  0.4  &  2010-01-17 &BL\,Lac - LSP\\
PKS\,0420$-$01   & 1FGLJ\,0423.2$-$0118 &  04 23 15.8 & $-$01 20 33 &  0.916 &  16.7 &    1.39 &  2726. &  5.6 $\pm$  0.5  &  2009-08-27 &FSRQ - LSP\\
PKS\,0426$-$380  & 1FGLJ\,0428.6$-$3756 &  04 28 40.4 & $-$37 56 19 &  1.030 &  16.3 &    0.42 &   753. & 25.7 $\pm$  1.0  &  2010-08-17 &BL\,Lac - LSP\\
PKS\,0454$-$234  & 1FGLJ\,0457.0$-$2325 &  04 57 03.1 & $-$23 24 52 &  1.003 &  17.9 &    0.91 &  1727. & 32.5 $\pm$  1.1  &  2010-02-25 &FSRQ - LSP\\
PKS\,0528+134 & 1FGLJ\,0531.0+1331 &  05 30 56.4 &  13 31 55 &  2.070 &  18.9 &    0.80 &  1556. &  4.0 $\pm$  0.5  &  2009-09-24 &FSRQ - LSP\\
PKS\,0537$-$441  & 1FGLJ\,0538.8$-$4404 &  05 38 50.3 & $-$44 05 08 &  0.892 &  16.0 &    2.10 &  3729. & 21.3 $\pm$  0.9  &  2010-03-03 &BL\,Lac - LSP\\
PKS\,0735+17   & 1FGLJ\,0738.2+1741 &  07 38 07.3 &  17 42 19 &  0.424 &  14.5 &    0.97 &  2258. &  4.4 $\pm$  0.5  &  2010-10-07  & BL\,Lac - ISP\\
S4\,0814+425  & 1FGLJ\,0818.2+4222 &  08 18 16.0 &  42 22 45 &  0.530 &  19.6 &    0.32 &  1091. &  8.7 $\pm$  0.6  &  2010-10-15  &BL\,Lac - LSP\\
OJ\,535       & 1FGLJ\,0825.0+5555 &  08 24 47.2 &  55 52 42 &  1.417 &  18.1 &    0.64 &  1449. &  0.9 $\pm$  0.3  &  2010-03-28 &FSRQ - LSP\\
PKS\,0851+202  & 1FGLJ\,0854.8+2006 &  08 54 48.8 &  20 06 30 &  0.306 &  14.4 &    1.72 &  1512. &  2.7 $\pm$  0.4  &  2010-04-10 &BL\,Lac - LSP\\
S4\,0917+44    & 1FGLJ\,0920.9+4441 &  09 20 58.4 &  44 41 54 &  2.190 &  19.2 &    1.04 &  1017. & 14.0 $\pm$  0.7  &  2009-10-29 &FSRQ - LSP\\
4C\,55.17      & 1FGLJ\,0957.7+5523 &  09 57 38.1 &  55 22 57 &  0.896 &  16.8 &    0.51 &  3079. & 10.5 $\pm$  0.6  &  2009-11-01  &FSRQ - LSP\\
4C\,01.28      & 1FGLJ\,1058.4+0134 &  10 58 29.6 &  01 33 58 &  0.888 &  17.6 &    1.08 &  3220. &  7.1 $\pm$  0.6  & 2009-12-03  &Uncertain - LSP\\
PKS\,1057$-$79   & 1FGLJ\,1058.1$-$8006 &  10 58 43.3 & $-$80 03 54 &  0.581 &  17.3 &    0.43 &   534. &  2.2 $\pm$  0.4  &  2010-08-30 &BL\,Lac -LSP\\
PKS\,1127$-$145  & 1FGLJ\,1130.2$-$1447 &  11 30 07.0 & $-$14 49 27 &  1.184 &  16.0 &    1.39 &  5622. &  2.4 $\pm$  0.4  & 2009-12-28 &FSRQ - LSP\\
PKS\,1144$-$379  & 1FGLJ\,1146.9$-$3812 &  11 47 01.3 & $-$38 12 11 &  1.048 &  15.7 &    0.91 &  1804. &  2.4 $\pm$ 0.4  &  2010-06-24 &BL\,Lac - LSP\\
4C\,29.45      & 1FGLJ\,1159.4+2914 &  11 59 31.8 &  29 14 44 &  0.729 &  16.4 &    0.84 &  2031. &  5.3 $\pm$  0.5  &  2010-05-28 &FSRQ - LSP\\
ON\,231       & 1FGLJ\,1221.5+2814 &  12 21 31.6 &  28 13 58 &  0.102 &  14.3 &    1.30 &   732. &  6.9 $\pm$  0.5  &  2009-12-10 &BL\,Lac - ISP\\
3C\,273        & 1FGLJ\,1229.1+0203 &  12 29 06.7 &  02 03 08 &  0.158 &  14.1 &   63.11 & 54991. &  9.6 $\pm$  0.6  &  2010-01-11 &FSRQ - LSP\\
PKS\,1244$-$255  & 1FGLJ\,1246.7$-$2545 &  12 46 46.8 & $-$25 47 49 &  0.635 &  16.7 &    1.30 &  1165. &  8.1 $\pm$  0.6  &  2010-01-25& FSRQ - LSP\\
3C\,279        & 1FGLJ\,1256.2$-$0547 &  12 56 11.1 & $-$05 47 21 &  0.536 &  15.0 &   20.90 &  9711. & 32.4 $\pm$  1.1  &  2010-01-15 &FSRQ - LSP\\
1Jy\,1308+326  & 1FGLJ\,1310.6+3222 &  13 10 28.6 &  32 20 43 &  0.997 &  19.6 &    0.53 &  1687. &  6.8 $\pm$  0.5  &  2009-12-12& Uncertain - LSP\\
PKS\,1502+106  & 1FGLJ\,1504.4+1029 &  15 04 24.9 &  10 29 39 &  1.839 &  19.5 &    0.16 &  1774. & 67.0 $\pm$  1.6  &  2010-07-29 &FSRQ - LSP\\
4C\,$-$05.64     & 1FGLJ\,1511.1$-$0545 &  15 10 53.5 & $-$05 43 07 &  1.191 &  16.9 &    1.16 &  3569. &  2.1 $\pm$  0.4  &       \tablefootmark{d} &  FSRQ - LSP \\
AP\,Lib        & 1FGLJ\,1517.8$-$2423 &  15 17 41.8 & $-$24 22 19 &  0.048 &  12.6 &    1.05 &  2042. &  5.6 $\pm$  0.5  &  2010-02-20 &BL\,Lac - LSP\\
OS\,$-$237.8     & 1FGLJ\,1625.7$-$2524 &  16 25 46.8 & $-$25 27 38 &  0.786 &  20.6 &     ... &  2521. &  4.4 $\pm$  0.6  &  2010-08-26 &Uncertain - LSP\\
4C\,38.41      & 1FGLJ\,1635.0+3808 &  16 35 15.4 &  38 08 04 &  1.814 &  17.3 &    0.17 &  2726. &  6.8 $\pm$  0.5  &  2010-03-07 &FSRQ - LSP\\
NRAO\,512     & 1FGLJ\,1642.5+3947 &  16 40 29.6 &  39 46 46 &  1.660 &  17.5 &    0.15 &   976. &  5.6 $\pm$  0.5  & 2010-08-06 &FSRQ - LSP\\
Mkn\,501       & 1FGLJ\,1653.9+3945 &  16 53 52.2 &  39 45 36 &  0.033 &   8.3 &   36.90 &  1558. &  8.3 $\pm$  0.6  & 2010-03-21 &BL\,Lac - HSP\\
OT\,081       & 1FGLJ\,1751.5+0937 &  17 51 32.8 &  09 39 00 &  0.322 &  17.0 &    1.18 &   623. &  6.4 $\pm$  0.6  &  2010-04-01 &Uncertain - LSP\\
S5\,1803+784   & 1FGLJ\,1800.4+7827 &  18 00 45.6 &  78 28 04 &  0.680 &  14.7 &    0.79 &  2223. &  3.0 $\pm$  0.4  &  2009-10-13 &BL\,Lac - LSP\\
2E\,1908.2$-$2011& 1FGLJ\,1911.2$-$2007 &  19 11 09.6 & $-$20 06 55 &  1.119 &  18.9 &    1.77 &  2714. &  4.5 $\pm$  0.5  &  2009-10-04  &FSRQ - LSP\\
PMNJ\,1923$-$2104& 1FGLJ\,1923.5$-$2104 &  19 23 32.1 & $-$21 04 33 &  0.874 &  16.6 &    0.77 &  3167. & 11.9 $\pm$  0.7  &  2010-09-30 &FSRQ - LSP\\
1Jy\,2005$-$489  & 1FGLJ\,2009.5$-$4849 &  20 09 25.3 & $-$48 49 53 &  0.071 &  11.0 &   33.24 &  1282. &  5.0 $\pm$  0.5  &  2009-10-05  &BL\,Lac - HSP\\
PKS\,2052$-$47   & 1FGLJ\,2056.3$-$4714 &  20 56 16.3 & $-$47 14 47 &  1.491 &  18.3 &    0.56 &  2223. &  4.6 $\pm$  0.5  &   2010-10-18   &  FSRQ - LSP \\
S3\,2141+17    & 1FGLJ\,2143.4+1742 &  21 43 35.5 &  17 43 48 &  0.213 &  14.4 &    0.63 &   651. &  4.9 $\pm$  0.5  &  2009-11-20 &FSRQ - LSP\\
1Jy\,2144+092  & 1FGLJ\,2147.2+0929 &  21 47 10.1 &  09 29 46 &  1.113 &  16.9 &    1.49 &   934. &  4.1 $\pm$  0.4  &      \tablefootmark{d} &  FSRQ - LSP  \\
BL\,Lac        & 1FGLJ\,2202.8+4216 &  22 02 43.2 &  42 16 40 &  0.069 &  12.5 &    1.58 &  6051. &  7.1 $\pm$  0.6  & 2009-12-23 &BL\,Lac - LSP\\
PKS\,2204$-$54   & 1FGLJ\,2207.8$-$5344 &  22 07 43.7 & $-$53 46 33 &  1.215 &  18.2 &    0.52 &  1526. &  1.6 $\pm$  0.3  &  2010-05-03&FSRQ - LSP\\
PKS\,2227$-$08   & 1FGLJ\,2229.7$-$0832 &  22 29 40.0 & $-$08 32 54 &  1.560 &  16.8 &    8.74 &   968. &  4.6 $\pm$  0.5  & 2009-11-19 &FSRQ - LSP\\
4C\,11.69      & 1FGLJ\,2232.5+1144 &  22 32 36.4 &  11 43 50 &  1.037 &  16.5 &    1.26 &  7202. &  4.1 $\pm$  0.4  &  2009-11-29 &FSRQ - LSP \\
3C\,454.3      & 1FGLJ\,2253.9+1608 &  22 53 57.7 &  16 08 53 &  0.859 &  13.2 &    7.80 & 12657. & 46.2 $\pm$  1.3  & 2009-12-14 &FSRQ - LSP \\
\hline
\end{tabular}
\label{tab:fermi_sample}
\tablefoot{
\tablefoottext{a}{843\,MHz flux is reported for sources with Dec $<-40^{\circ}$}; 
\tablefoottext{b}{Units of $10^{-12}$\,erg\,cm$^{-2}$\,s$^{-1}$}; 
\tablefoottext{c}{Units of $10^{-10}$\,ph\,cm$^{-2}$\,s$^{-1}$}\\
\tablefoottext{d}{\swift~simultaneous observation not available.}
\tablefoottext{e}{Optical spectrum completely featureless or not available, redshift unknown}
} 
\normalsize
\end{center}
\end{table*}

Details are presented in Table~\ref{tab:fermi_sample}, where column 1 gives the source common name, column 2 gives the \fermi-LAT~name as it appears in \cite{abdo2010a}, columns 3 and 4 give the source position in equatorial coordinates, columns 5, 6, and 7 give the redshift, $V$ magnitude, and X-ray flux (0.1--2.4\,keV) from BZCAT \citep{bzcat10}, column 8 gives the 1.4\,GHz or 843\,MHz flux density from NVSS \citep{nvss} or 
from SUMSS \citep{sumss} with Dec $<-40\degr$, column 9 gives the \gr~flux from \cite{abdo2010a} \footnote{We give average \gr\ fluxes from the 1-year \fermi\ catalog rather than the three-month fluxes that were used to define the sample for consistency with Tables \ref{tab:bat_sample} and \ref{tab:rass_sample}.}, and column 10 gives the date of the \swift\ ToO observation made when the source was within the \planck\ field of view.

\subsection{The \swift/BAT (hard X-ray flux-limited)  sample}

We defined our hard X-ray flux-limited sample starting from the \swift-BAT 54 month source catalog\footnote{\url{http://www.asdc.asi.it/bat54/}} \citep{batpa}, and selecting all the sources identified with blazars with X-ray flux $>10^{-11}$\,erg\,cm$^{-2}$\,s$^{-1}$ in the 15--150\,keV energy band. The BAT catalog includes 70 known blazars that satisfy the X-ray flux condition, but many of them are too faint to be detected at millimetre wavelengths by \planck. Therefore, although a pure X-ray selection would be preferable, we have decided to add a mild radio flux-density constraint ($S_{5\,\rm GHz}>100$\,mJy, with $S_{5\,\rm GHz}$ taken from BZCAT) to select only sources that can be detected by \planck\ or for which \planck\ will be able to provide meaningful upper limits, leaving enough sources 
to build a statistically sizable sample.
The list of the 34 sources included in this sample is given in Table~\ref{tab:bat_sample}. The column description is the same as for Table~\ref{tab:fermi_sample}.

\begin{table*}[t]
\caption{The \swift-BAT (hard X-ray flux-limited) sample.}
\begin{center}
\scriptsize
\begin{tabular}{llcrcD{.}{.}{2.2}D{.}{.}{2.2}D{.}{}{5.0}ccl}
\hline
\hline
      &                    &         &         &   &  \multicolumn{1}{c}{}    & \multicolumn{1}{c}{X-ray Flux}  & \multicolumn{1}{c}{Flux Density} & 
\fermi-LAT~Flux & &\\
      &                    & R.A.    & Dec.~~~~     &   &  \multicolumn{1}{c}{}    & \multicolumn{1}{c}{0.1--2.4\,keV} & \multicolumn{1}{c}{1.4\,GHz \tablefootmark{a}} & 
1--100\,GeV && \\
Source name & \fermi-LAT~name  & (J2000) & (J2000)~~ & $z$ & \multicolumn{1}{c}{Rmag} & \multicolumn{1}{c}{\tablefootmark{b}} & \multicolumn{1}{c}{mJy}
 & \tablefootmark{c} & Swift obs. date & Blazar type\\
 \hline
III\,ZW\,2    &      ...         &   00 10 31.0 &   10 58 29 &  0.089 &    13.9 &    6.14 &       98. &      ...       &   2010-07-08 & FSRQ - LSP\\
S5\,0014+813   &      ...         &   00 17 08.4 &   81 35 08 &  3.387 &    15.9 &    0.77 &      693. &      ...       &   2010-09-21 &FSRQ - ISP\\
1ES\,0033+595  & 1FGLJ\,0035.9+5951 &   00 35 52.6 &   59 50 03 &   \tablefootmark{e} &    17.2 &    5.41 & 147. & 3.2 $\pm$  0.5 & \tablefootmark{d} &  BL\,Lac - HSP    \\
Mkn\,348    &      ...         &   00 48 47.1 &   31 57 25 &  0.015 &     9.3 &    0.12 &      292. &      ...       &      \tablefootmark{d}    &BL\,Lac - LSP\\
1Jy\,0212+735  & 1FGLJ\,0217.8+7353 &   02 17 30.8 &   73 49 32 &  2.367 &    18.8 &    0.54 &     2272. & 1.0 $\pm$  0.4 &   2010-09-11 &Uncertain - LSP\\
NGC\,1275      & 1FGLJ\,0319.7+4130 &   03 19 48.1 &   41 30 42 &  0.018 &    12.3 &  197.92 &    22830. &17.3 $\pm$  0.8 &   2010-08-09&FSRQ -LSP \\
NRAO\,140     & 1FGLJ\,0334.2+3233 &   03 36 30.1 &   32 18 29 &  1.259 &    16.6 &    2.80 &     2677. & 1.0 $\pm$  0.4 &   2010-08-25 &Uncertain - LSP\\
3C\,120        &      ...         &   04 33 11.0 &   05 21 15 &  0.033 &    13.8 &   22.68 &     3440. &      ...       &   2010-02-25 &Uncertain - LSP\\
PKS\,0521$-$36   & 1FGLJ\,0522.8$-$3632 &   05 22 57.9 & $-$36 27 30 &  0.055 &    11.6 &   10.42 &    15620. & 2.9 $\pm$  0.4 &   2010-03-05 &Uncertain - LSP \\
PKS\,0528+134 & 1FGLJ\,0531.0+1331 &   05 30 56.4 &   13 31 55 &  2.070 &    18.9 &    0.80 &     1556. & 4.0 $\pm$  0.5 &   2009-09-24 &FSRQ - LSP\\
1Jy\,0537$-$286  & 1FGLJ\,0539.1$-$2847 &   05 39 54.2 &  $-$28 39 55 &  3.104 &    19.0 &    0.83 &      862. & 0.9 $\pm$  0.0 &   2010-03-12 &FSRQ - LSP\\
PKS\,0548$-$322 &      ...         &   05 50 40.6 &  $-$32 16 20 &  0.069 &    13.1 &   26.34 &      344. &      ...       &   2010-03-12 &BL\,Lac - HSP \\
B2.2\,0743+25 & 1FGLJ\,0746.6+2548 &   07 46 25.8 &   25 49 02 &  2.979 &    19.2 &    0.38 &      417. & 0.7 $\pm$  0.2 &    2010-10-15 & FSRQ - LSP\\  
4C\,71.07      & 1FGLJ\,0842.2+7054 &   08 41 24.3 &   70 53 42 &  2.218 &    16.8 &    5.52 &     3823. & 1.2 $\pm$  0.3 &   2010-03-21 &FSRQ - LSP\\
Mkn\,421       & 1FGLJ\,1104.4+3812 &   11 04 27.3 &   38 12 31 &  0.030 &     8.3 &  180.94 &      767. &26.1 $\pm$  1.0 &   2009-11-17 &BL\,Lac - HSP\\
PKS\,1127$-$145  & 1FGLJ\,1130.2$-$1447 &   11 30 07.0 &  $-$14 49 27 &  1.184 &    16.0 &    1.39 &     5622. & 2.4 $\pm$  0.4 &   2009-12-28 &FSRQ - LSP\\
PKS\,1219+04  & 1FGLJ\,1222.5+0415 &   12 22 22.5 &   04 13 15 &  0.967 &    17.1 &    0.91 &      800. & 0.9 $\pm$  0.3 &   2010-07-17 &FSRQ - LSP\\
3C\,273        & 1FGLJ\,1229.1+0203 &   12 29 06.7 &   02 03 08 &  0.158 &    14.1 &   63.11 &    54991. & 9.6 $\pm$  0.6 &   2010-01-11 &FSRQ - LSP\\
3C\,279        & 1FGLJ\,1256.2$-$0547 &   12 56 11.1 &  $-$05 47 21 &  0.536 &    15.0 &   20.85 &     9711. &32.4 $\pm$  1.1 &   2010-01-15 &FSRQ - LSP\\
AP\,Lib        & 1FGLJ\,1517.8$-$2423 &   15 17 41.8 &  $-$24 22 19 &  0.048 &    12.6 &    1.05 &     2042. & 5.7 $\pm$  0.5 &   2010-02-20 &BL\,Lac - LSP\\
Mkn\,501       & 1FGLJ\,1653.9+3945 &   16 53 52.2 &   39 45 36 &  0.033 &     8.3 &   36.93 &     1558. & 8.3 $\pm$  0.6 &   2010-03-21 &BL\,Lac - HSP\\
ARP\,102B     &      ...         &   17 19 14.4 &   48 58 49 &  0.024 &     9.4 &    0.59 &      145. &      ...       &  2010-03-31 &Uncertain - ISP\\
PKSB\,1830$-$210 & 1FGLJ\,1833.6$-$2103 &   18 33 39.8 &  $-$21 03 39 &  2.507 &    16.6 &    0.69 &    10896. &10.7 $\pm$  0.8 &   2010-09-23 &FSRQ - LSP\\
OV\,$-$236       & 1FGLJ\,1925.2$-$2919 &   19 24 51.0 &  $-$29 14 30 &  0.352 &    17.3 &    2.42 &    13387. & 1.4 $\pm$  0.4 &   2010-09-30 &FSRQ - LSP\\
1ES\,1959+650  & 1FGLJ\,2000.0+6508 &   19 59 59.8 &   65 08 54 &  0.047 &    11.9 &   35.28 &     250.  & 6.0 $\pm$  0.5 &   2009-09-26 &BL\,Lac - HSP\\
1Jy\,2126$-$158  &      ...         &   21 29 12.1 &  $-$15 38 41 &  3.268 &    16.5 &    1.54 &     590.  &      ...       &   2010-05-03 &FSRQ - ISP\\
4C\,06.69      & 1FGLJ\,2148.5+0654 &   21 48 05.4 &   06 57 38 &  0.999 &    15.1 &    1.46 &    2589.  & 0.7 $\pm$  0.3 &   2009-11-21 &FSRQ - LSP\\
PKS\,2149$-$307  &      ...         &   21 51 55.5 &  $-$30 27 53 &  2.345 &    17.4 &    4.80 &    1243.  &      ...       &   2010-05-11 &FSRQ - LSP\\
BL\,Lac        & 1FGLJ\,2202.8+4216 &   22 02 43.2 &   42 16 40 &  0.069 &    12.5 &    1.57 &    6051.  & 7.1 $\pm$  0.6 &   2009-12-23 &BL\,Lac - LSP\\
4C\,31.63      &      ...         &   22 03 14.9 &   31 45 38 &  0.295 &    14.3 &    3.22 &    2878.  &      ...       &   2009-11-27 &FSRQ - LSP\\
NGC\,7213     &      ...         &   22 09 16.2 &  $-$47 10 00 &  0.006 &    10.3 &   35.34 &      98.  &      ...       &   2010-10-23  &    Uncertain \\
4C\,11.69      & 1FGLJ\,2232.5+1144 &   22 32 36.4 &   11 43 50 &  1.037 &    16.5 &    1.26 &    7202.  & 4.1 $\pm$  0.4 &   2009-11-29&FSRQ - LSP\\
3C\,454.3      & 1FGLJ\,2253.9+1608 &   22 53 57.7 &   16 08 53 &  0.859 &    13.2 &    7.80 &   12657.  &46.2 $\pm$  1.3 &   2009-12-14 &FSRQ - LSP\\
PKS\,2325+093  & 1FGLJ\,2327.7+0943 &   23 27 33.5 &   09 40 09 &  1.843 &    18.8 &    0.73 &     741.  & 3.0 $\pm$  0.4 &   2010-07-18 &FSRQ - LSP\\
\hline
\end{tabular}
\label{tab:bat_sample}
\tablefoot{
\tablefoottext{a}{843\,MHz flux is reported for sources with Dec $<-40^{\circ}$.}
\tablefoottext{b}{Units of $10^{-12}$\,erg\,cm$^{-2}$\,s$^{-1}$}; 
\tablefoottext{c}{Units of $10^{-10}$\,ph\,cm$^{-2}$\,s$^{-1}$}\\
\tablefoottext{d}{\swift~simultaneous observation not available.}
\tablefoottext{e}{Optical spectrum completely featureless or not available, redshift unknown}
}
\normalsize
\end{center} 
\end{table*}


\subsection{The \ROSAT/RASS (soft X-ray flux-limited)  sample} 
The soft X-ray flux-limited sample was defined starting from the RASS  catalog (1RXS) \citep{rass}, and selecting all the sources identified with blazars with count rates higher than 0.3 counts/s in the 0.1--2.4\,keV energy band, and radio flux densities (taken from BZCAT) of $S_{5\,\rm GHz} > 200$\,mJy. The reasons for using an additional radio flux constraint are the same as for the hard X-ray
flux-limited sample, where, however, we chose $200$\,mJy to reduce the size of the sample to be
comparable to those of  the \gr~and hard X-ray samples.  We realize that this is a stringent cut that removes about two thirds of the sources from the purely soft X-ray selected sample. However, all the 
sources below the radio threshold {\it are HSP BL Lacs}, thus implying that the subsample of LBL sources remains purely X-ray flux-limited and, consequently, that high \nupS\, objects are strongly underrepresented.
The list of the 43 sources included in this sample is given in Table~\ref{tab:rass_sample}. The column description is the same as for Tables~\ref{tab:fermi_sample} and~\ref{tab:bat_sample}.

\begin{table*}[t]
\caption{The \ROSAT/RASS (soft X-ray flux-limited) sample.}
\begin{center}
\scriptsize
\begin{tabular}{llcrcD{.}{.}{2.2}D{.}{.}{2.2}D{.}{}{5.0}ccl}
\hline
\hline
      &                    &         &         &   &  \multicolumn{1}{c}{}    & \multicolumn{1}{c}{X-ray Flux}  & \multicolumn{1}{c}{Flux Density} & 
\fermi-LAT~Flux & &  \\
      &                    & R.A.    & Dec.~~~~     &   &  \multicolumn{1}{c}{}    & \multicolumn{1}{c}{0.1--2.4\,keV} & \multicolumn{1}{c}{1.4\,GHz \tablefootmark{a}} & 
1--100\,GeV &&  \\
Source name & \fermi-LAT~name  & (J2000) & (J2000)~~ & $z$ & \multicolumn{1}{c}{Rmag} & \multicolumn{1}{c}{ \tablefootmark{b}} & \multicolumn{1}{c}{mJy}
 &  \tablefootmark{c}  & Swift obs. date & Blazar type\\
 \hline
III\,ZW\,2       &      ...         & 00 10 31.0 &  10 58 29 &  0.089 &  13.9 &    6.14 &       98. &      ...       &    2010-07-08 & FSRQ - LSP\\
GB6J\,0214+5145   &      ...         & 02 14 17.9 &  51 44 52 &  0.049 &  16.5 &    4.58 &      294. &      ...       &   \tablefootmark{d}  &    BL\,Lac - HSP  \\  
3C\,120           &      ...         & 04 33 11.0 &  05 21 15 &  0.033 &  13.8 &   22.68 &     3440. &   ...          &    2010-02-25 & Uncertain - LSP\\
PKS\,0521$-$36      & 1FGLJ\,0522.8$-$3632 & 05 22 57.9 & $-$36 27 30 &  0.055 &  11.6 &   10.42 &    15620. & 2.9 $\pm$  0.4 &    2010-03-05 & Uncertain - LSP\\
PKS\,0548$-$322    &      ...         & 05 50 40.6 & $-$32 16 20 &  0.069 &  13.1 &   26.34 &      344. &      ...       &      2010-03-12 &BL\,Lac - HSP\\
IRAS-L\,06229$-$643  &      ...         & 06 23 07.6 & $-$64 36 20 &  0.129 &  13.7 &    5.34 &      274. &      ...         &   2010-08-18 &FSRQ - LSP\\ 
4C\,71.07         & 1FGLJ\,0842.2+7054 & 08 41 24.3 &  70 53 42 &  2.218 &  16.8 &    5.52 &     3823. & 1.2 $\pm$  0.3 &    2010-03-21 &FSRQ - LSP\\
B2\,0912+29      & 1FGLJ\,0915.7+2931 & 09 15 52.4 &  29 33 24 &   \tablefootmark{e} &  15.0 &    6.25 &      342. & 2.1 $\pm$  0.3   &   2010-10-28  & BL\,Lac -HSP \\
PKS\,0921$-$213    &      ...         & 09 23 38.8 & $-$21 35 47 &  0.053 &  12.8 &    4.80 &      268. &      ...       &      2010-05-02 &Uncertain - LSP\\
1H\,1013+498     & 1FGLJ\,1015.1+4927 & 10 15 04.1 &  49 26 00 &  0.212 &  15.1 &   13.23 &      378. & 8.7 $\pm$  0.6 &     2010-04-24&BL\,Lac - HSP\\ 
1RXSJ\,105837.5+562816   & 1FGLJ\,1058.6+5628 & 10 58 37.7 &  56 28 11 &  0.143 &  14.0 &    3.13 &      228. & 5.7 $\pm$  0.5   &   2010-04-18 &BL\,Lac - HSP\\
PKS\,1124$-$186     & 1FGLJ\,1126.8$-$1854 & 11 27 04.3 & $-$18 57 17 &  1.048 &  19.2 &    5.33 &      536. & 2.4 $\pm$  0.4  &   2010-06-10 &FSRQ - LSP\\
B2\,1128+31      &      ...         & 11 31 09.4 &  31 14 05 &  0.289 &  15.8 &    5.02 &      370. &      ...       &     2009-11-28 &FSRQ - HSP\\
S5\,1133+704     & 1FGLJ\,1136.6+7009 & 11 36 26.4 &  70 09 27 &  0.045 &  11.0 &   35.08 &      327. & 1.7 $\pm$  0.3 &     2009-10-27& BL\,Lac - HSP\\ 
4C\,49.22         &      ...         & 11 53 24.4 &  49 31 08 &  0.334 &  16.9 &    3.31 &     1572. &      ...       &      200911-17 &FSRQ - LSP\\
ON\,325          & 1FGLJ\,1217.7+3007 & 12 17 52.0 &  30 07 00 &  0.130 &  14.5 &   24.90 &      572. & 6.7 $\pm$  0.6 &     2009-12-03 & BL\,Lac - HSP\\
PKS\,1217+02     &      ...         & 12 20 11.8 &  02 03 42 &  0.241 &  15.6 &    2.78 &      672. &      ...       &      2010-06-24 &FSRQ - LSP\\
3C\,273           & 1FGLJ\,1229.1+0203 & 12 29 06.7 &  02 03 08 &  0.158 &  14.1 &   63.11 &    54991. & 9.6 $\pm$  0.6 &    2010-01-11 &FSRQ - LSP\\ 
PG\,1246+586     & 1FGLJ\,1248.2+5820 & 12 48 18.7 &  58 20 28 &   \tablefootmark{e} &  14.5 &    3.99 &      245. & 4.5 $\pm$  0.4 &     2010-05-20 &BL\,Lac - ISP \\
3C\,279           & 1FGLJ\,1256.2$-$0547 & 12 56 11.1 & $-$05 47 21 &  0.536 &  15.0 &   20.85 &     9711. &32.4 $\pm$  1.1 &    2010-01-15 &FSRQ - LSP\\
1Jy\,1302$-$102     &      ...         & 13 05 33.0 & $-$10 33 19 &  0.286 &  14.4 &    4.20 &      711. &      ...       &    \tablefootmark{d}    &  FSRQ - ISP \\   
GB6B\,1347+0955   &      ...         & 13 50 22.1 &  09 40 10 &  0.133 &  13.6 &    3.74 &      293. &      ...       &      2010-07-18 &Uncertain - ISP\\ 
1WGAJ\,1407.5$-$2700&      ...         & 14 07 29.7 & $-$27 01 04 &  0.022 &   9.7 &   15.28 &      646. &      ...         &    2010-02-12 & Uncertain - HSP\\
3C\,298.0        &      ...         & 14 19 08.1 &  06 28 34 &  1.437 &  16.4 &    0.00 &     6100. &      ...        &    2010-07-30 &Radio Galaxy\\
BZQJ\,1423+5055        &      ...         & 14 23 14.1 &  50 55 37 &  0.286 &  15.1 &    3.35 &      178. &      ...        &   2010-07-13 &FSRQ - HSP\\
PG\,1424+240     & 1FGLJ\,1426.9+2347 & 14 27 00.3 &  23 48 00 &   \tablefootmark{e} &  14.5 &    3.57 &      430. &10.2 $\pm$  0.6 &   2010-01-22 &BL\,Lac - ISP\\ 
1RXSJ\,145603.4+504825   &      ...         & 14 56 03.6 &  50 48 25 &  0.479 &  18.1 &   13.02 &      220. &      ...       &      2009-12-25 &BL\,Lac - HSP\\
BZQJ\,1507+0415   &      ...         & 15 07 59.7 &  04 15 12 &  1.701 &  19.0 &    6.11 &      167. &      ...       &      2010-08-05 &FSRQ - LSP\\
PG\,1553+113     &      ...         & 15 55 43.0 &  11 11 24 &   \tablefootmark{e} &  14.0 &   17.85 &      312. &      ...       &      2010-02-05 &BL\,Lac - ISP\\
WE\,1601+16W3    &      ...         & 16 03 38.0 &  15 54 02 &  0.110 &  13.4 &    4.14 &       97. &      ...        &    2010-08-14 &Uncertain - HSP\\ 
3C\,345           & 1FGLJ\,1642.5+3947 & 16 42 58.8 &  39 48 37 &  0.593 &  15.0 &    2.52 &     7099. & 5.6 $\pm$  0.5 &     2010-03-06 &FSRQ - LSP\\
Mkn\,501          & 1FGLJ\,1653.9+3945 & 16 53 52.2 &  39 45 36 &  0.033 &   8.3 &   36.93 &     1558. & 8.3 $\pm$  0.6   &   2010-03-21 &BL\,Lac - HSP\\
1ES\,1741+196     & 1FGLJ\,1744.2+1934 & 17 43 57.8 &  19 35 09 &  0.084 &  12.7 &    4.23 &      301. & 1.1 $\pm$  0.3 &    \tablefootmark{d} &  BL\,Lac - ISP  \\  
PKS\,1833$-$77      &      ...         & 18 40 38.4 & $-$77 09 28 &  0.018 &   8.3 &    5.93 &     1108. &      ...       &      2010-03-11 &Uncertain - ISP\\ 
1ES\,1959+650     & 1FGLJ\,2000.0+6508 & 19 59 59.8 &  65 08 54 &  0.047 &  11.9 &   35.28 &      250. & 6.0 $\pm$  0.5 &     2009-09-26 &BL\,Lac - HSP\\
1Jy\,2005$-$489     & 1FGLJ\,2009.5$-$4849 & 20 09 25.3 & $-$48 49 53 &  0.071 &  11.0 &   33.24 &     1282. & 5.0 $\pm$  0.5 &    2009-10-05  &BL\,Lac - HSP\\
PKS\,2149$-$307     &      ...         & 21 51 55.5 & $-$30 27 53 &  2.345 &  17.4 &    4.80 &     1243. &      ...       &     2010-05-11 & FSRQ - LSP\\
NGC\,7213        &      ...         & 22 09 16.2 & $-$47 10 00 &  0.006 &  10.3 &   35.34 &       98. &      ...       &     2010-10-23       & Uncertain - HSP \\ 
PKS\,2227$-$399    &      ...         & 22 30 40.2 & $-$39 42 52 &  0.318 &  16.0 &    4.23 &      369. &      ...         &  2010-05-09 &Uncertain - ISP\\
3C\,454.3         & 1FGLJ\,2253.9+1608 & 22 53 57.7 &  16 08 53 &  0.859 &  13.2 &    7.80 &    12657. &46.2 $\pm$  1.3   &  2009-12-14 &FSRQ - LSP\\
PKS\,2300$-$18     &      ...         & 23 03 02.9 & $-$18 41 25 &  0.129 &  15.5 &    5.16 &      861. &      ...         &     2010-05-30 &Uncertain - ISP\\
PKS\,2331$-$240     &      ...         & 23 33 55.2 & $-$23 43 40 &  0.048 &  11.4 &    3.92 &      782. &      ...        &   2010-06-05 &Uncertain - ISP\\ 
1ES\,2344+514     & 1FGLJ\,2347.1+5142 & 23 47 04.8 &  51 42 17 &  0.044 &  10.7 &    7.71 &      250. & 1.4 $\pm$  0.3 &   2010-01-17 & BL\,Lac - HSP\\
\hline
\end{tabular}
\label{tab:rass_sample}
\tablefoot{
\tablefoottext{a}{843\,MHz flux is reported for sources with Dec $<-40^{\circ}$.}
\tablefoottext{b}{Units of $10^{-12}$\,erg\,cm$^{-2}$\,s$^{-1}$}; 
\tablefoottext{c}{Units of $10^{-10}$\,ph\,cm$^{-2}$\,s$^{-1}$}\\
\tablefoottext{d}{\swift~simultaneous observation not available.}
\tablefoottext{e}{Optical spectrum completely featureless or not available, redshift unknown}
}
\normalsize
\end{center} 
\end{table*}

\subsection{The radio flux-density limited  sample}
The radio flux-density limited sample is presented in the companion \planck\ paper  \citep{planck2011-6.3a}, where all the observational details are given.
The sample consists of 104 bright northern and equatorial radio-loud AGN characterized by $S_{37\,{\rm GHz}} > 1$\,Jy as measured with the Mets\"ahovi 
radio telescope. 

\medskip
Although the samples are defined by different criteria, four sources are common to all samples.  These are the well-known objects 3C\,273, 3C\,279, Mkn\,501, and 3C\,454.3, which are
among the brightest objects across the entire electromagnetic spectrum.
A summary of the number of sources common to more than one sample is given in Table~\ref{tab:samples}.

\begin{table*}[htbp]
  \caption{Summary of the samples, blazar types, and selection methods considered in this paper.}
  \label{tab:samples}
  \begin{center}
  \begin{tabular}{l l c c c c c c c c}
    \hline
    \hline
     & Selection  & No. of & Blazars & &  \multicolumn{5}{c}{Sources in common} \\
   Sample   & band   &    sources           & FS/BL/Unc. &Other AGN & RASS   &  BAT   & \fermi-LAT   &  Radio  & All  \\
    \hline
    RASS   & Soft X-ray &43 &15/16/11 & 1 & ...  & 12 & 5 & 9  & 4 \\
    BAT    & Hard X-ray &34 &21/7/6 & ... & 12 & ...  & 9 & 16 & 4 \\
    \fermi-LAT $^*$  & \gr & 50 &28/16/6 & ... & 5  & 9  & ... & 23 & 4 \\
    \fermi-LAT FL$^{**}$ & \gr & 40 &27/8/5 & ... & 3  & 7  & ... & 19 & 3 \\
    \hline
    Total this paper &   & 105 & 52/32/20 & 1 & ... & ...   & ...  & ...   & ...\\
    \hline
   Radio   & radio & 104 & 73/18/10 & 3 & 9  &  16 & 23 & ... & 4 \\
    \hline   
\end{tabular}
   \end{center}
      $^{ *}$ Total  \fermi-LAT sample (TS limited), \\
     $^{**}$ Flux-limited  \fermi-LAT sample $F(E >100\,{\rm MeV}) > 8\times 10^{-8}$ ph~cm$^{-2}$~s$^{-1}$
\end{table*}

\section{Data analysis}

\subsection{Ground-based follow-up data}

Following the launch of  \planck, several follow-up programs with ground-based facilities started collecting
simultaneous radio and optical data. 
In this paper, we used data from the observatories listed in
Table~\ref{tab:radioobs}.

\begin{table*}[htbp]
  \caption{Optical and radio observatories participating in the \planck~multi-frequency campaigns.}
  \label{tab:radioobs}
  \begin{center}
  \begin{tabular}{l l}
    \hline
    \hline
  Radio observatory &     Frequencies [GHz]       \\
    \hline
  APEX, Chile        &    345      \\
  \multirow{2}{*}{ATCA, Australia}        &   4.7, 5.2, 5.8, 6.3, 8.2, 8.7, 9.3, 9.8, 17.2, 17.7, 18.3, 18.8 \\
                         &   23.2, 23.7, 24.3, 24.8, 32.2, 32.7, 33.3, 33.8, 38.2, 38.7, 39.3, 39.8  \\
  Effelsberg, Germany  &   2.64, 4.85, 8.35, 10.45, 14.60, 23.05, 32.00, 43.00         \\
  IRAM, Spain        &     86.2, 142.3         \\
  Medicina, Italy    &   5, 8     \\
  Metsahovi, Finland   &     37      \\
  OVRO, USA        &     15      \\
  RATAN, Russia       &   1.1, 2.3, 4.8, 7.7, 11.2, 21.7  \\
  UMRAO, USA       &   4.8, 8.0, 14.5  \\
  VLA, USA         &   5, 8, 22, 43      \\
    \hline \hline
    Optical observatory &   Band \\
    \hline
    KVA, Spain   & R \\
    Xinglong, China & I \\
    \hline
  \end{tabular}
  \end{center}
\end{table*}

\subsubsection{APEX}

Some sources from our sample were observed
in the submillimetre domain with the 12-m Atacama
Pathfinder Experiment (APEX) in Chile. The observations were
made using the LABOCA bolometer array centered at 345\,GHz.
Data were taken at two epochs: September 3--4, 2009,
and November 12, 2009. The data were reduced using the script
package {\it minicrush}\footnote{\url{http://www.submm.caltech.edu/~sharc/crush/}},
version 30-Oct-2009. Uranus was used as a calibrator of the flux densities.

\subsubsection{ATCA-PACO}

The \planck-ATCA Co-eval Observations (PACO) project \citep{massardi11, bonavera11}
 observed  480 sources selected from the Australia Telescope 20\,GHz catalogue
(AT20G, \citealt{massardi10}), with the Australia Telescope Compact Array (ATCA)
in the frequency range  4.5--40\,GHz, at several epochs close in time to the \planck\ observations
in the period July 2009 to August 2010. The PACO sample is  a complete, flux-density limited, and spectrally selected
sample of southern sources, with the exception of the region with Galactic latitude $|b|<5\degr$. 
A total of 147 PACO point-like sources have at least one observation within ten days of the \planck~observations.

\subsubsection{Effelsberg and IRAM}

Quasi-simultaneous cm/mm radio spectra for a larger number of \planck\ blazars were obtained
within the framework of a \fermi\ monitoring program of $\gamma$-ray blazars \citep[F-GAMMA:][]{fuhrmann07,angelakis08} on the Effelsberg 100-m and IRAM 30-m telescopes. The frequency range was 2.64--142\,GHz.

The Effelsberg measurements were conducted with the secondary focus heterodyne 
receivers at 2.64, 4.85, 8.35, 10.45, 14.60, 23.05, 32.00, and 43.00\,GHz. 
The observations were performed quasi-simultaneously with cross-scans, 
that is by slewing over the source position in azimuth and elevation 
with the number of sub-scans chosen to reach the desired 
sensitivity \citep[for details, see][]{fuhrmann08,angelakis08}.
Pointing offset, gain, atmospheric opacity, and sensitivity corrections were applied to the data. 

The IRAM 30-m observations were carried out with calibrated cross-scans 
using the EMIR horizontal and vertical polarization receivers operating 
at 86.2\,GHz and 142.3\,GHz. The opacity-corrected intensities were converted 
into the standard temperature scale and finally corrected for small remaining 
pointing offsets and systematic gain-elevation effects. Conversion to a standard 
flux density scale was based on frequent observations of primary calibrators (Mars, Uranus) and secondary
calibrators (W3(OH), K3-50A, NGC\,7027).
 
From this program, radio spectra measured quasi-simultaneously with the 
\planck\ observations were collected for a total of 37 \planck\ blazars 
during the period August 2009 to June 2010. Results are reported in Table~\ref{tab:effelsbergdata} and~\ref{tab:iramdata}.
\addtocounter{table}{1}
\addtocounter{table}{1}

\subsubsection{Medicina}

The Simultaneous Medicina \planck\ Experiment 
(SiMPlE, \citealt{procopio11}) used  the 32-m Medicina
single dish to make almost simultaneous observations at 5\,GHz and 8.3\,GHz of the 263 sources of 
the NEWPS sample \citep{massardi09} with Dec $>0^{\circ}$,
and partially overlapping with the PACO observations for $-10^{\circ}<$ Dec $<0^{\circ}$.
The project began in June 2010 and  finished in June 2011, 
observing our sample several times throughout two complete \planck\ surveys. It does not overlap with the \planck\ first survey.

\subsubsection{Mets\"ahovi}
The 37\,GHz observations were made with the 13.7-m
Mets\"ahovi radio telescope using
a 1\,GHz bandwidth, dual-beam receiver centered at 36.8\,GHz. We performed
ON-ON observations, by alternating between the source and
the sky in each feed horn. A typical integration time for obtaininig
one flux density data point was 1200--1400\,s. The telescope detection limit
at 37\,GHz was $\sim$0.2\,Jy under optimal conditions. 
Data points with a signal-to-noise ratio (S/N) smaller than four
are handled as non-detections. The flux-density scale was set by
observations of DR\,21. Sources NGC\,7027, 3C\,274, and 3C\,84
were used as secondary calibrators. A detailed description of the
data reduction and analysis is given in \citep{terasranta98}.
The error estimate in the flux density includes the contribution
from the measurement rms and the uncertainty in the absolute
calibration.

\subsubsection{OVRO}
Some of the sources in our samples were monitored at 15\,GHz
using the 40-metre telescope of the Owens Valley Radio
Observatory as part of a larger monitoring program 
\citep{richards2010}. The flux density of each source was measured 
approximately twice weekly, with occasional gaps due to poor
weather or instrumental problems. The telescope was equipped
with a cooled receiver installed at prime focus, with two 
symmetric off-axis corrugated horn feeds that are sensitive to left circular
polarization. The telescope and receiver combination produces
a pair of approximately Gaussian beams (157\,arcsec FWHM),
separated in azimuth by 12.95\,arcmin. The receiver has a 
central frequency of 15.0\,GHz, a 3.0\,GHz bandwidth, and a 
noise-equivalent reception bandwidth of 2.5\,GHz. Measurements were
made using a Dicke-switched dual-beam system, with a second
level of switching in azimuth where we alternated between source and sky in each
of the two horns. Our calibration is referred to 3C\,286, for which a
flux density of 3.44\,Jy at 15\,GHz is assumed \citep{baars77}.
Details of the observations, calibration, and analysis are given
by \citep{richards2010}.

\subsubsection{RATAN}
A six-frequency broadband radio spectrum was obtained
with the RATAN-600 radio telescope in transit mode by observing 
simultaneously at 1.1, 2.3, 4.8, 7.7, 11.2, and 21.7\,GHz
\citep{parijskij93,berlin96}.
Data were reduced using the
RATAN standard software FADPS (Flexible Astronomical Data
Processing System) reduction package \citep{verkhodanov97}.
The flux density measurement procedure at RATAN-600 is
described by \citep{aliakberov85}.

\subsubsection{UMRAO}

Centimetre-band observations were obtained with the
University of Michigan 26-m prime focus paraboloid
equipped with radiometers operating at central frequencies of
4.8, 8.0, and 14.5\,GHz. Observations at all three frequencies
utilized rotating polarimeter systems permitting both total flux
density and linear polarization to be measured. A typical
measurement consisted of 8 to 16 individual measurements over a
20--40 minute time period. Frequent drift scans were made
across stronger sources to verify the telescope pointing 
correction curves; and observations of program sources were 
intermixed with observations of a grid of calibrator sources to 
correct for temporal changes in the antenna aperture efficiency. The
flux scale was based on observations of Cassiopeia A \citep{baars77}.
Details of the calibration and analysis 
techniques are described by \cite{aller85}.

\subsubsection{VLA}
The Very Large Array (VLA) and, since Spring 2010, the Expanded VLA (EVLA),
 observed a subset of the sources as 
simultaneously as possible. Most of the VLA and EVLA runs were performed in one to two hour chunks of time. 
We observed during a one-hour chunk of time, in addition to flux calibrators and phase calibrators, typically 
5--8 \planck\ sources. In many cases, VLA flux-density and
phase calibrators were themselves of interest, since they were
bright enough to be detected by \planck. For these bright sources,
the integration times could be extremely short. 

Integration times were about 30\,s at 4.86\,GHz and 8.46\,GHz, 100\,s at
22.46\,GHz, and 120\,s at 43.34\,GHz. All VLA/EVLA flux 
density measurements were calibrated using standard values for one
or both of the primary calibrator sources used by NRAO, 3C\,48
or 3C\,286, and the $u$-$v$ data were flagged, calibrated, and imaged
using standard NRAO software (AIPS or CASA). It is important to
bear in mind that the VLA and EVLA were in different 
configurations at different times in the several months duration of the
observations. As a consequence, the angular resolution changed, becoming, for a given configuration, 
much higher at higher frequencies. For that reason, sources that appeared to be resolved in 
any VLA configuration or at any VLA frequency were carefully flagged.

\subsubsection{KVA}

Optical observations were made with the 35 cm KVA
(Kunliga Vetenskapsakademiet) telescope at La Palma, Canary
islands. All observations were made through the $R$-band filter
($\lambda_{\rm eff} = 640$\,nm) using a Santa Barbara ST-8 CCD camera with a
gain factor of 2.3\,$e^-/{\rm ADU}$ and readout noise of 14 electrons. 
Pixels were binned 2$\times$2 pixels giving a plate scale of 0.98\,arcsec/pixel.
We obtained 3--6 exposures of 180\,s per target.
The images were reduced in the standard way of subtracting
the bias and dark frames and dividing by twilight flat-fields.
The fluxes of the target and 3--5 stars in the target field were
measured with aperture photometry and the magnitude difference
between the target and a primary reference star in the same
field was determined. The use of differential mode makes the observations
insensitive to variations in atmospheric transparency
and accurate measurements can be obtained even in partially
cloudy conditions. The $R$-band magnitude of the primary reference
star was determined from observations made on photometric nights,
using comparison stars in known blazar fields as
calibrators \citep{fiorucci96,fiorucci98,raiteri98,villata98,nilsson07}
and taking into account the color term of the $R$-band filter employed. After
the $R$-band magnitude of the primary reference star was determined,
the object magnitudes were computed from the magnitude differences.
At this phase we assumed $V-R = 0.5$ for the
targets. Several stars in the field were used to check the quality 
of the photometry and stability of the primary reference. The
uncertainties in the magnitudes include the contribution from both
measurement and calibration errors.

\subsubsection{Xinglong}
The monitoring at Xinglong Station, National Astronomical
Observatories of China, was performed with a 60/90 cm f/3
Schmidt telescope. The telescope is equipped with a 4096$\times$4096
E2V CCD, which has a pixel size of 12\,$\mu$m and a spatial 
resolution of 1\farcs3 pixel$^{-1}$ . The observations were made with an
intermediate-band filter, the $I$ filter. Its central wavelength and
passband width are 6685\,\AA\  and 514\,\AA, respectively. The exposure times were mostly 120\,s but 
ranged from 60\,s to 180\,s, depending on weather and lunar phase.

\subsection{\planck\ \mw~data}

\planck\ \citep{tauber2010a,planck2011-1.1} is the third generation space mission
to measure the anisotropy of the cosmic \mw~ background (CMB).  It observes the
sky in nine frequency bands covering 30--857\,GHz with high sensitivity 
and angular resolution from 31\arcm\ to 5\arcm. Full sky coverage is attained in about 
seven months. The Low Frequency Instrument (LFI; \citealt{Mandolesi2010,
Bersanelli2010, Planck2011-1.4}) covers the 30, 44, and 70\,GHz bands with
amplifiers cooled to 20\,\hbox{K}.  The High Frequency Instrument (HFI;
\citealt{Lamarre2010, Planck2011-1.5}) covers the 100, 143, 217, 353, 545,  and
857\,GHz bands with bolometers cooled to 0.1\,\hbox{K}.  Polarization is measured in
all but the highest two bands 
\citep{Leahy2010, Rosset2010}.  A combination of radiative cooling and three
mechanical coolers produces the temperatures needed for the detectors and optics
\citep{Planck2011-1.3}.  Two data processing centers (DPCs) check and calibrate the
data and make maps of the sky \citep{Planck2011-1.7, Planck2011-1.6}.  \planck's
sensitivity, angular resolution, and frequency coverage make it a powerful
instrument for Galactic and extragalactic astrophysics as well as cosmology.  
Early astrophysics results are given in Planck Collaboration, 2011h--z.

The Early Release Compact Source Catalog (ERCSC, \citealt{Planck2011-1.10})
contains all sources, both Galactic and extragalactic, detected with high confidence
over the full sky during the period between August 12, 2009 and June 6, 2010
(corresponding to \planck\  operational days 91 to 389).
The ERCSC only contains average intensity information for the
sources. However, many of the sources were observed more than once during the time
period spanned by the ERCSC, and some of the \swift~observations used for this paper
were carried out between June and October 2010. Therefore, to have simultaneous data,
we produced independent maps for the first (OD 91--274), 
the second (OD 275--456), and the beginning of the third \planck\ survey (OD 457--550) through the
LFI and HFI pipelines described in \cite{Planck2011-1.6} and \cite{Planck2011-1.7}, and we 
extracted the flux densities from each map using IFACMEX, which is 
an implementation of the Mexican Hat Wavelet 2 (MHW2) algorithm available at the LFI DPC.
The MHW2 tool has been extensively
used to detect point-like objects in astronomical images, both with
simulations from various experiments and data from the \wmap, \planck,
and Herschel satellites \citep{gonzaleznuevo06,lopezcaniego06,lopezcaniego07,massardi09}.
This wavelet is defined as the fourth derivative of the two-dimensional Gaussian function, where
the scale of the filter $R$ is optimized to look for the maximum in the S/N of the sources in the filtered map. In
practice, the IFCAMEX code, our implementation of the MHW algorithm,
deals with flexible image transport system (FITS) maps in Healpix format \citep{Healpix} and can be used to detect
sources in the whole sky or at the position of known objects. For this
analysis, we looked for objects above the $4\sigma$ level at the
positions corresponding to the 105 sources of our sample. For objects with S/N
smaller than four we adopted the $4\sigma$ level as an upper limit.
The results of the analysis of \planck\ simultaneous data are reported in Table~\ref{tab:planckdata}, where 
columns 1 and 2 give the source name, columns 3 and 4 give the observation start and end times, and columns 5--13 give the flux densities in units of Jy 
at 30, 44, 70, 100, 143, 217, 353, 545, and 857\,GHz.
\addtocounter{table}{1}

Owing to source variability,  we do not expect these simultaneous flux densities to
be the same as the time-averaged  ERCSC measurements, except in the case of the sources that were observed only once during the ERCSC time 
range and for which we estimated the \planck\ flux densities, measured simultaneously with the \swift~observation, in the same period.
We verified that, for the sources fulfilling these requirements, the flux densities extracted for this paper
are in good agreement with those of the ERCSC.

In addition to simultaneous \planck\ data, we also used ERCSC flux densities in both our analysis of flux correlations (Sect.~\ref{mwvsgr}) and the 
SED plots described in Sect.~\ref{seds}.

\subsection{\swift~optical, UV, and X-ray data}

The {\it Swift} Gamma-Ray-Burst (GRB) Explorer \citep{gehrels2004} is a multi-frequency space observatory
devoted to the discovery and rapid follow-up of GRBs. There are three instruments on board the spacecraft:
the UV and Optical Telescope (UVOT, \citealt{roming2005}), the X-Ray Telescope (XRT, \citealt{burrows2005})
sensitive to the 0.3--10.0\,keV band, and the Burst Alert Telescope (BAT, \citealt{barthelmy2005}) sensitive
to the 15--150\,keV band. Although the primary scientific goal of the satellite is the observation
of GRBs, the wide frequency coverage is suitable for blazar studies, because it covers the region where
the transition between synchrotron and inverse Compton emission usually occurs.

When not engaged in GRB observations, {\it Swift} is available for target of opportunity (ToO) requests, and the
{\it Swift} team decided to devote an average of three ToO observations per week to this project for simultaneous
observations of blazars.

\subsubsection{UVOT}

The \swift~UVOT telescope can produce images in each of
its six lenticular filters (V, B, U, UVW1, UVM2, and UVW2). However, in an effort to reduce the 
use of mechanical parts after several years of orbital operations, observations are  carried out using
only one filter, unless specifically requested by the user. Thus images are not always available for all filters.

The photometry analysis of all our sources was performed using the standard UVOT
software distributed within the HEAsoft 6.8.0 package and the calibration
included in the latest release of the ``Calibration Database''.
A specific procedure was developed at the ASI Science Data Center (ASDC) to process all ToO observations requested
for the blazar sample. Counts were extracted from apertures of 5\arcsec\ radius for 
all filters and converted to fluxes using the standard zero points \citep{poole08}. The fluxes were 
then de-reddened using the appropriate values of $E(B-V)$ for each source taken 
from \citep{schlegel1998} with $A_{\lambda}/E(B-V)$ ratios calculated for UVOT 
filters using the mean Galactic interstellar extinction curve from \citep{Fitzpatrick1999}. 
No variability was detected within single exposures in any filter.
The processing results were carefully validated including checks for possible
contamination by nearby objects within the source and background apertures.
Some sources, such as 3C\,273 and NGC\,1275, needed special analysis, and results for some other sources
had to be discarded.

The results of the UVOT data analysis are summarized in Table~\ref{tab:uvotdata},
where columns 1 and 2 give the source name, columns 3 and 4 give the observation date and the \swift~observation ID, and  the remaining columns give the 
magnitudes in the six UVOT filters with errors.
\addtocounter{table}{1}

\subsubsection{XRT}

The \swift\ XRT is usually operated in ``auto-state'' mode, which automatically adjusts the CCD read-out mode
to the source brightness, in an attempt to avoid pile-up \citep{burrows2005, hill2004}. 
As a consequence, some of the data were collected
using the most sensitive photon counting (PC) mode, while windowed timing (WT) mode
was used for bright sources.

The XRT data were processed with the XRTDAS software package (v. 2.5.1, \cite{capalbi2005})
developed at the ASI Science Data Center (ASDC) and distributed by the NASA High Energy Astrophysics Archive
Research Center (HEASARC) within the HEASoft package (v. 6.9). Event files were calibrated
and cleaned with standard filtering criteria using the {\it xrtpipeline} task and the latest calibration
files available in the \swift\ CALDB. Events in the energy range 0.3--10\,keV with grades 0--12 (PC
mode) and 0--2 (WT mode) were used for the analysis.

Events for the spectral analysis were selected within a circle of 20 pixels ($\sim 47\arcsec$) radius,
which encloses about 90\% of the point spread function (PSF)  at $1.5$\,keV \citep{moretti2005}, centered on the source
position.
When the source count rate is above $\sim0.5$\,counts/s, the  PC mode data are
significantly affected by pile-up in the inner part of the PSF. In these 
cases, and after comparing the observed PSF profile with the analytical model derived by \citep{moretti2005},
we removed pile-up effects by excluding events detected within up to 6 pixels from the
source position, and used an outer radius of 30 pixels. The value of the inner radius was evaluated
individually for each observation affected by pile-up, in a way that depended on the observed source count rate.

Ancillary response files were generated with the xrtmkarf task by applying corrections for the PSF
losses and CCD defects. Source spectra were binned to ensure a minimum of 20 counts per bin when 
utilizing the $\chi^2$ minimization fitting technique.
We fitted the spectra adopting an absorbed power-law model with photon index $\Gamma_x$. When
deviations from a single power-law model were found, we adopted a log-parabolic law of the form
$F (E) = KE^{(a+b\log E)}$ \citep{massaro2004}, which has been shown to fit the X-ray spectrum
of blazars of the HSP type well \citep[e.g.][]{giommi2005, tramacere2009}. This spectral model is described by only two
parameters: $a$, the photon index at $1$\,keV, and $b$, the curvature of the parabola. For both models,
the amount of hydrogen-equivalent column density ($N_{\rm H}$) was fixed to the Galactic value along the
line of sight \citep{kalberla2005}. For a fraction of the sources fitted with a power-law model ($\sim$15\%),
we found evidence of an absorption excess at low energies and the hydrogen column density $N_{\rm H}$ parameter was left free.

The results of the spectral fits with a power-law model and Galactic $N_{\rm H}$ are shown in Table~\ref{tab:xrtdata}, 
where columns 1 and 2 give the source name,
column 3 gives the \swift~observation date, column 4 gives the best-fit photon index $\Gamma_x$, column 5 gives the
Galactic $N_{\rm H}$, columns 6 and 7 give the 0.1--2.4 and 2--10\,keV X-ray fluxes, column 8 gives the value of the reduced $\chi^2$, and column 9
gives the number of degrees of freedom.
\addtocounter{table}{1}

In Table~\ref{tab:xrtdatalogpar}, we report data obtained using a log-parabola to describe the spectrum model.
Columns 1 and 2 give the source name, column 3 gives the \swift~observation date, columns 4 and 5 give the log parabola
parameters $a$ and $b$, column 6 gives the
Galactic $N_{\rm H}$, columns 7 and 8 give the 0.1--2.4 and 2--10\,keV X-ray fluxes, column 9 gives the value of the reduced $\chi^2$, and column 10
gives the number of degrees of freedom.
\addtocounter{table}{1}

Finally, in Table~\ref{tab:xrtdatanhfree} we report the results of the spectral fits that were performed leaving the hydrogen
column density $N_{\rm H}$ to vary as a free parameter. Columns 1 and 2 give the source name,
column 3 gives the \swift~ observation date, column 4 gives the best-fit photon index $\Gamma_x$, column 5 gives the
estimated $N_{\rm H}$, columns 6 and 7 give the 0.1--2.4 and 2--10\,keV X-ray fluxes, column 8 gives the value of the reduced $\chi^2$, and column 9
gives the number of degrees of freedom.
\addtocounter{table}{1}

\subsection{{\em Fermi}-LAT \gr~data}
\label{sec:Fermi}

The Large Area Telescope (LAT) on-board \fermi\ is an electron-positron pair conversion telescope sensitive to $\gamma$-rays of energies from 20\,MeV to $>$ 300\,GeV. The Fermi-LAT consists of a high-resolution silicon microstrip tracker, a CsI hodoscopic electromagnetic calorimeter, and an anticoincidence detector for charged particle background identification. A full description of the instrument and its performance can be found in \cite{atwood2009}. The large field of view ($\sim$2.4\,sr) allows the LAT to observe the full sky in survey mode every 3\,hr. The LAT point spread function (PSF) depends strongly on both the energy and the conversion point in the tracker, but less on the incidence angle. 

The LAT $\gamma$-ray spectra of all AGN sources are studied in \cite{abdo2010a} based on 11 months of Fermi-LAT data. Here we derived the $\gamma$-ray spectra of the blazars for which we built the simultaneous SEDs, integrating for two weeks  encompassing the whole duration of the \planck\ observations.

The \fermi-LAT data considered for this analysis cover the period from August 4, 2008 to November 4, 2010  and were analyzed using the standard \fermi-LAT ScienceTools software package\footnote{\url{http://fermi.gsfc.nasa.gov/ssc/data/analysis/documentation/Cicerone}}  (version v9r16 and selecting for each source only photons 
of energies above 100~MeV belonging to the diffuse class (Pass6 V3 IRF; \citealt{atwood2009}), which have the lowest background contamination. 
For each source, we selected only photons within a 15$^{\circ}$ region of interest (RoI) centered around the source itself. 
To avoid background contamination from the bright Earth limb, time intervals where the Earth entered the LAT Field of View (FoV) were excluded from the data sample. In addition, we excluded observations in which the source under study was viewed at zenith angles larger than 105$^{\circ}$, where Earth's atmospheric $\gamma$-rays increase the background contamination.
The data were analyzed with a binned maximum likelihood technique \citep{mattox96} using the analysis software {\em gtlike} developed by the LAT team$\footnote{\url{http://fermi.gsfc.nasa.gov/ssc/data/analysis/documentation/Cicerone/Cicerone\_Likelihood}}$. 
A model accounting for the diffuse emission and nearby $\gamma$-ray sources was included in the fit. 

The diffuse foreground, including Galactic interstellar emission, extragalactic $\gamma$-ray emission, and residual CR background, was modeled with $ gll$\_$iem$\_$v02$\footnote{\url{http://fermi.gsfc.nasa.gov/ssc/data/access/lat/BackgroundModels.html}} for the Galactic diffuse emission and isotropic$\_$iem$\_$v02 for the extragalactic isotropic emission.
Each source under study was modeled with a power-law function 
\begin{equation}
 \frac{dN}{dE} = \frac{N(\Gamma +1)E^{\Gamma}}{E_{\rm max}^{\Gamma +1}-E_{\rm min}^{\Gamma +1}}
\end{equation}
where both the normalization factor $N$ and the spectral index $\Gamma$ were allowed to vary in the model fit.
The model also includes all the sources within a 20$^{\circ}$ RoI included in Fermi-LAT one-year catalog \cite{Abdo1FGL} and modeled using power-law functions. If a source included in the model is a pulsar belonging to the \fermi-LAT pulsar catalog \cite{abdopsrcat}, we modeled the spectrum with a power-law with exponential cut-off using the spectral parameters in the pulsar catalog.


For the evaluation of the $\gamma$-ray SEDs, the whole energy range from 100\,MeV to 300\,GeV was divided into two equal logarithmically spaced bins per decade. In each energy bin, the standard {\em gtlike} binned analysis was applied assuming power-law spectra with photon index $=-2.0$  for all the point sources in the model.
Assuming that in each energy bin the spectral shape can be approximated by a power-law, the flux of the source in all selected energy bins was evaluated, requiring in each energy bin a TS greater than ten.  If the TS is lower than ten, an upper limit (UL) was evaluated in that energy bin. Only the statistical errors in the fit parameters are shown in the plots. Systematic errors due  mainly to uncertainties in the LAT effective area derived from the on-orbit estimations, are \ $<5\%$ near 1\,GeV, 10$\%$ below 0.1\,GeV, and 20$\%$ above 10\,GeV. 

For each source, we considered the three different integration periods for the \gr~data:

\begin{itemize}
 \item {\it Simultaneous observations}: data accumulated during the period of \planck\ observation of the source. As the \planck\ instruments point in slightly different directions and the field of view depends on the frequency of observation, a typical observation covering all \planck\ channels~takes about one week, the exact integration time depending on the position of the source. 
 \item  {\it Quasi-simultaneous observations}: data integrated over a period of two months centered on the \planck\ observing period of the source.
 \item  {\it Twenty-seven month \fermi-LAT~integration}: data integrated over a period of  27 months from August 4, 2008 to November 4, 2010, i.e., the entire \fermi-LAT data set available for this paper.
\end{itemize}

Tables~\ref{tab:fermisummary} and~\ref{tab:fermifsrqsbllacs} give a summary of the \gr~detections ($\textrm{TS} >25$) in all our samples. The fraction of  sources detected by \fermi-LAT during 
the simultaneous integrations is not very large and varies from $\sim 40\%$ in the  \fermi-LAT sample to just $\sim20\%$ in the soft X-ray selected sample. 
We note that even considering all the \fermi-LAT data available at the time of writing (27 month integration), a sizable fraction of the blazars in the radio and both soft and hard X-ray selected samples were not detected. 

\begin{table}
\begin{center}
\caption{Summary of \gr~detections with significance $\textrm{TS}>25$.}
  \begin{tabular}{lcccc}
\hline
\hline
      &  \multicolumn{3}{c}{No. of detected sources}& Sources    \\    
    Sample      & simult.    & 2 months   &27 months  &in sample   \\                 
\hline  
\fermi-LAT & 18 (36\%) &40 (80\%) & 50 (100\%) & 50 \\   
\swift-BAT  & ~9 (26\%) &12 (35\%) & 27 (79\%)& 34 \\       
{\it ROSAT}/RASS & 10 (23\%) &15 (35\%) & 24 (56\%)&43 \\     
{Radio}& 22 (21\%) &38 (37\%) & 78 (75\%)&104 \\  
\hline
\label{tab:fermisummary}
 \end{tabular}
\end{center}
\end{table}

\begin{table}
\begin{center}
\caption{Statistics of \gr~detections ($\textrm{TS}>25$) in the 27 month \fermi-LAT data set.}
  \begin{tabular}{lccc}
\hline
\hline
     & \multicolumn{3}{c}{No. of detected sources} \\       
    Sample   &  FSRQs & BL Lacs &Uncertain   \\        
\hline  
\fermi-LAT & 28 (100\%) &14 (100\%) & 8 (100\%) \\   
\swift-BAT  & 17 (63\%) &7 (100\%)  &3 (50\%)\\       
{\it ROSAT}/RASS & 8 (53\%) &14 (88\%)& 2 (17\%)\\     
{Radio}& 48 (72\%) &16 (100\%) & 9 (64\%) \\  
\hline
\label{tab:fermifsrqsbllacs}
 \end{tabular}
\end{center}
\end{table}

Detailed results of the \fermi-LAT analysis are given in Tables~\ref{tab:fermidatasim2bin}--\ref{tab:fermidata27m1bin}, where the observed fluxes or upper limits are given in six or three energy bands depending on the source brightness.
\addtocounter{table}{6}

Two sources (PKS\,0548$-$322 and NGC\,7213) appear as significant \gr~detections in our 27-month data set, although they were not included in any of the  
\fermi-LAT catalogues published so far \citep[][and \footnote{\url{http://heasarc.gsfc.nasa.gov/W3Browse/all/fermilpsc.html}}]{abdoAGNpaper,Abdo1FGL}. 
These should therefore be considered as new \gr\ detections.

\section{The importance of simultaneity}

Blazars are, by definition, highly variable sources. It is therefore important to use simultaneous multi-frequency data to build SEDs for comparison with theoretical models.  
In this section, we compare our measurements with data taken from the literature in order to derive an estimate of the uncertainties introduced by the use of non-simultaneous data 
in different parts of the spectrum.

\begin{figure}
 \centering
  \includegraphics[width=7.5cm,angle=-90]{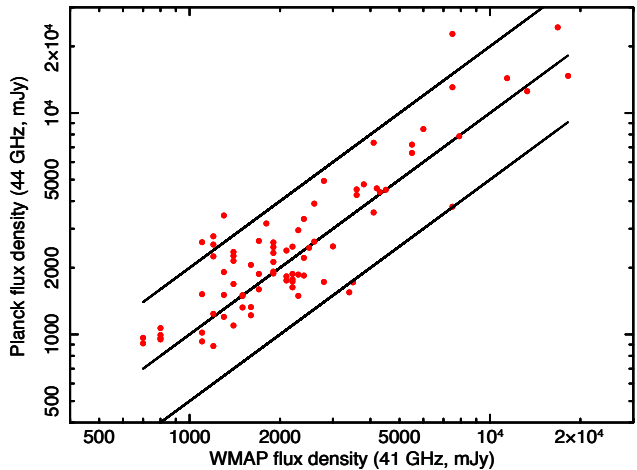}
 \caption{The \planck\ 44\,GHz flux density of the sources in our sample is plotted against the 41\,GHz flux density from the \wmap\ five-year catalog (81 sources). The three solid lines represent equal flux densities (i.e., no variation) and a factor of two variability above or below the equal flux level. Almost all the points lie between the factor of two variability lines.}
 \label{fig.planckwmapvariability}
\end{figure}

Figure \ref{fig.planckwmapvariability} plots the \planck\ flux density at 44\,GHz presented in this paper versus the \wmap~flux density at 41\,GHz from the \wmap~point source catalogs \citep{bennett03,wright09}.  Some scatter is present, but most of the points lie between the two solid lines indicating a factor of two variability.

\begin{figure}
 \centering
 \includegraphics[width=7.5cm,angle=-90]{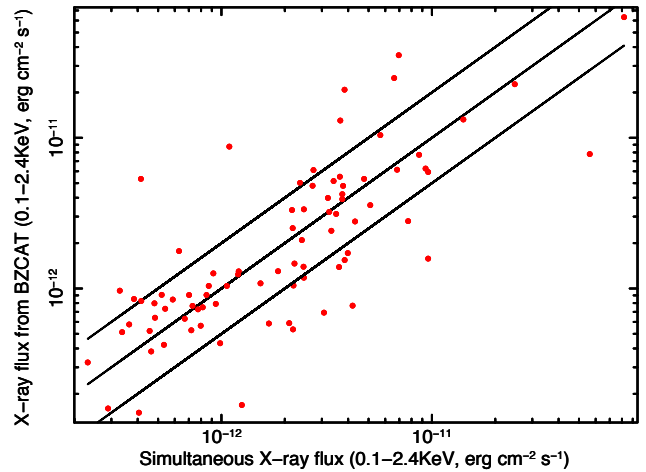}
 \caption{The \swift~X-ray (0.1--2.4\,keV) flux of the sources in our sample measured simultaneously with \planck\ is plotted against the 0.1--2.4\,keV flux reported in the BZCAT catalog (83 sources).The three solid lines represent 
 equal fluxes (i.e., no variation) and a factor of two variability above or below the equal flux level. Note that several points are outside the factor of two variability lines, revealing variability of up to about a factor ten.}
 \label{fig.fxvariability}
\end{figure}

Figure \ref{fig.fxvariability} plots the X-ray fluxes of the sources observed by \swift\ simultaneously with \planck\ (see Table \ref{tab:xrtdata}) against the X-ray fluxes of the same sources 
from the BZCAT catalog \citep{bzcat09,bzcat10}. In this case, a large scatter is present, with variations of over a factor of ten. 

\begin{figure}
 \centering
 \includegraphics[width=7.5cm,angle=-90]{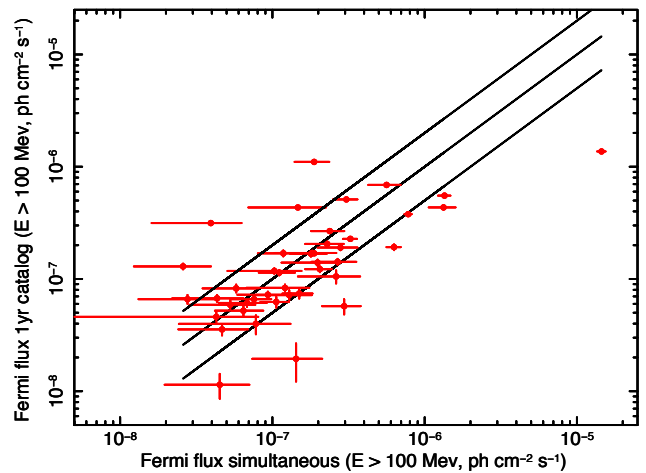}
 \caption{The \fermi-LAT \gr~flux of the sources in our samples detected during the simultaneous integration with \planck\ is plotted against the flux reported in the \fermi-LAT 1-year catalog. The three solid lines represent equal fluxes (i.e., no variation) and a factor of two variability above or below the equal flux level. Note that several points are outside the factor of two variability lines, revealing variability of up to about a factor ten.}
 \label{fig.fermivariability}
\end{figure}

Figure \ref{fig.fermivariability}  shows the \fermi-LAT \gr\ fluxes of our sources measured simultaneously with \planck\ plotted against their \gr\ fluxes in the \fermi-LAT 1FGL catalog \citep{Abdo1FGL}. As in the X-ray sample,  a scatter with variations larger than a factor of ten  is observed.


We conclude that SEDs built with non-simultaneous data suffer from uncertainties in the \mw\ region that are relatively modest and generally limited to about a factor of two, while the high energy part of the spectrum (X-ray and \gr) is much more affected, with uncertainties caused by flux variations of up to a factor of ten or more. The same uncertainties,
of course, apply when searching for correlations in non-simultaneous multi-frequency data.

\section{Spectral energy distributions}
\label{seds}

We constructed the SEDs of all the blazars in our samples from the simultaneous multi-frequency data described above using the ASDC SED Builder, an on-line service developed at the ASI Science Data Center (ASDC)\footnote{\url{http://tools.asdc.asi.it/SED/}} \citep{SEDtool}. This is a WEB-based tool that allows users to build multi-frequency 
SEDs combining data from local catalogs and external services (e.g., NED, SDSS, USNO) with the user's own data.
The tool converts observed fluxes or magnitudes into de-reddened fluxes at a given frequency using standard recipes that take into account the instrument response and assumed average spectral slopes. The SED builder can display SEDs both in flux and in luminosity (if redshift information is given); it also provides useful tasks such as the overlay of templates for blazar host galaxies and nuclear optical emission (blue-bump), and allows users to compare the SED with models including one or more SSC components.

The SEDs of all the sources in our samples are shown in Figs.~\ref{fig:firstSED}--\ref{fig:lastSED}. In these figures, red  points  represent strictly simultaneous multi-frequency data, green points represent \gr\ data integrated over a period of two months centered on the times of the \swift/ \planck\ observations, ground-based data taken quasi-simultaneously, and \planck-ERCSC flux densities, and blue points represent \gr\ data integrated over the full period of 27 months. In the few cases where no \swift~simultaneous observations could be obtained, we plot only \planck, \fermi-LAT, and ground-based data.  Two-$\sigma$ upper limits are indicated by arrows.

\subsection{Distinguishing the non-thermal/jet-related radiation from QSO accretion and host galaxy emission}
\label{ThermalContamination}

\begin{figure}
 \centering
 \includegraphics[width=8.0cm,angle=-90]{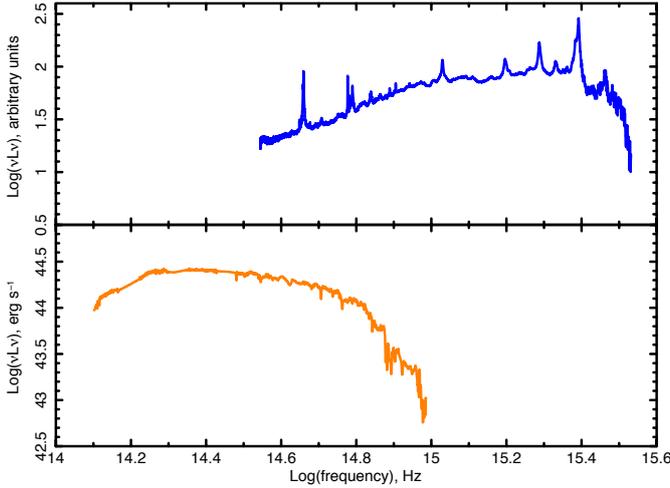}
 \caption{Top panel: The SDSS template of \citealt{vandenberk01} for the broad-line and thermal emission from a QSO.
 Bottom panel :  The giant elliptical galaxy template of \citealt{man01}. See text for  details.}
 \label{fig.GalQSOtemplates}
\end{figure}

We used the simultaneous SEDs of  Figs.\ref{fig:firstSED}--\ref{fig:lastSED} to determine some parameters that can constrain the physical mechanisms powering blazars. However, before doing so we had to identify and separate the radiation that is unrelated to the non-thermal, relativistically amplified emission from the jet of the blazars;  that is, radiation from accretion onto the central black hole and from the host galaxy (see, e.g., \citealt{perlman08}).
To do so, we estimated the contamination by the host galaxy assuming that all blazars are hosted by giant elliptical  galaxies
\citep[e.g.,][]{kotilainen98,nilsson03,leontavares10} with absolute magnitude of $M_{\rm R} = -23.7$. As  \cite{scarpa00} and \cite{urryscarpa00} demonstrated, this value is within
one magnitude of the observed values in a sample of over 100 BL Lacs observed with the \textit{Hubble} Space Telescope.    
For the spectral shape, we used the elliptical galaxy template of \citealt{man01} (bottom panel of Fig.~\ref{fig.GalQSOtemplates}), which is based on low spectral resolution observations of a number of nearby galaxies in the wavelength range $0.12$ --$ 2.4$\,\micron~and is a good match to  the predictions of spectrophotometric models for giant ellipticals \citealt{man01}.

The radiation produced by accretion was estimated from the composite optical spectrum built by \cite{vandenberk01}  using over 2200 optical spectra of radio-quiet QSOs taken from the SDSS database \citep{york00} (top panel of Fig.\ref{fig.GalQSOtemplates}), and the expected soft X-ray emission of radio quiet AGN from \citealt{grupe10}.

The ratio of optical to soft X-ray light has been known to be a function of optical luminosity since the early obervations of the {\it Einstein}
observatory \citep{avnitananbaum}. More recently, this dependence has been confirmed using simultaneously acquired optical and soft X-ray data
from \swift\ \citep{grupe10} and {\it XMM-Newton} \citep{vagnetti10}. To assess the presence of a possible thermal component in the X-ray emission, 
we used the relationship given by \citet{grupe10}
\begin{equation}
\alpha_{\rm UV-X(Radio-quietQSO)} = 0.114 \log(L_{2500\,\AA})-1.177,
\label{eq:aoxbb}
\end{equation}
where $L_{2500\,\AA}$ is the rest-frame luminosity of the thermal emission at 2500\,\AA\ in units of W\,Hz$^{-1}$ and \aoxbb\ is
the usual slope between the UV (2500\,\AA) and the soft X-ray (2\,keV) flux in radio quiet QSOs \citep[e.g.,][]{vagnetti10}.

\begin{figure*}[t]
 \centering
 \includegraphics[width=6.4cm,angle=-90]{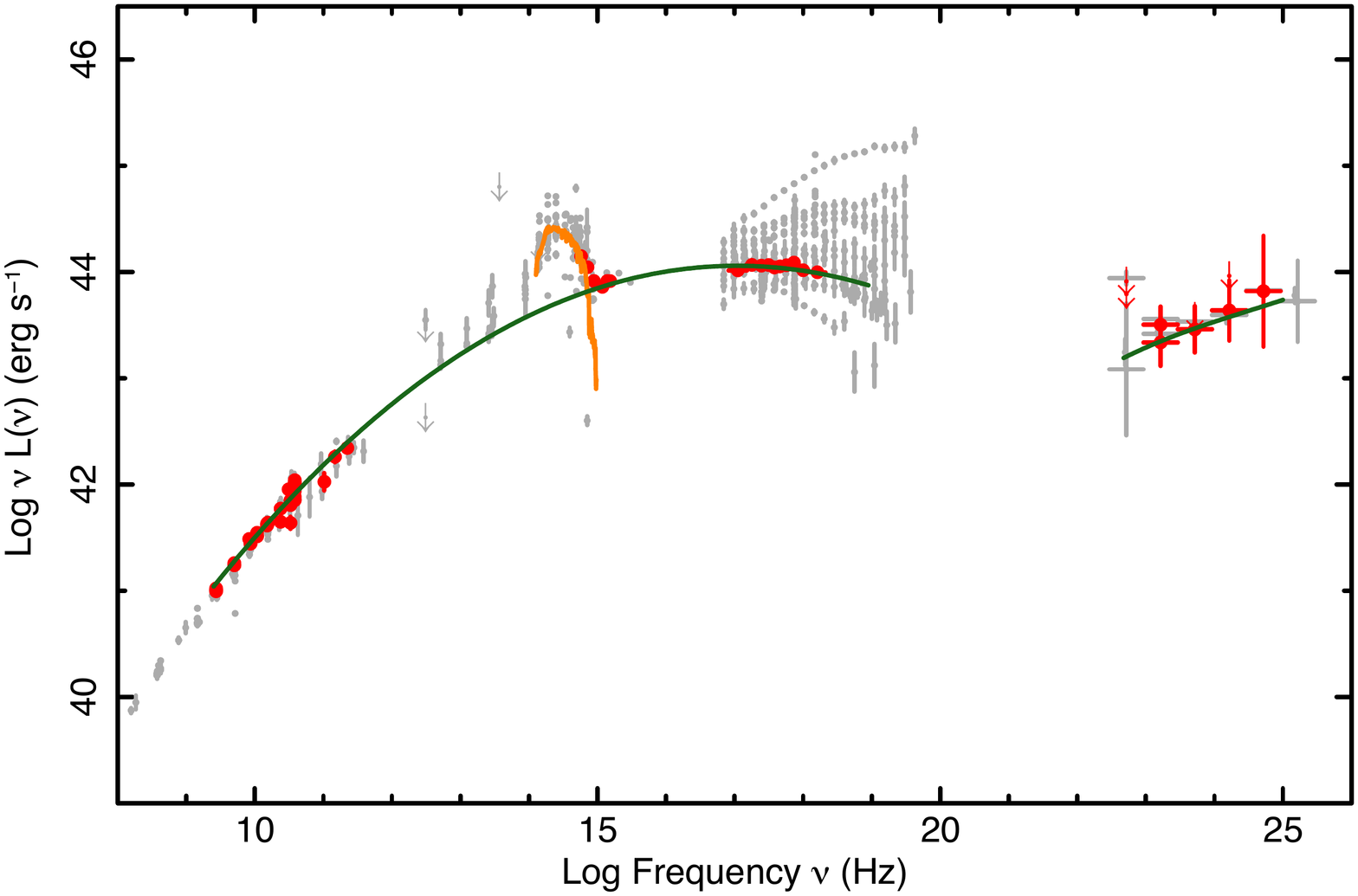}
 \includegraphics[width=6.4cm,angle=-90]{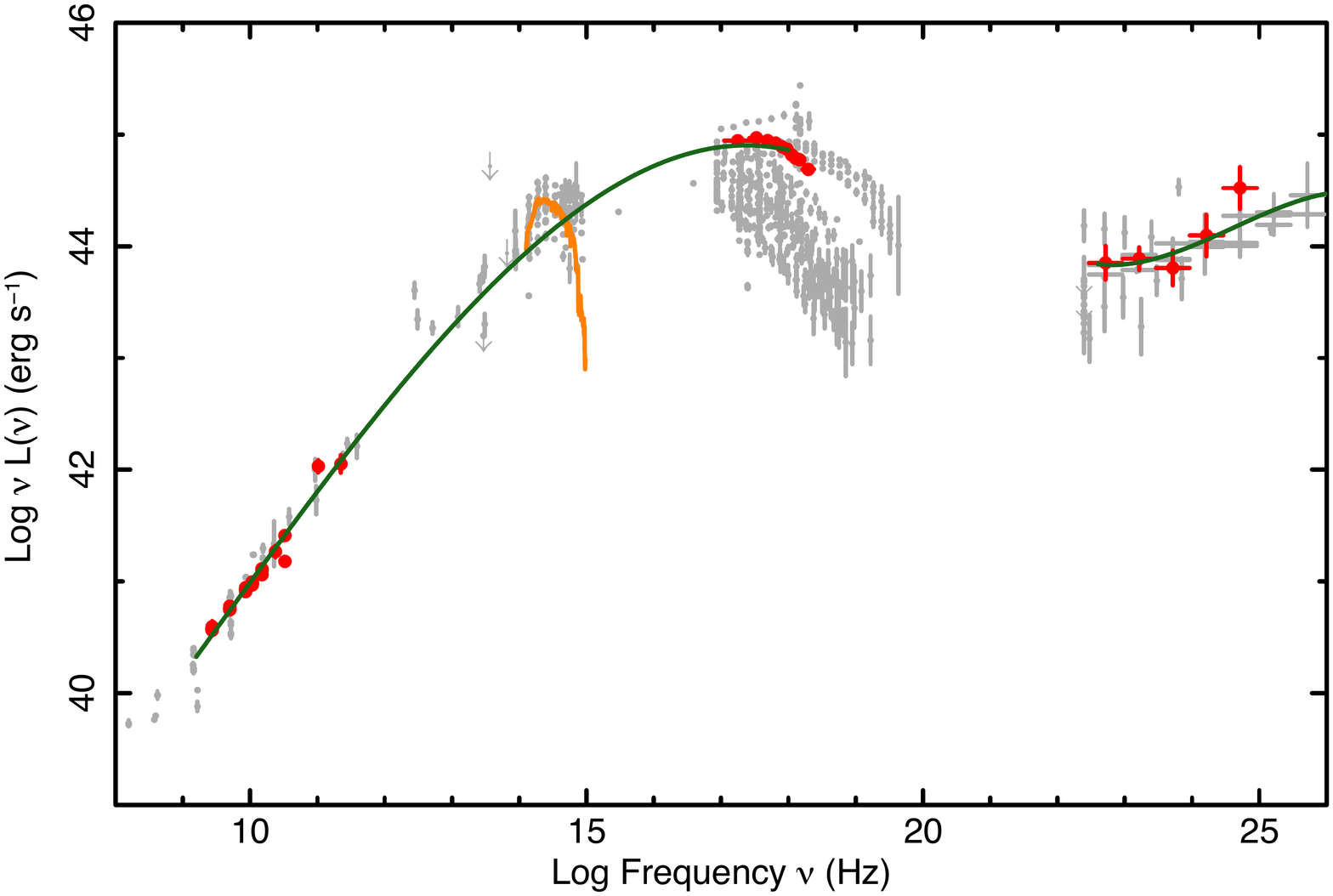}
 \caption{The SEDs of Mkn\,501 ({\it left}) and  Mkn\,421 ({\it right}) showing the expected emission from the host galaxy (giant elliptical) as an orange line. The green lines are the best-fit to the simultaneous non-thermal data using a third degree polynomial function. See text for details.}
 \label{fig.mkn501mkn421}
\end{figure*}

Examples of the emission from these components unrelated to the jet  are shown in Fig.~\ref{fig.mkn501mkn421}, which 
shows the SEDs of  Mkn\,501 and Mkn\,421 where the optical light is dominated by the host galaxy.
\begin{figure*}[ht]
 \centering
 \includegraphics[width=6.4cm,angle=-90]{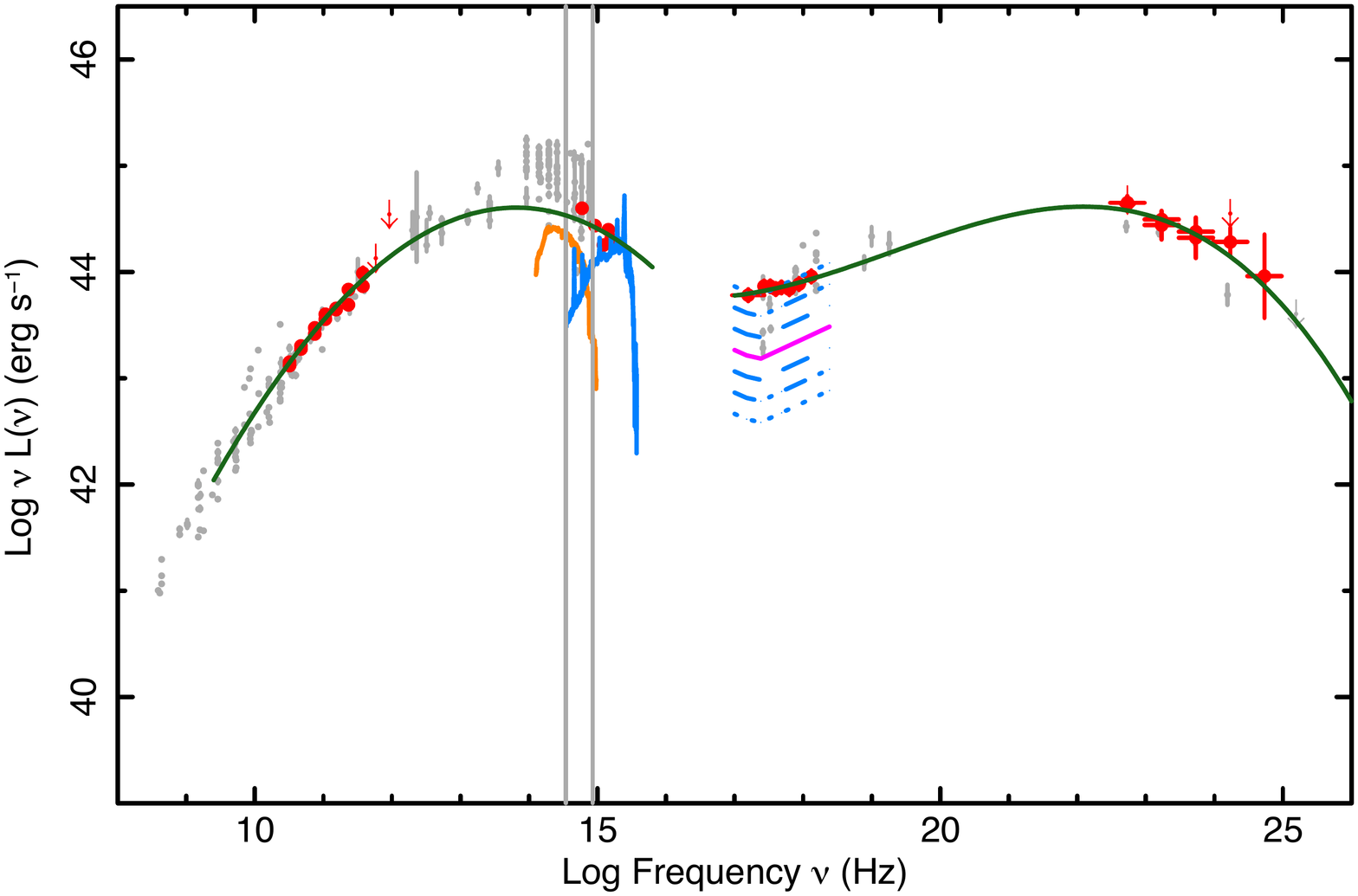}
 \includegraphics[width=6.4cm,angle=-90]{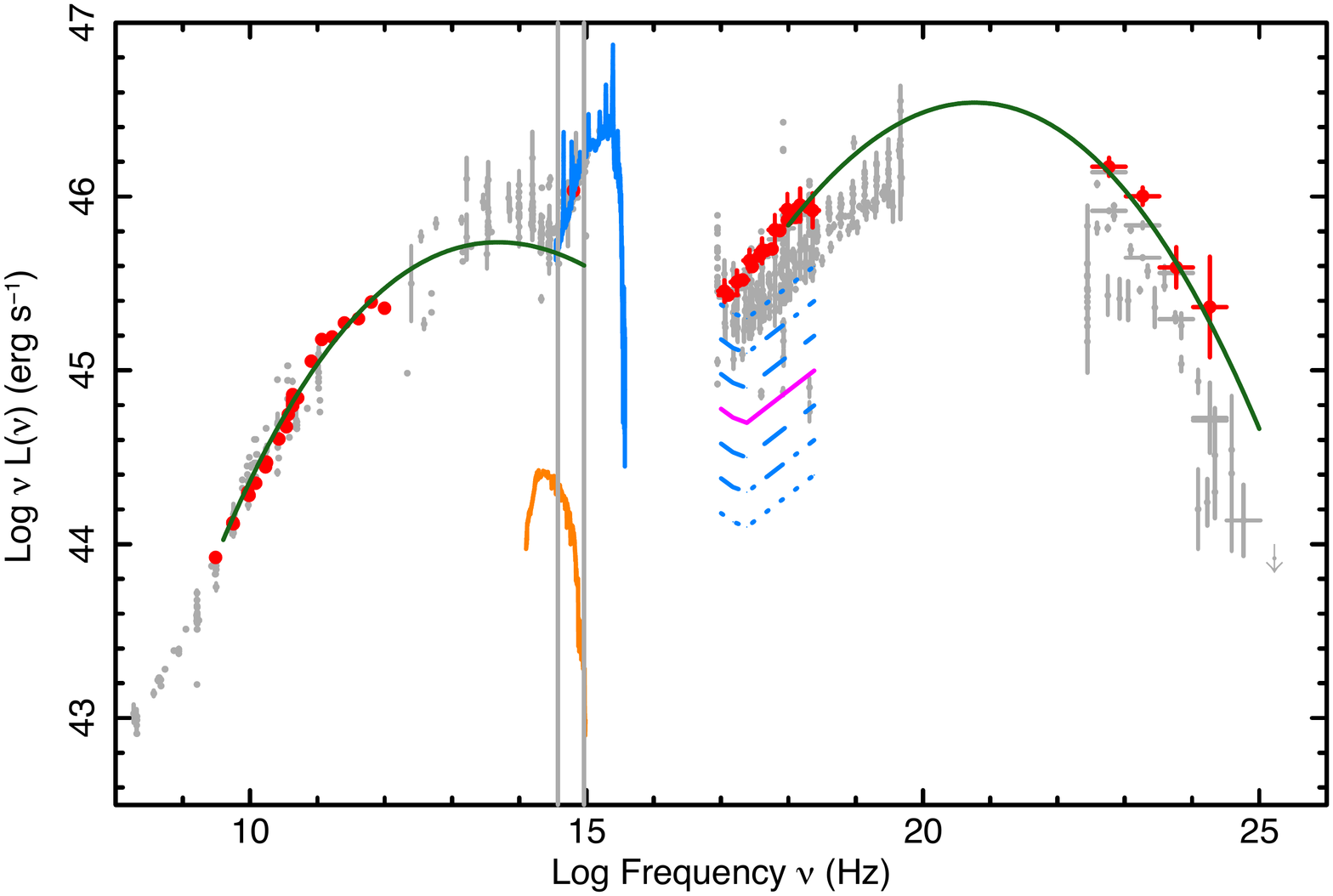}
 \caption{{\it Left}: the SED of BL Lacertae  showing the expected emission from the host galaxy (just below the observed non-thermal radiation, giant elliptical, orange line) and the blue bump emission \citep[blue line, see also][]{raiteri09}. {\it Right}: the SED of 3C\,273 showing the thermal emission from the blue bump and the expected X-ray emission from accretion including 1, 2, and 3$\sigma$ bands (purple and blue lines) derived from Eq.~\ref{eq:aoxbb}. The vertical parallel lines represent the optical  window (4\,000--10\,000\,\AA). The green lines are the best-fit to the simultaneous non-thermal data using a third degree polynomial function.}
 \label{fig.lowzWithTemplate}
\end{figure*}

\begin{figure*}[ht]

\centering
\includegraphics[width=6.4cm,angle=-90]{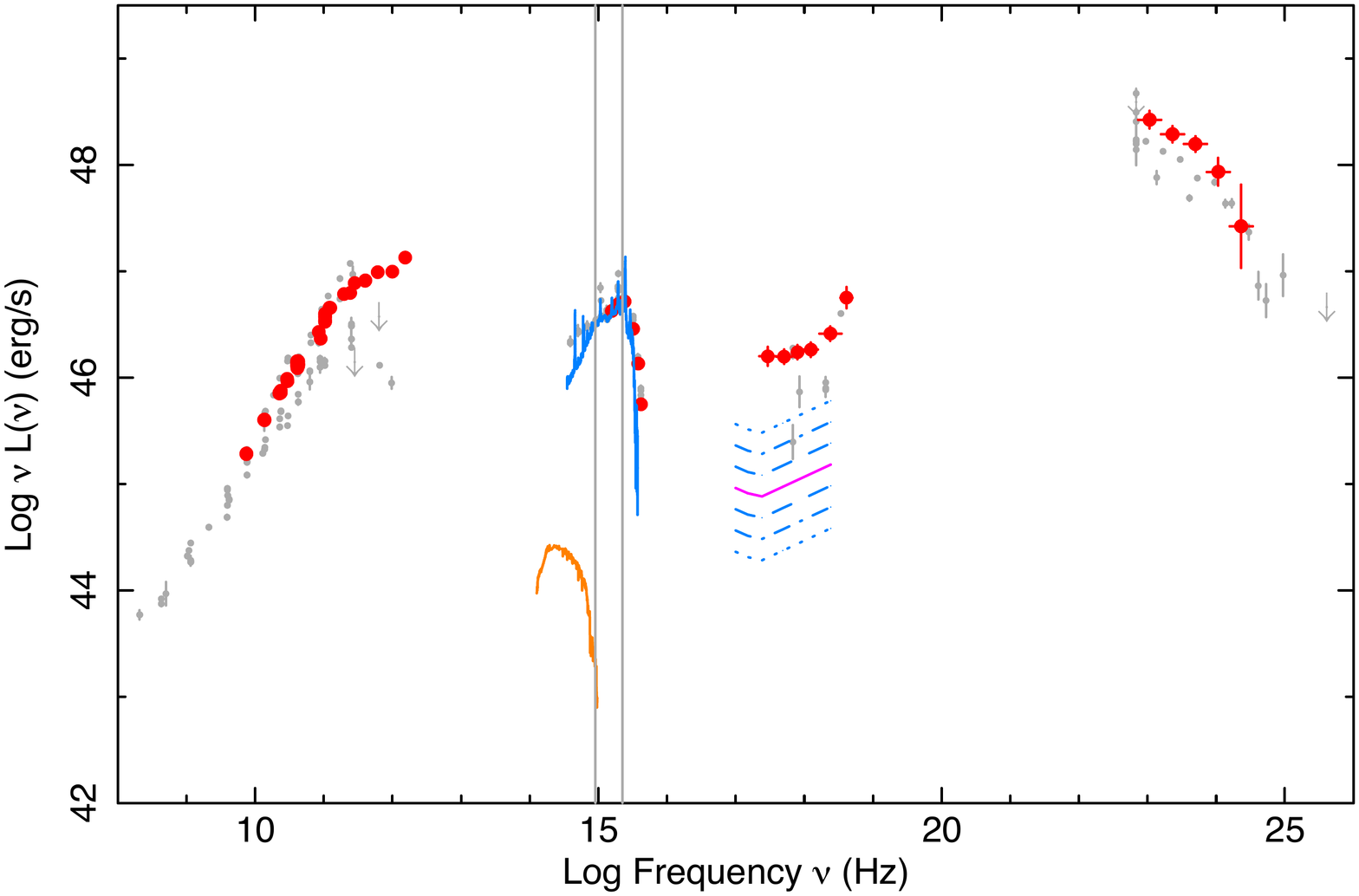}
\includegraphics[width=6.4cm,angle=-90]{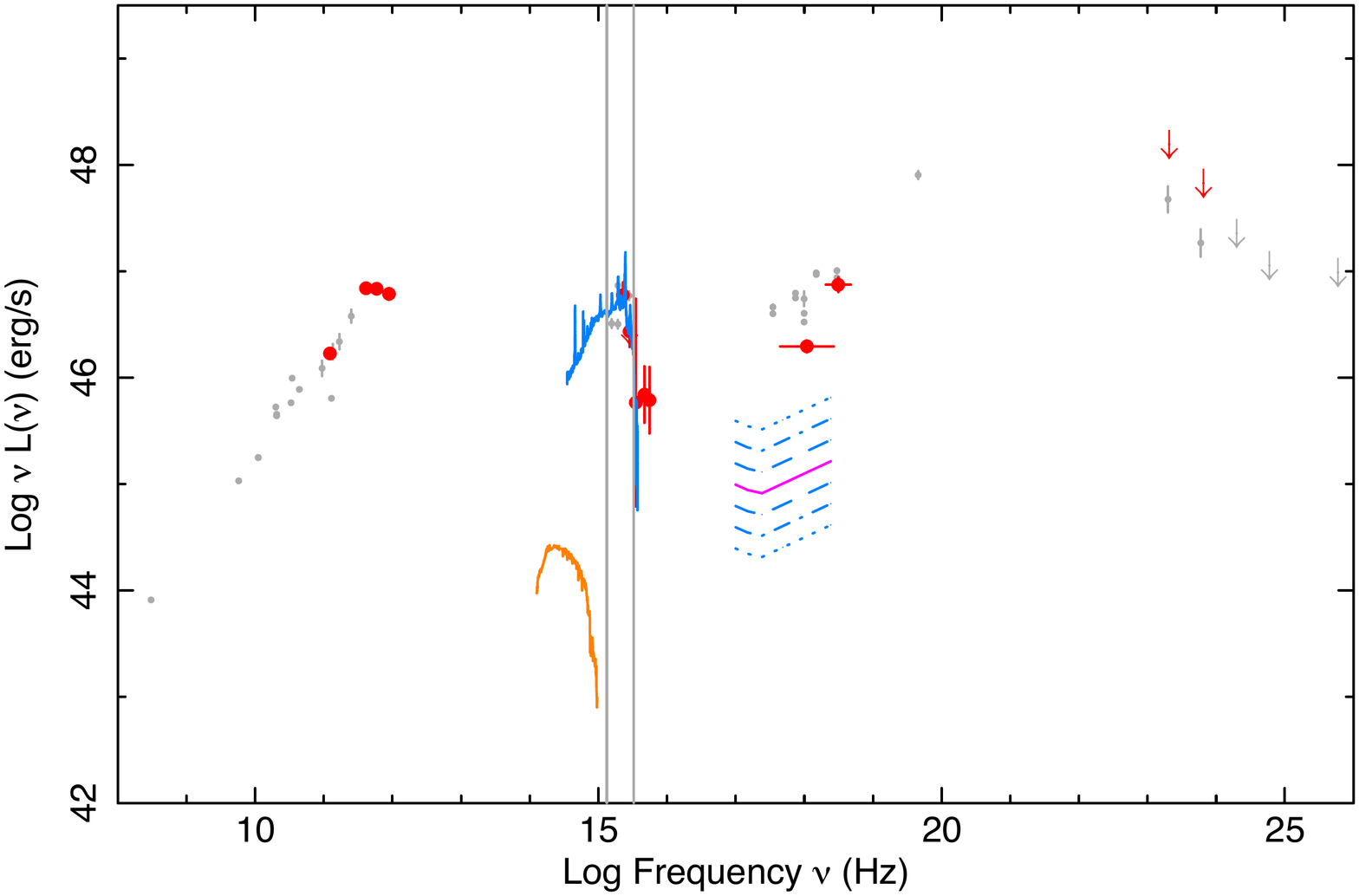}
\caption{The SEDs of 4C\,38.41 ($z = 1.814$; {\it left}) and 1Jy\,0537$-$286 ($z = 3.104$; {\it right}).
Simultaneous data are shown in red; non-simultaneous literature or archival data are shown in light gray. 
Note that the UVOT data of these medium and high redshift objects matches quite well the QSO template of \cite{vandenberk01} that we use to 
estimate the thermal emission from the blue bump ({\it blue line}). The emission from the host galaxy ({\it orange line}) is very low compared to other components. 
The observed X-ray emission from these sources is more than $3\sigma$ above the expected emission 
from accretion derived from Eq.~\ref{eq:aoxbb}. In both cases, the optical light is dominated by radiation from the accretion while the X-rays 
originate from the non-thermal component.}
 \label{fig.highzWithTemplate}
\end{figure*}

Figs. ~\ref{fig.lowzWithTemplate} and \ref{fig.highzWithTemplate} show the SEDs of BL Lacertae where part of the UV light is thought  to originate in 
disk emission \citep{raiteri09}, the nearby FSRQ 3C\,273 ($z=0.158$), and the high-redshift FSRQs  4C\,38.41 ($z = 1.814$) and 1Jy\,0537$-$286 ($z = 3.104$) 
where the optical/UV light is heavily, or completely, contaminated by radiation coming from accretion onto the central black hole.

We compared the amplitude of the optical thermal emission to non-thermal radiation using the parameter \arbb, defined as the spectral 
slope between the 5\,GHz radio flux density and the 5000\,\AA~optical flux density that can be attributed to the blue-bump/disk/thermal emission. 
This quantity depends on both the relativistic amplification factor and the intrinsic ratio of non-thermal/jet radiation to disk emission. 
In the case of FSRQs, the optical spectrum displays emission lines by definition, therefore we were able to constrain \arbb\ by adjusting the optical thermal emission 
to the same level as the data. When the thermal blue bump was seen directly in the optical/UV part of the SED (e.g., Figs.~\ref{fig.lowzWithTemplate} and \ref{fig.highzWithTemplate}), we fit the template of \cite{vandenberk01} to the observed data;
when the disk emission was not obviously visible we set the intensity of the blue-bump template to just below the observed optical/UV emission
where the broad emission lines had been detected.

The BL Lacs do not show emission lines in their optical spectrum, so for this class of objects we could only set
upper limits on \arbb. We did that by assuming that the template for the optical thermal emission is at least one order of
magnitude below the observed data, that is sufficiently low to hamper any broad line detection. 
The estimation of \arbb~relies strongly on the quality of the optical data available, in particular on the simultaneous UVOT data.
Therefore we define an ``optical data quality'' flag as follows:

\begin{itemize}
 \item 0: no simultaneous data available;
 \item 1: poor quality  (e.g., only one UVOT filter available);
 \item 2: good quality (e.g., two or three UVOT filters available);
 \item 3: excellent quality  (all UVOT filters available).
\end{itemize}

\begin{table*}[htbp]
  \caption{Contamination of the X-ray emission from a thermal component and $\langle \alpha_{\rm R-O(Thermal)} \rangle$ estimated for FSRQs in the whole sample and for each sample
independently, for various thresholds on the quality of the optical data.  We have considered as contaminated sources all those with X-ray thermal contamination flag $\le 2$ (see the text for a description of the flags).}
  \label{tab:xthem}
  \begin{center}
  \begin{tabular}{c c c c c c c c c c c c c}
    \hline
    \hline
  &     \multicolumn{3}{c}{ALL}     & \multicolumn{3}{c}{{\it Fermi}-LAT} &  \multicolumn{3}{c}{{\it Swift}-BAT} &
   \multicolumn{3}{c}{{\it ROSAT}/RASS}       \\
Opt Data   & cont/& \% & $\langle \alpha_{\rm R-O(Thermal)} \rangle$ & cont/& \% & $\langle \alpha_{\rm R-O(Thermal)} \rangle$ & 
        cont/ & \% & $\langle \alpha_{\rm R-O(Thermal)} \rangle$ &cont/ & \% & $\langle \alpha_{\rm R-O(Thermal)} \rangle$ \\
     quality   & total & & &total & & & total & & & total & & \\
\hline
ALL    & 10/47 &    21 &  $-0.69\pm$0.02 &     3/25  &   12 &   $-0.72\pm$0.02 &  2/21 &    10  &  $-0.70\pm$0.03 &   6/13 &   46 & $-0.65\pm$0.04 \\
$\ge$2 & 10/44 &    23 &  $-0.69\pm$0.02 &   3/22  &   14 &  $-0.73\pm$0.02   & 2/20 &    10  &  $-0.70\pm$0.03 &   6/13 &   46 & $-0.65\pm$0.04 \\
3      & 9/40 &    23 &   $-0.69\pm$0.02 &  3/22  &   14 &   $-0.73\pm$0.02 &  1/16 & 6 & $-0.69\pm$0.04  &    6/12 &   50 & $-0.64\pm$0.05 \\
\hline
\end{tabular}
\end{center}
\end{table*}

\begin{figure}
 \centering
 \includegraphics[width=7.8cm,angle=-90]{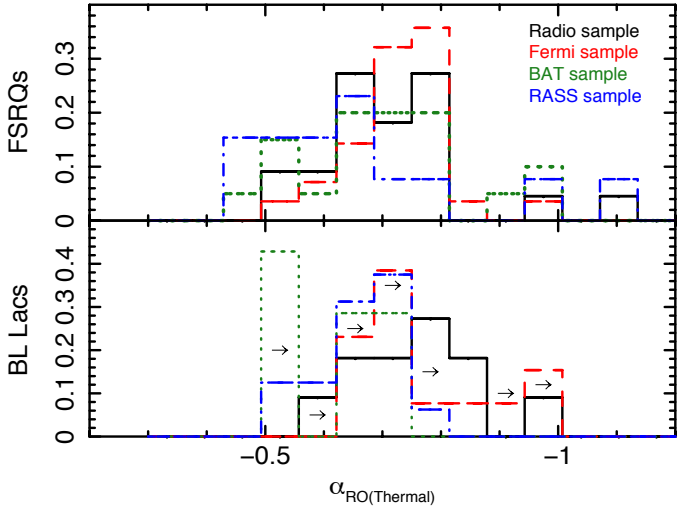}
 \caption{Distribution of \arbb\ for all the sources and for each blazar sample. Only SEDs with good or excellent optical data (flag = 2 or 3 , see text) were used. {Black solid histograms}: radio sample; {red dashed histograms}: \fermi~sample; {green dot-dashed histograms}: \swift~BAT sample; {blue dotted histograms}: RASS sample. Distributions for BL Lac objects refer to upper limits only.}
 \label{fig:arobb}
\end{figure}

In Figure~\ref{fig:arobb} (upper panel), we show the distribution of \arbb\ for all the FSRQs with excellent optical data for the whole sample and each of our samples independently. The distribution of the \arbb\ upper limits for BL Lac objects is shown in the bottom panel. The results suggest 
a possible difference in the \arbb\ distribution between different samples. As reported in Table~\ref{tab:xthem}, when only sources with excellent optical data are considered we obtain $\langle \alpha_{\rm R-O(Thermal)} \rangle = -0.64\pm0.05$ for the {\it ROSAT}/RASS sample, to be compared with  $\langle \alpha_{\rm R-O(Thermal)} \rangle = -0.73\pm0.02$ for the \fermi-LAT sample.

To assess the possible presence of a thermal component in the X-ray emission, for each source we compare the predicted thermal emission
from accretion with the actual X-ray spectrum, and we also include the uncertainties in the parameters of Eq.~\ref{eq:aoxbb}
in determining the uncertainties given by 1, 2, and $3\sigma$ bands. We therefore define the following X-ray thermal contamination flags: 
\begin{itemize}
 \item 0: no X-ray data available;
 \item 1: X-ray emission mostly or entirely due to accretion/reflection (data agree with the expectations for accretion emission within $1\sigma$);
 \item 2: X-ray emission probably contaminated by the accretion component (data agree with with the expectations for accretion emission within $2\sigma$);
 \item 3: X-ray emission mostly of non-thermal origin (data are at $2$--$3\sigma$ from the expectations for radio quiet QSOs);
 \item 4: X-ray emission certainly of non-thermal origin (data are more than $3\sigma$ away from the expectations for radio quiet QSOs).
\end{itemize}

\begin{figure}
 \centering
 \includegraphics[width=7.8cm,angle=-90]{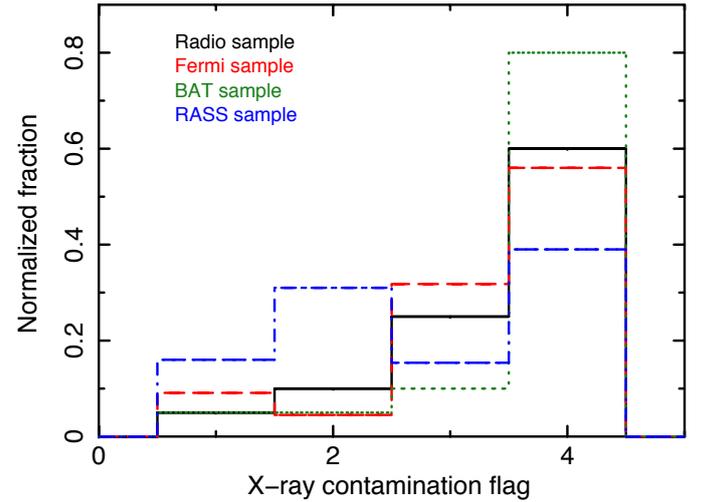}
 \caption{Distribution of the thermal emission flag for all the FSRQs with good or excellent optical data flag (see text for details). Black solid histograms: radio sample; red dashed histograms: \fermi-LAT sample; green dot-dashed histograms: \swift~BAT sample; and blue dotted histograms: RASS sample. As  the X-ray emission of all BL Lacs  was 
not contaminates, the BL Lac data are not plotted here.}
 \label{fig:xcontamination}
\end{figure}

Results for all the FSRQs with good or excellent optical data are shown in Figure~\ref{fig:xcontamination}, and summarized in 
Table~\ref{tab:xthem} for the whole sample and each sample independently. We considered as contaminated all the sources with X-ray thermal contamination flag $\le$\,2.
There is a large difference between the \fermi-LAT and \swift-BAT samples, where $\lsim 15\%$ of the sources
have a thermal component in their X-ray emission, and the {\it ROSAT}/RASS sample, where $\sim 50\%$ of the sources 
are contaminated. This demonstrates 
that the thermal component in the X-ray emission of blazars cannot be neglected, even in bright sources.

\subsection{SED parameter estimation}

We used  the SEDs of all the objects in our samples to estimate the values of important physical parameters such as  \nupS,  \nupS $F$(\nupS), \nupIC, and \nupIC $F$(\nupIC) (see Table~\ref{tab:peaks})  taking into account only the non-thermal radiation and fitting third-degree polynomials as described in \cite{abdoseds} (see Figs.~\ref{fig.mkn501mkn421}, \ref{fig.lowzWithTemplate}, 
and \ref{fig.highzWithTemplate} for examples). 

\begin{table*}[htbp]
  \caption{Rest-frame $\langle $\nupS $\rangle$ and  $\langle $\nupIC $\rangle$ values for different samples and classes.}
  \label{tab:peaks}
  \begin{center}
  \begin{tabular}{l l c c c c}
    \hline
    \hline
 &  &     \multicolumn{2}{c}{Synchrotron peak}     & \multicolumn{2}{c}{inverse Compton peak}  \\
Sample    &  Class   & No. of sources & $\langle \log(\nu_{Peak})\rangle $ & No. of sources & $\langle\log(\nu_{Peak}) \rangle $ \\
\hline
\multirow{3}{*}{\fermi-LAT}           & ALL     &    45 &     13.34$\pm$0.12 &       42  &  22.19$\pm$0.10     \\
                             & FSRQs     &    24 &     13.07$\pm$0.07   &    23  &  22.21$\pm$0.14    \\
                             & BL Lacs     &    13 &     14.11$\pm$0.32   &    11  &  22.35$\pm$0.24    \\
\hline
\multirow{3}{*}{\swift-BAT}           & ALL     &   26  &    13.65$\pm$0.26   &   22   & 21.66$\pm$0.28     \\
                             & FSRQs     &    18 &     13.16$\pm$0.09  &    16  &  21.45$\pm$0.17    \\
                             & BL Lacs     &    5 &     15.75$\pm$0.81   &    3  &  23.41$\pm$1.52     \\
\hline
\multirow{3}{*}{\textit{ROSAT}/RASS} & ALL     &   25  &    14.27$\pm$0.31    &   20   &  22.15$\pm$0.38   \\      
                                & FSRQs     &    11 &     13.02$\pm$0.17   &    11  &  21.43$\pm$0.30     \\
                             & BL Lacs     &    11 &     15.79$\pm$0.28   &    6  &  24.24$\pm$0.37     \\  
\hline
\multirow{3}{*}{Radio}                & ALL     &   94  &    13.20$\pm$0.06    &   77   &  21.99$\pm$0.10   \\      
                                & FSRQs     &    64 &     13.08$\pm$0.05   &    49  &  21.99$\pm$0.12     \\
                             & BL Lacs     &    14 &     13.89$\pm$0.27   &    13  &  22.26$\pm$0.20     \\     
\hline
\end{tabular}
\end{center}
\end{table*}

\begin{figure}[htbp]
 \centering
 \includegraphics[width=6cm,angle=-90]{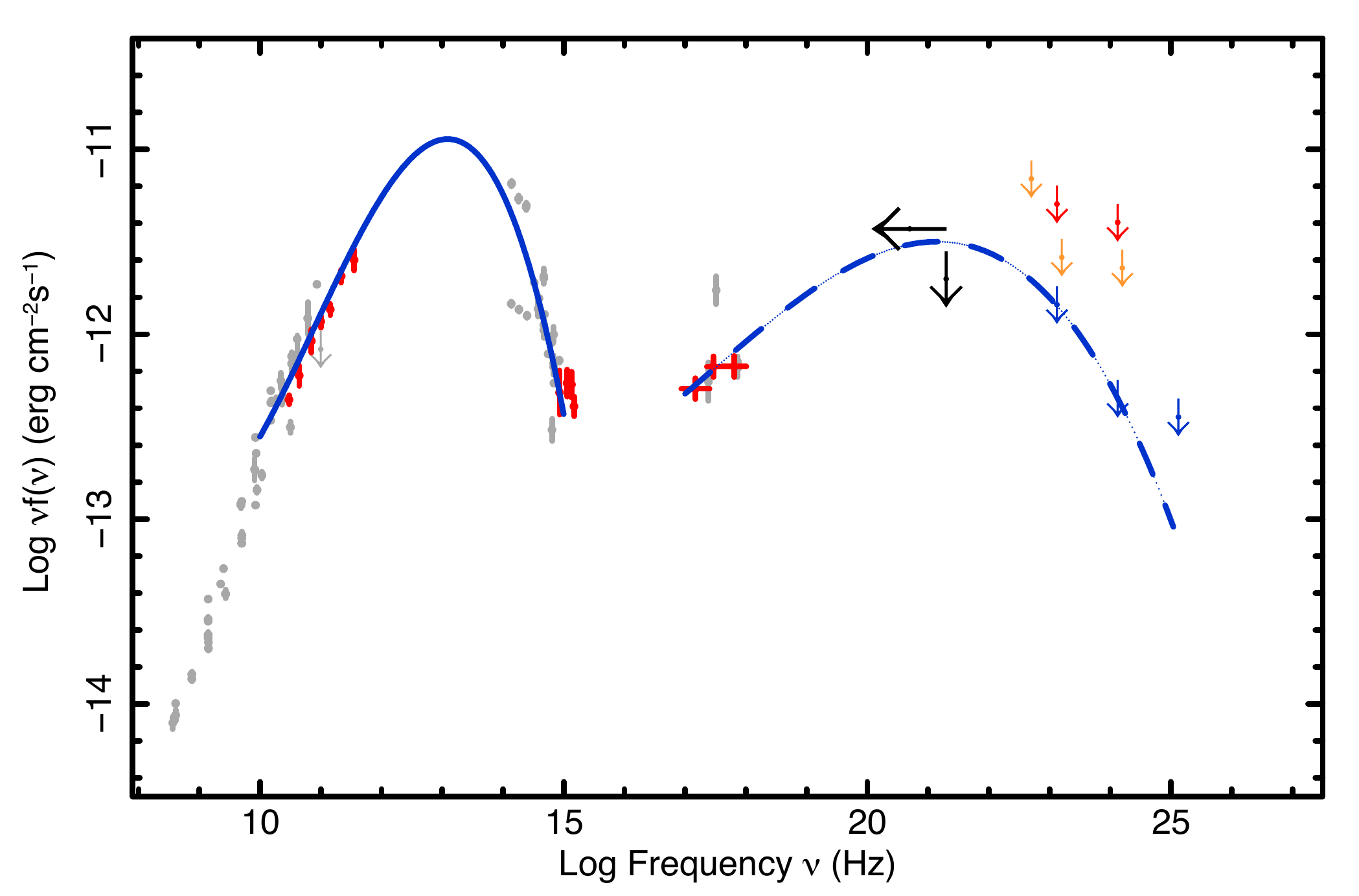}
 \caption{The SED of  the blazar PKS\,0003$-$066, illustrating the estimation of the upper limits to \nupIC and to \nupIC $F$(\nupIC) by combining the X-ray data with the 27-month \fermi-LAT 
 upper limits.} 
 \label{fig.ICUpperLimit}
\end{figure}

For the sources that were not detected by \fermi-LAT even in the 27-month integration, we estimated limits of \nupIC and \nupIC $F$(\nupIC) by constraining the polynomial in the 
high-energy part of the SED with the 27-month \fermi-LAT upper limits, as shown in Fig.~\ref{fig.ICUpperLimit}. 

In some HSP BL Lacs with particularly high \nupS~values (see, e.g., Figs.~\ref{fig:sed19}, 
\ref{fig:sed22}, \ref{fig:sed37}, and \ref{fig:sed43}) the \fermi-LAT data alone are insufficient to ensure a good measure of  \nupIC, as the spectra are still rising at the highest \fermi-LAT energies 
and no simultaneous TeV data are available. The  \nupIC~values for these sources should therefore be considered as lower limits.

\section{The spectral slope of blazars in the radio--\mw\ region}

While \wmap~results are consistent with a single flat spectral index in the relatively narrow frequency range $23-94$\,GHz \citep{wright2009,gold2010},
the blazar spectrum must steepen at frequencies closer to the synchrotron peak. 
Adding {\it Planck} data and  simultaneous ground-based observations at centimetre wavelengths to the \wmap\ data improves the spectral coverage, and allows us to probe the spectral shape of blazars over the much wider frequency range $\sim1$\,GHz to $\sim1$\,THz.

\begin{figure*}
\centering
\includegraphics[width=6.8cm,angle=-90]{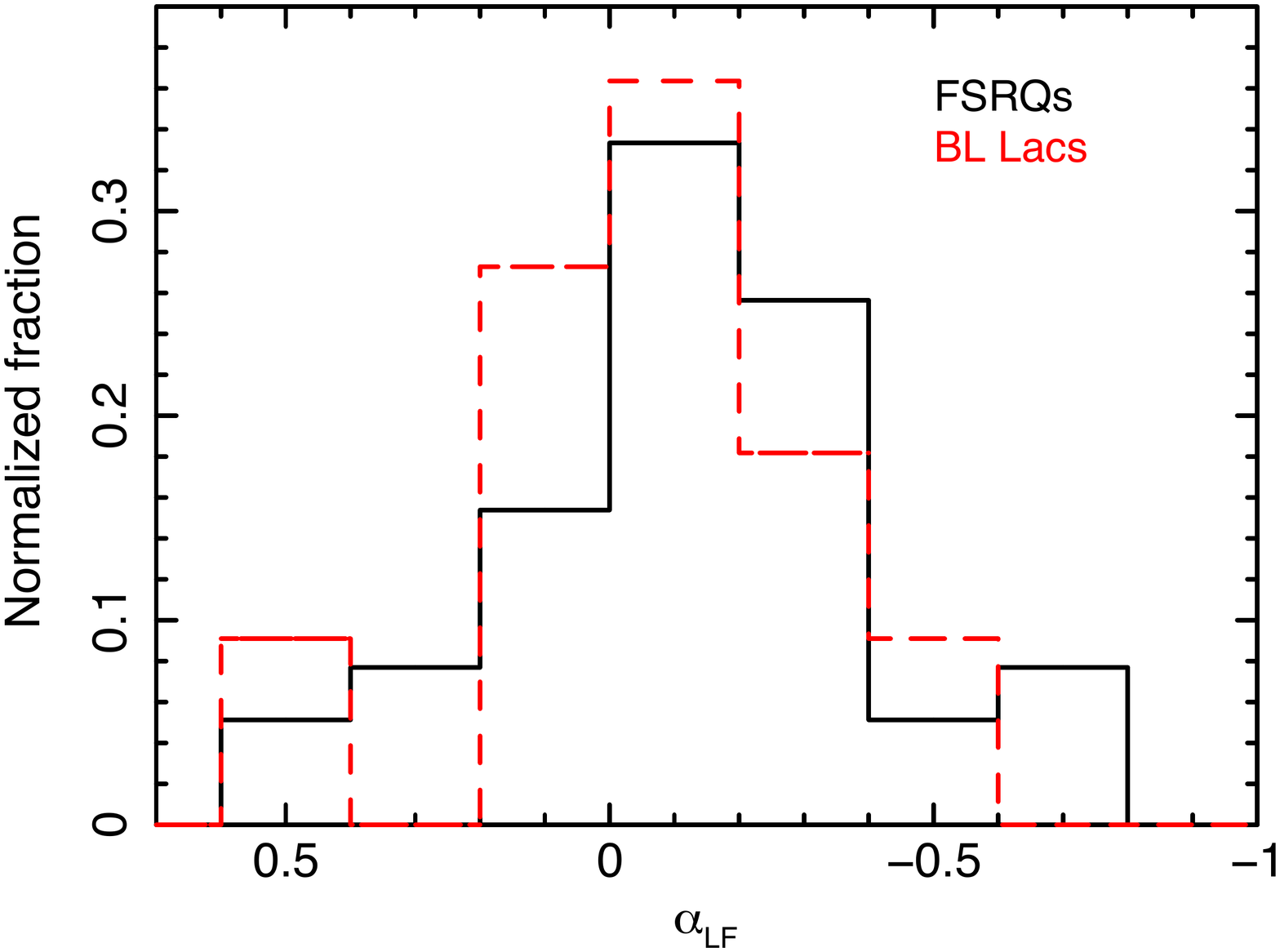}
\includegraphics[width=6.8cm,angle=-90]{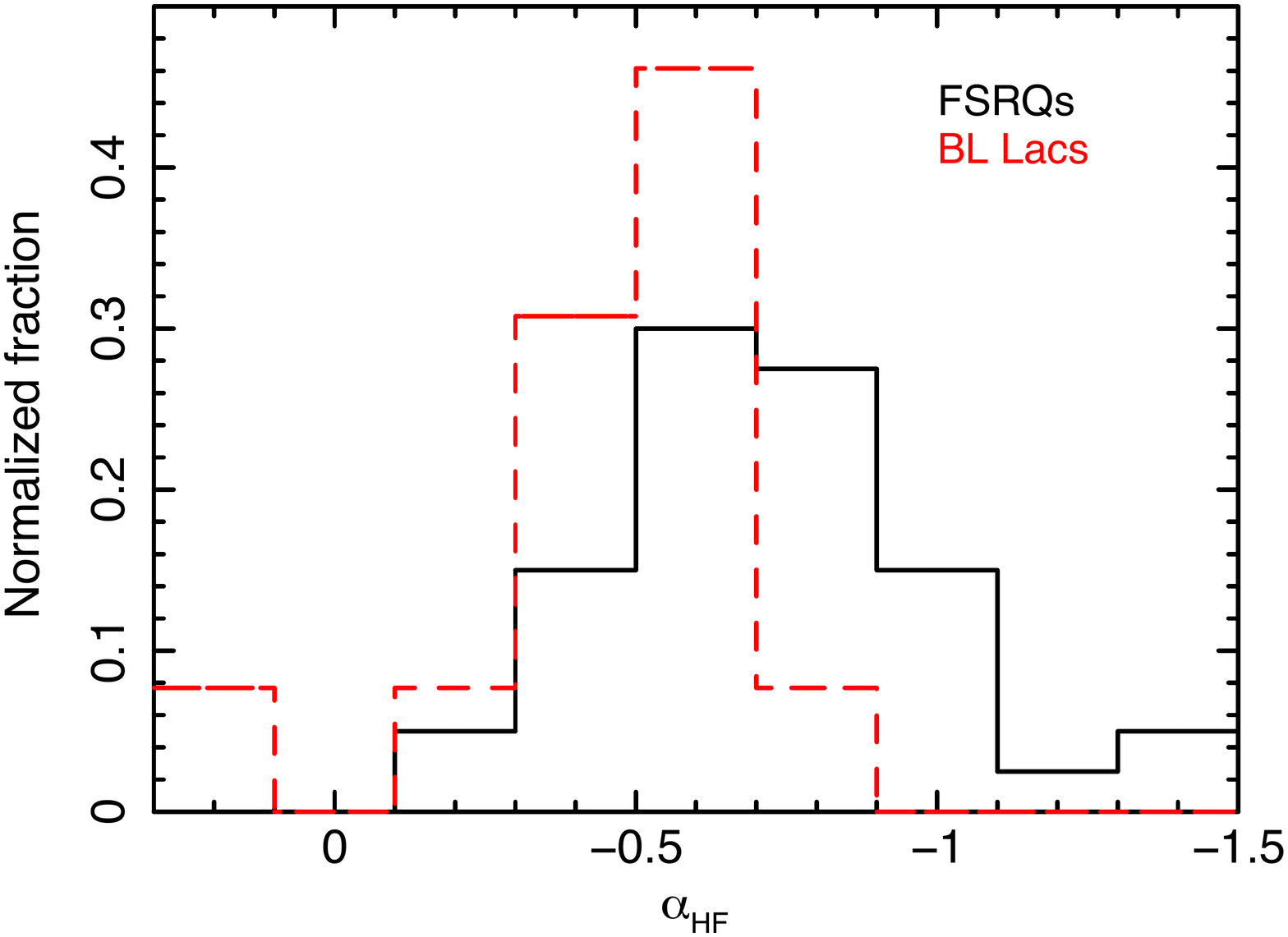}
\caption{Radio/\mw~low frequency ($\nu < 70$ GHz, LF; left side) and high frequency ($\nu > 70$ GHz, HF; right side) spectral index distributions of FSRQs and BL Lacs in our samples for the case $\nu_{Break}=70$\,GHz. While the distributions of low frequency spectral indices are very similar for both types of blazars with $\langle \alpha_{\rm LF} \rangle \sim -0.1$, the distributions of the high frequency slopes indicate that  the spectra of FSRQs ($\langle \alpha_{\rm HF} \rangle=-0.73 \pm0.04$) might be somewhat steeper than 
those of BL Lacs ($\langle \alpha_{\rm HF} \rangle=-0.51\pm 0.07$).  A Kolmogorov Smirnov (KS) test gives a probability of less than 3\% that the HF distributions for FSRQs and BL Lacs come from the same parent distribution.}
\label{fig:radioslopes}
\end{figure*}

We studied the low frequency (LF) and the high frequency (HF) regions of the centimetre to sub-millimetre blazar spectra separately
to search for differences in the spectral index $\alpha$ and determine the frequency at which the spectral index changes.
We fitted the two frequency regions independently with power-laws to estimate the spectral indices at both low frequencies ($\alpha_{\rm LF}$, for $\nu \le \nu_{\rm Break}$)
and high frequencies ($\alpha_{\rm HF}$, for $\nu > \nu_{\rm Break}$), assuming a range of break frequency values, $\nu_{\rm Break}$, from 30\,GHz to 100\,GHz. In Fig.~\ref{fig:radioslopes}, we show the distributions of $\alpha_{\rm LF}$ and
$\alpha_{\rm HF}$ for $\nu_{\rm Break}= 70$\,GHz  and 100\,GHz; the blazar spectra steepen from $\alpha_{\rm LF}\sim 0$ to $\alpha_{\rm HF}\sim -0.65$.

To verify the robustness of the results we repeated the analysis by imposing the two minimum numbers of independent frequencies needed to perform the fit, 
namely 3 and 5, and the results are consistent (Table~\ref{tab:radioslope}).

\begin{table*}[ht]
 \caption{Radio LF and HF spectral index distributions for different values of the break frequency $\nu_{Break}$ and of the minimum number of frequency bands considered for the fit.}
 \label{tab:radioslope}

 \begin{center}
 \begin{tabular}{ccccccccccc}
   \hline
   \hline
   & & &\multicolumn{3}{c}{Low Frequency} &  \multicolumn{3}{c}{High Frequency} \\
$\nu_{Break}$ & n. of frequencies & Class & n. of sources & $\langle \alpha_{\rm LF} \rangle$ & $\sigma_{\rm LF}$ & n. of sources  & $\langle \alpha_{\rm HF} \rangle$ & $\sigma_{\rm HF}$ \\
\hline
 30 &  3 & ALL       &  47 &    $~~0.00 \pm$     0.04 & 0.27 &  69 &  $-0.51 \pm$  0.04 &  0.30 \\
 30 &  5 & ALL       &  46 &    $~~0.00 \pm$     0.04 & 0.27 &  57 &  $-0.49 \pm$  0.03 &  0.26 \\
 44 &  3 & ALL       &  47 &    $-0.02 \pm$     0.04 & 0.25 &  66 &  $-0.64 \pm$  0.03 &  0.26 \\
 44 &  5 & ALL       &  47 &    $-0.02 \pm$     0.04 & 0.25 &  47 &  $-0.62 \pm$  0.03 &  0.22 \\
 70 &  3 & ALL       &  63 &    $-0.08 \pm$     0.03 & 0.28 &  66 &  $-0.67 \pm$  0.03 &  0.28 \\
 70 &  5 & ALL       &  47 &    $-0.02 \pm$     0.04 & 0.25 &  35 &  $-0.65 \pm$  0.04 &  0.22 \\
100 &  3 & ALL       &  65 &    $-0.11 \pm$     0.03 & 0.27 &  48 &  $-0.56 \pm$  0.04 &  0.26 \\
100 &  5 & ALL       &  47 &    $-0.02 \pm$     0.04 & 0.24 &  15 &  $-0.52 \pm$  0.05 &  0.19 \\
\hline
 70 &  3 & FSRQ      &  39 &   $-0.11 \pm$     0.04 & 0.28 &  40 &  $-0.73 \pm$  0.04 &  0.27 \\
 70 &  3 & BL Lac    &  12 &    $-0.08 \pm$     0.08 & 0.29 &  14 &  $-0.51 \pm$  0.07 &  0.27 \\ 
 70 &  3 & Unc. Type &  12 &  $-0.02 \pm$     0.08 & 0.29 &  12 &  $-0.64 \pm$  0.07 &  0.25 \\
\hline
\end{tabular}
\end{center}
\end{table*}

We also analyzed the different classes of blazars --flat spectrum radio quasars (FSRQ), BL Lac objects (BL Lac), and radio-loud AGN of unknown classification (Uncertain Type) -- separately, and we 
have found a difference at a level of about $3\sigma$ in $\alpha_{\rm HF}$, which is $\sim -0.7$ for FSRQ but $\sim -0.5$ for BL Lacs.

\begin{figure}[htbp]
\centering
\includegraphics[width=7.8cm,angle=-90]{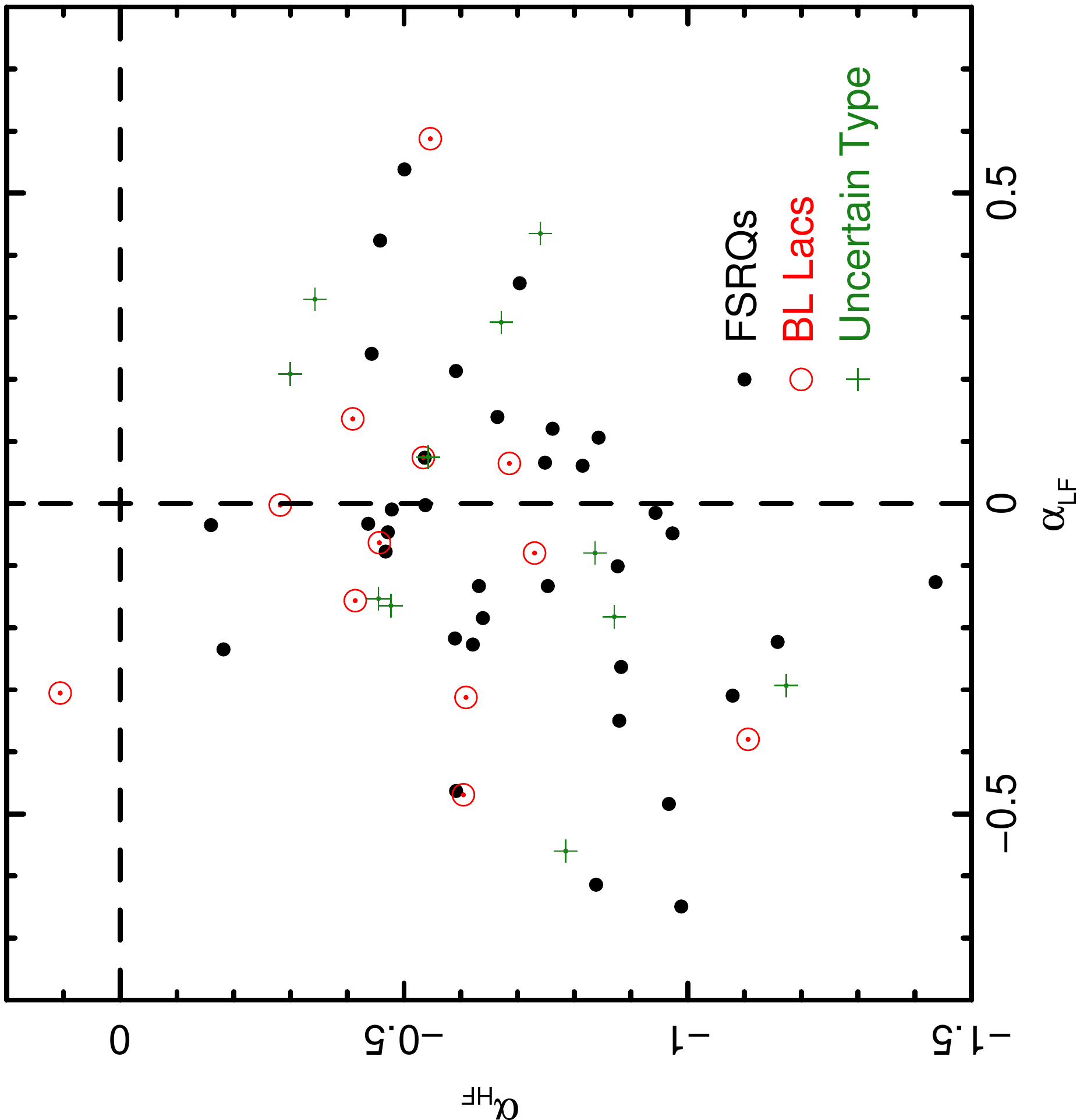}
\caption{$\alpha_{\rm HF}$ vs. $\alpha_{\rm LF}$ diagram for $\nu_{Break}=70$\,GHz for different blazar classes: flat spectrum radio quasars (FSRQ),
BL Lac objects (BL Lac), and blazars of uncertain classification.}
\label{fig:colorcolor}
\end{figure}

The spectral steepening and the difference in  $\alpha_{\rm HF}$  are  visible in Figure~\ref{fig:colorcolor}, where we
show $\alpha_{\rm LF}$ versus (vs.) \ $\alpha_{\rm HF}$ for the sources that meet the requirements for the minimum amount of
independent data at both low and high frequencies. 

Our results are in complete agreement with the findings of  \citealt{planck2011-6.3a} for the radio selected sample, i.e., flat
spectral index at low frequency and $\alpha_{\rm HF}\sim -0.6$ at high frequency with a break frequency $\sim70$\,GHz,
suggesting that this is a general feature of all blazars regardless of the selection criteria.
There is also general agreement with the results presented in \citealt{planck2011-6.1} for all
the sources in the \planck\ ERCSC catalogs, and the slight discrepancy in the spectral index estimated at low frequency
can be explained by the fact that, unlike \citealt{planck2011-6.1}, we have included the ground-based
observations at 5--30\,GHz.

\section{Searching for correlations between fluxes in different energy bands}

We used our  simultaneous and average multi-frequency data to identify possible correlations between fluxes measured in different energy 
bands. As some of the blazars in our samples were not detected  by either \fermi-LAT or  \planck, we estimated the significance of the correlations using the ASURV code Rev1.2 \citep{92ASURV}, which takes account of upper limits as described in \citet{86isobe}. We search for possible correlations using fluxes or flux-densities
rather than in luminosity-luminosity space because this allows us to use all objects in the samples, including those with no redshift (and consequently luminosity) information. 

\subsection{Microwave vs. X-ray}

\begin{figure}
\centering
\includegraphics[width=8.4cm,angle=0]{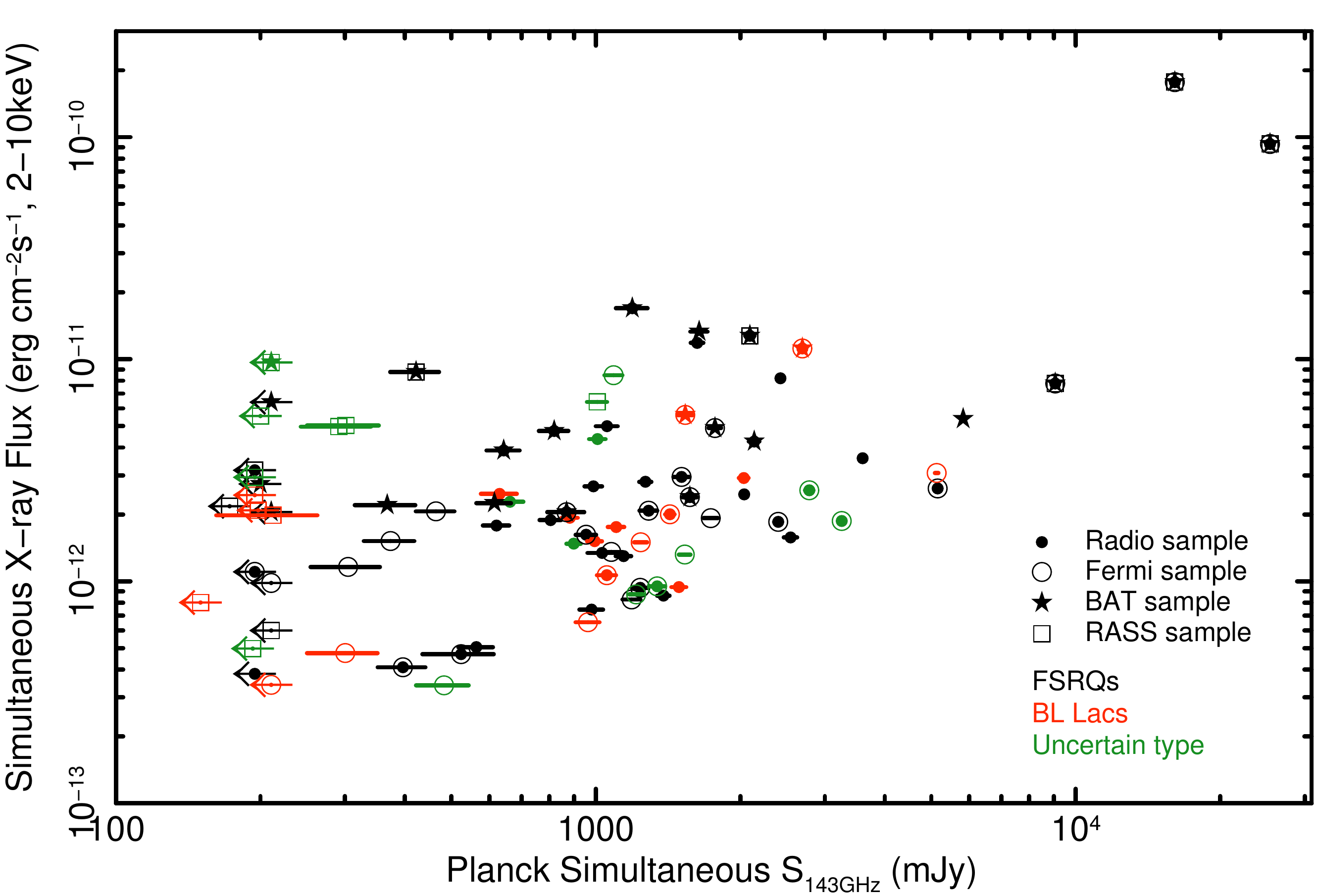}
\caption{The \planck\ 143\,GHz flux density plotted vs. the simultaneous \swift\ XRT X-ray flux for all sources where the X-ray flux is expected to be due to the inverse Compton component.
Blazars with significant X-ray contamination due to accretion have been excluded from the plot. }
\label{fig.fxvs143ghz}
\end{figure}

Figure \ref{fig.fxvs143ghz} shows the \planck\ 143\,GHz flux density versus the simultaneous \swift-XRT X-ray flux for all the sources where the X-rays are expected
to be due to the inverse Compton component (that is all LSP blazars) and not significantly contaminated by X-ray emission that is unrelated to the jet, such as that 
produced by the accretion process (see Sect.~\ref{ThermalContamination} for details).  Sources that were not detected by \planck\  are plotted as  upper limits; all the sources were detected in the X-ray band.

A correlation, although with some scatter, is clearly present. The Spearman rank coefficients ($\rho$) and the corresponding probabilities that the observed correlation
is the result of chance are given in Table~\ref{tab:Xcorr}. Results obtained using \planck\ flux densities at other frequencies from 30\,GHz to 217\,GHz are similar and 
not shown here.

\begin{table*}[htbp]
\caption{Spearman correlation parameters for 143\,GHz flux density vs.\  X-ray flux.}
\label{tab:Xcorr}
 \begin{center}
 \begin{tabular}{ c c c c c c c c c c c c c}
   \hline
   \hline
 Class & \multicolumn{3}{c}{\fermi-LAT sample}  &     \multicolumn{3}{c}{\swift-BAT  sample}  &   \multicolumn{3}{c}{{\textit{ROSAT}}/RASS sample}  &   
\multicolumn{3}{c}{Radio sample} \\
           & No. of  & $\rho$   &  $P(\%)$ & No. of  & $\rho$   &  $P(\%)$ & No. of & $\rho$   &  $P(\%)$ & No. of  & $\rho$   &  $P(\%)$ \\
                   & sources &  &  & sources &   & & sources &  & & sources &  &  \\
                 & (a)/(b) &  &  &  (a)/(b) &   & &  (a)/(b) &  & &  (a)/(b) &  &  \\
\hline
 All       &    35/3 &  0.72 &  $<$0.01  &    18/4  &  0.50  &    2.34 &    9/10  &  0.70  &    0.30  &    49/3  &  0.51  &    0.03 \\ 
 FSRQs     &    20/2 &  0.75 &  0.06  &    16/3  &  0.58  &    1.40 &    5/3  &  0.89  &    1.85      &    33/3  &  0.57  &    0.08 \\ 
 BL Lacs   &    8/1 &  0.95 &  0.72 &    2/0  &  ...$^*$  &    ... &   1/3   & ...   &  ...     &    9/0  &  0.28  &    42.29  \\

\hline
\end{tabular}
\end{center}
$^{(a)}$ Number of \planck\ detections. $^{(b)}$Number of \planck\ upper limits.\\
$^*$ The number of detections is too low to allow a reliable estimation of $\rho$. 
\end{table*}

\subsection{Microwave vs. \gr}
\label{mwvsgr}
The relationship between radio or \mw~and \gr~fluxes is a topic that has been addressed several times in the literature.
A positive correlation between the radio and  \gr~fluxes is generally found, though with a large scatter
\citep[e.g.,][]{kovalev09, leon10,giroletti10,ghirlanda10,mahony10,peel10,linford11,ackerman11,tavares11}. 
However, in most cases radio and {\it non-simultaneous} \gr~data for sources {\it detected in both energy bands} are compared. For the first time, we present 
simultaneous \mw~and \gr~data and take into account upper limits.

\begin{figure}
  \centering
 \includegraphics[width=8.4cm,angle=0]{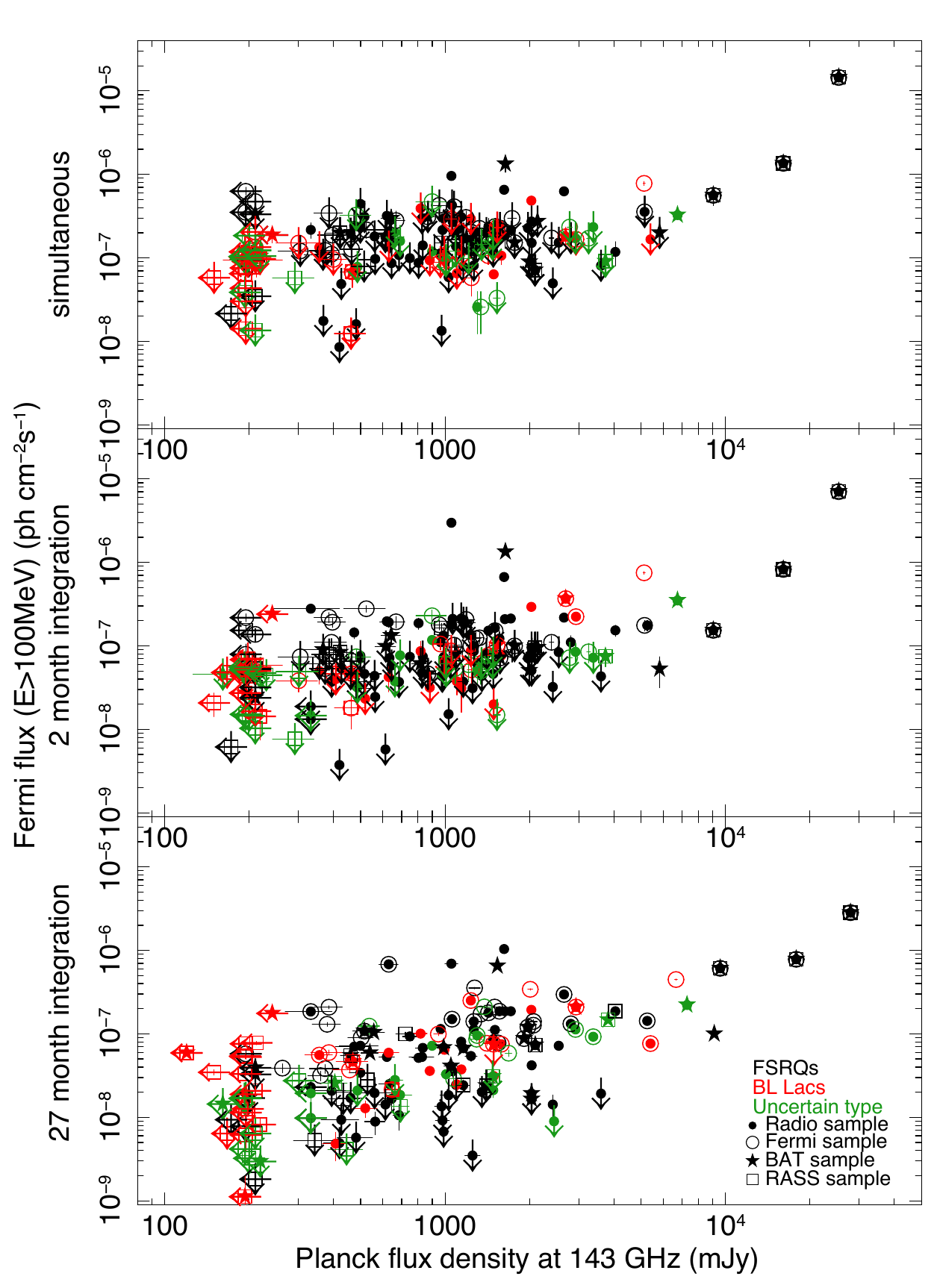}
 \caption{Top panel: The \planck\ flux density at 143\,GHz is plotted against the simultaneous \fermi-LAT flux. 
 Middle panel : The \planck\ flux density at 143\,GHz is plotted against the \fermi-LAT flux integrated over the 2-month period centered on the \planck-\swift~observations.
 Bottom panel: The \planck\ ERCSC flux density at 143\,GHz (flux averaged over more than one \planck\ survey) is plotted versus the \gr~fluxes averaged over the entire 27 month \fermi-LAT observing period.}
 \label{fig:PFcorrelations}
\end{figure}

The top panel of Figure~\ref{fig:PFcorrelations} shows the simultaneous \fermi-LAT \gr~ flux plotted versus the \planck\ flux density at 143\,GHz.
Sources with $\textrm{TS} < 25$ in the \fermi-LAT data and below 4$\sigma$ in \planck\ maps are shown as upper limits.  The plot shows a clear trend but
from Table~\ref{tab:gammacorr}, which gives the Spearman's $\rho$ correlation parameter and the probability $P$ that the observed level of 
correlation is caused by chance, we see that the level of significance of the correlation is never very high ($P$ is of the order of a few percent and never  lower than 0.05\%), 
especially in the hard X-ray and \gr~selected samples and for  BL Lacs. 
This result is partly due to the large number of upper limits to the simultaneous \gr~flux, which was estimated typically over a period of one week. 
To improve the statistics, in the middle panel of Fig.~\ref{fig:PFcorrelations} we show the same plot using the \fermi-LAT \gr~ flux integrated over a two-month period centered on the 
time  of the \planck\ observations. 
From Table~\ref{tab:gammacorr}, we see that, although the significance of the correlation in the various samples and blazar classes substantially increases, it is never very high and the scatter remains. 

This result may be due to the limited flux dynamic range in the \gr~band for the \fermi-LAT sample and to the small number of objects (in the case of BL Lacs). However, 
a weak correlation could occur if  the \mw~flux density represents the  superposition of multiple synchrotron components, 
while the simultaneous \gr~flux represents the emission from a single dominant component that may be active for only a short time. For this reason, 
in the bottom panel of Fig.~\ref{fig:PFcorrelations}, we plot the \planck\ ERCSC flux densities (which in most cases represent the flux density averaged over more than one \planck\ observation) versus the \fermi-LAT flux averaged over the entire 27-month period. In this case the correlation is highly significant, although, the many upper limits in both energy bands clearly imply that the dispersion is very large.

\begin{table*}[htbp]
 \caption{Microwave (143\,GHz) vs. \gr~flux correlation parameters.}
 \label{tab:gammacorr}
 \begin{center}
 \begin{tabular}{c c c c c c c c c c c c c c}
   \hline
   \hline
\gr  & Class & \multicolumn{3}{c}{\fermi-LAT sample}  &     \multicolumn{3}{c}{\swift-BAT  sample}  &   \multicolumn{3}{c}{{\textit{ROSAT}}/RASS sample}  &   
\multicolumn{3}{c}{Radio sample} \\
  integration              &            & No. of  & $\rho$   &  $P(\%)$ & No. of  & $\rho$   &  $P(\%)$ & No. of & $\rho$   &  $P(\%)$ & No. of  & $\rho$   &  $P(\%)$ \\
period                       &            & sources &  &  & sources &   & & sources &  & & sources &  &  \\
                                   &            & (a)/(b)/(c) &  &  &  (a)/(b)/(c) &   & &  (a)/(b)/(c) &  & &  (a)/(b)/(c) &  &  \\
\hline
\multirow{3}{*}{Simult.} & All       &    16/5/32 &  0.22 &  12.4  &    8/6/20  &  0.42  &    2.49 &    6/20/25  &  0.42  &    1.42  &    22/2/71  &  0.35  &    0.07 \\ 
                                         & FSRQs     &    10/3/16&  0.23 &  23.4  &    5/2/12  &  0.39  &    9.92 &    4/3/6  &  0.65  &    3.97      &    15/1/46  &  0.33  &    1.19 \\ 
                                         & BL Lacs   &    5/2/9 &  0.55 &  4.76 &    2/2/2   &  ...$^*$  &    ...  &   2/11/4   & ...   &   ...    &    4/0/10  &  0.38  &    17.6  \\

\hline
\multirow{3}{*}{2 months} & All  &    36/5/9 &  0.29 &  4.18      &    11/7/20 &  0.61  &    0.07  &    8/21/18 &  0.31  &    6.00    &   37/3/56   &  0.40  &   0.01 \\
                                              & FSRQs     &    19/3/6 &  0.17 &  38.7  &    6/3/15  &  0.59  &    0.89   &    4/4/9 &  0.51  &    6.83      &    23/2/42  &  0.36  &    0.39 \\
                                              & BL Lacs    &    11/2/1 &  0.72 &  0.92  &    3/3/2   &  ...  &    ...    &    3/11/5  & ...   &    ...     &    10/0/6  &  0.42  &    10.2  \\

\hline
\multirow{3}{*}{27 months}  & All		         &    47/3/0 &  0.47 &  0.10  &    22/16/7 &  0.67  &    $<$0.01  &    12/23/18 &  0.40  &    0.99    &   75/4/16   &  0.48  &   $<$0.01 \\
                    & FSRQs     &    25/3/0 &  0.48 &  1.27  &    16/3/4  &  0.54  &    1.59   &    8/3/7 &  0.85  &    0.15      &    47/2/19  &  0.48  &    0.01 \\
                    & BL Lacs    &    14/0/0 &  0.72 &  0.91  &    3/4/0   &  ...  &    ...    &    2/14/2  & ...  &   ...     &    16/0/0  &  0.55  &    3.22  \\

\hline
\end{tabular}
\end{center}
$^{(a)}$ Number of sources detected both by \planck\ and \fermi-LAT. $^{(b)}$Number of \planck\ upper limits. $^{(c)}$Number of \fermi-LAT~upper limits.\\
$^*$ The number of detections is too low to allow a reliable estimation of $\rho$.
\end{table*}

\subsection{X-ray vs. \gr}

\begin{figure}
\centering
\includegraphics[width=6.6cm,angle=-90]{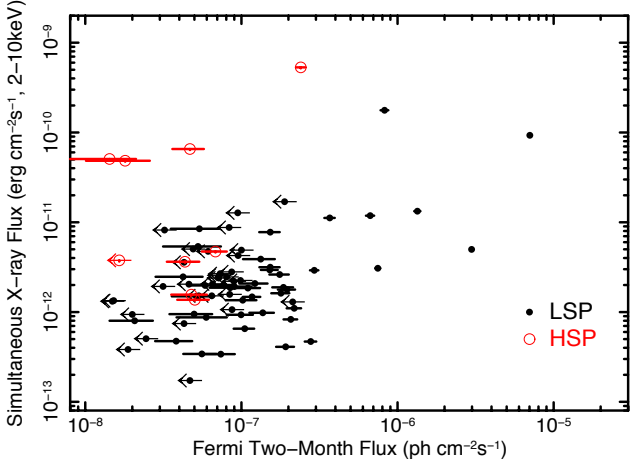}
\caption{The \swift~XRT flux is plotted against the two-month \fermi-LAT flux for the sources included in the three high-energy flux-limited samples. 
Open red circles represent HSP sources, i.e., high synchrotron-peaked BL Lacs where the X-ray flux is dominated by synchrotron radiation. Filled black circles represent LSP sources, i.e., blazars with low synchrotron peak, for which the X-ray flux is dominated by inverse Compton radiation.}
\label{fig.fxvsgamma}
\end{figure}

Figure \ref{fig.fxvsgamma} plots the 2-month \gr~flux versus the simultaneous X-ray flux for all sources observed by \swift~that do not show signatures of thermal contamination
in the X-ray spectrum. 
Open red circles represent HSP sources, where the X-ray flux is due to the tail of the synchrotron emission, while black filled circles are LSP and ISP sources 
for which the X-ray flux is related to the inverse Compton radiation. This distinction is needed in order to properly compare sources where the emission is produced by 
the same mechanism. We therefore compute the \gr~vs. X-ray correlation coefficient only for LSP and ISP sources. The Spearman rank test shows 
moderate evidence of a correlation that is less significant in the longer integrations: $P=2.55\%$ for simultaneous data, $P=6.9\%$ for a 2-month integration, or $P=17.4\%$ for 
a 27-month integration.

\begin{figure}
\centering
\includegraphics[width=11.0cm,angle=0]{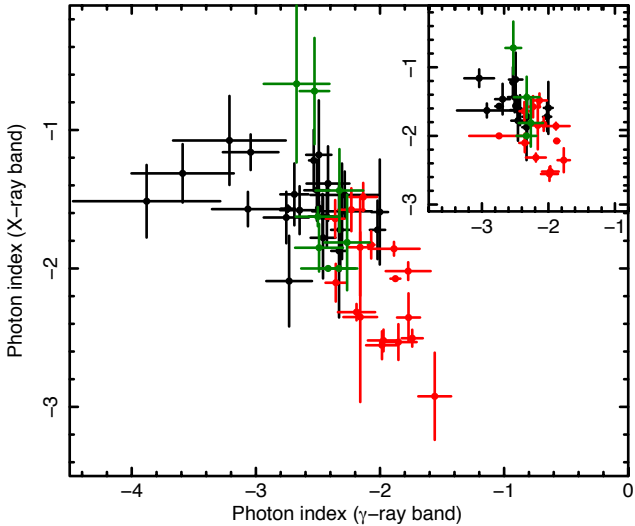}
\caption{The power-law spectral index  in the \swift~XRT energy band (0.3--10\,keV) is plotted versus the slope in the \fermi-LAT \gr~band (100\,MeV--100\,GeV) derived
using \fermi-LAT data integrated over a period of 2 months. The same plot built with simultaneous \gr~data is shown in the inset at the upper   right.
A clear anti-correlation is present; the Spearman test gives a probability of less than 0.01\% that the correlation is due to chance, even in the simultaneous data.}
\label{fig.xgammaslopes2m}
\end{figure}

Figure \ref{fig.xgammaslopes2m} plots the power-law spectral index in the \swift~XRT energy band (0.3--10\,keV) versus the spectral slope in the \fermi-LAT \gr~band (100\,MeV--100\,GeV) derived using the 2-month \gr~data. 
A significant correlation is present, confirming the results of \cite{abdoseds}. 
The top right corner of Fig.  \ref{fig.xgammaslopes2m} shows the same plot built with simultaneous \gr~data; although the number of objects is smaller, the correlation is still present.   

\section{Discussion and conclusions}

We have collected simultaneous \planck, \swift, \fermi-LAT, and ground-based multi-frequency data for 105 blazars included in three statistically well-defined samples characterized by flux limits in the soft X-ray (0.1--2.4\,keV, \textit{ROSAT}), hard X-ray (15--150\,keV, \swift-BAT), and \gr~($E > 100$\,MeV, \fermi-LAT) energy bands, with the addition of a cut to 
the radio 5\,GHz flux density to ensure a high probability of detection by \planck. This study complements a similar study of 104 radio-bright AGN (f$_{37GHz} > 1 $Jy) \citep{planck2011-6.3a}. Altogether, the four samples contains a total of 175 distinct objects.
The acquisition of this unprecedented multi-frequency/multi-satellite data set was possible thanks to cooperation between the \planck, \swift, and \fermi-LAT teams, who agreed to share data and organize an extensive program of multi-frequency observations involving over 160 \swift~ToO pointings scheduled when the blazars were within the field of view of the \planck\ instruments. 

We have used this unique multi-frequency dataset to build well-sampled, simultaneous SEDs of all the blazars included in our high-energy selected samples. This collection of SEDs is an improvement over  previous compilations \citep[e.g., ][]{abdoseds} because: a) the SEDs presented here are strictly simultaneous, while in  \cite{abdoseds} the multi-frequency data were collected over a period of up to nine months centered on the first three months of \fermi-LAT operations; b) the sources were selected according to different statistical criteria allowing us to probe the blazar parameter space from widely different viewpoints; and c)  we took care to identify and separate radiation components unrelated to the emission from the jet, such as the light from the blazar's host galaxy and the radiation produced by the accretion onto the central black hole, which often contaminate the non-thermal blazar spectrum in the optical, UV, and X-ray bands.

Our findings are broadly consistent with those of \cite{abdoseds}. However, our analyses of larger samples selected in different parts of the electromagnetic spectrum, wider wavelength coverages, different level of simultaneity, and the ability to separate the emission components, have allowed us to make significant progress in several areas. 
The use of four widely different samples has allowed us to investigate the consequences of selection effects on the estimation of critical parameters such as $\langle$\nupS$\rangle$, average Compton dominance, especially for the case of BL Lacs.
 
Some of our sources have been observed simultaneously by \swift, \planck, and \fermi-LAT during more than one \planck\ survey. In these cases, we have presented only the data collected during the first observation. Multiple simultaneous observations of a subset of our blazars and detailed model fitting of the SEDs will be the subject of future papers. 
The main results of this work are discussed below.



\subsection{\fermi-LAT detection statistics and the effects of variability}

The percentage of \fermi-LAT detected sources during the simultaneous integrations, typically lasting about one week, ranges from $\lsim 40\%$ in the  \fermi-LAT sample to 20--25\% in the radio and X-ray selected samples (see Table~\ref{tab:fermisummary}). When 2-month integrations centered on the \planck-\swift~observations are considered, these percentages grow to 80\% in the \fermi-LAT sample and to $\sim$ 35\% in the other samples. 
However, even when using data from the 27-month \fermi-LAT integrations available at the time of writing, many of the blazars belonging to the radio, and both soft and hard X-ray selected samples remain undetected.  

We note that the detection rate is quite different for FSRQs and BL Lacs: if we exclude the \fermi-LAT sample where all the objects have been detected (by definition), the percentage of detections is 95\% for BL Lacs and only $\sim$60\% for FSRQs, with values ranging from 
72\% in the radio sample to just 53\% in the RASS sample (see Table~\ref{tab:fermifsrqsbllacs}).

A comparison of our simultaneous data with published and archival measurements shows that the use of non-simultaneous data in the SED of blazars typically introduces a scatter of about a factor of two in the \mw~band, and a factor of up to ten or more in the X-ray and \gr~ bands.

\subsection{The spectral slope in the radio--sub-millimetre region}

We confirm that the energy spectrum of blazars in the radio--\mw~spectral region is quite flat, with an average slope of $\langle\alpha\rangle \sim 0$  ($f(\nu)\propto\nu^{~\alpha}$) up to about 70\,GHz, above which it steepens to  $\langle\alpha\rangle \sim -0.6$.  This behaviour is very similar to that observed in the sample of radio bright blazars considered by \cite{planck2011-6.3a}  \citep[see also][]{tucci11} confirming the findings of \cite{abdoseds} that the radio to microwave part of the spectrum is approximately the same in all blazars (FSRQs and BL Lacs) independently of the selection band. 
However, the spectral slope of BL Lacs  above  $\sim$70\,GHz is flatter than that of FSRQs with $\langle \alpha_{\rm HF}\rangle = -0.51 \pm 0.07$ for BL Lacs compared to  
$\langle \alpha_{\rm HF}\rangle = -0.73 \pm 0.04$ for FSRQs (see Table~\ref{tab:radioslope} and Fig.~\ref{fig:radioslopes}). A KS test, performed on the subsamples of FSRQs and BL Lacs with 
radio flux density larger than 1 Jy,  gives a probability of less than 3\% that the two samples come from the same parent population. 
This difference in the high frequency spectral index may reflect that the radio-submm  part of the spectrum is closer to \nupS\ in LBL than in HSP sources.

\subsection{Synchrotron self-absorption}

We searched for signatures of synchrotron self-absorption, which in simple homogeneous SSC models is expected to cause strong spectral flattening below $\sim$100\,GHz, but we found no evidence of any common behaviour; indeed, the average spectrum in that region steepens instead of flattening.  Possible cases where some evidence of synchrotron self-absorption may be present are
PKS\,0454$-$234, PKS\,0521$-$36 (Fig.~\ref{fig:sed10}), S4\,0917+44 (Fig.~\ref{fig:sed16}), PKS\,1127$-$145 (Fig.~\ref{fig:sed22}), and 3C\,454.3 (Fig.~\ref{fig:sed49}).

\subsection{Non-thermal versus disk radiation}
In several blazars, the optical/UV light is contaminated significantly by thermal/disk  radiation (known as the blue bump, see Figs.~\ref{fig.lowzWithTemplate} and~\ref{fig.highzWithTemplate}), while the soft X-ray flux is contaminated by radiation produced in the accretion process in approximately 25\% of the blazars in our samples (see Fig.~\ref{fig:xcontamination}). In some of the closest sources, the optical light is instead either contaminated or dominated by the emission from the host galaxy.
Ignoring this contamination may cause an overestimate of the position of both \nupS and  \nupIC  by 0.5 dex or more.

We investigated the relationship between the radiation produced by accretion and the jet, which co-exist in most FSRQs, using the parameter \arbb, defined 
as the spectral slope between the 5\,GHz radio (non-thermal) flux and the 5000\,\AA~optical flux that can be attributed to the blue-bump/disk emission. 
In the blazar paradigm, this quantity depends on both the amount of relativistic amplification (which only affects the non-thermal radiation from the jet) 
and the intrinsic ratio of non-thermal/jet radiation to disk emission. We estimated the value of \arbb~in all sources for which we had optical data of good quality.
Since BL Lacs do not display broad lines, only lower limits to  \arbb~can be derived. 

Figure~\ref{fig:arobb} shows the distribution of \arbb~for all our samples, which range from $\sim$0.4 to just above 1.0 and have peak values between 0.6 and 0.8. 
The distributions are all similar, with the largest difference being between the \fermi-LAT and the RASS samples (a KS test gives a probability of 3.6\% that the two distributions 
originate from the same parent population), possibly reflecting differences in either the amplification factor or the ratio of accretion to jet emission between 
\gr~and soft X-ray selected blazars.  
We note that the distribution of the \arbb~limits for BL Lacs is consistent with intrinsic values of \arbb~for BL Lacs that are within the range of values observed in FSRQs. 

\subsection{The distribution of rest-frame peak energies}
\begin{figure}[ht]
 \centering
 \includegraphics[width=7.5cm,angle=-90]{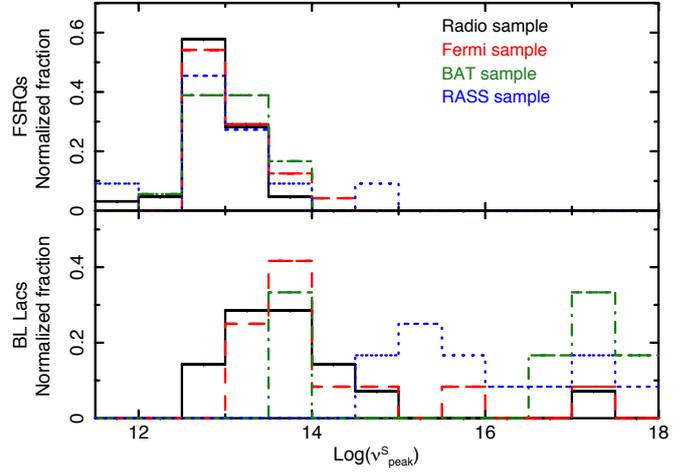}
 \caption{The rest-frame synchrotron \nup~distributions of FSRQs and BL Lacs in different samples. Black solid histograms: radio sample; red dashed histograms: \fermi-LAT sample; green dot-dashed histograms: \swift~BAT sample; blue dotted histograms: RASS sample.}
 \label{fig:nupeaks_distr_sy}
\end{figure}

\begin{figure}[ht]
 \centering
 \includegraphics[width=7.5cm,angle=-90]{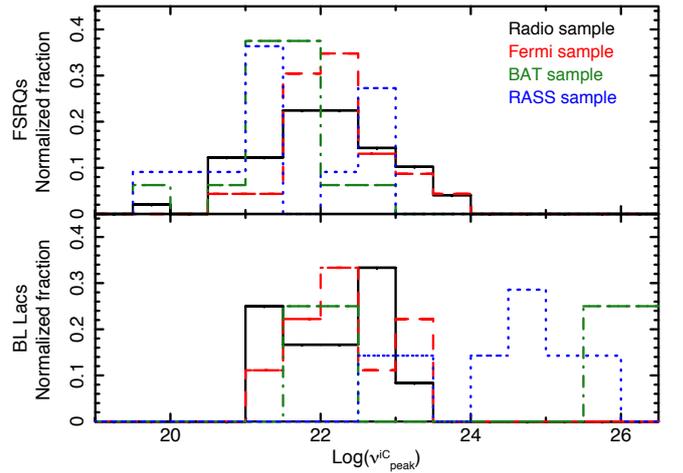}
 \caption{ The rest-frame inverse Compton \nup~distributions of FSRQs and BL Lacs in different samples. Black solid histograms: radio sample; red dashed histograms: \fermi-LAT sample; green dot-dashed histograms: \swift~BAT sample; blue dotted histograms: RASS sample.}
 \label{fig:nupeaks_distr_ic}
\end{figure}

The distribution of rest-frame synchrotron peak energies (\nupS) of FSRQs is very similar in all our samples with a strong peak at $ \approx 10^{12.5}$\,Hz, an average of  $\langle$\nupS$\rangle 10^{13.1\pm 0.1}$\,Hz, and a dispersion of only $\sim$ 0.5\,dex.
However, for BL Lacs  the value of $\langle$\nupS$\rangle$ is at least one order of magnitude larger than that of FSRQs, the exact value depending considerably on the selection method (see Table~\ref{tab:peaks} and Fig.~\ref{fig:nupeaks_distr_sy}). Since all the sources that are below the radio flux density cut in the RASS and BAT samples are ISP or HSP blazars, their inclusion would increase the difference between the \nupS\ distributions.
The distributions of \nupIC\ for FSRQs and BL Lacs also differ, but not as much as those of \nupS\ (see Fig.~\ref{fig:nupeaks_distr_ic}).  The majority of the sources in all the samples (both FSRQs and BL Lacs) peak between $10^{21}$\,Hz and $10^{23}$\,Hz, with a few extreme HSP BL Lacs reaching  $\sim~10^{26}$\,Hz.

\subsection{Correlations between fluxes and other blazar parameters} 
Despite the strict simultaneity of our  data, plots of fluxes in different spectral regions (\mw~ vs.\ X-ray, \mw~ vs.\ \gr, and X-ray vs.\ \gr) still have a large scatter (see Figs.~\ref{fig.fxvs143ghz}, \ref{fig:PFcorrelations}, and~\ref{fig.fxvsgamma}). This is somewhat surprising, as positive correlations between the radio and \grs\ have been reported \citep[e.g.][]{kovalev09, leon10,giroletti10,ghirlanda10,mahony10,peel10,linford11,ackerman11}. 
The difference might be due to the different synchrotron peak energies of the objects in the samples. Even in simple SSC scenarios, this introduces a scatter in the correlation between the fluxes 
(e.g., for the same radio flux, an object with higher \nupS\ is expected to produce more \grs\ than one with a smaller \nupS). 
The large scatter present in Fig.~\ref{fig:PFcorrelations} could also imply that \gr~ emission is due to components not always directly related to radiation in other energy bands, e.g., multiple SSC components \citep[see also][]{abdoseds}.  A good correlation is however present between the X-ray and \gr~spectral slopes (see Fig.~\ref{fig.xgammaslopes2m}). 

\begin{figure}[ht]
 \centering
 \includegraphics[width=8.0cm,angle=-90]{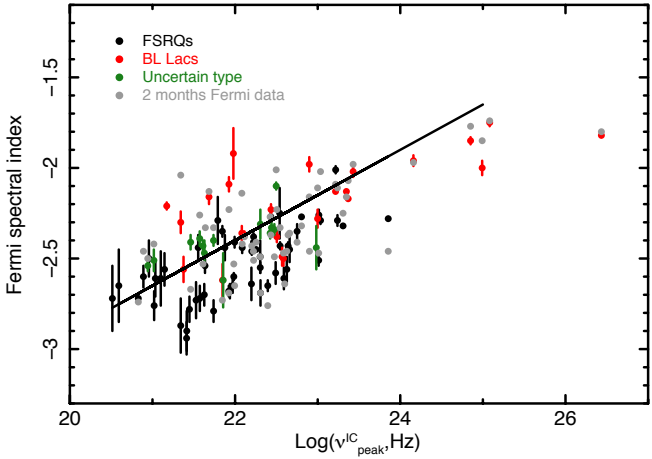}
 \caption{The \gr~spectral index, estimated from the entire 27-month \fermi-LAT data set of all sources in our samples is plotted against 
 log(\nupIC). Gray points, plotted without statistical errors to avoid confusion, represent the 
 \fermi-LAT spectral slopes estimated using the 2-month integrations. The solid line is the best-fit given by \cite{abdoseds}}
 \label{fig:slopeVsnupIC}
\end{figure}

We confirm the correlations between the \fermi-LAT spectral index and the SED peak energies \nupS\ and \nupIC\ found by \cite{abdoseds}. As an illustration of the agreement, Fig.~\ref{fig:slopeVsnupIC} plots the \fermi-LAT spectral slope estimated using the full 27-month data set as a function of \nupIC. The gray points represent the \gr~spectral slopes estimated using the quasi-simultaneous two-month dataset. These points, plotted without the much larger error bars to avoid confusion, clearly cluster around the 27-month data, confirming the correlation. The solid line represents the best-fit obtained by \cite{abdoseds}.

\subsection{Comparison with the expectation of simple SSC models}
 As discussed in \cite{abdoseds}, simple SSC models  predict that in the Thomson regime  the peak frequency of the synchrotron (\nupS)  and inverse Compton (\nupIC) components are related by
\begin{equation}
{\nu_{\rm peak}^{\rm IC} \over \nu_{\rm peak}^{S}} \simeq \frac{4}{3}\left( \gamma^{\rm SSC}_{\rm peak} \right)^2,
\label{nupSSC}
\end{equation}
where $\gamma^{\rm SSC}_{\rm peak}$ is the Lorentz factor of the electrons radiating at the peak energy. This is related to the observed peak frequency of the 
observed photon spectrum by
\begin{equation}
\gamma^{\rm SSC}_{\rm peak} \propto \left( \frac{\nu_{\rm peak}^{\rm S}}{B \delta} \right)^{1/2} ,
\label{nupS}
\end{equation}
where $\nu_{\rm peak}^{\rm S}$ is the synchrotron peak frequency in the rest-frame of the  emitting region, {\it B}  is the magnetic field, and $ \delta$ is the usual Doppler factor \citep[e.g.][]{UP95}.

\begin{figure}[ht]
 \centering
 \includegraphics[width=8.4cm,angle=-90]{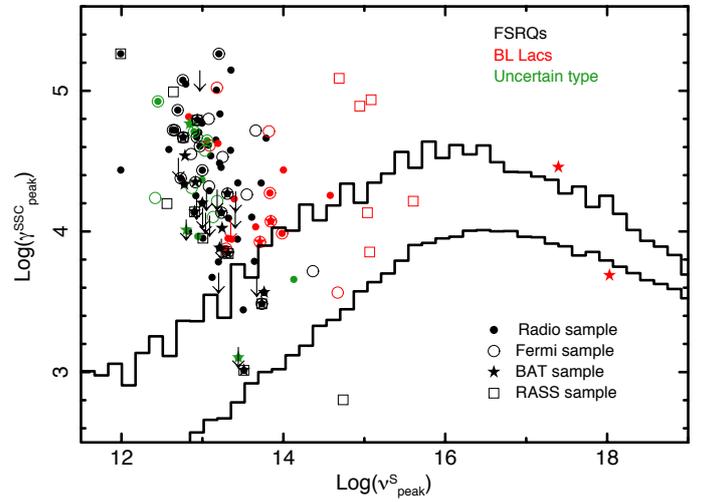}
 \caption{log($\gamma_{peak}^{\rm SSC}$), obtained from Eq.~\ref{nupSSC} for all the objects in our samples for which we could obtain \nupS\ and \nupIC\, is plotted against log(\nupS) in the rest-frame of the blazars. The two black lines delimit the area predicted by simple homogeneous SSC models obtained through extensive Monte Carlo simulations 
 \citep[see][for details]{abdoseds}.}
 \label{fig:gammaevsnupeak}
\end{figure}

In objects where  \nupS is higher than $\approx 10^{15}$\,Hz, the Thomson approximation is no longer valid and the inverse Compton scattering 
occurs under the Klein--Nishina (KN) regime. Using Monte Carlo simulations, \cite{abdoseds} estimated the area covered by SSC models in the plane  log(\nupS)--log($\gamma^{SSC}_{\rm peak}$).
This area is delimited by the solid contour lines shown in Fig.~\ref{fig:gammaevsnupeak}  where we plot  the $\log(\gamma^{\rm SSC}_{\rm peak}$) of our sources, calculated from the observed values of \nupS\ and \nupIC\  using eq.~\ref{nupSSC}, versus the rest-frame log(\nupS).

As in the case of the bright \gr~blazars considered by  \cite{abdoseds}, only a few objects are inside or close to the SSC area, implying that simple SSC models cannot explain the SED of many blazars in our samples. This conclusion is supported by the lack of a strong correlation between radio and \gr\ fluxes.

However, the $\gamma^{\rm SSC}_{\rm peak}$ of blazars that were not detected by \fermi-LAT and for which we could only infer a limit to \nupIC\ (from 30\% to 40\% of the FSRQs in 
the radio and X-ray selected samples; see  Fig.~\ref{fig.ICUpperLimit} for
one example) are plotted as upper limits in Fig.~\ref{fig:gammaevsnupeak}; many of these limits are close to or inside the SSC area and therefore the SEDs of these objects are likely consistent with simple SSC emission.

\subsection{The Compton dominance of blazars} 
The Compton dominance (CD, defined as the ratio of the inverse Compton to synchrotron peak luminosities) is a crucial parameter for the study of blazar physics,
as it is strictly related to the location of the maximum power output in the energy spectrum of a blazar. 

\begin{figure}
\centering
\includegraphics[width=8.4cm,angle=-90]{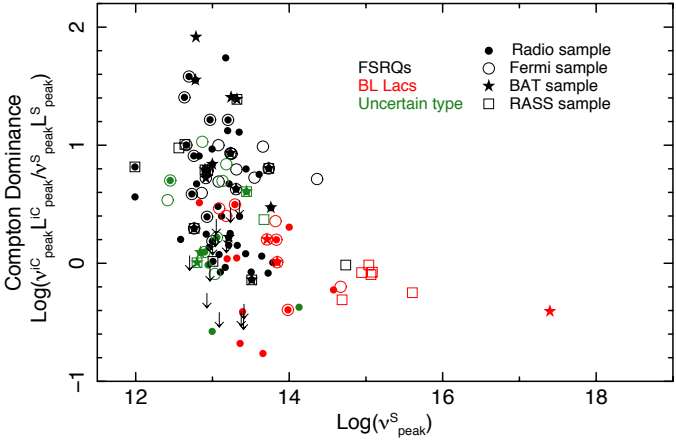}
\caption{The logarithm of the Compton dominance is plotted as a function of log(\nupS) for all sources for which \nupS\ and \nupIC\ could be reliably determined.}
\label{fig:comptdomvsnupeak}
\end{figure}

\begin{figure}
\centering
\includegraphics[width=7.5cm,angle=-90]{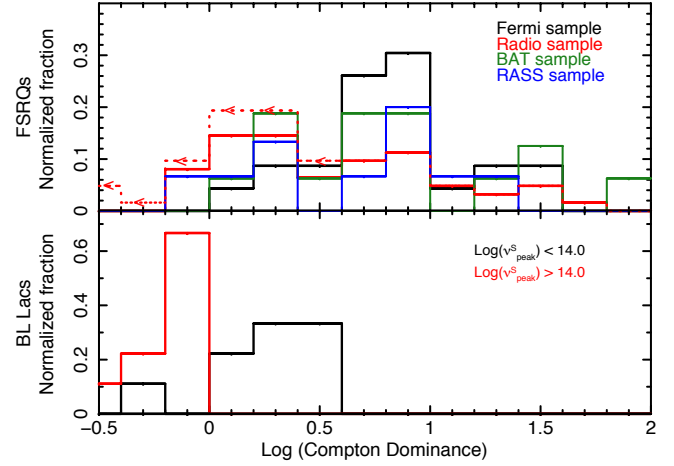}
\caption{Distributions of the Compton dominance for FSRQs and BL Lacs of the LSP and HSP type. A significant fraction of FSRQs in the radio, RASS, and BAT samples have not been
detected in the \gr~band and therefore only limits to the Compton dominance (shown as a dashed histogram;  only the radio sample to avoid confusion)  can be calculated. 
The large difference between the CD distribution of FSRQs in the \fermi-LAT and the other samples illustrates the strong bias that \gr~selection induces.}
\label{fig:cddistr}
\end{figure}

Figure \ref{fig:comptdomvsnupeak} plots the CD values, estimated from our SEDs, as a function of  \nupS, showing that  $\log({\rm CD})$ ranges from about $-0.5$ to about 2.
The larger values are always associated with LSP objects, while HSP sources always have values of $\log({\rm CD})$ lower than $\approx$ 0.5.
In this figure, the two blazar subclasses appear to be quite different, with BL Lacs having significantly smaller CD values, even when their  \nupS\ values are equal to those of FSRQs. To better  understand this difference, in Fig.~\ref{fig:cddistr} we plot the CD distributions of FSRQs and BL Lacs for different samples and \nupS\ intervals.

The FSRQs included in the \fermi-LAT sample, 
which are \gr~bright by definition, show a CD distribution peaking at large values. We note that in this sample we also applied a radio flux-density limit of 1\,Jy,  hence 
the sources below the radio cut must be on average more Compton-dominated than those in our sample. This implies that the distribution of purely \gr~selected blazars 
must be even more strongly peaked at high CD values than that of the \fermi-LAT sample. 
Considering instead FSRQs selected in the radio and the X-ray bands, we get quite a different picture, with a broader distribution extending to values of less than 1. 
Moreover, about 30--45\% of FSRQs  in the radio, soft X-ray, and hard X-ray selected samples are not detected by \fermi-LAT and therefore they must populate the part of the CD distribution
with low CD values. This is shown by the dotted red histogram, which also includes upper limits to the CD estimated  
as the ratio of the upper limit to \nupIC $F$(\nupIC) and  \nupS $F$(\nupS) where limits to \nupIC\  and \nupIC $F$(\nupIC) are obtained by fitting the X-ray data together with the  27-month \fermi-LAT upper limits as shown in Fig.~\ref{fig.ICUpperLimit}.

\subsection{The blazar sequence} 
\begin{figure}[ht]
 \centering
 \includegraphics[width=12.5cm,angle=-90]{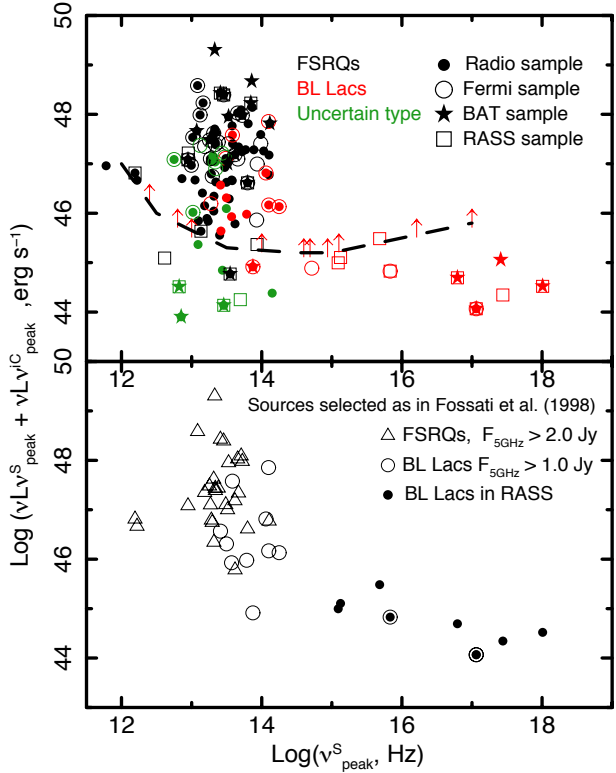} \\
 \caption{Top panel: The bolometric luminosity (represented by the sum of synchrotron and inverse Compton peak luminosities) is plotted against 
 \nupS.
The non-thermal optical light of blazars with no strong emission lines above the dashed line would be bright enough to swamp the emission from the host galaxy making the source appear featureless thus hampering any redshift measurement. 
The lower limits, representing  BL Lacs  with no known redshift, {\and} are estimated assuming that the non-thermal light is ten times brighter than that of a giant elliptical host galaxy (see text for details). 
The bottom panel shows the same plot for the subsample of sources satisfying the selection criteria of \cite{fossati98}.}
 \label{fig:BlazarSequence}
\end{figure}

\begin{figure*}
\centering
\includegraphics[width=6.5cm,angle=-90]{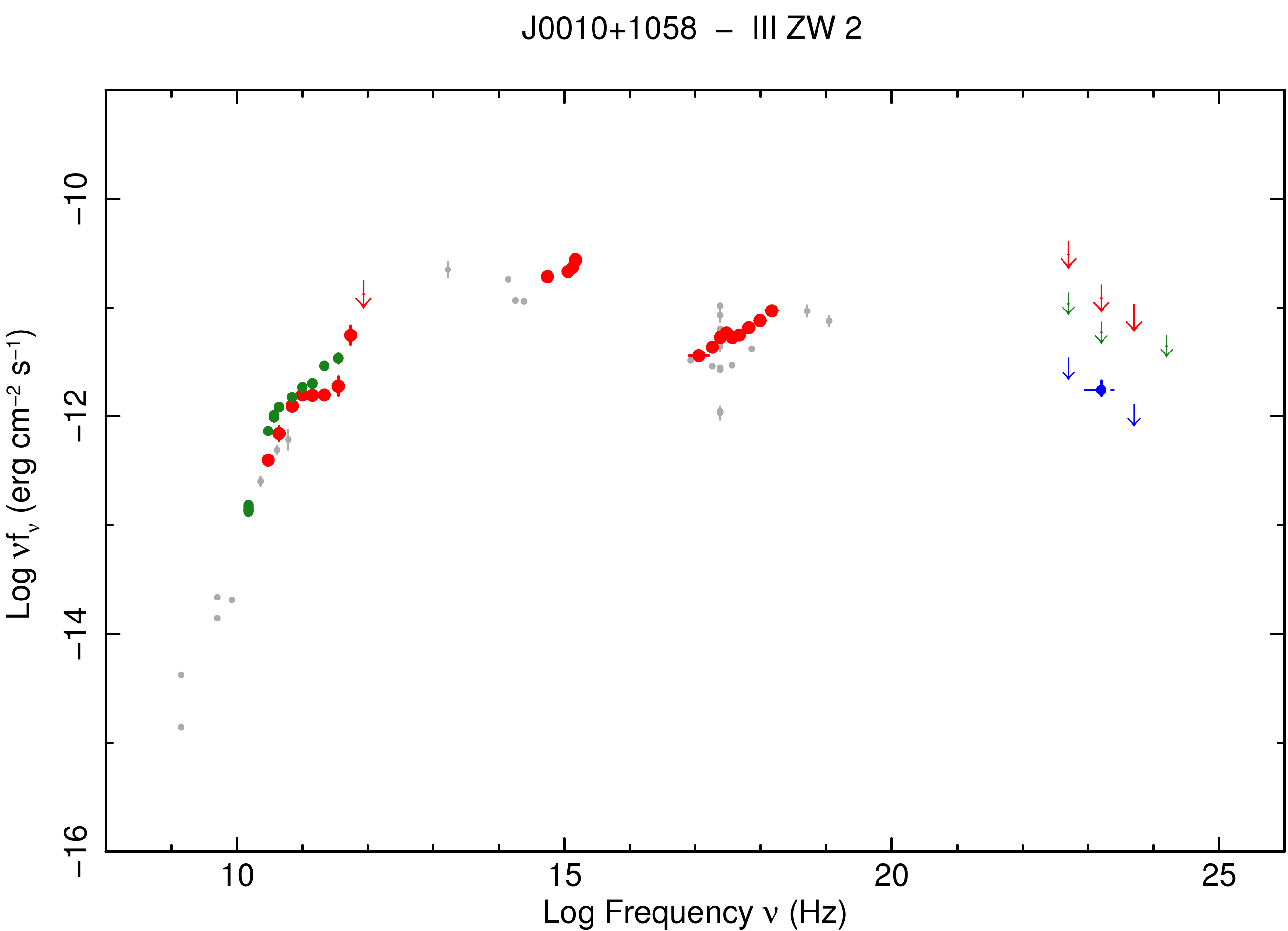}
\includegraphics[width=6.5cm,angle=-90]{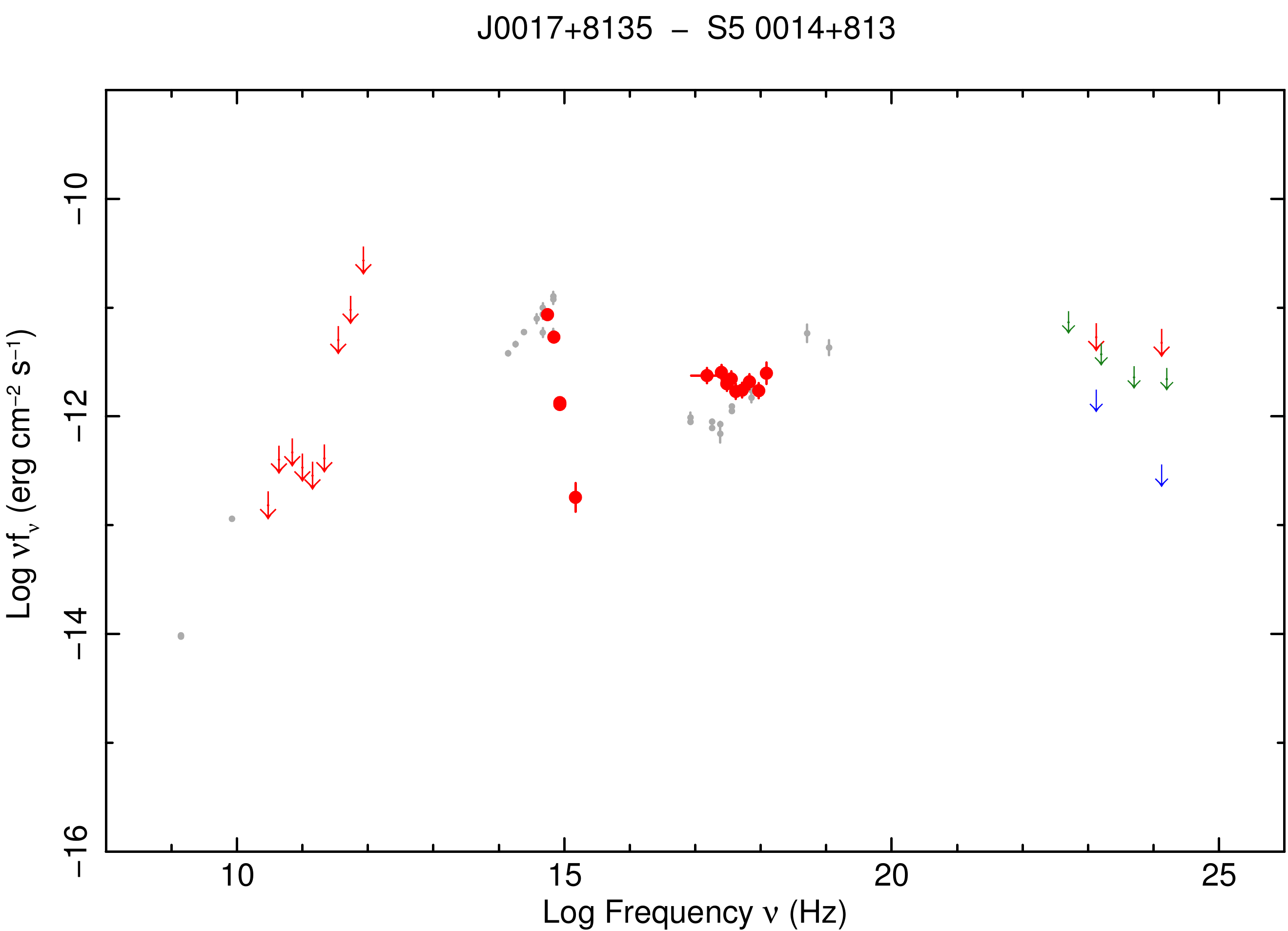}
\includegraphics[width=6.5cm,angle=-90]{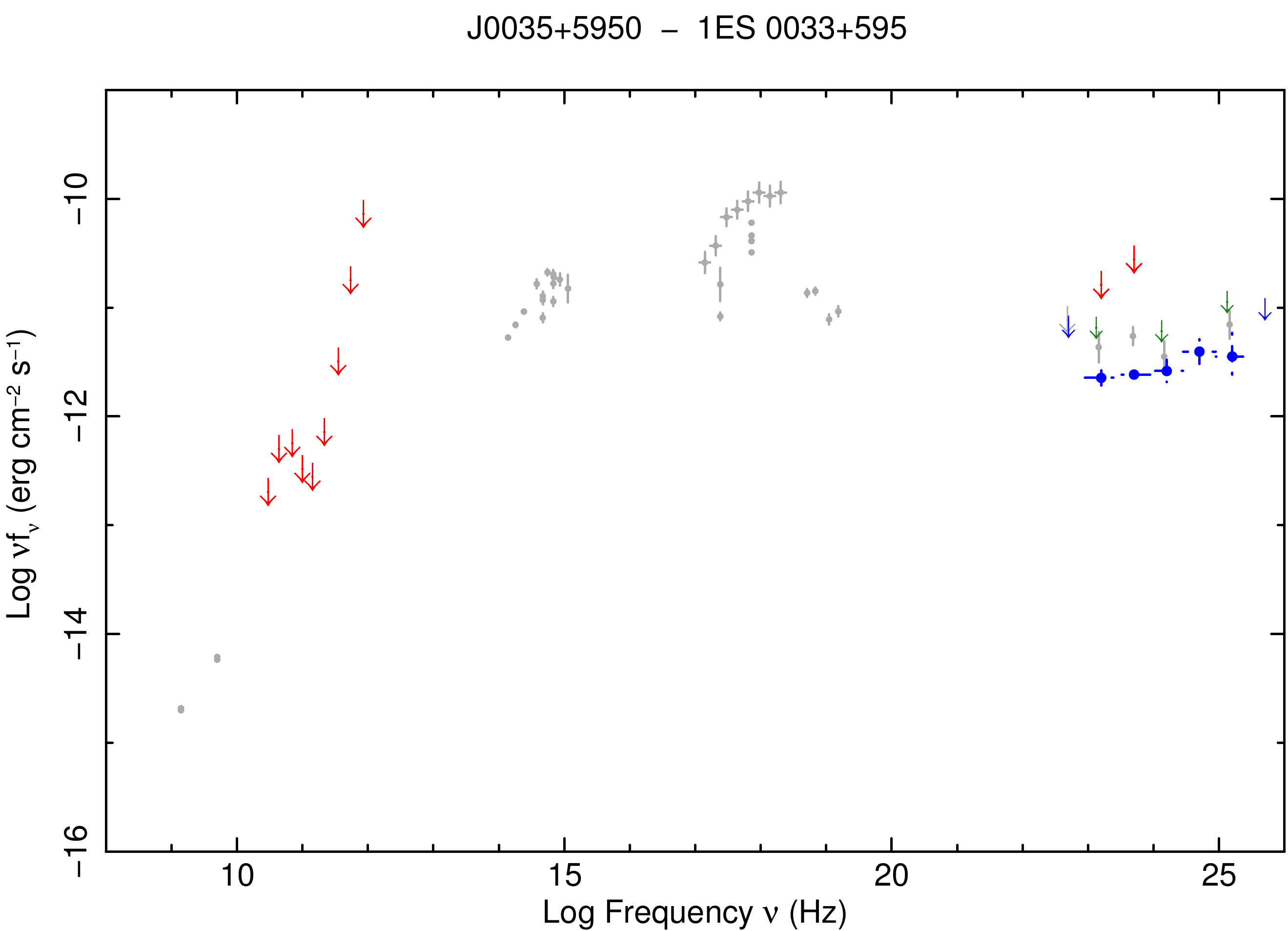}
\includegraphics[width=6.5cm,angle=-90]{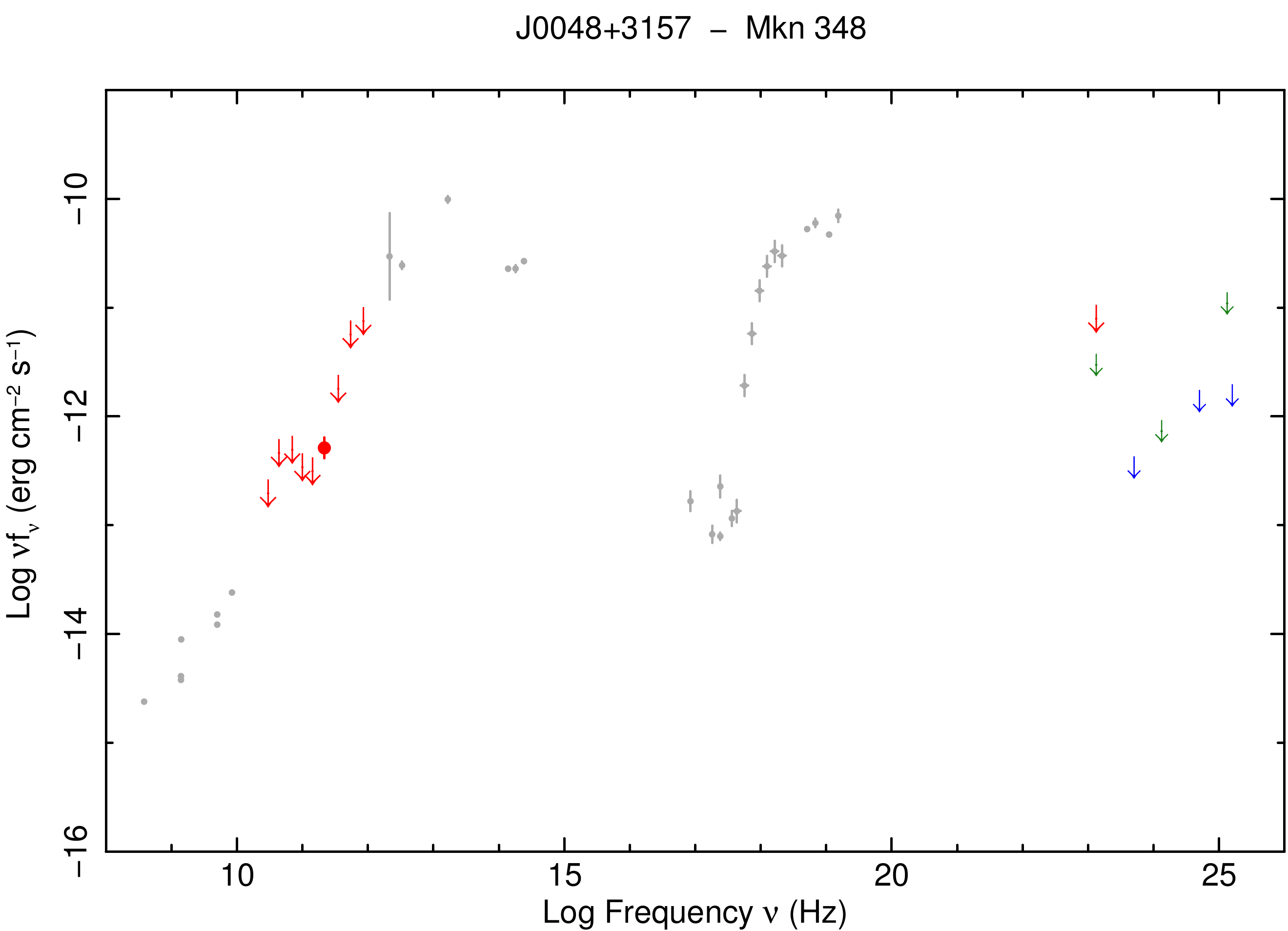}
\includegraphics[width=6.5cm,angle=-90]{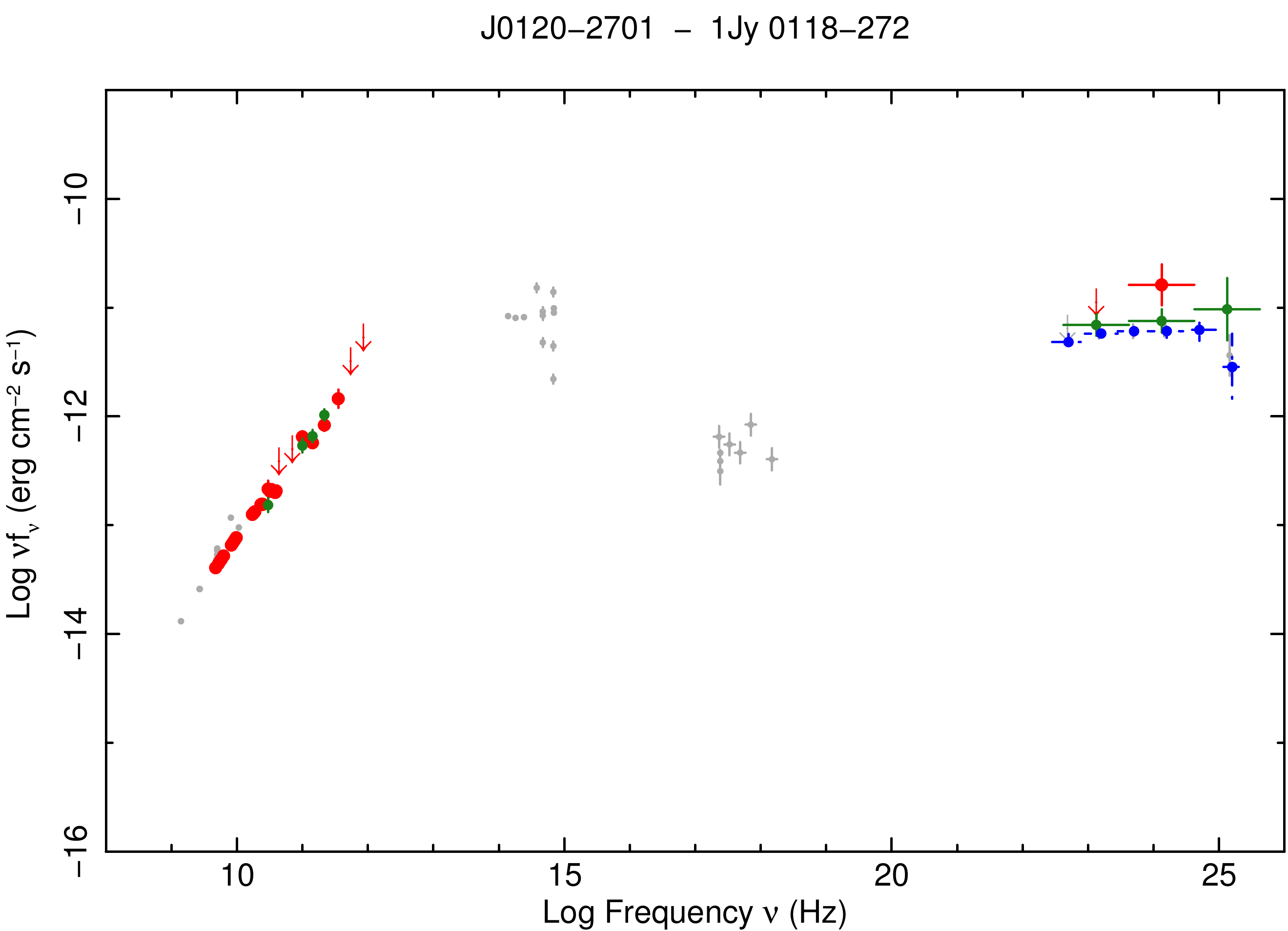}
\includegraphics[width=6.5cm,angle=-90]{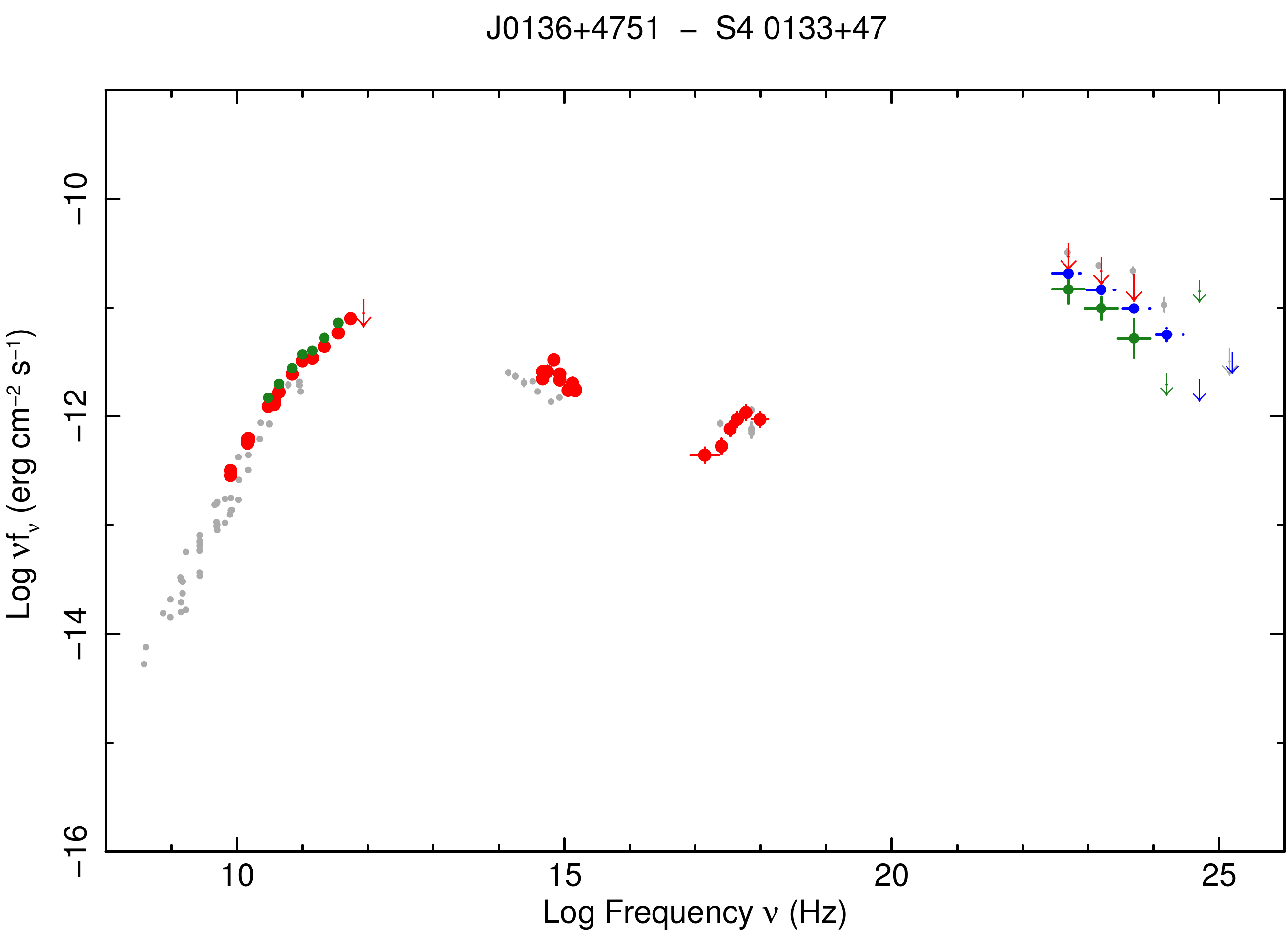}
\caption{The SED of III\,ZW\,2 (J0010+1058, top left), S5\,0014+813 (J0017+8135, top right),
1ES\,0033+595 (J0035+5950, middle left), Mkn\,348 (J0048+3157, middle right),
1Jy\,0118$-$272 (J0120$-$2701, bottom left), and S4\,0133+47 (J0136+4751, bottom right). 
Simultaneous data are shown in red; quasi-simultaneous data, i.e. {\it Fermi} data
integrated over 2 months, {\it Planck} ERCSC and non-simultaneous ground based observations
are shown in green; {\it Fermi} data integrated over 27 months are shown in blue;
literature or archival data are shown in light gray.}
\label{fig:firstSED}
\end{figure*}

\begin{figure*}
\centering
\includegraphics[width=6.5cm,angle=-90]{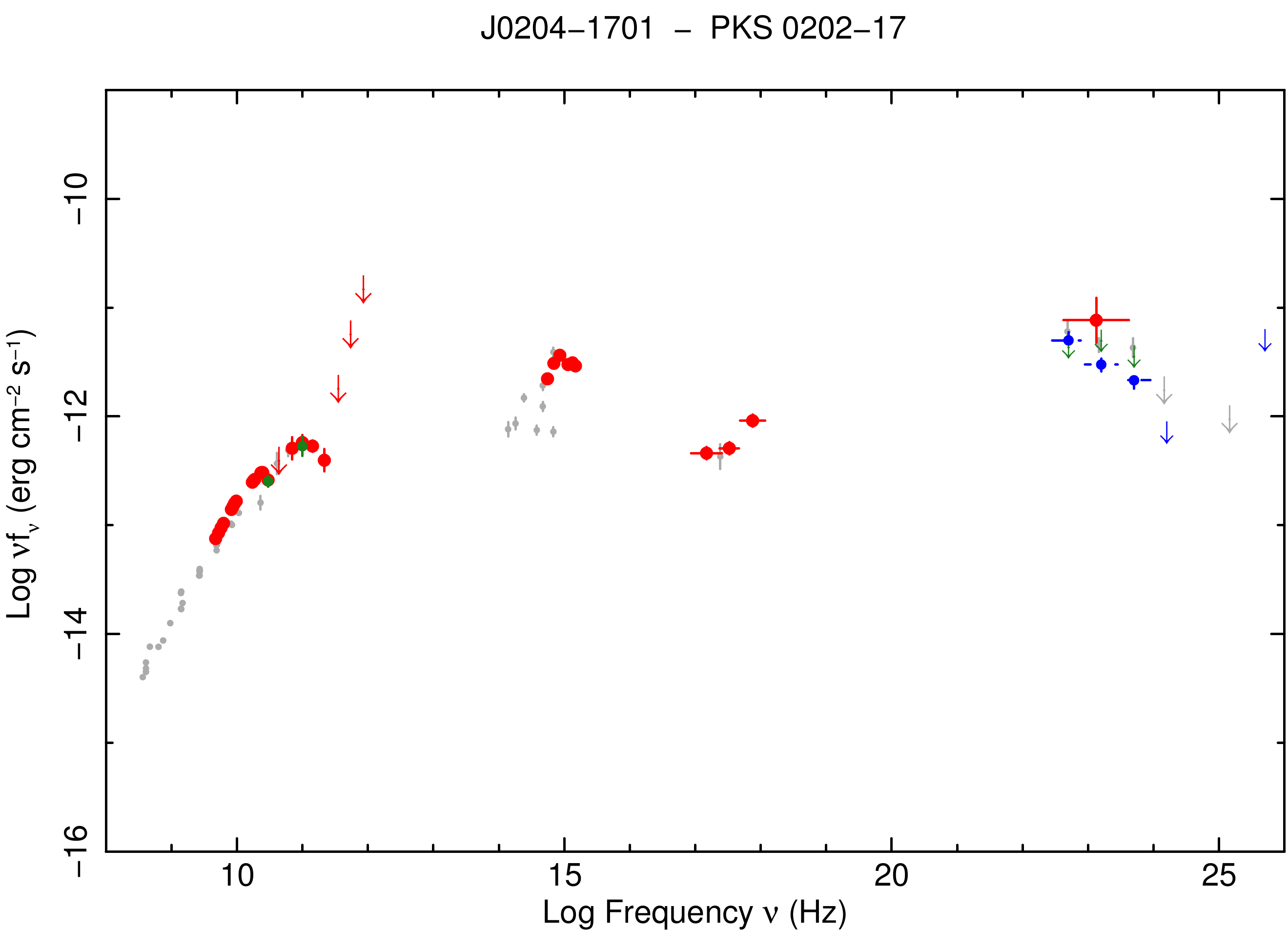}
\includegraphics[width=6.5cm,angle=-90]{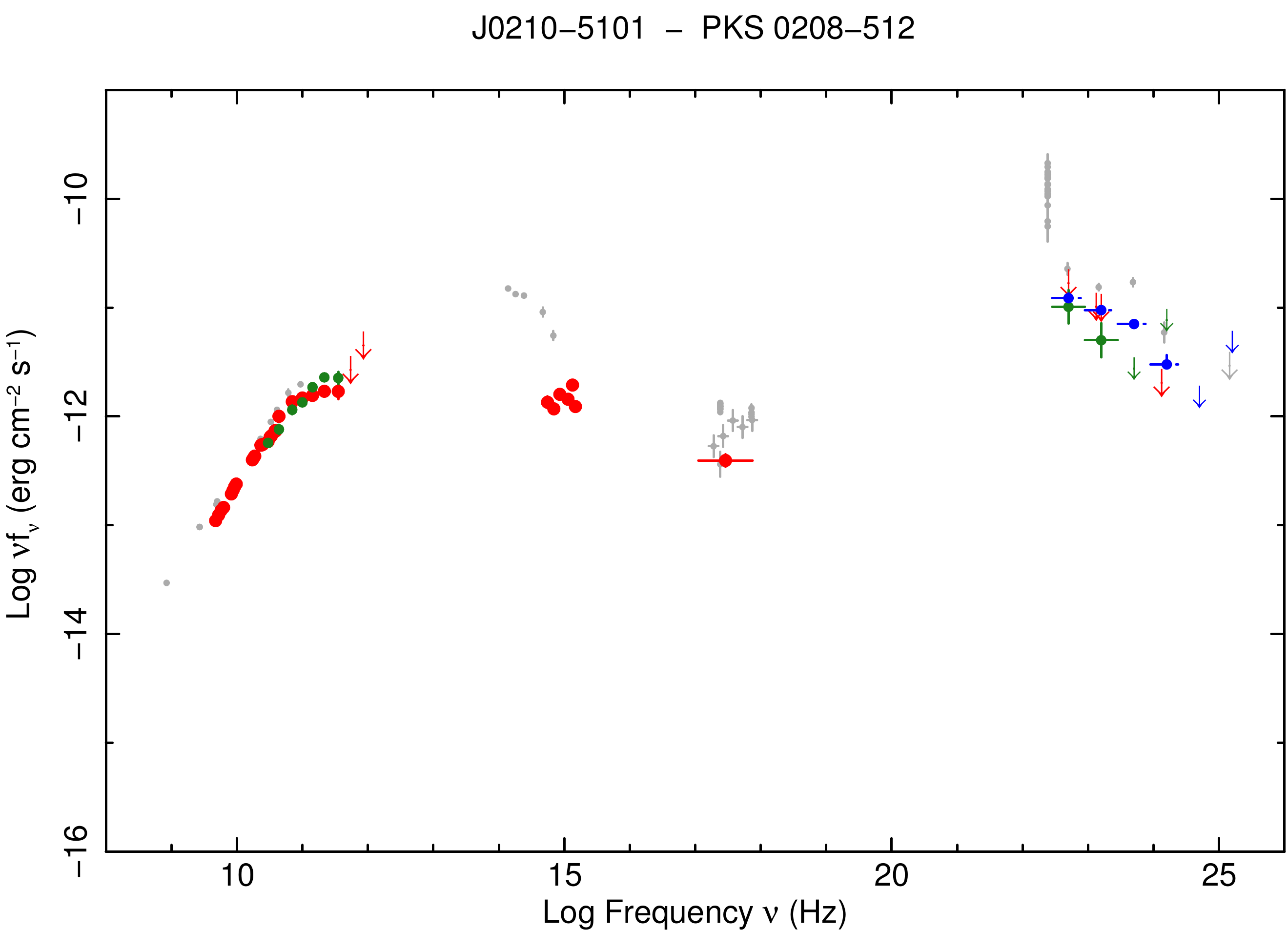}
\includegraphics[width=6.5cm,angle=-90]{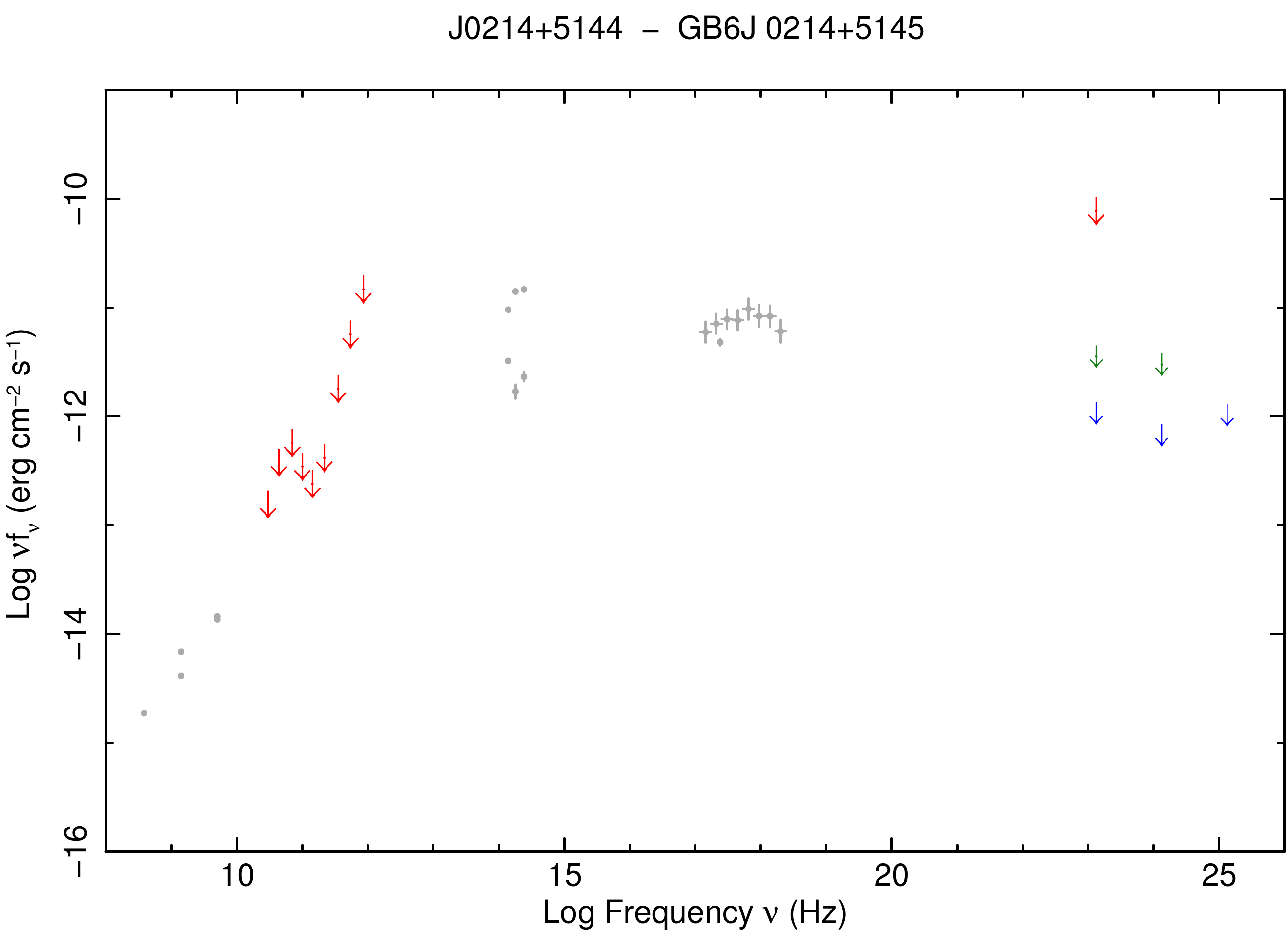}
\includegraphics[width=6.5cm,angle=-90]{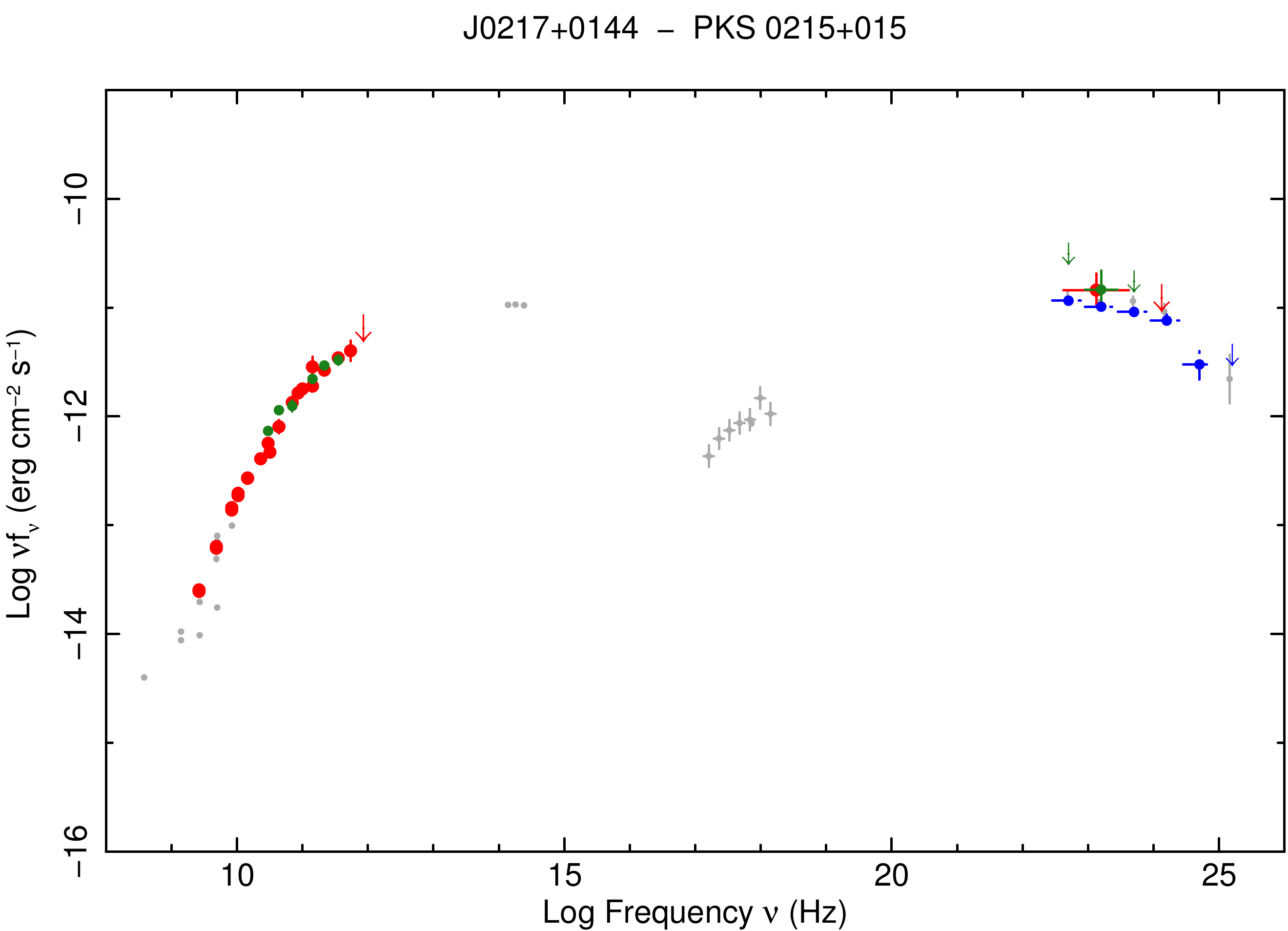}
\includegraphics[width=6.5cm,angle=-90]{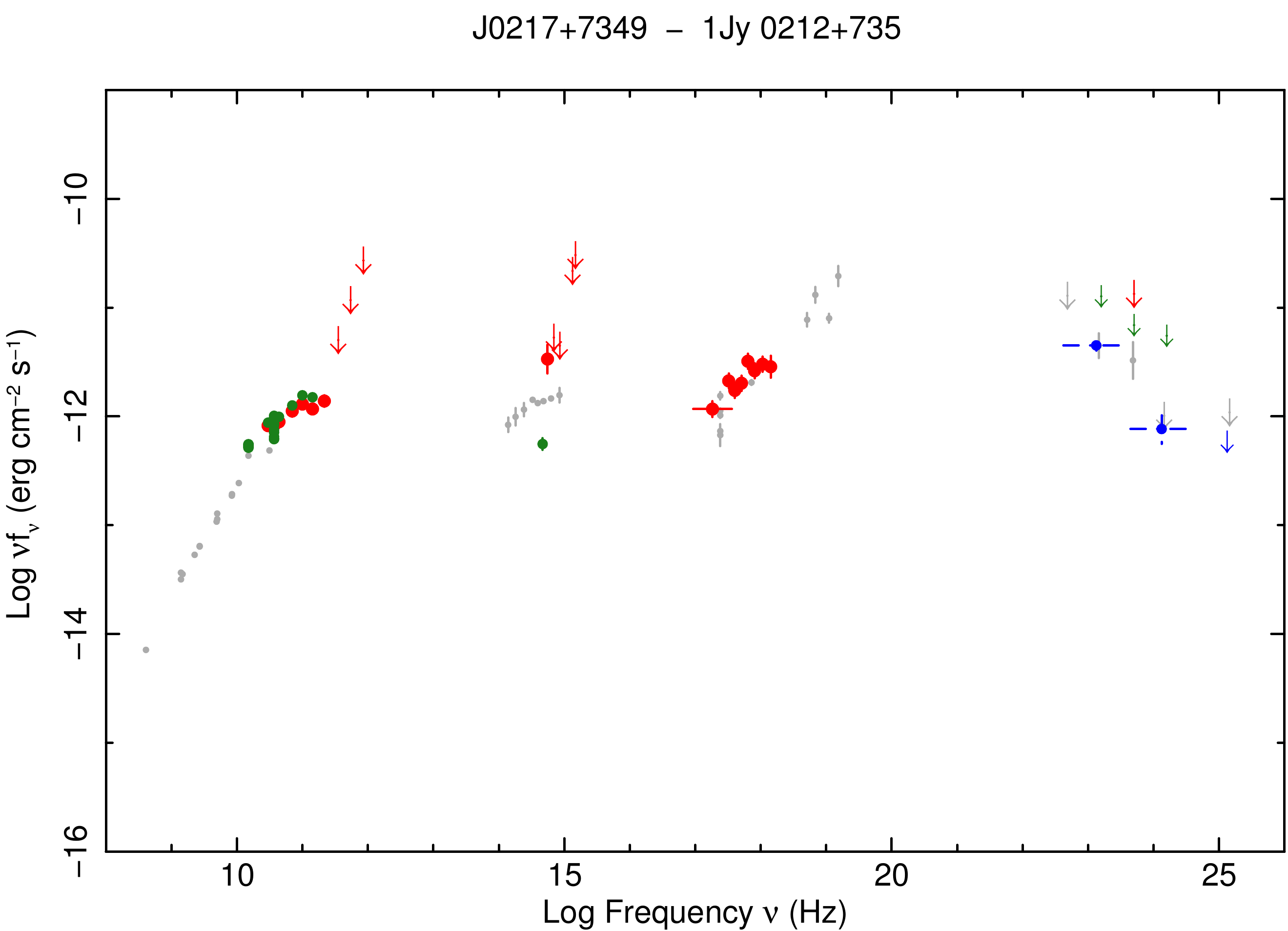}
\includegraphics[width=6.5cm,angle=-90]{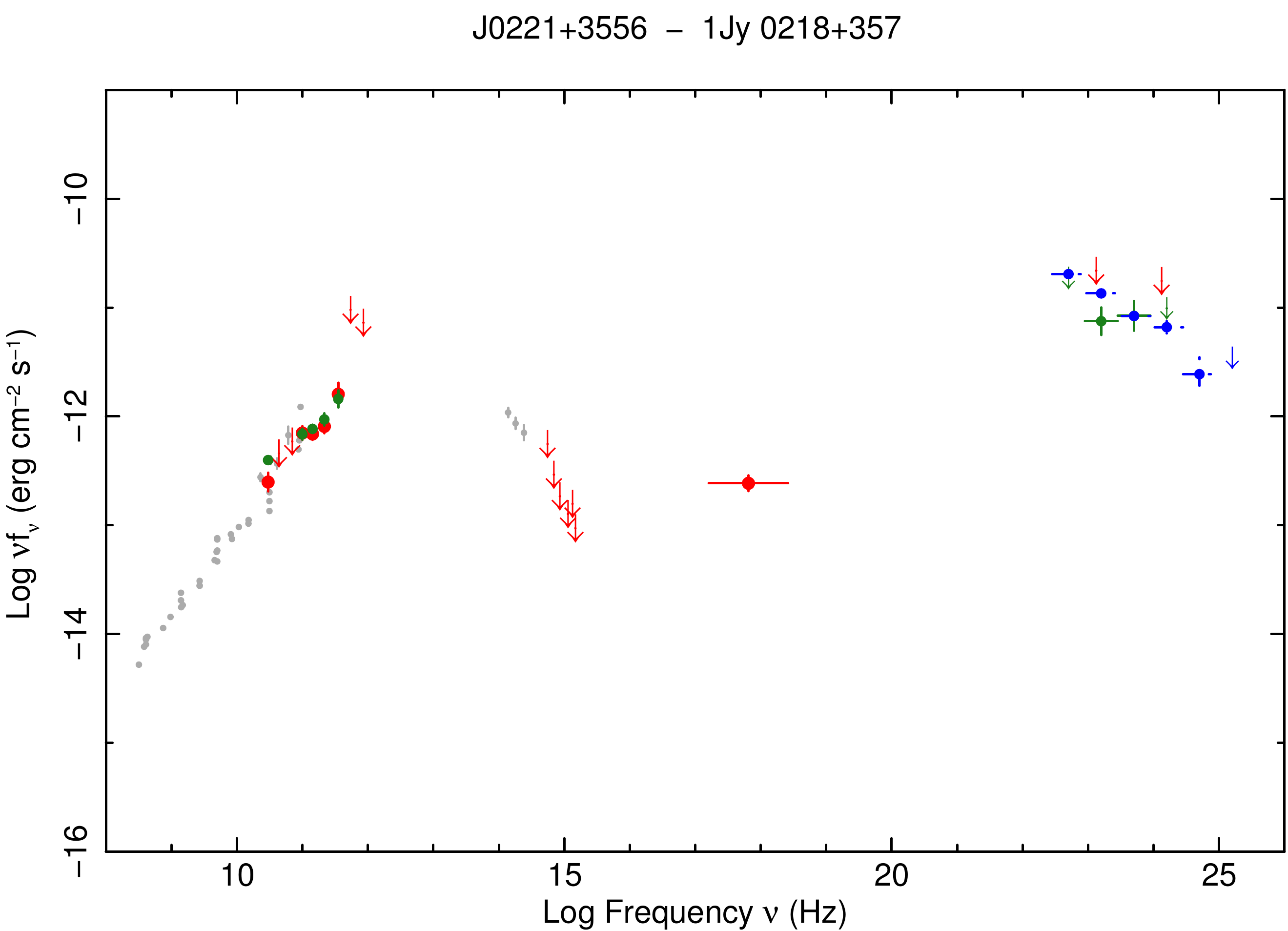}
\caption{The SED of PKS\,0202$-$17 (J0204$-$1701, top left), PKS\,0208$-$512 (J0210$-$5101, top right),
GB6J\,0214+5145 (J0214+5144, middle left), PKS\,0215+015 (J0217+0144, middle right),
1Jy\,0212+735 (J0217+7349, bottom left), and 1Jy\,0218+357 (J0221+3556, bottom right). 
Simultaneous data are shown in red; quasi-simultaneous data, i.e. {\it Fermi} data
integrated over 2 months, {\it Planck} ERCSC and non-simultaneous ground based observations
are shown in green; {\it Fermi} data integrated over 27 months are shown in blue;
literature or archival data are shown in light gray.}
\label{fig:sed4}
\end{figure*}

\begin{figure*}
\centering
\includegraphics[width=6.5cm,angle=-90]{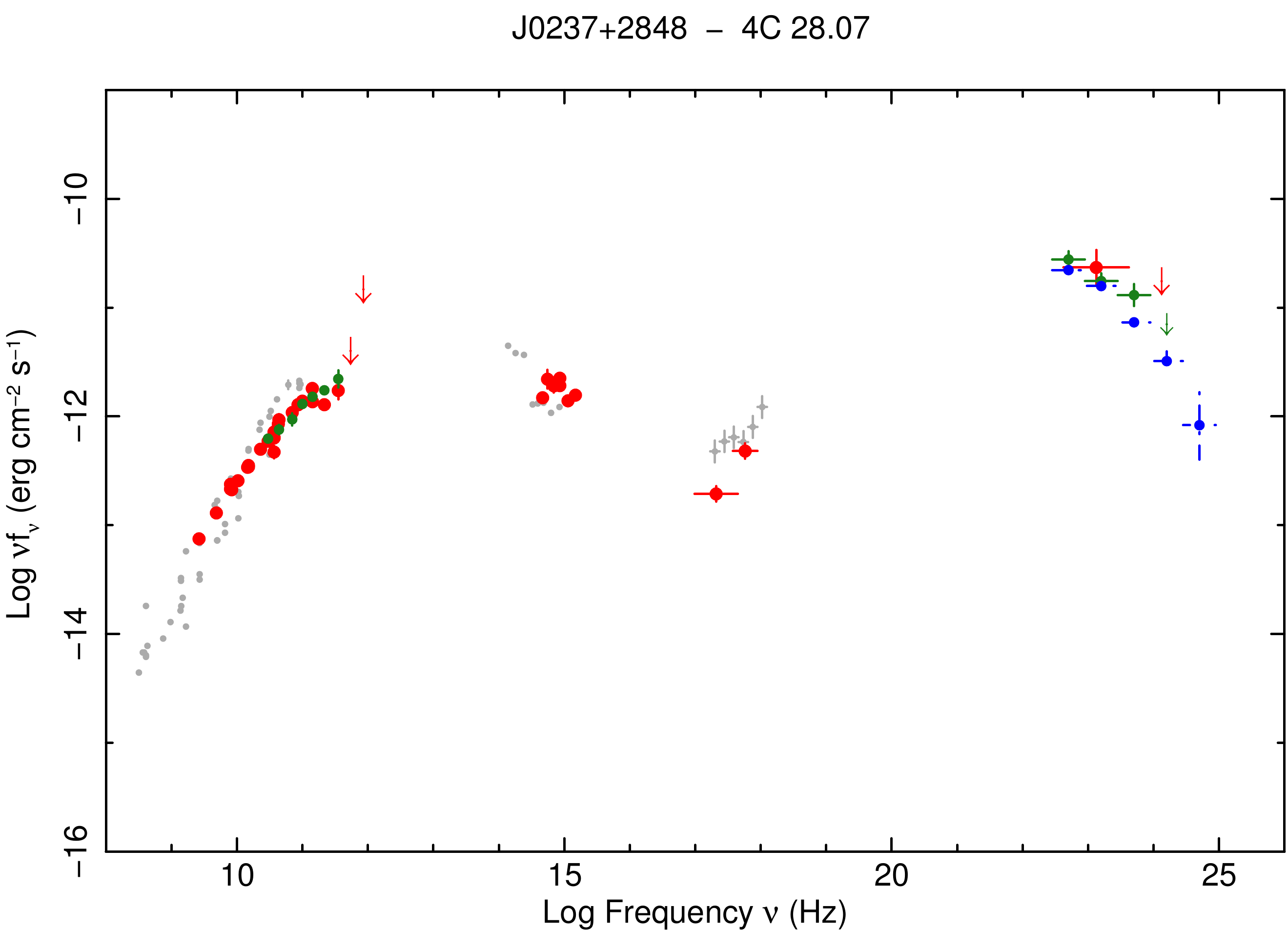}
\includegraphics[width=6.5cm,angle=-90]{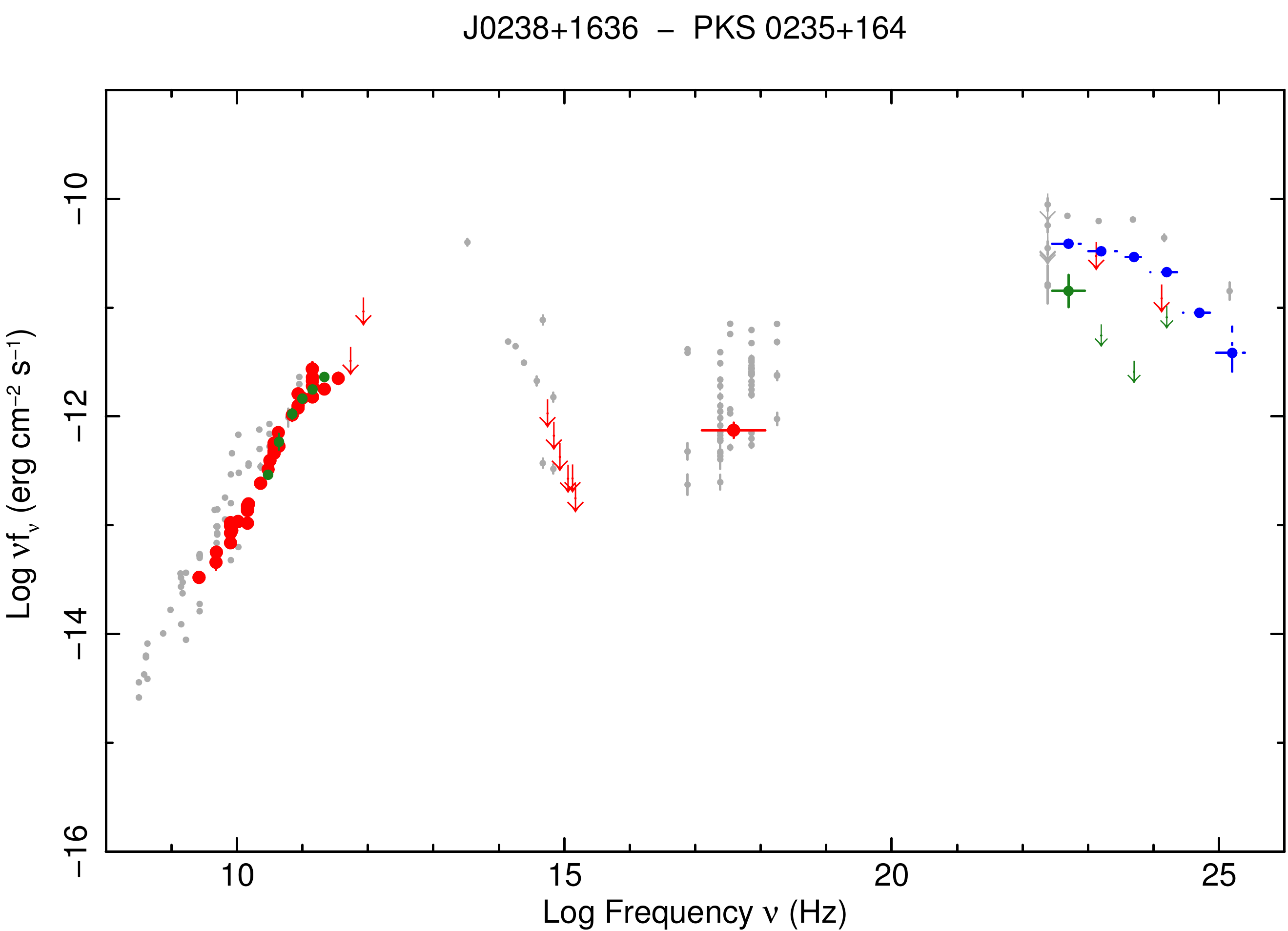}
\includegraphics[width=6.5cm,angle=-90]{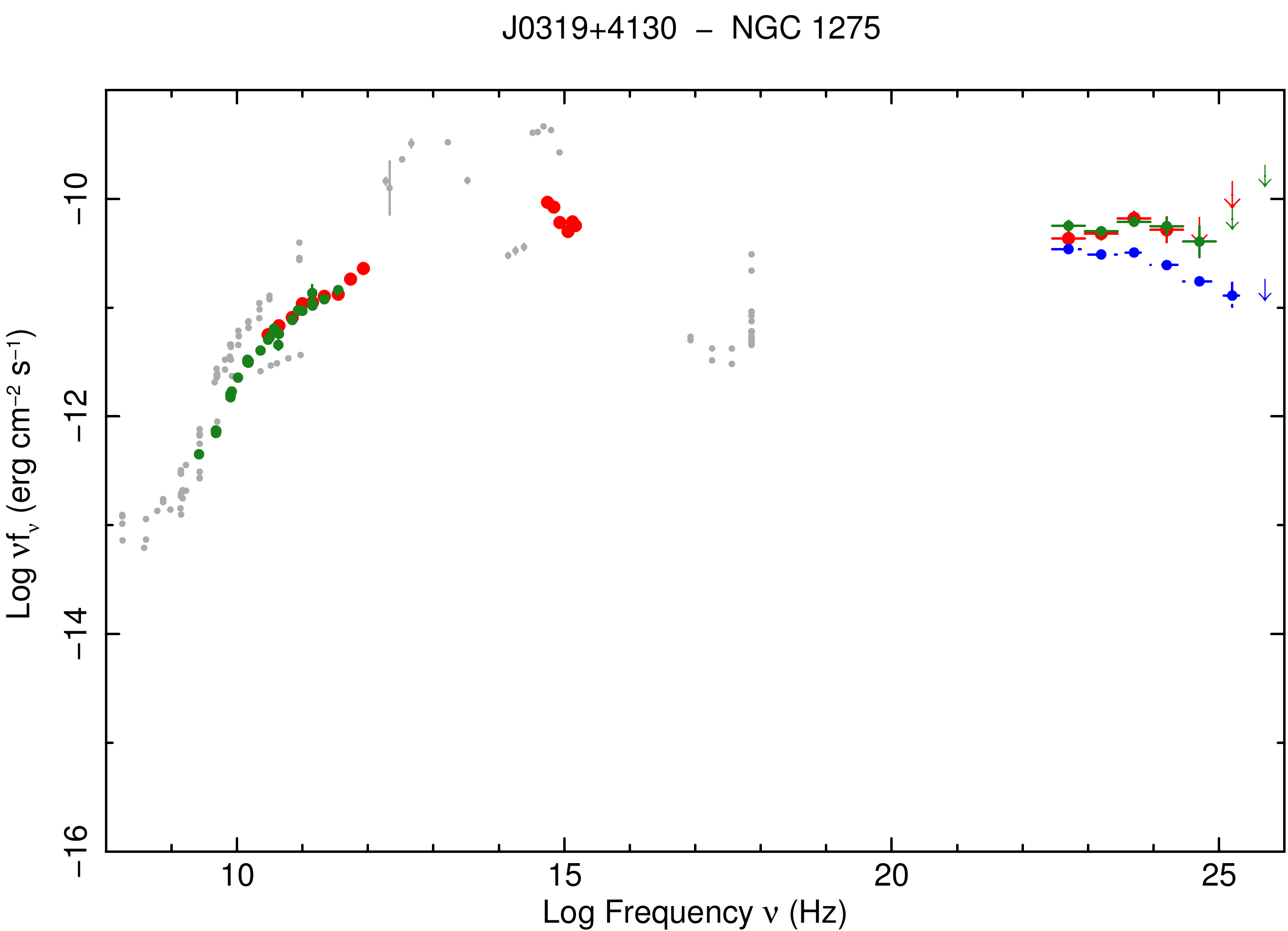}
\includegraphics[width=6.5cm,angle=-90]{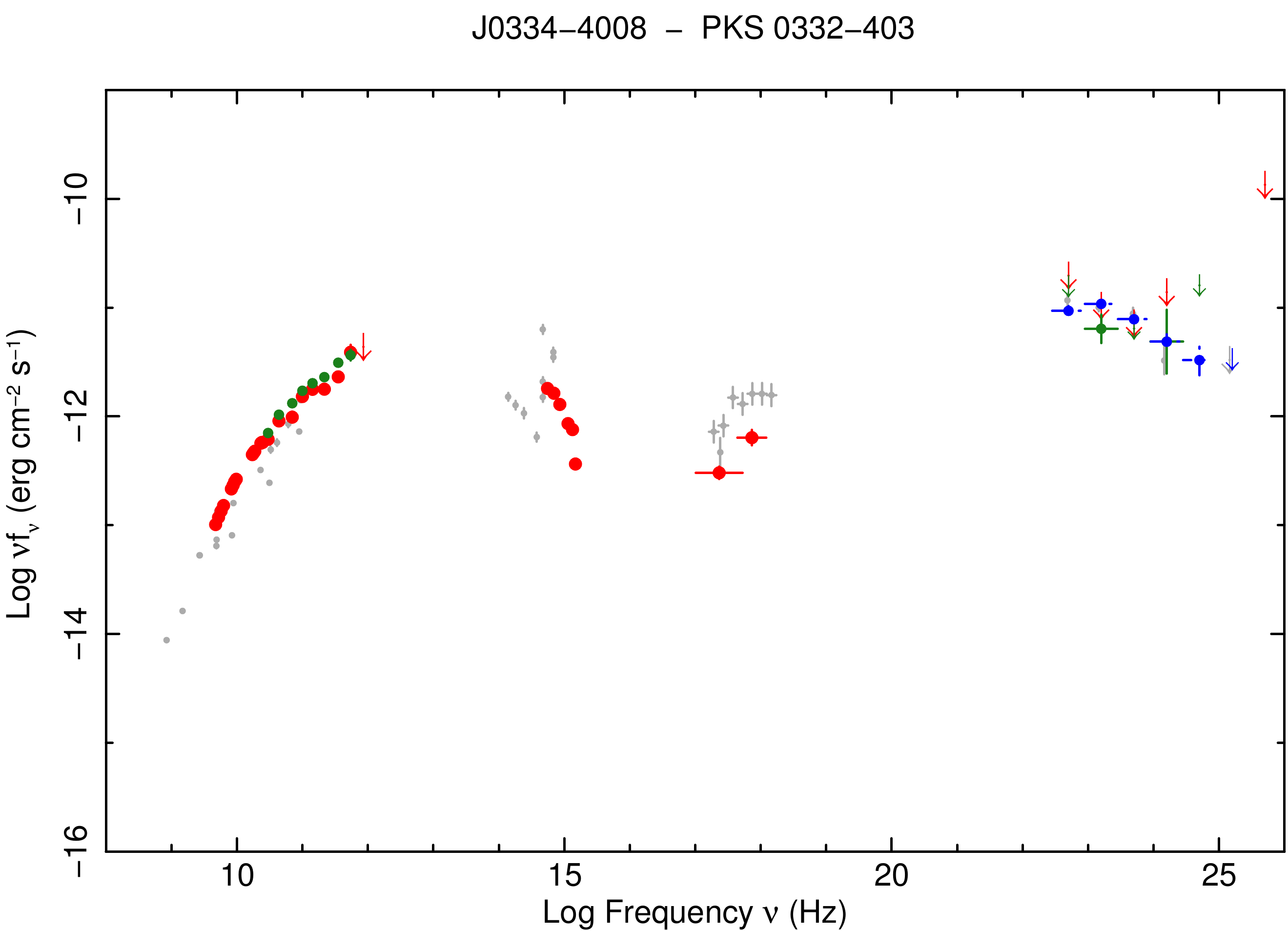}
\includegraphics[width=6.5cm,angle=-90]{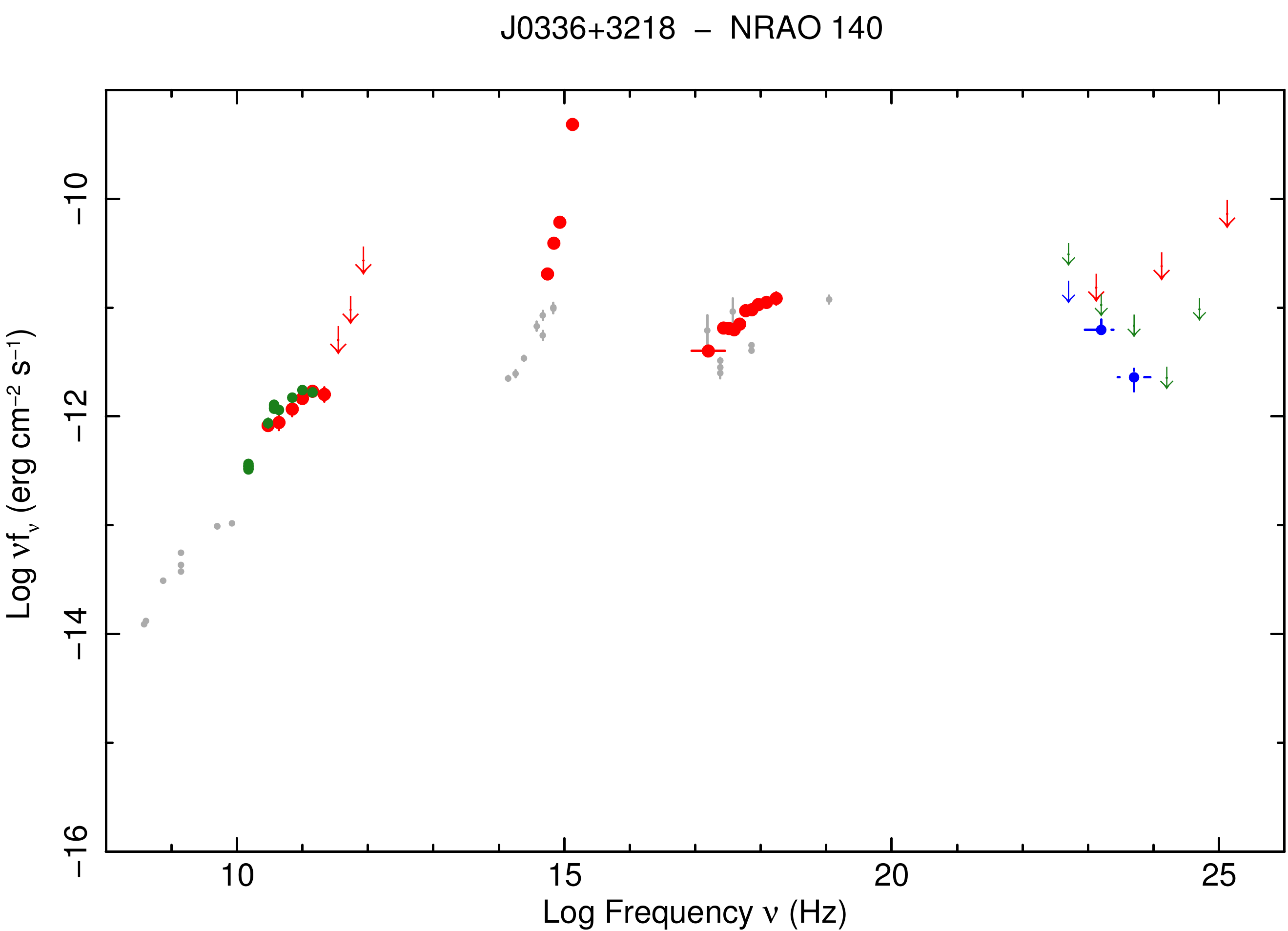}
\includegraphics[width=6.5cm,angle=-90]{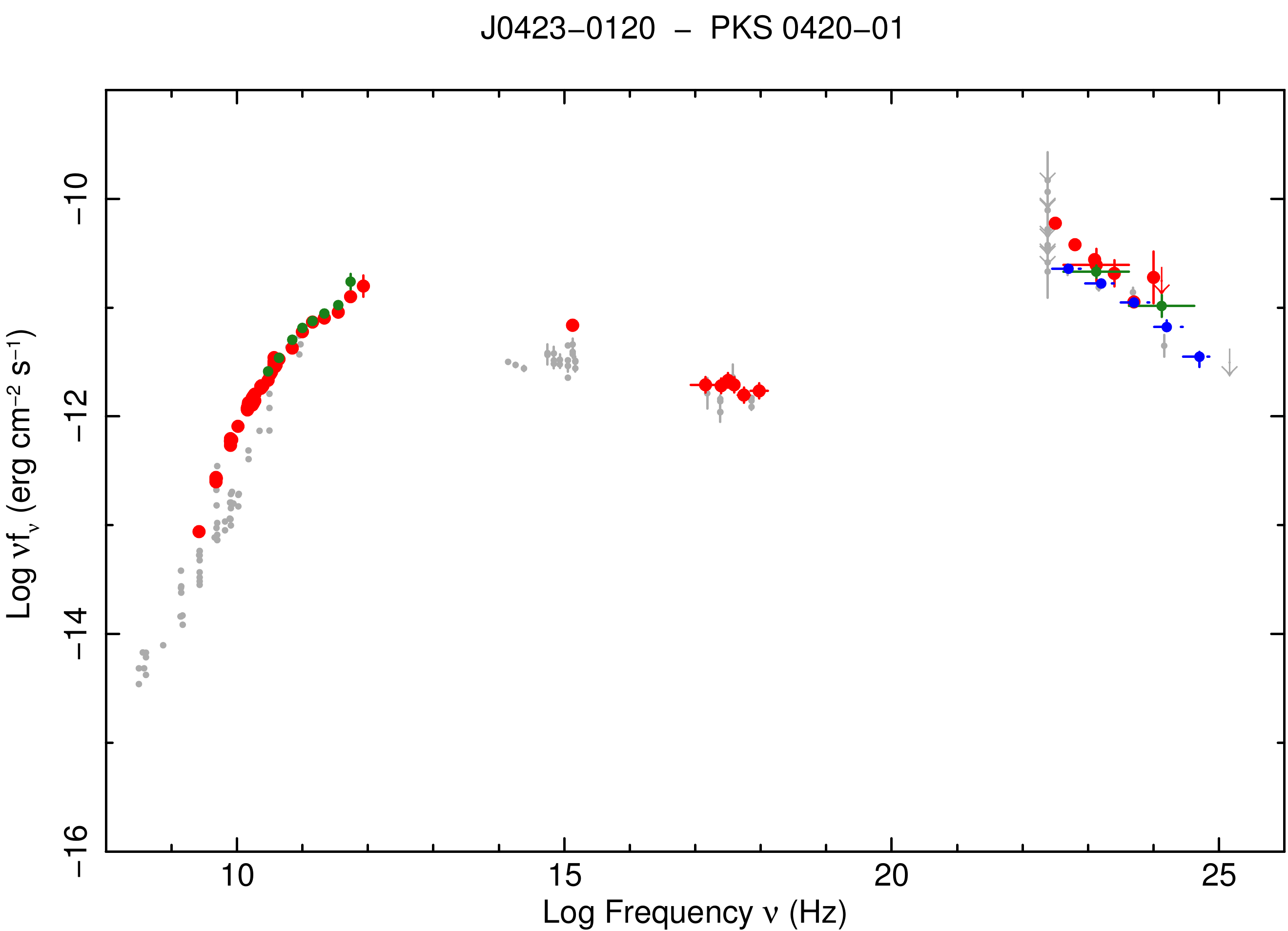}
\caption{The SED of 4C\,28.07 (J0237+2848, top left), PKS\,0235+164 (J0238+1636, top right),
NGC\,1275 (J0319+4130, middle left), PKS\,0332$-$403 (J0334$-$4008, middle right),
NRAO\,140 (J0336+3218, bottom left) and PKS\,0420$-$01 (J0423$-$0120, bottom right). 
Simultaneous data are shown in red; quasi-simultaneous data, i.e. {\it Fermi} data
integrated over 2 months, {\it Planck} ERCSC and non-simultaneous ground based observations
are shown in green; {\it Fermi} data integrated over 27 months are shown in blue;
literature or archival data are shown in light gray.}
\label{fig:sed7}
\end{figure*}

\begin{figure*}
\centering
\includegraphics[width=6.5cm,angle=-90]{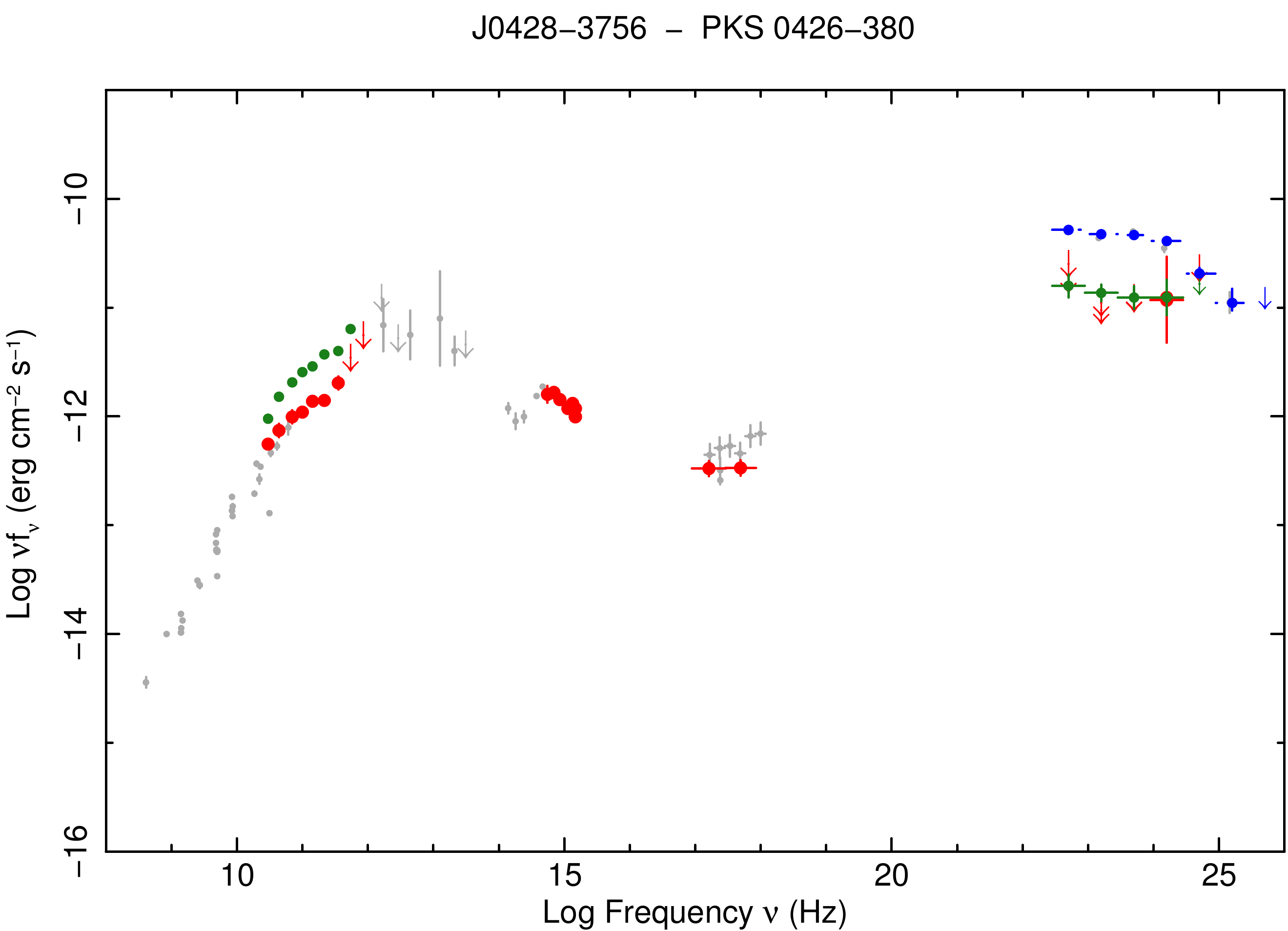}
\includegraphics[width=6.5cm,angle=-90]{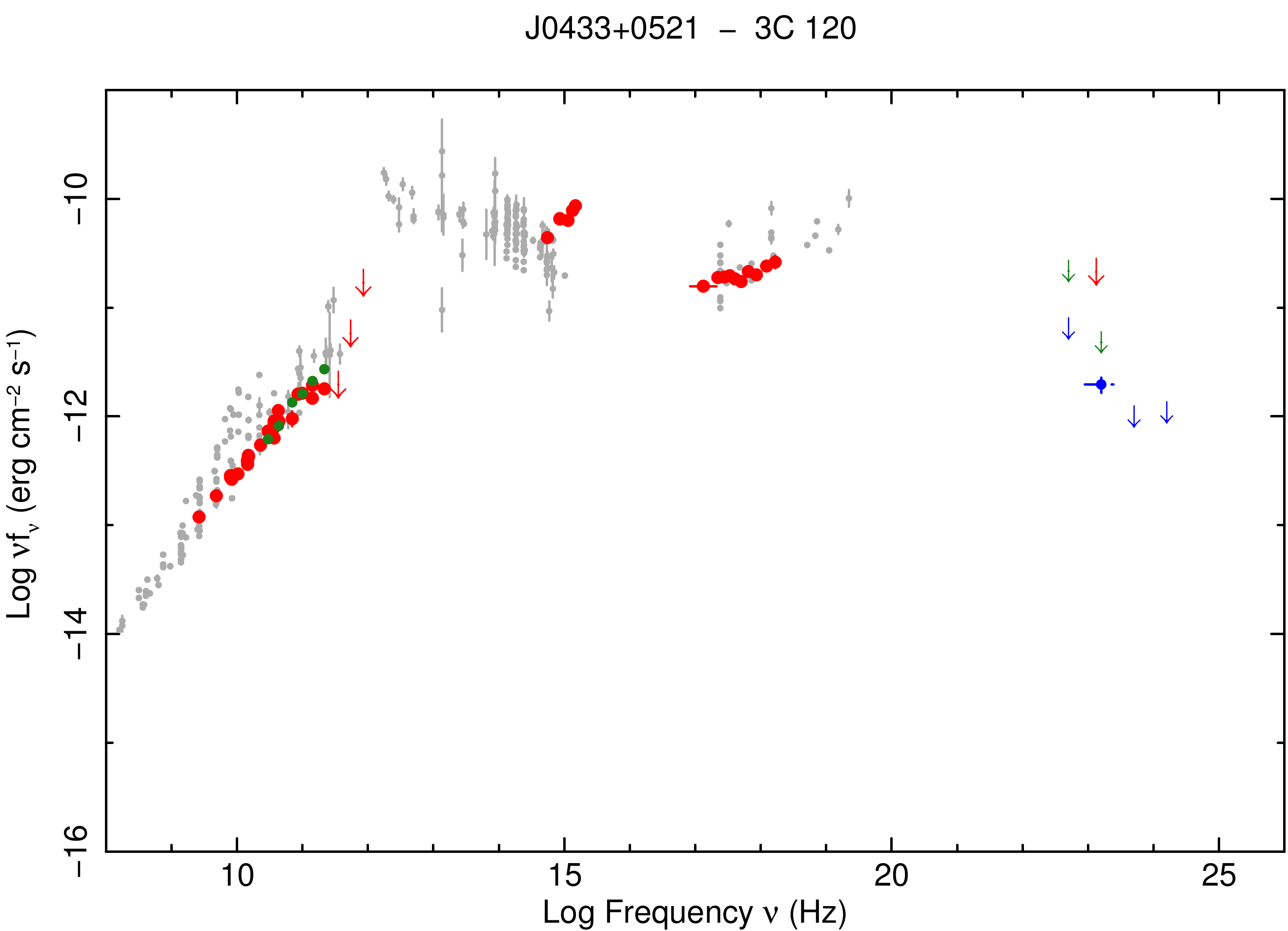}
\includegraphics[width=6.5cm,angle=-90]{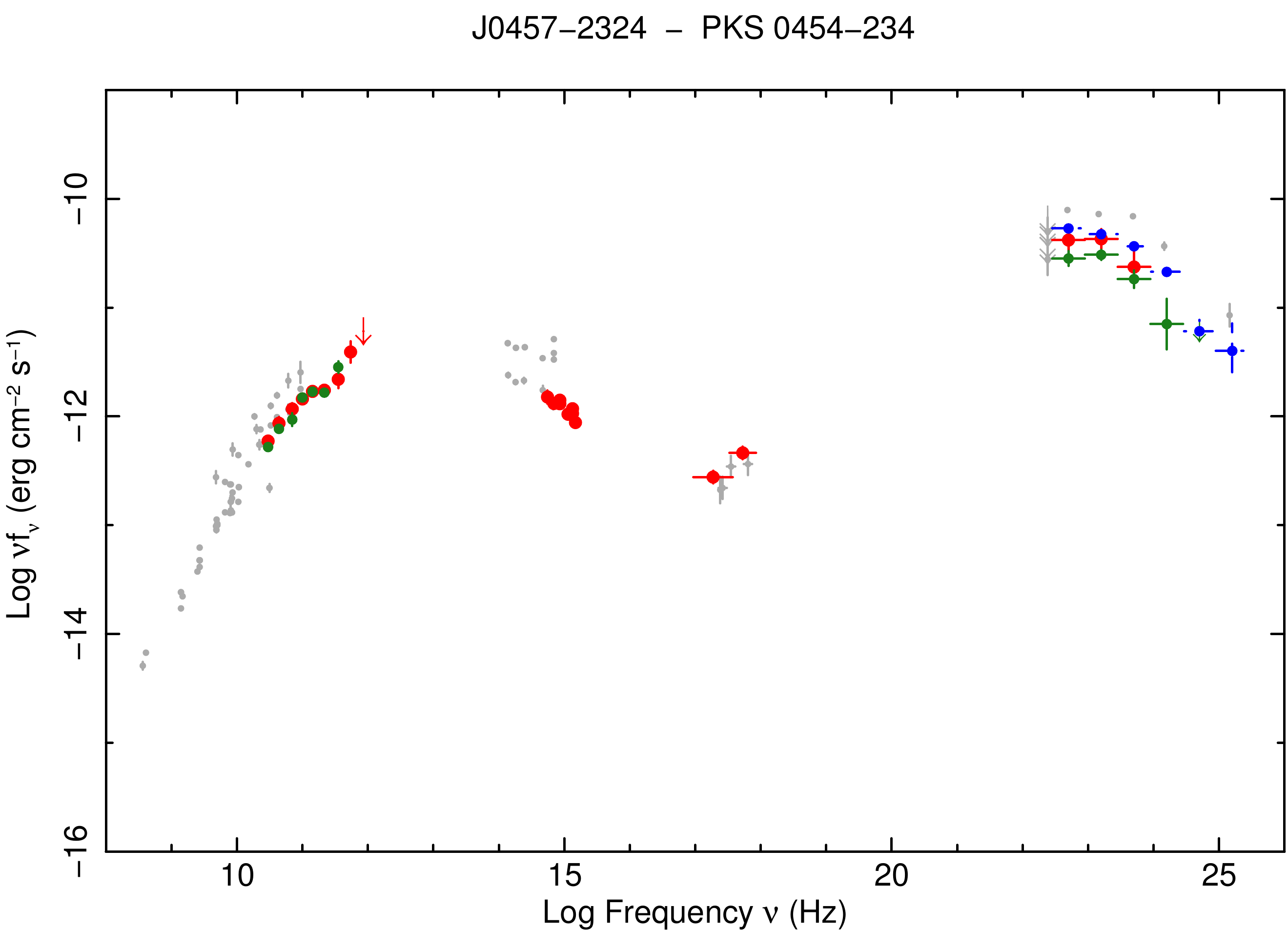}
\includegraphics[width=6.5cm,angle=-90]{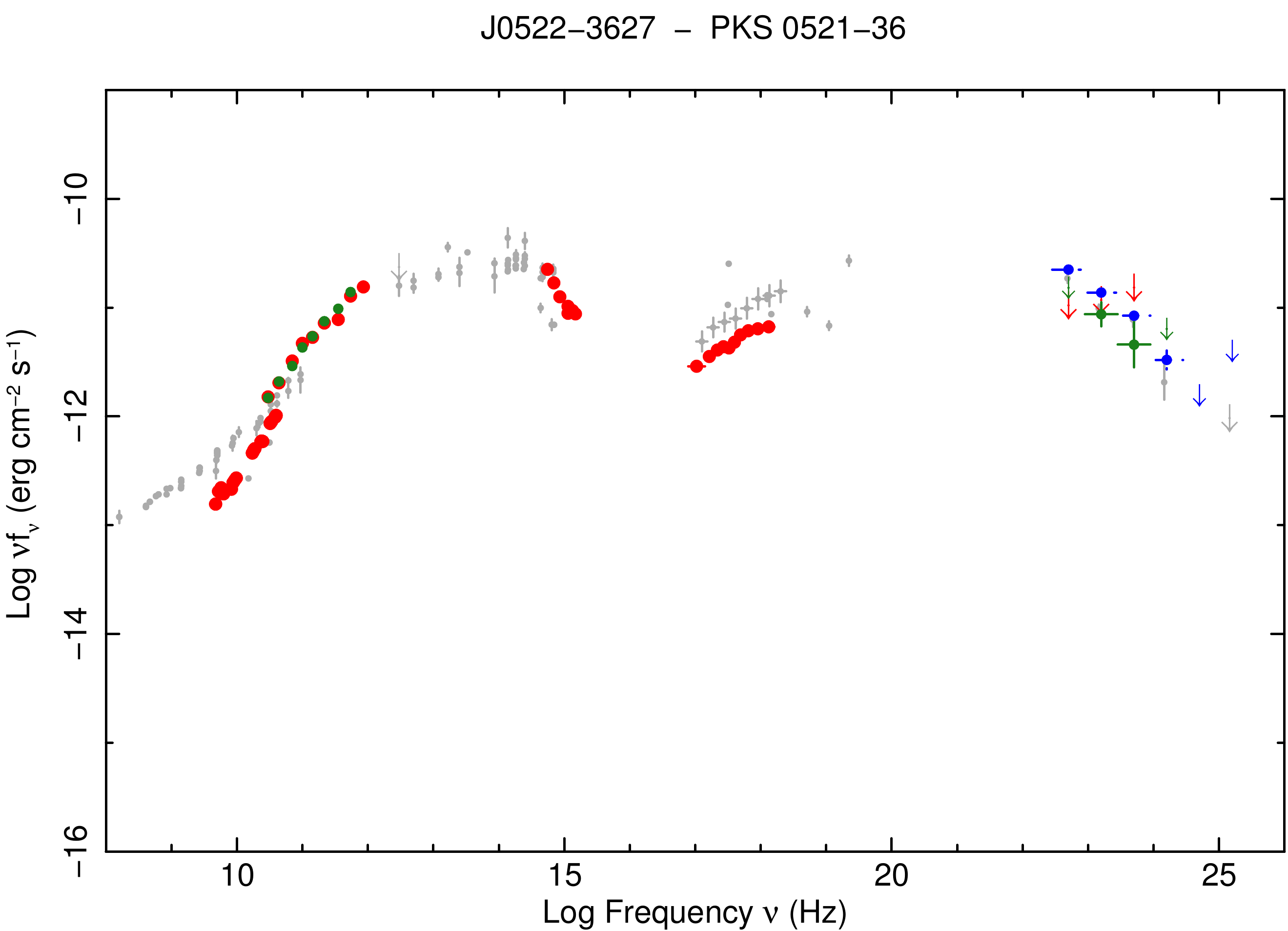}
\includegraphics[width=6.5cm,angle=-90]{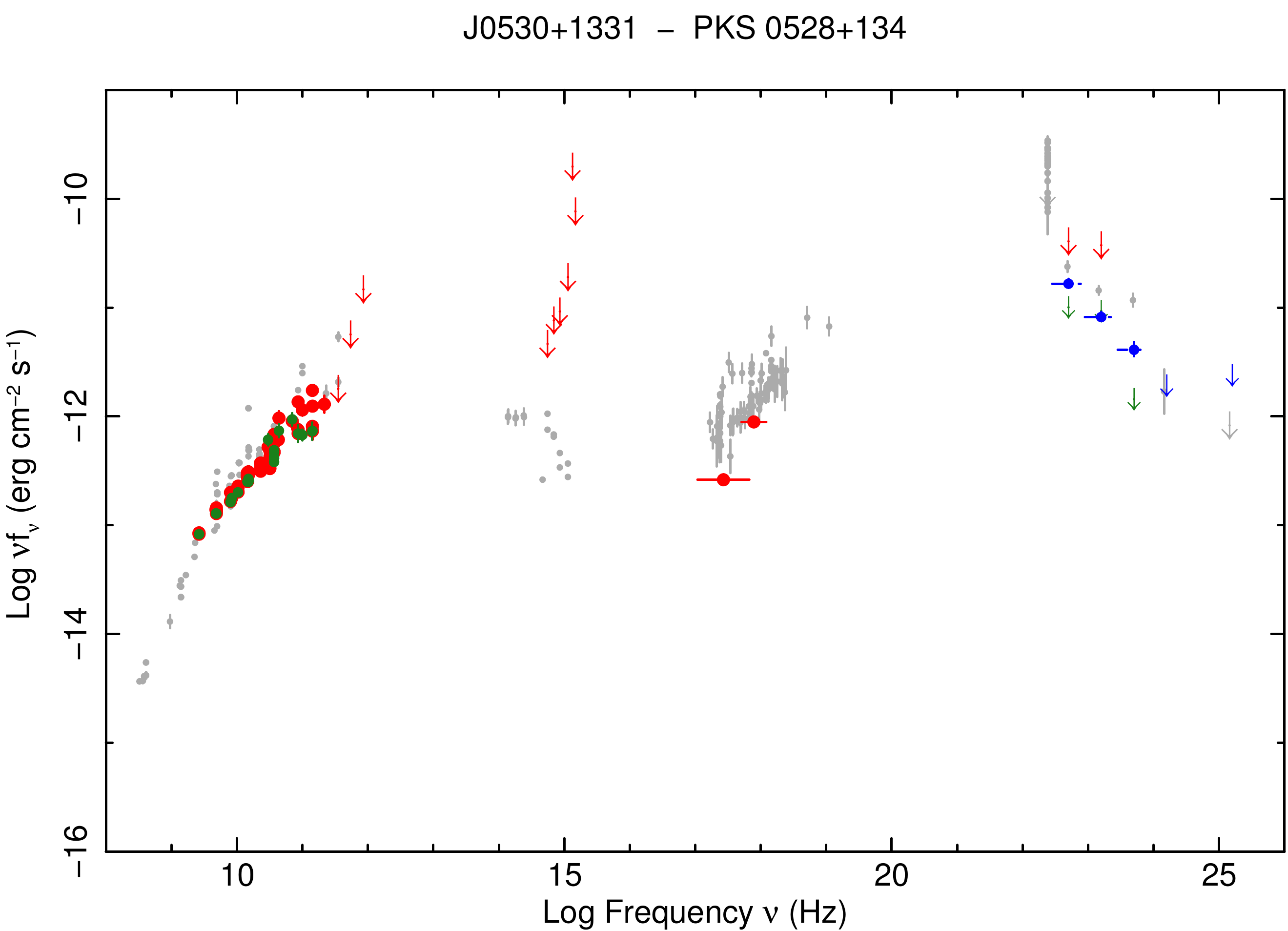}
\includegraphics[width=6.5cm,angle=-90]{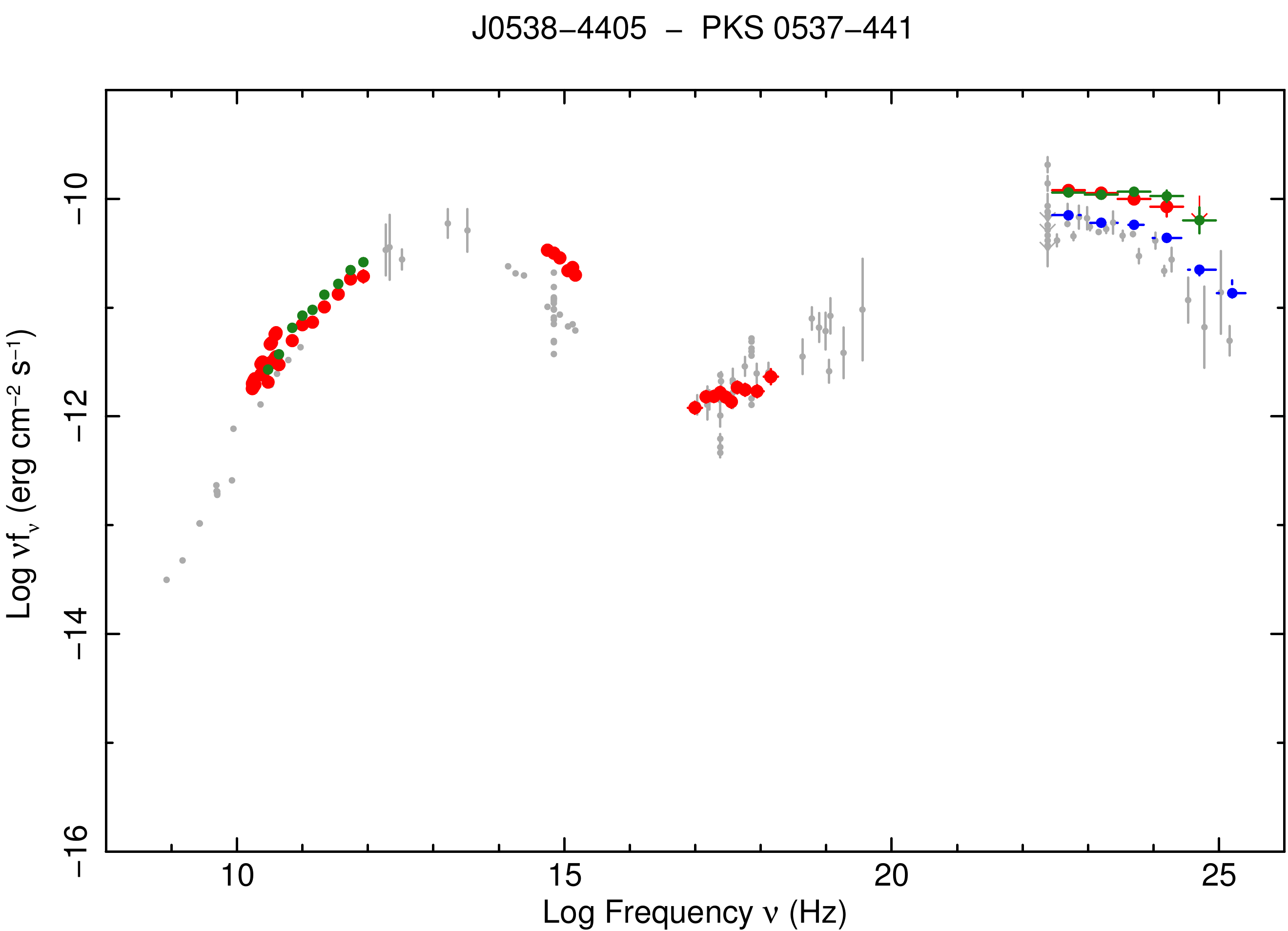}
\caption{The SED of PKS\,0426$-$380 (J0428$-$3756, top left), 3C\,120 (J0433+0521, top right),
PKS\,0454$-$234 (J0457$-$2324, middle left), PKS\,0521$-$36 (J0522$-$3627, middle right),
PKS\,0528+134 (J0530+1331, bottom left), and PKS\,0537$-$441 (J0538$-$4405, bottom right). 
Simultaneous data are shown in red; quasi-simultaneous data, i.e. {\it Fermi} data
integrated over 2 months, {\it Planck} ERCSC and non-simultaneous ground based observations
are shown in green; {\it Fermi} data integrated over 27 months are shown in blue;
literature or archival data are shown in light gray.}
\label{fig:sed10}
\end{figure*}

\begin{figure*}
\centering
\includegraphics[width=6.5cm,angle=-90]{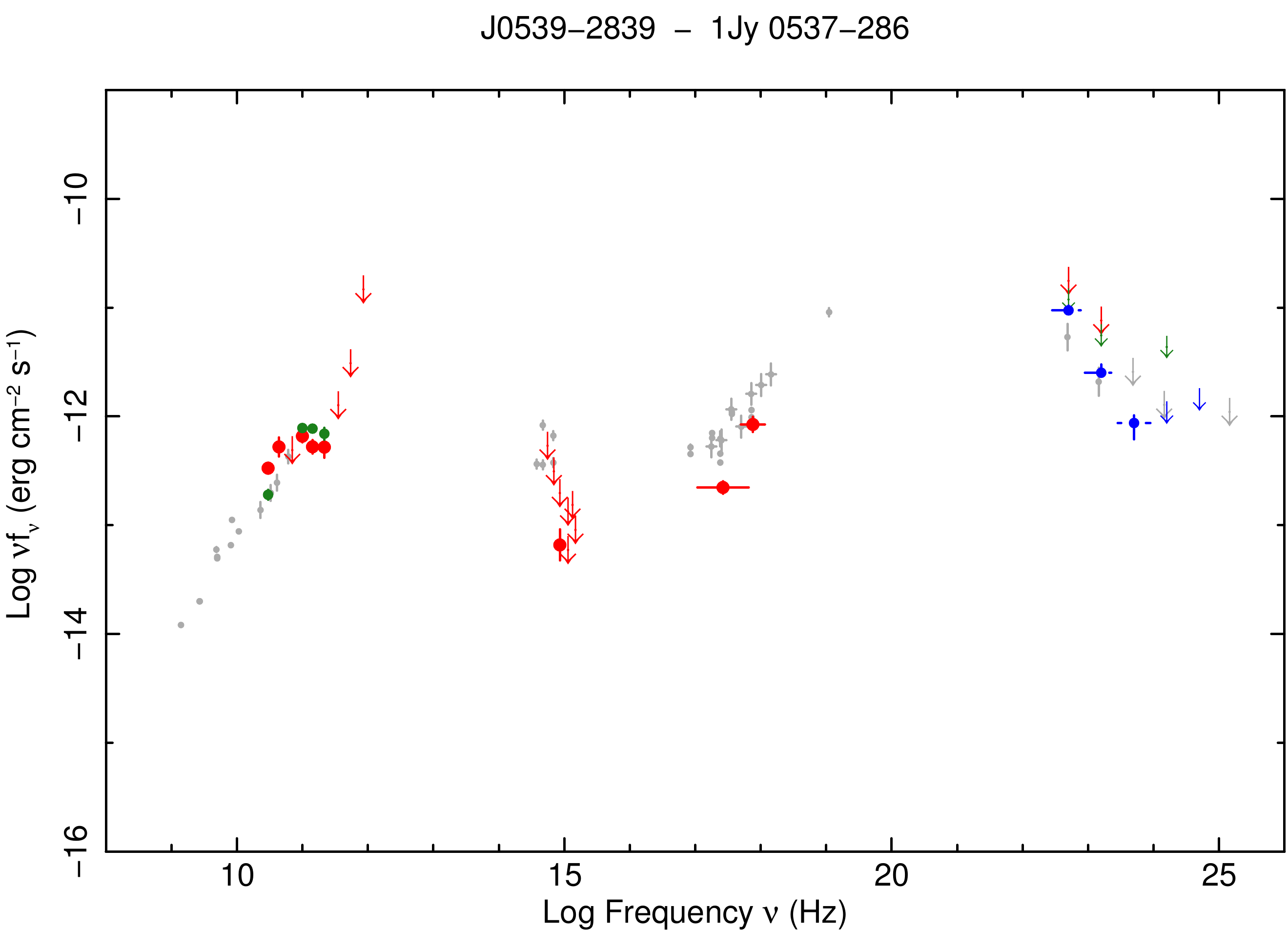}
\includegraphics[width=6.5cm,angle=-90]{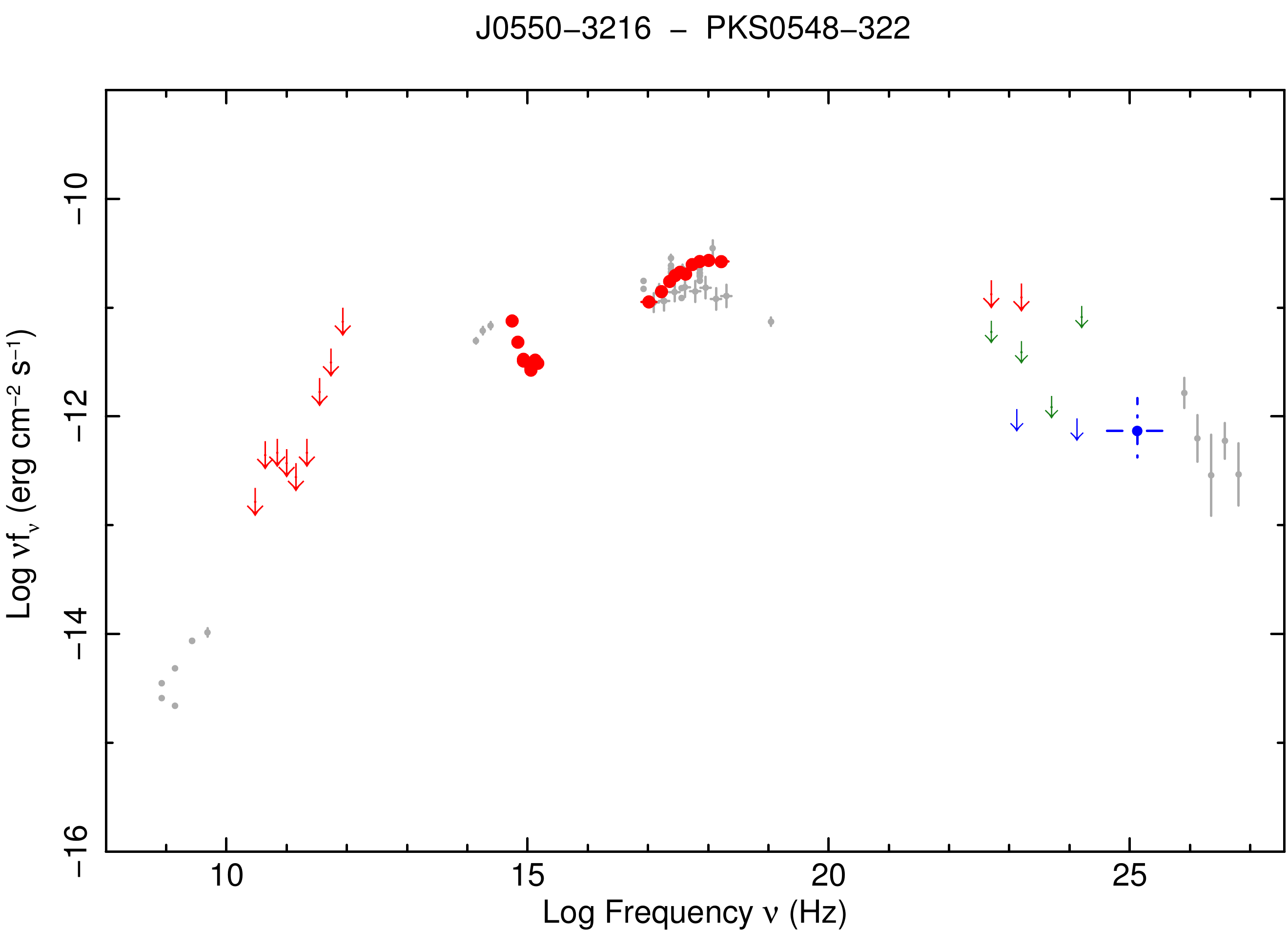}
\includegraphics[width=6.5cm,angle=-90]{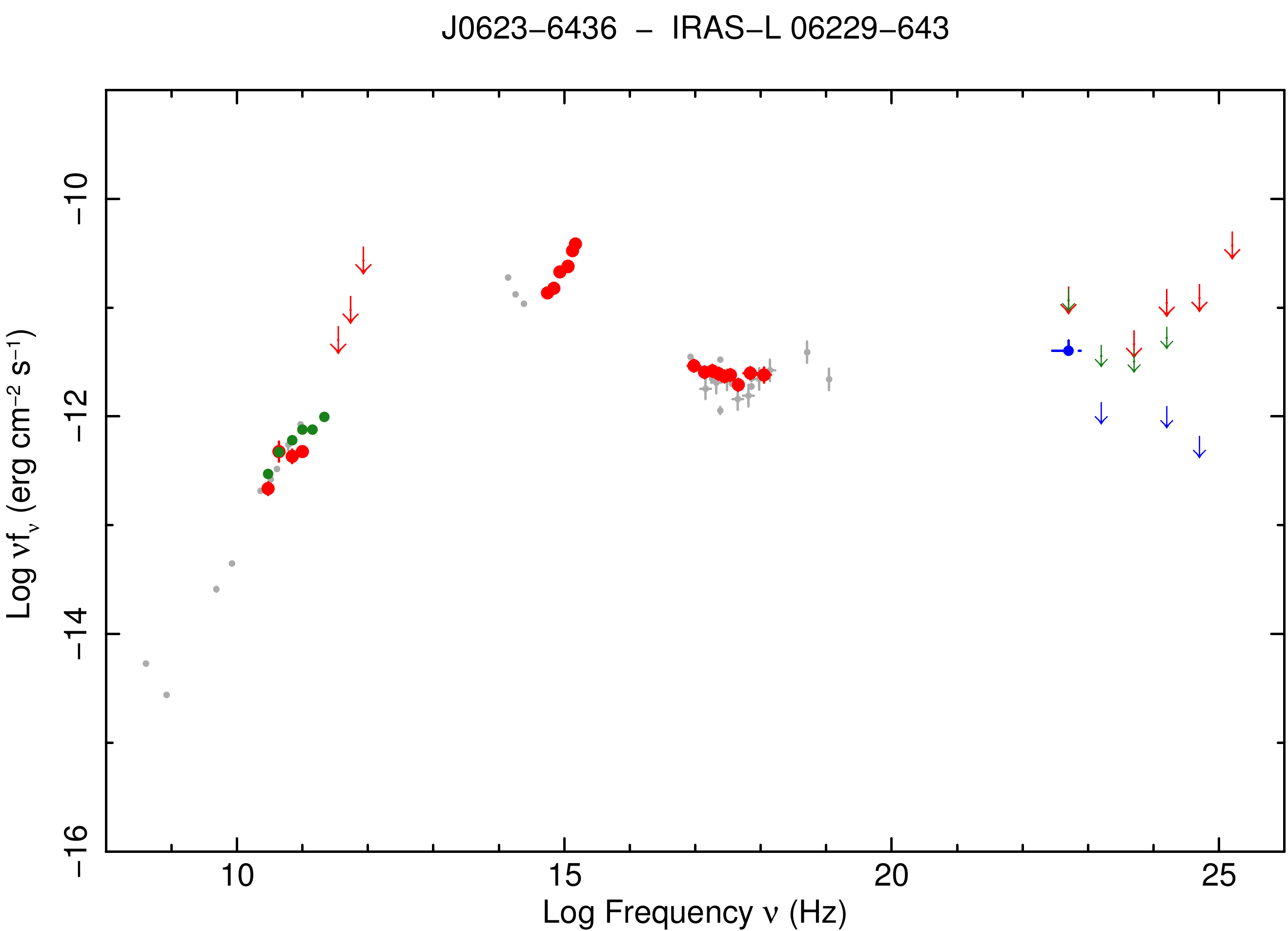}
\includegraphics[width=6.5cm,angle=-90]{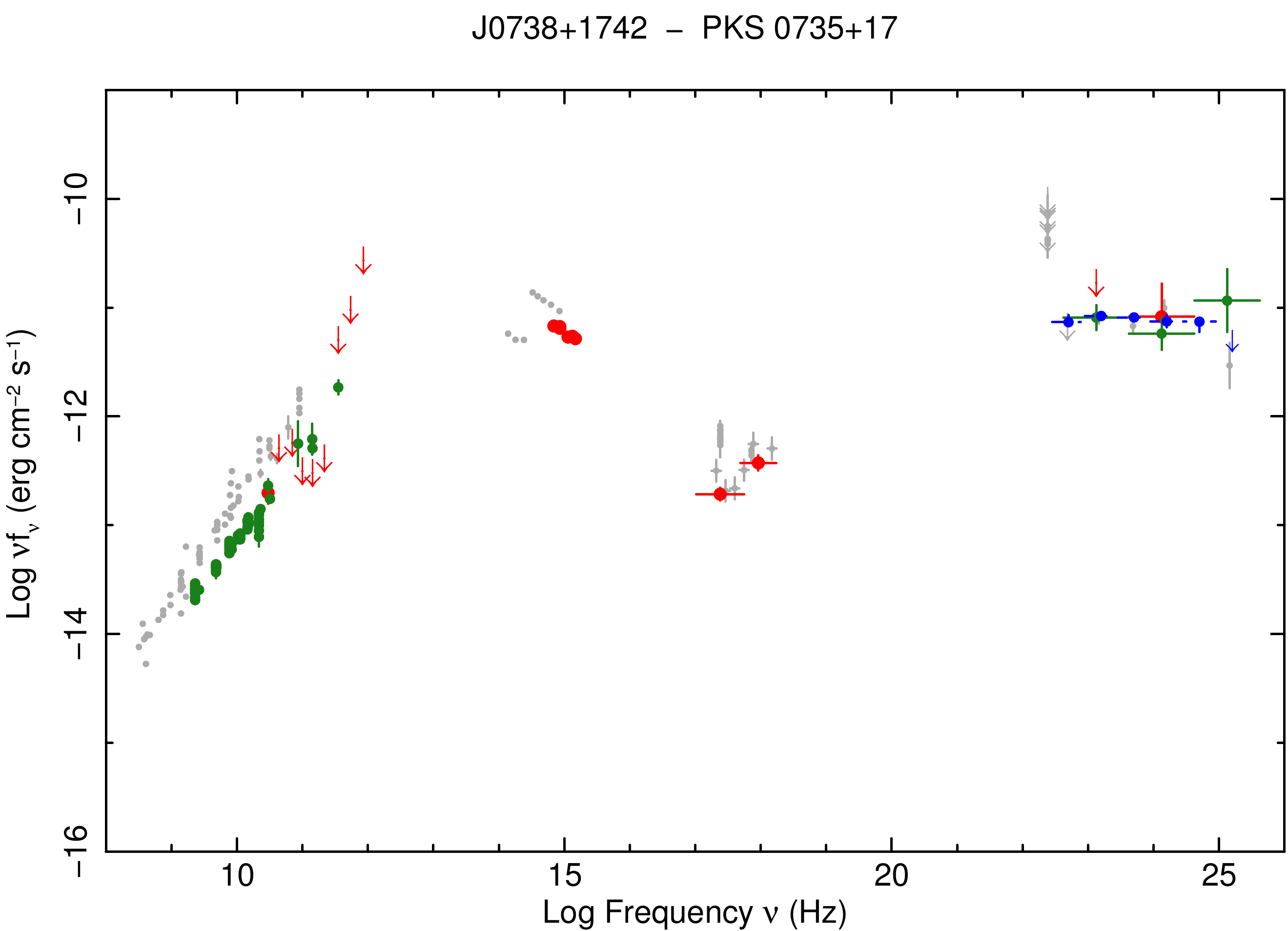}
\includegraphics[width=6.5cm,angle=-90]{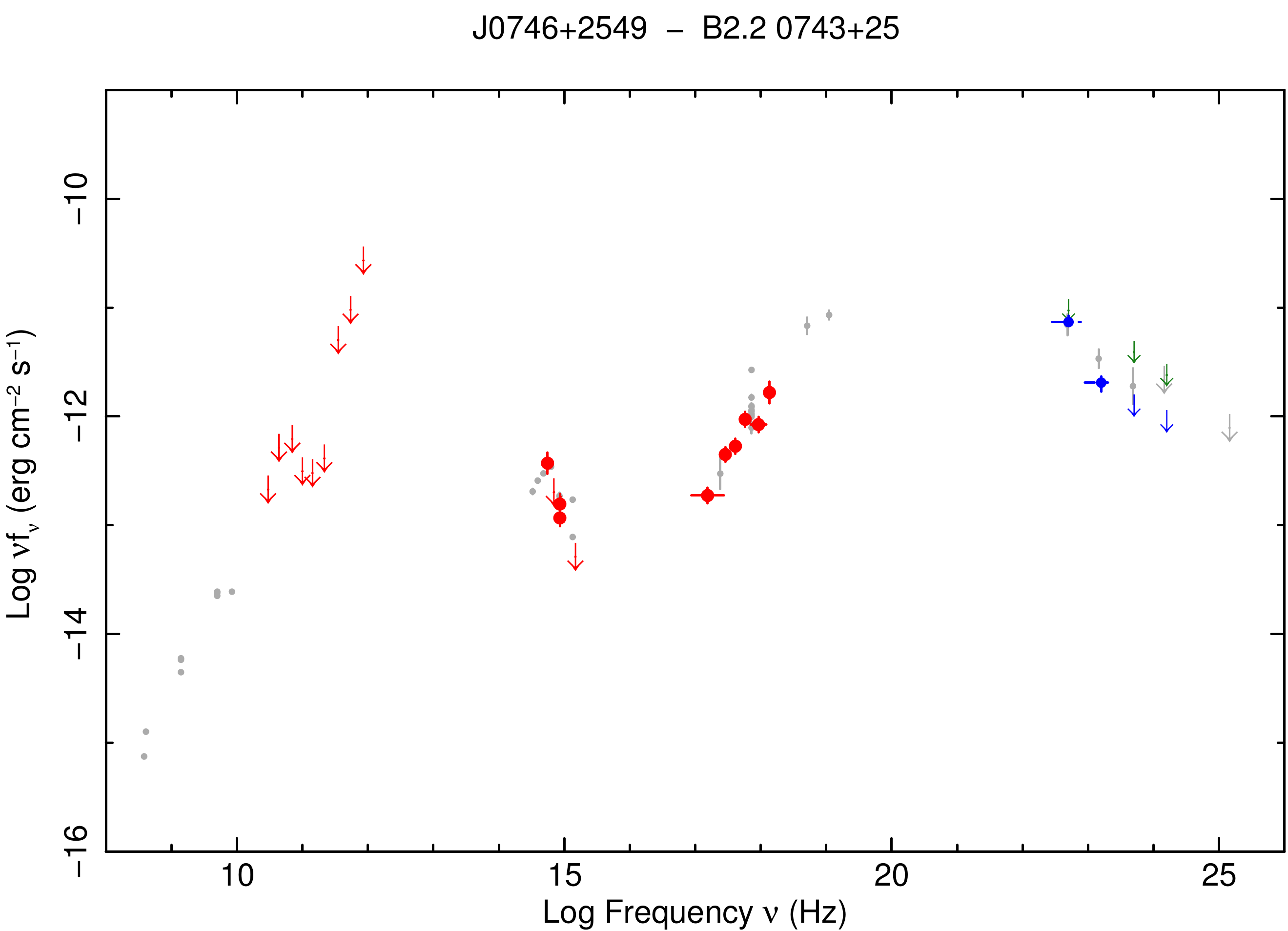}
\includegraphics[width=6.5cm,angle=-90]{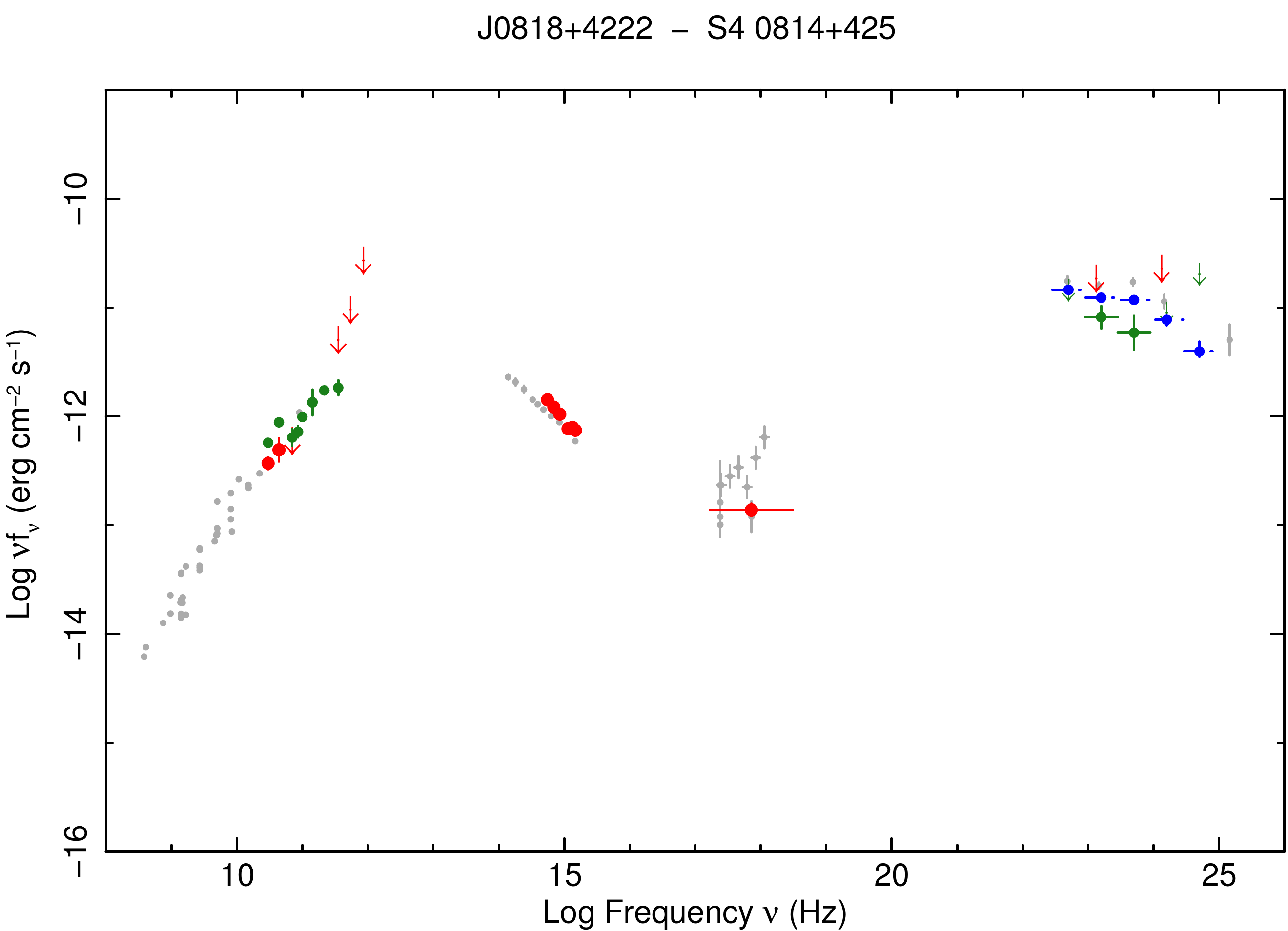}
\caption{The SED of 1Jy\,0537$-$286 (J0539$-$2839, top left), PKS\,0548$-$322 (J0550$-$3216, top right),
 IRAS-L\,06229$-$643 (J0623$-$6436, middle left), PKS\,0735+17 (J0738+1742, middle right),
 B2.2\,0743+25 (J0746+2549, bottom left), and S4\,0814+425 (J0818+4222, bottom right). 
Simultaneous data are shown in red; quasi-simultaneous data, i.e. {\it Fermi} data
integrated over 2 months, {\it Planck} ERCSC and non-simultaneous ground based observations
are shown in green; {\it Fermi} data integrated over 27 months are shown in blue;
literature or archival data are shown in light gray.}
\label{fig:sed13}
\end{figure*}

\begin{figure*}
\centering
\includegraphics[width=6.5cm,angle=-90]{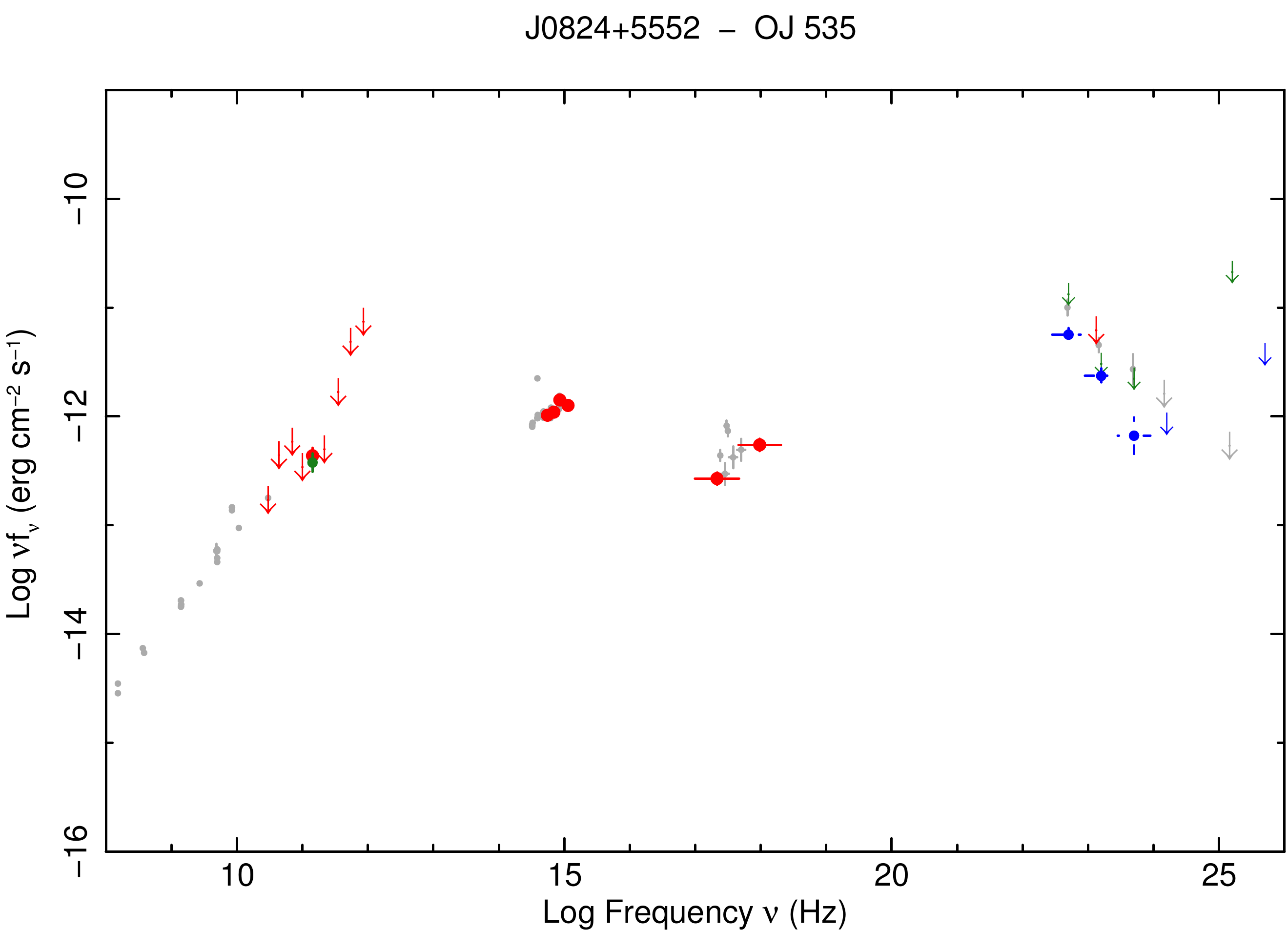}
\includegraphics[width=6.5cm,angle=-90]{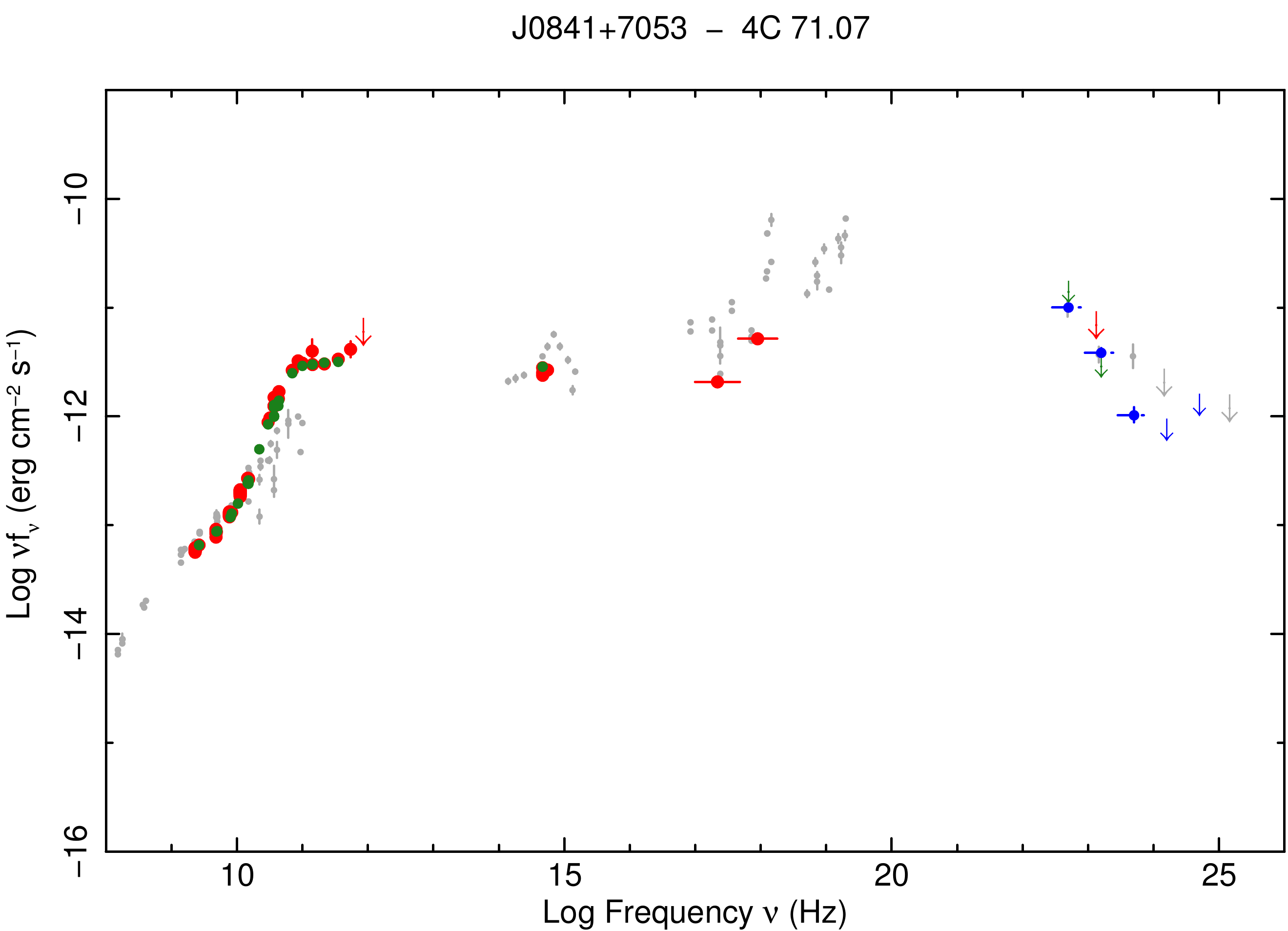}
\includegraphics[width=6.5cm,angle=-90]{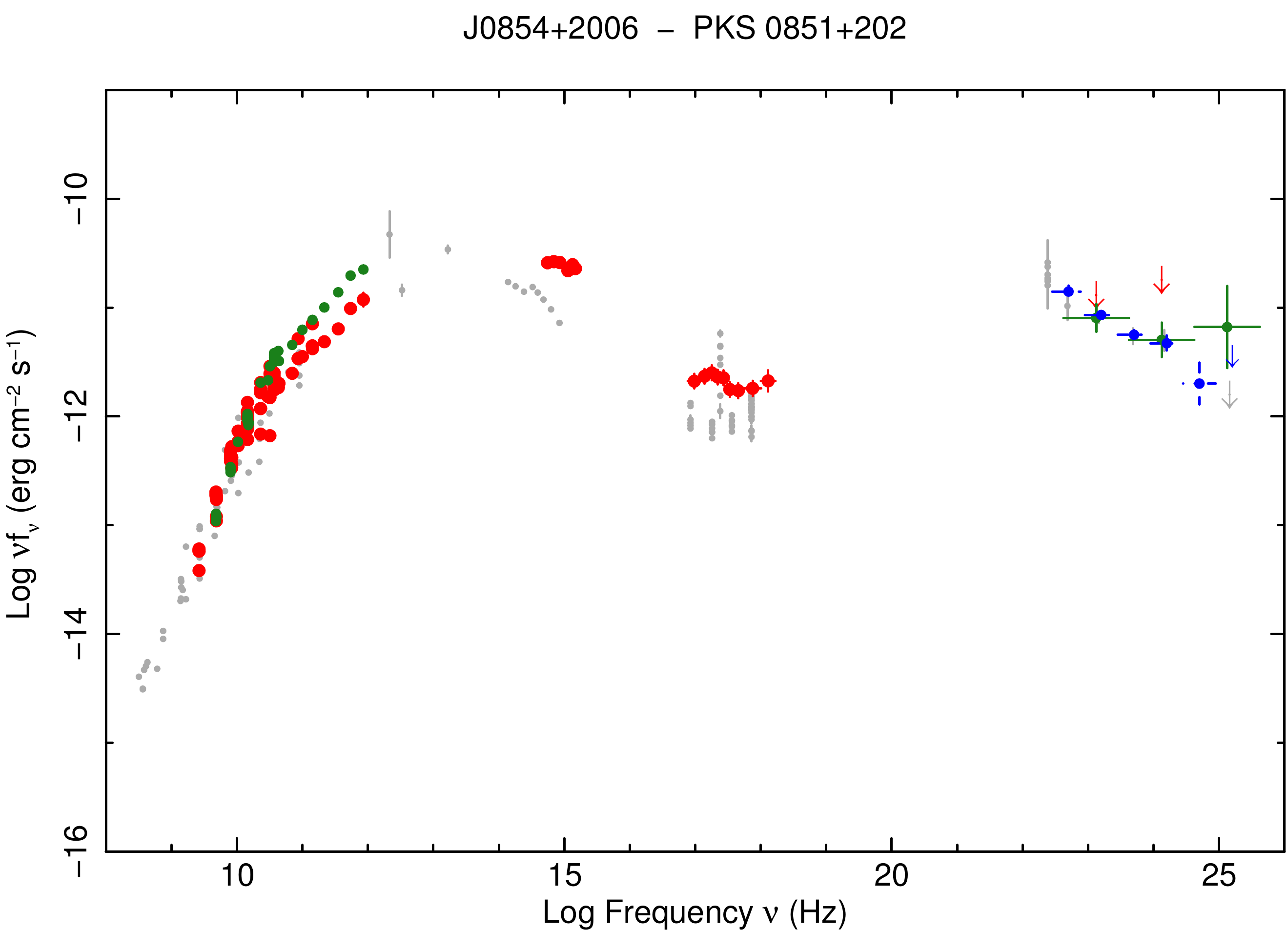}
\includegraphics[width=6.5cm,angle=-90]{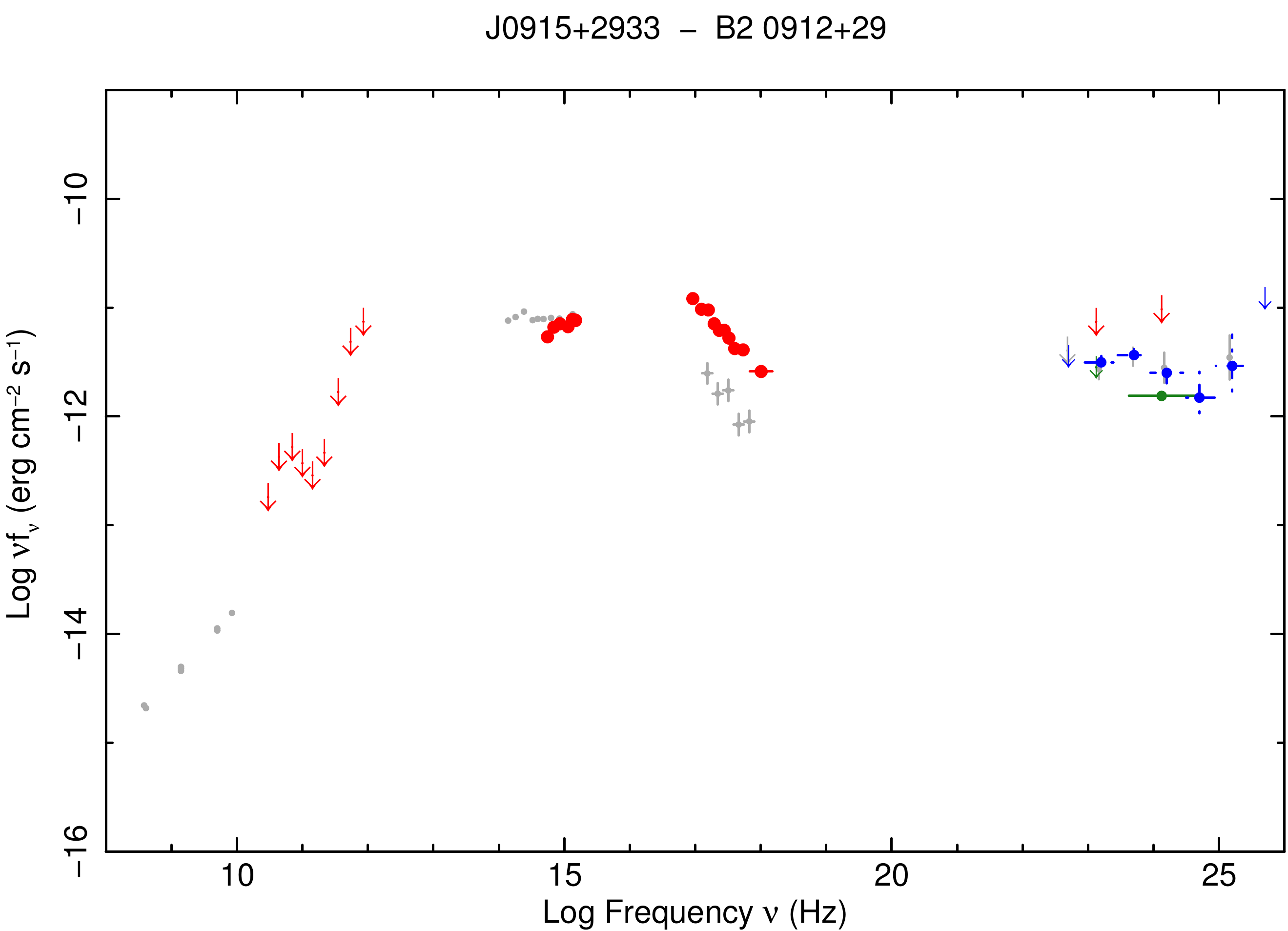}
\includegraphics[width=6.5cm,angle=-90]{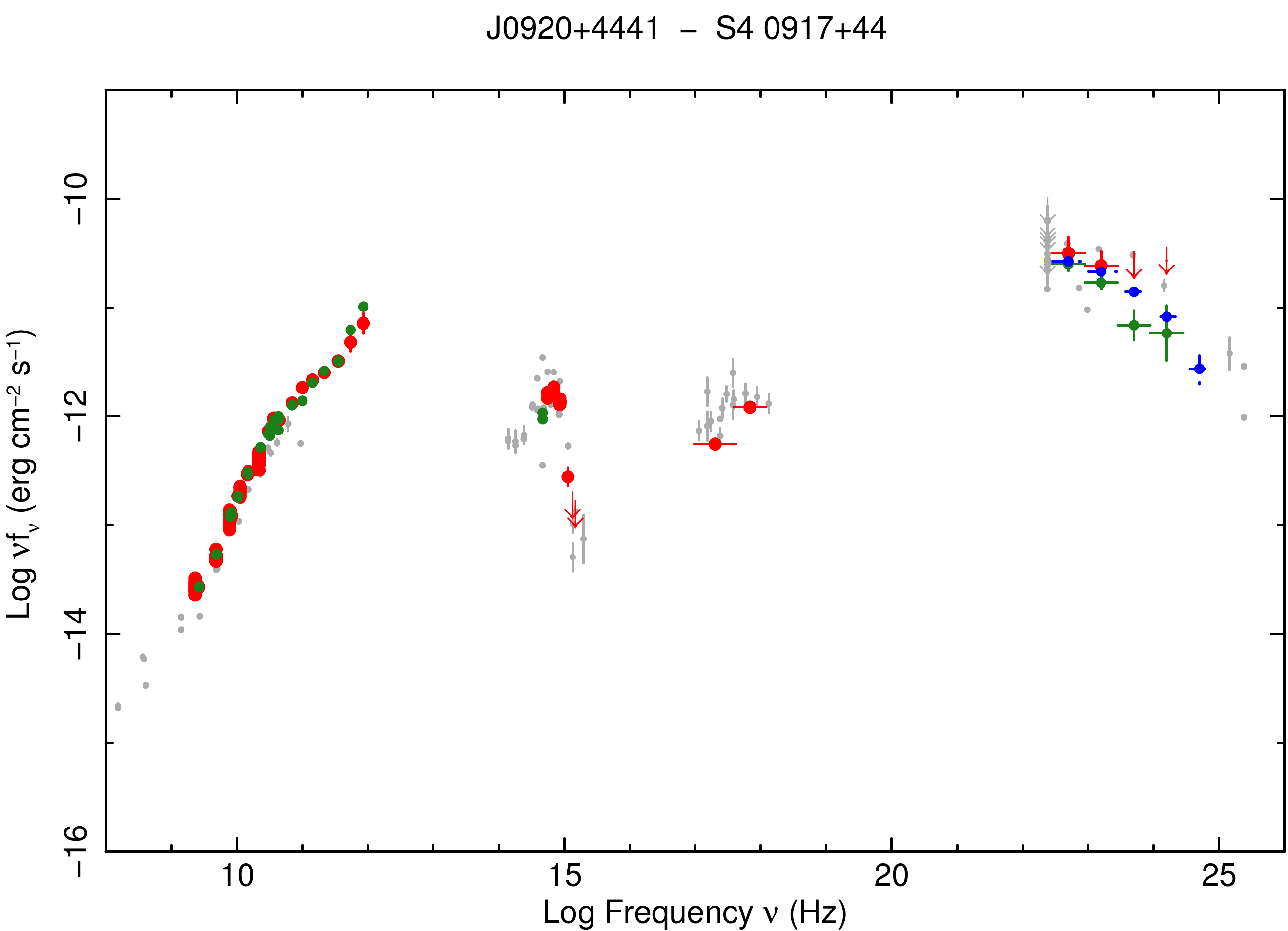}
\includegraphics[width=6.5cm,angle=-90]{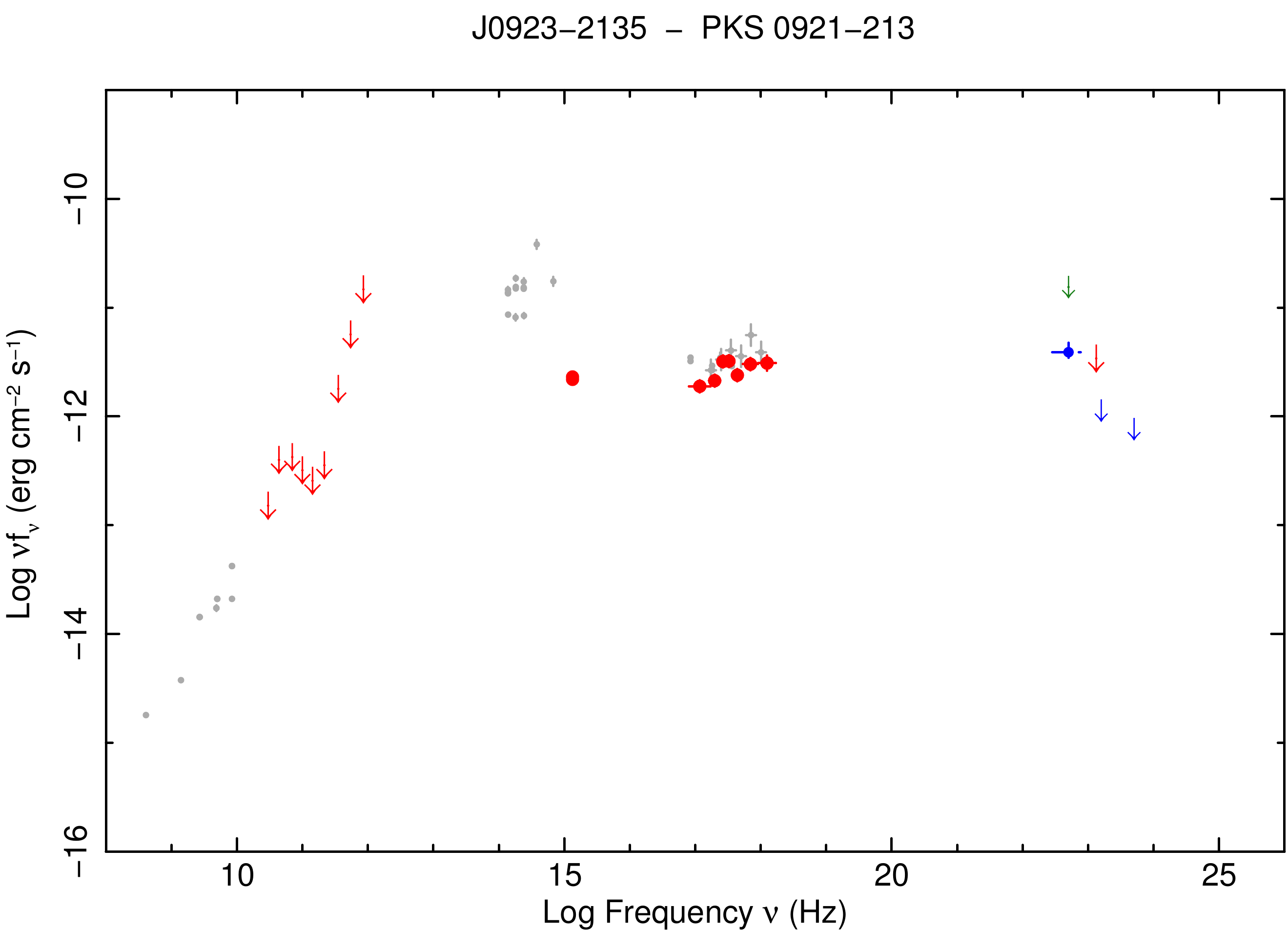}
\caption{The SED of OJ\,535 (J0824+5552, top left), 4C\,71.07 (J0841+7053, top right),
PKS\,0851+202 (J0854+2006, middle left), B2\,0912+29 (J0915+2933, middle right),
S4\,0917+44 (J0920+4441, bottom left), and PKS\,0921$-$213 (J0923$-$2135, bottom right). 
Simultaneous data are shown in red; quasi-simultaneous data, i.e. {\it Fermi} data
integrated over 2 months, {\it Planck} ERCSC and non-simultaneous ground based observations
are shown in green; {\it Fermi} data integrated over 27 months are shown in blue;
literature or archival data are shown in light gray.}
\label{fig:sed16}
\end{figure*}

\begin{figure*}
\centering
\includegraphics[width=6.5cm,angle=-90]{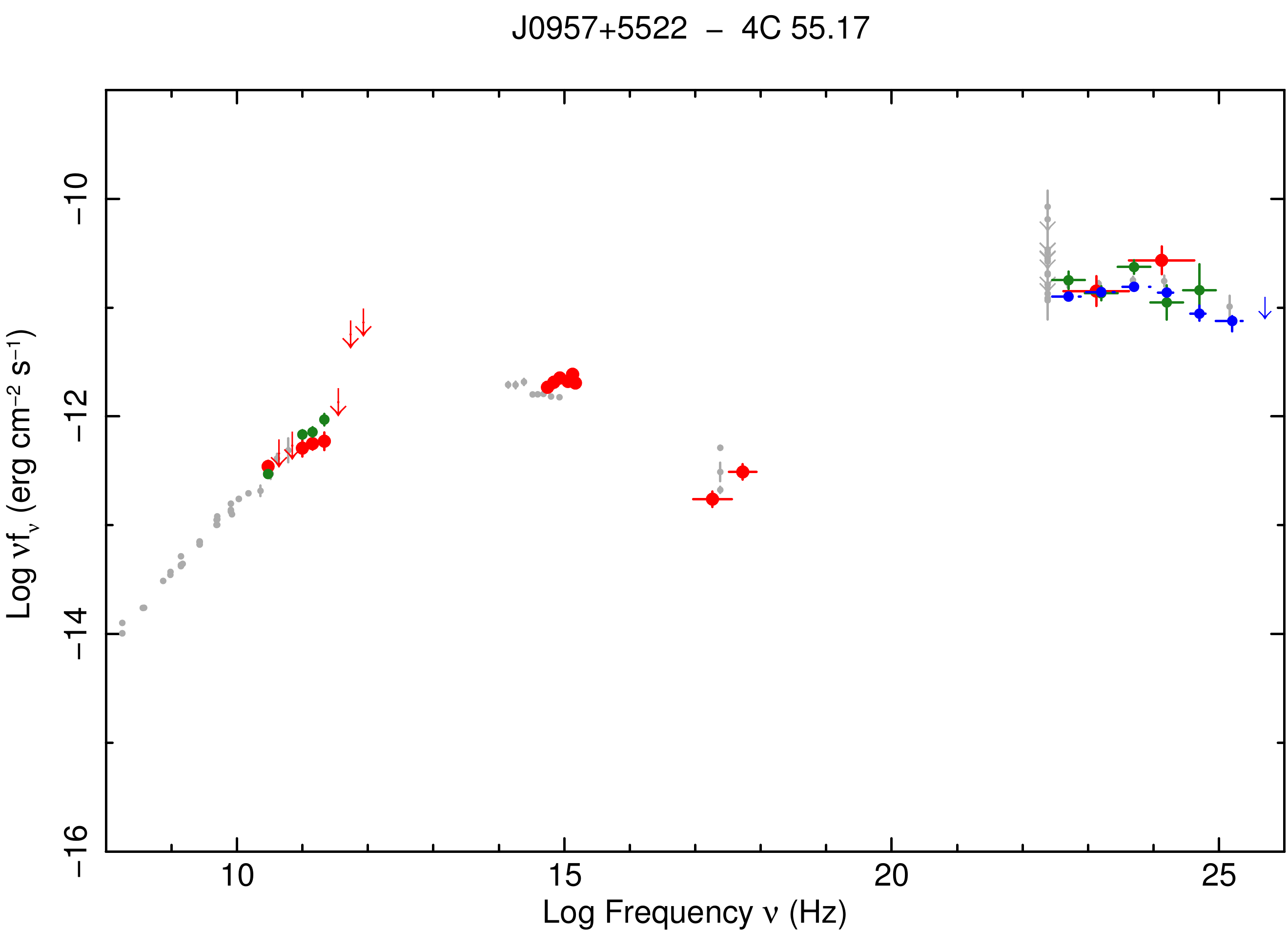}
\includegraphics[width=6.5cm,angle=-90]{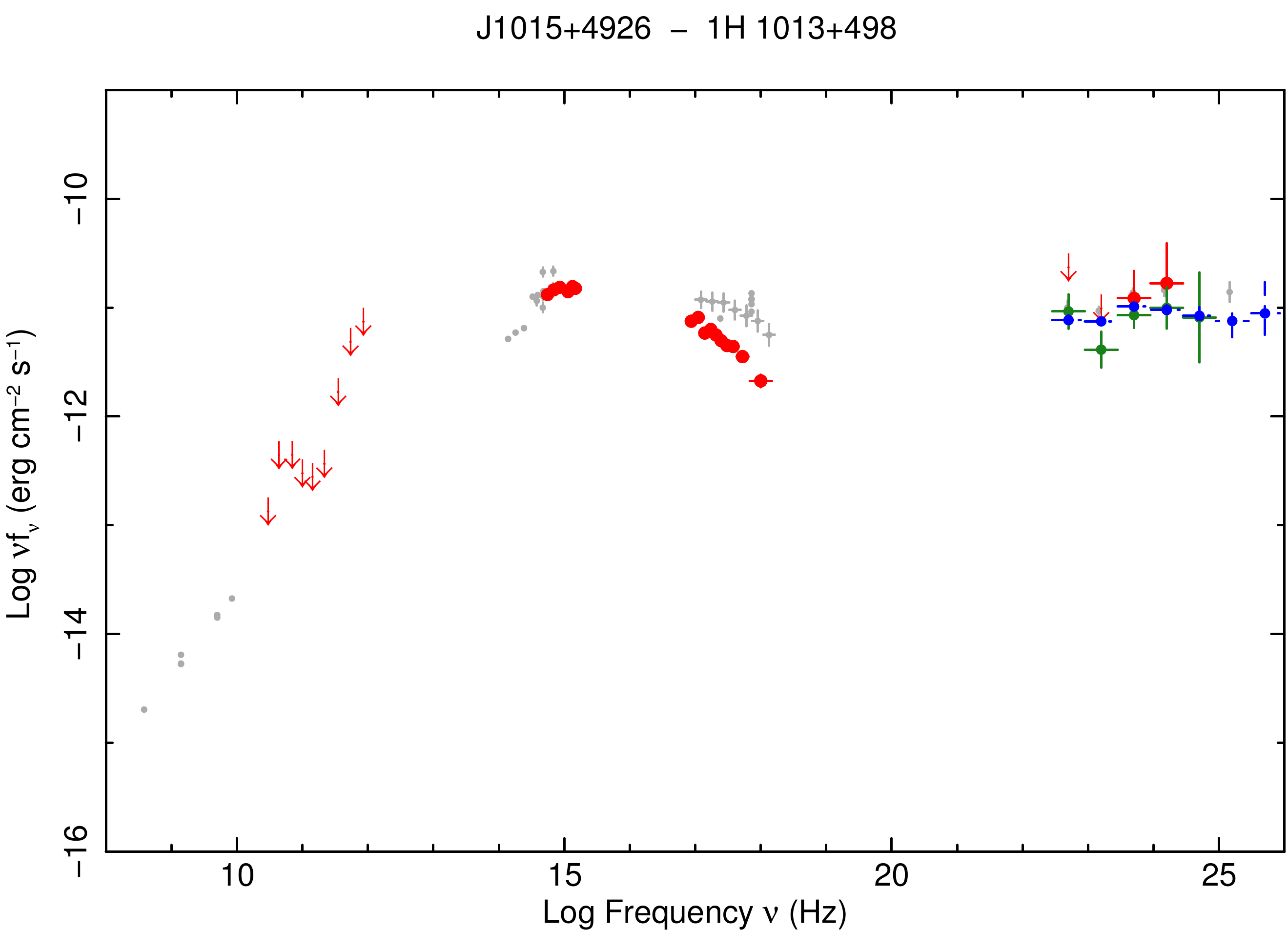}
\includegraphics[width=6.5cm,angle=-90]{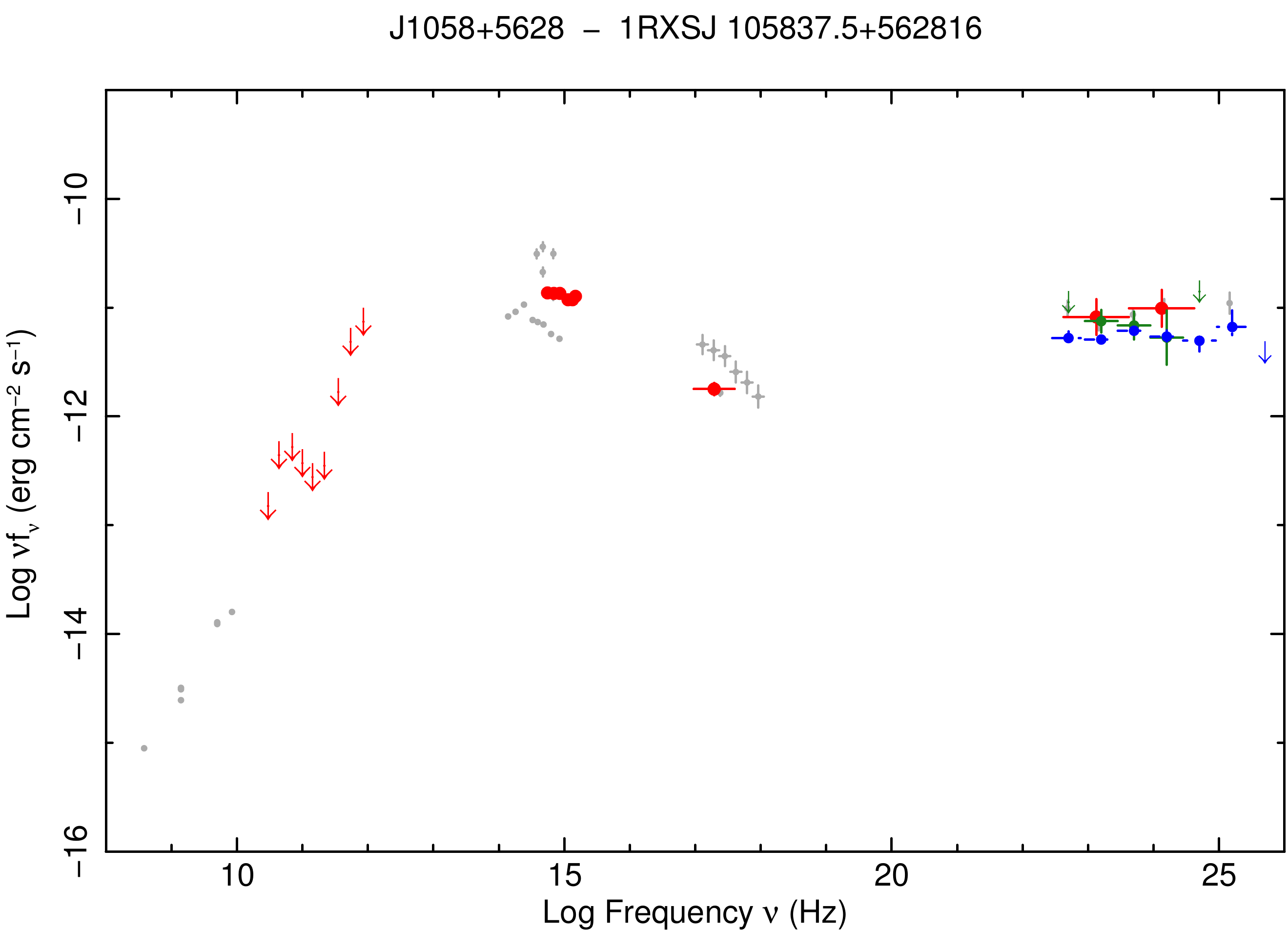}
\includegraphics[width=6.5cm,angle=-90]{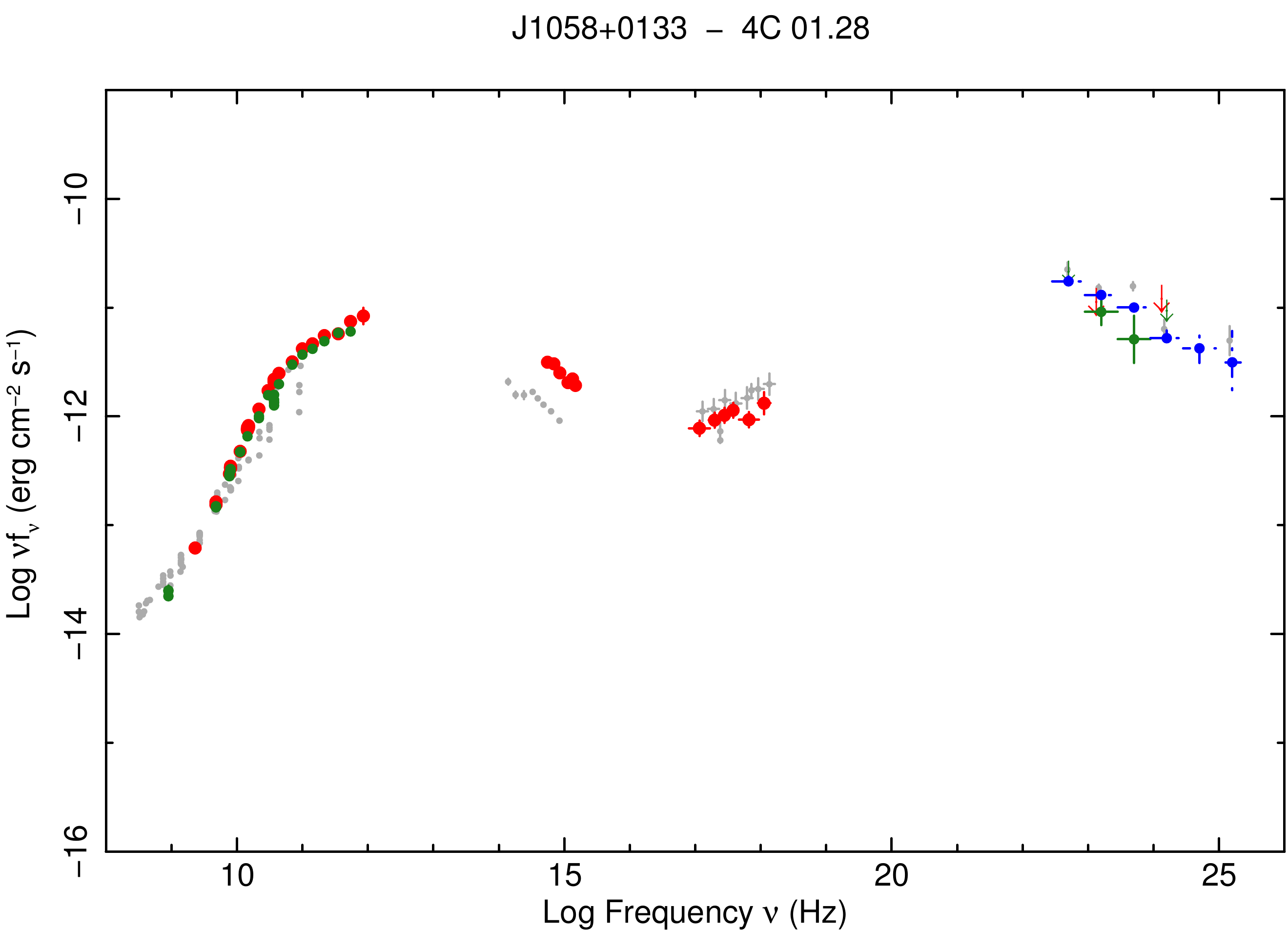}
\includegraphics[width=6.5cm,angle=-90]{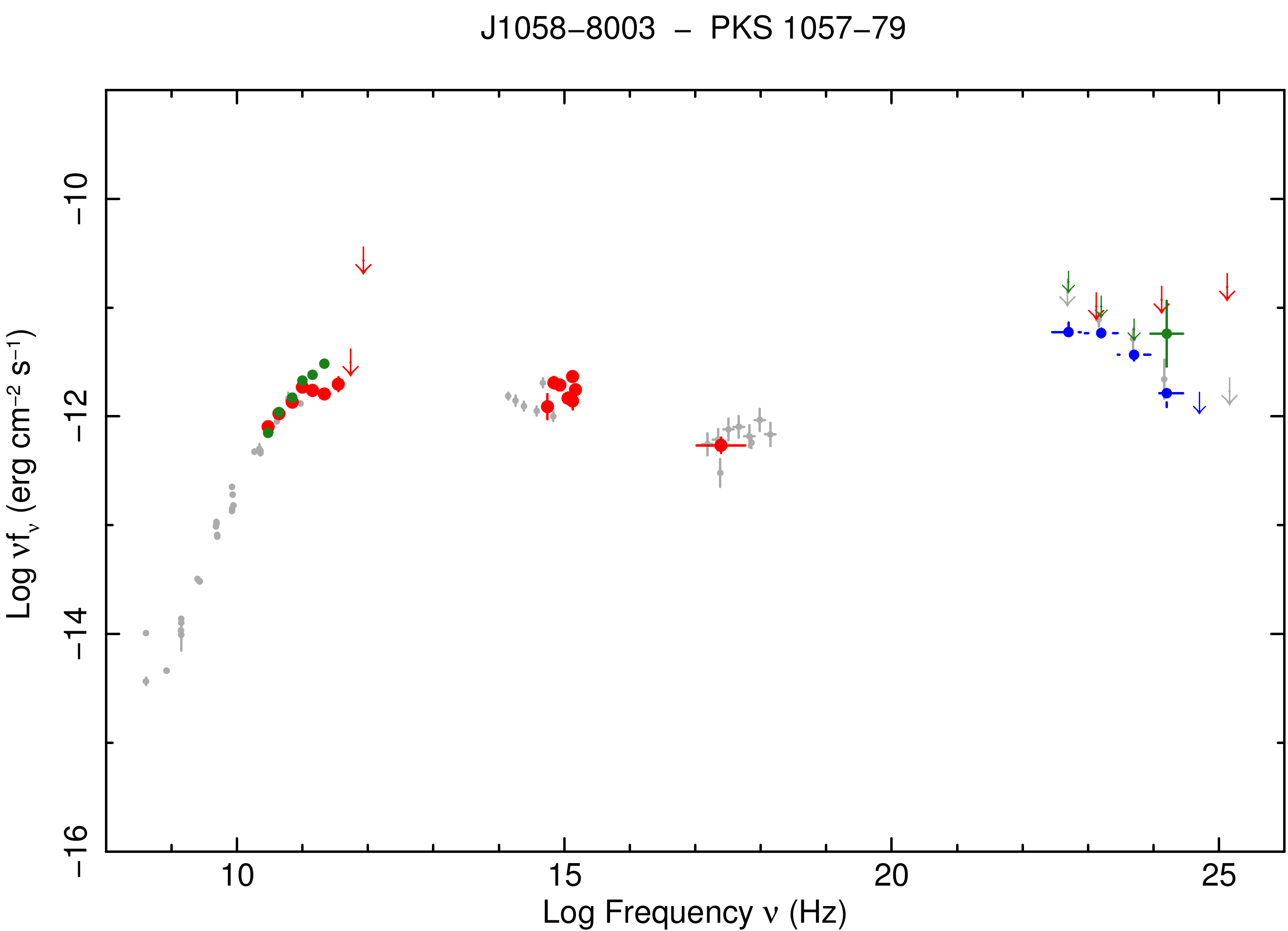}
\includegraphics[width=6.5cm,angle=-90]{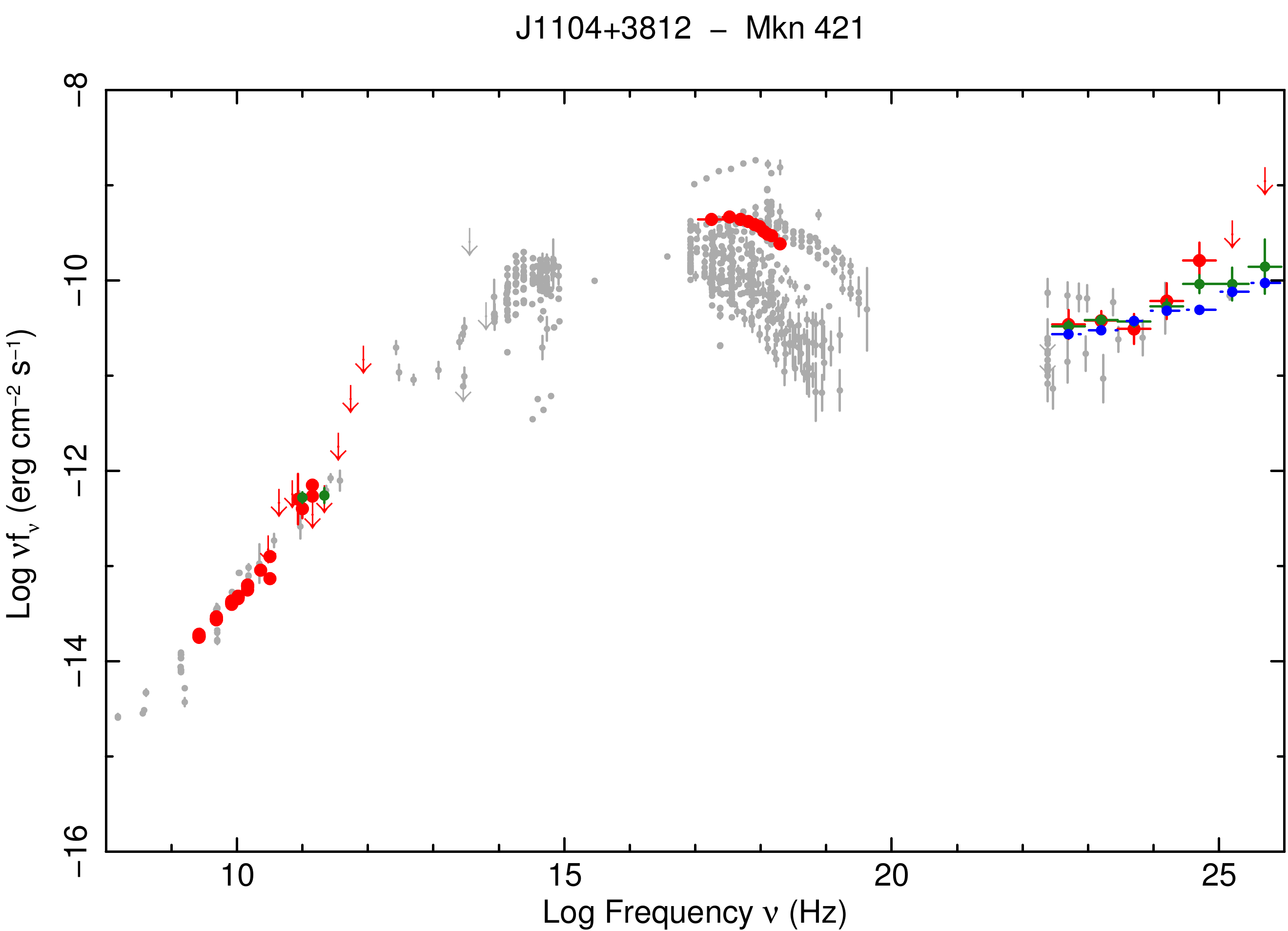}
\caption{The SED of 4C\,55.17 (J0957+5522, top left), 1H\,1013+498 (J1015+4926, top right),
1RXSJ\,105837.5+562816 (J1058+5628, middle left),  4C\,01.28 (J1058+0133, middle right),
PKS\,1057$-$79 (J1058$-$8003, bottom left), and Mkn\,421 (J1104+3812, bottom right). 
Simultaneous data are shown in red; quasi-simultaneous data, i.e. {\it Fermi} data
integrated over 2 months, {\it Planck} ERCSC and non-simultaneous ground based observations
are shown in green; {\it Fermi} data integrated over 27 months are shown in blue;
literature or archival data are shown in light gray.}
\label{fig:sed19}
\end{figure*}

\begin{figure*}
\centering
\includegraphics[width=6.5cm,angle=-90]{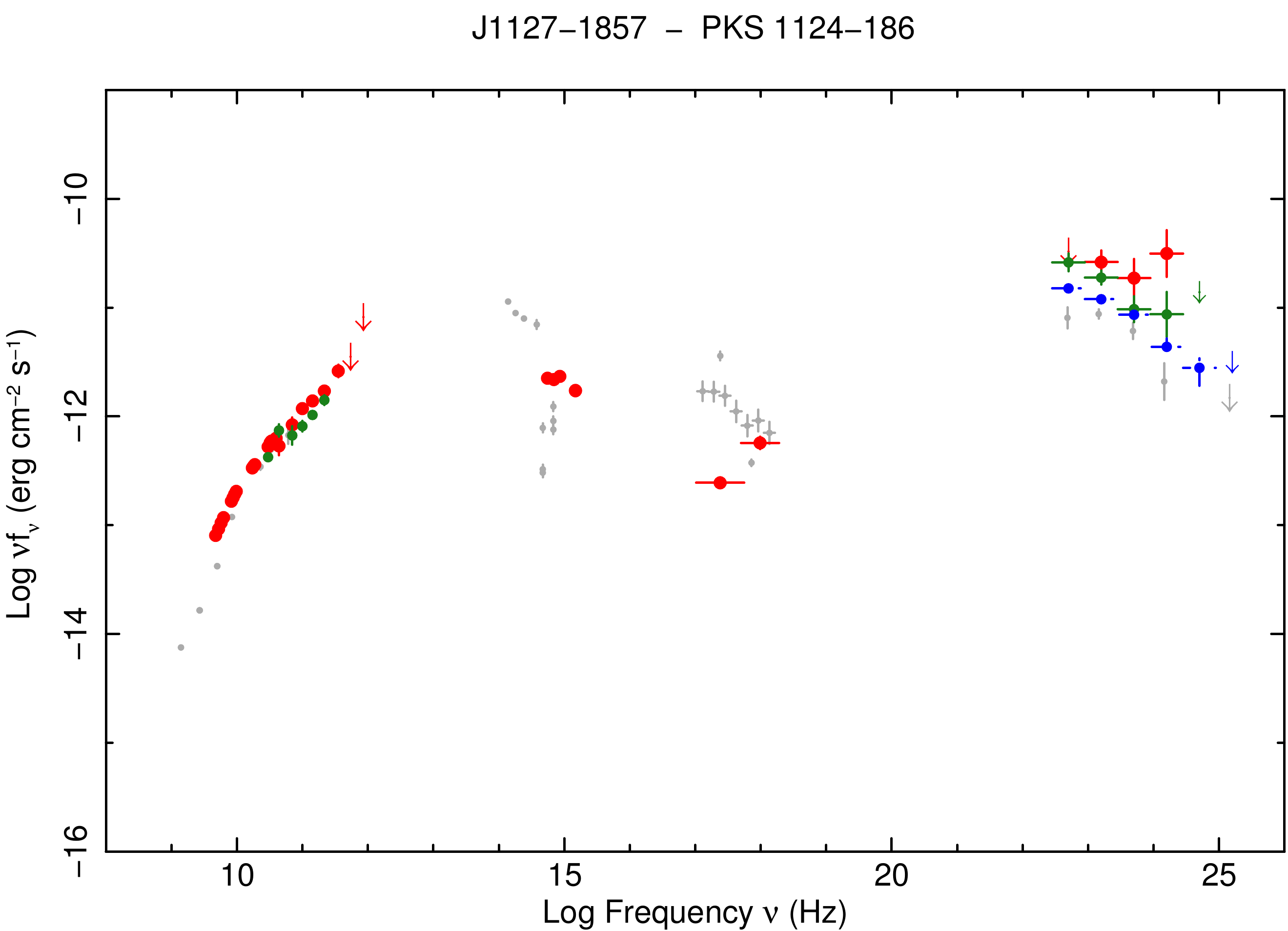}
\includegraphics[width=6.5cm,angle=-90]{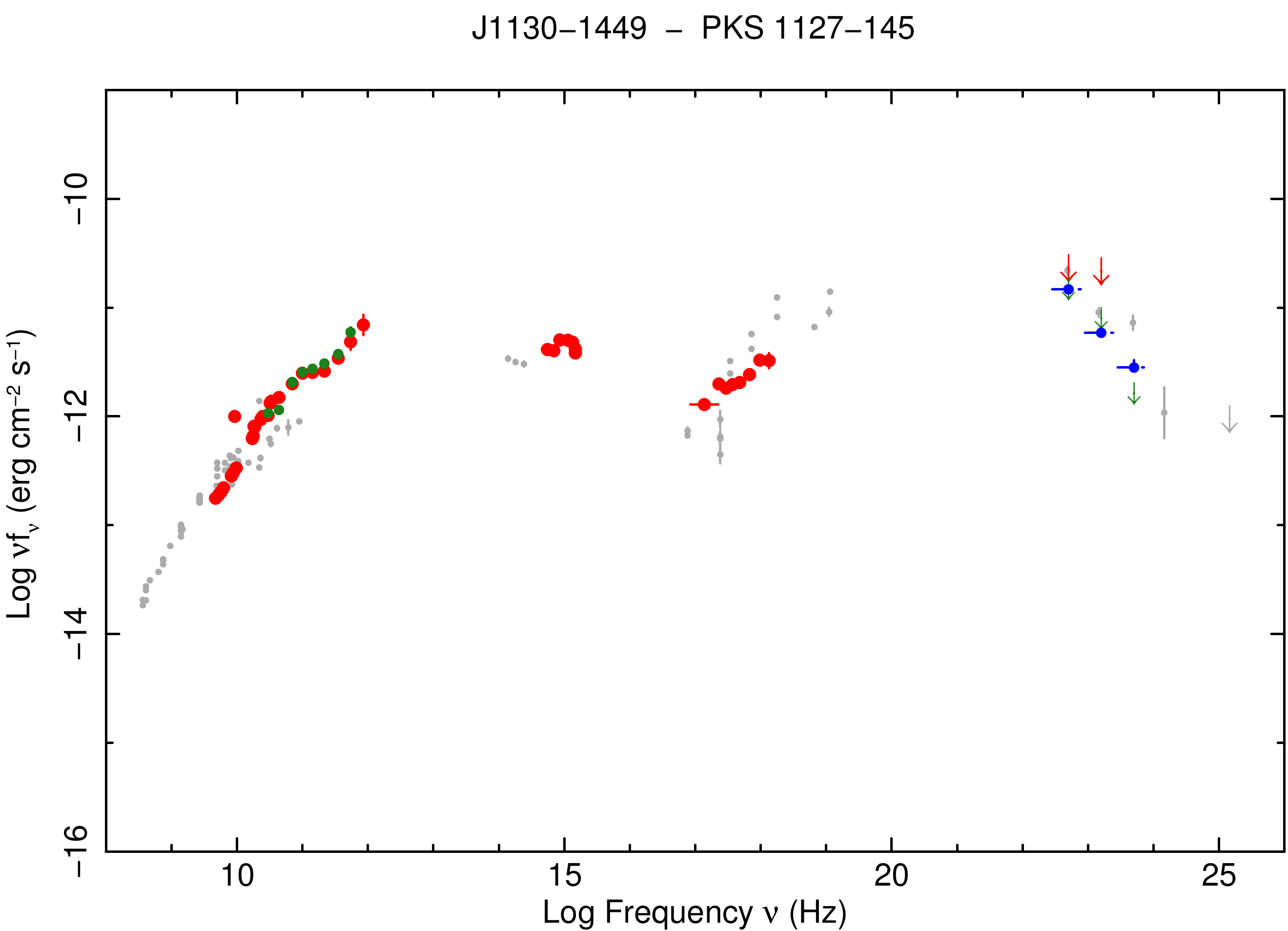}
\includegraphics[width=6.5cm,angle=-90]{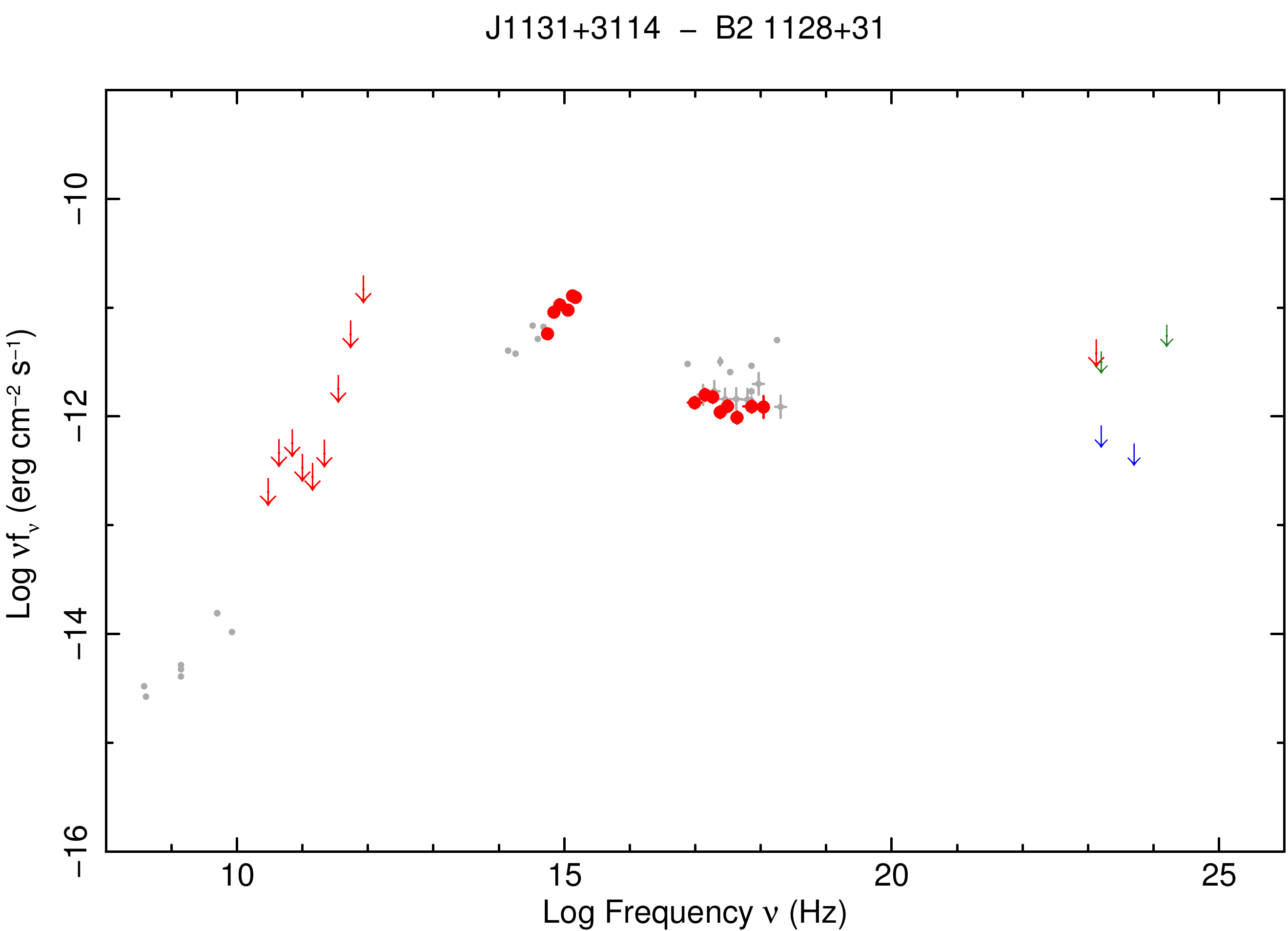}
\includegraphics[width=6.5cm,angle=-90]{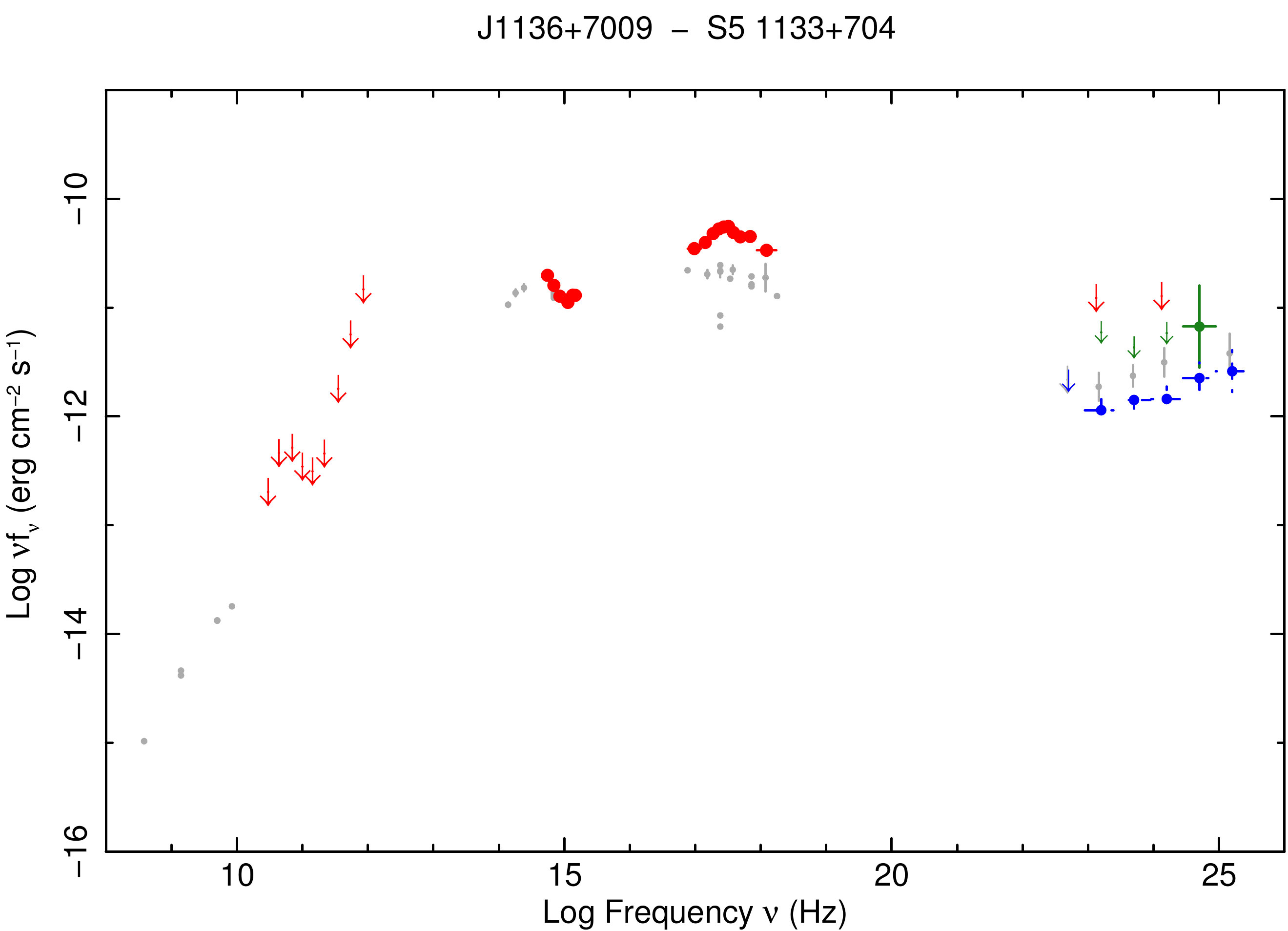}
\includegraphics[width=6.5cm,angle=-90]{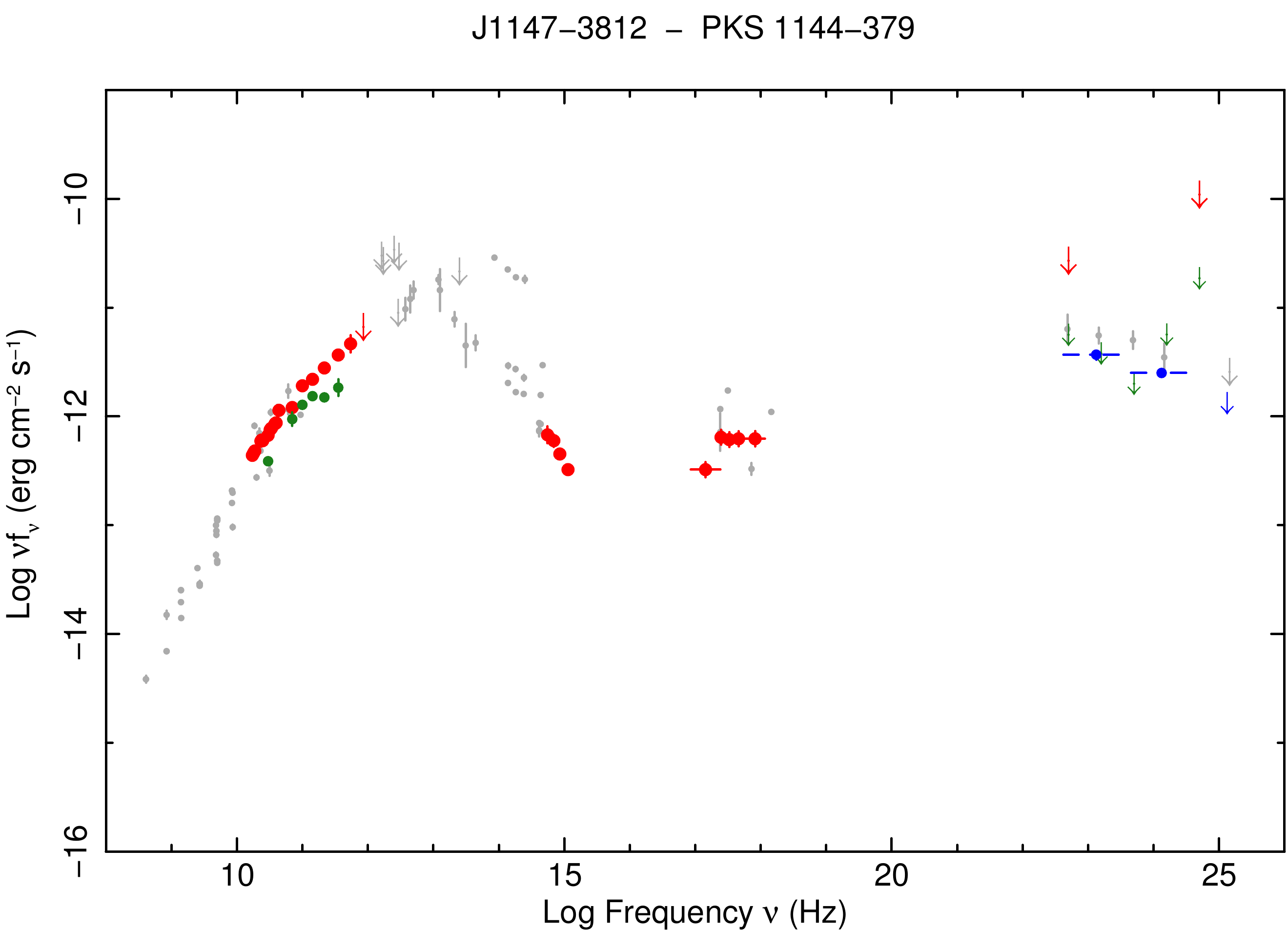}
\includegraphics[width=6.5cm,angle=-90]{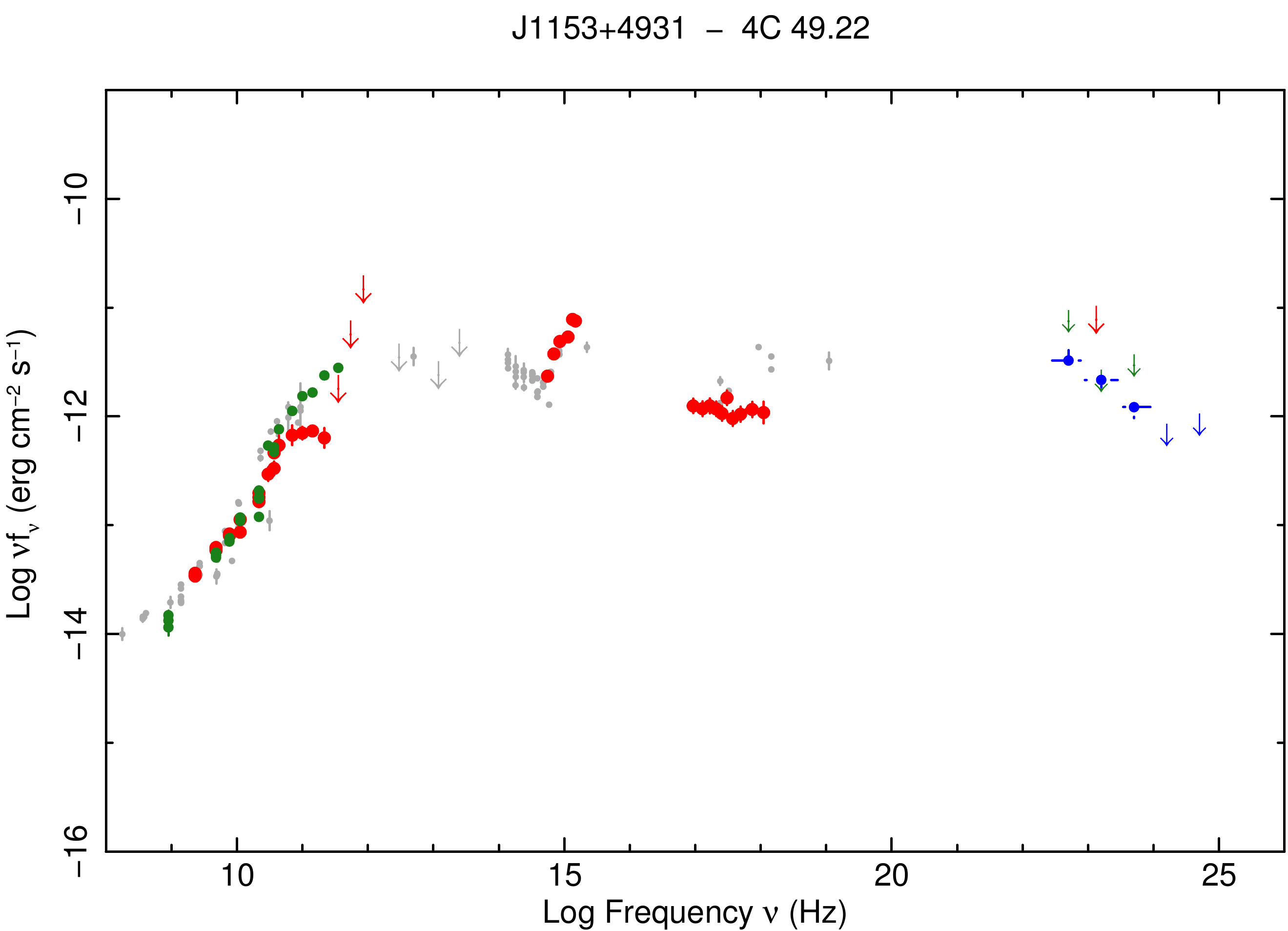}
\caption{The SED of PKS\,1124$-$186 (J1127$-$1857, top left), PKS\,1127$-$145 (J1130$-$1449, top right),
B2\,1128+31 (J1131+3114, middle left), S5\,1133+704 (J1136+7009, middle right),
PKS\,1144$-$379 (J1147$-$3812, bottom left), and 4C\,49.22 (J1153+4931, bottom right). 
Simultaneous data are shown in red; quasi-simultaneous data, i.e. {\it Fermi} data
integrated over 2 months, {\it Planck} ERCSC and non-simultaneous ground based observations
are shown in green; {\it Fermi} data integrated over 27 months are shown in blue;
literature or archival data are shown in light gray.}
\label{fig:sed22}
\end{figure*}

\begin{figure*}
\centering
\includegraphics[width=6.5cm,angle=-90]{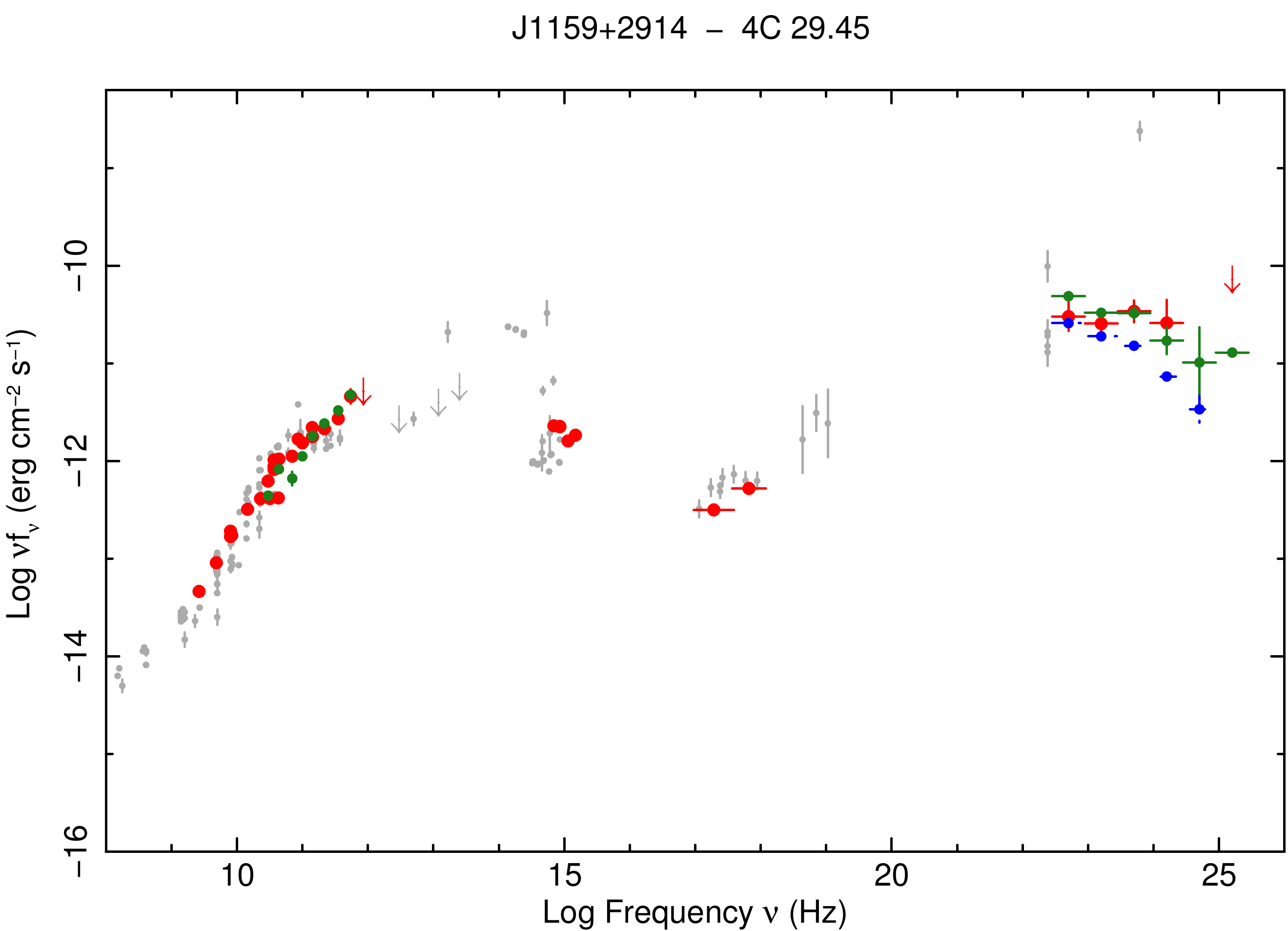}
\includegraphics[width=6.5cm,angle=-90]{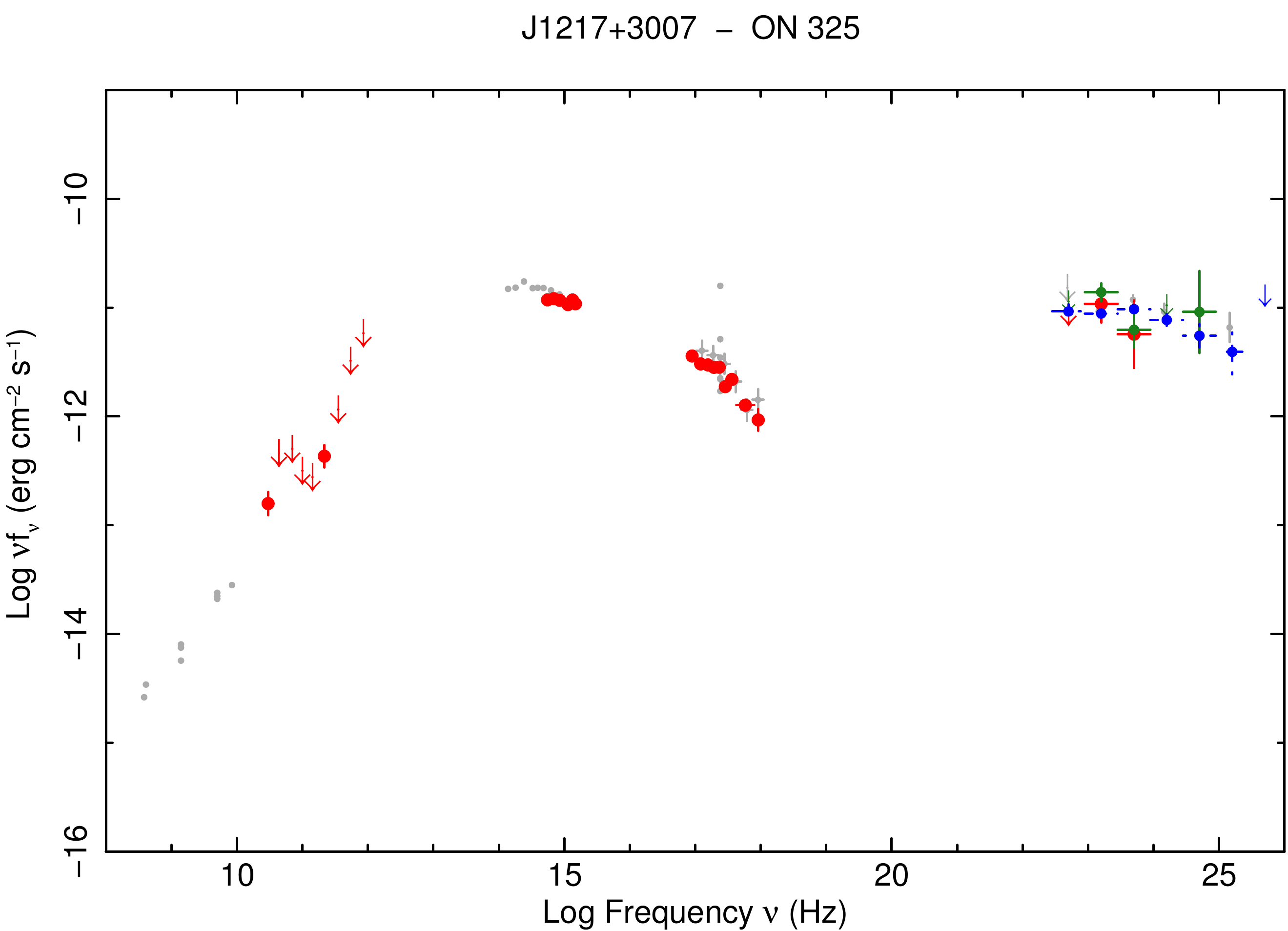}
\includegraphics[width=6.5cm,angle=-90]{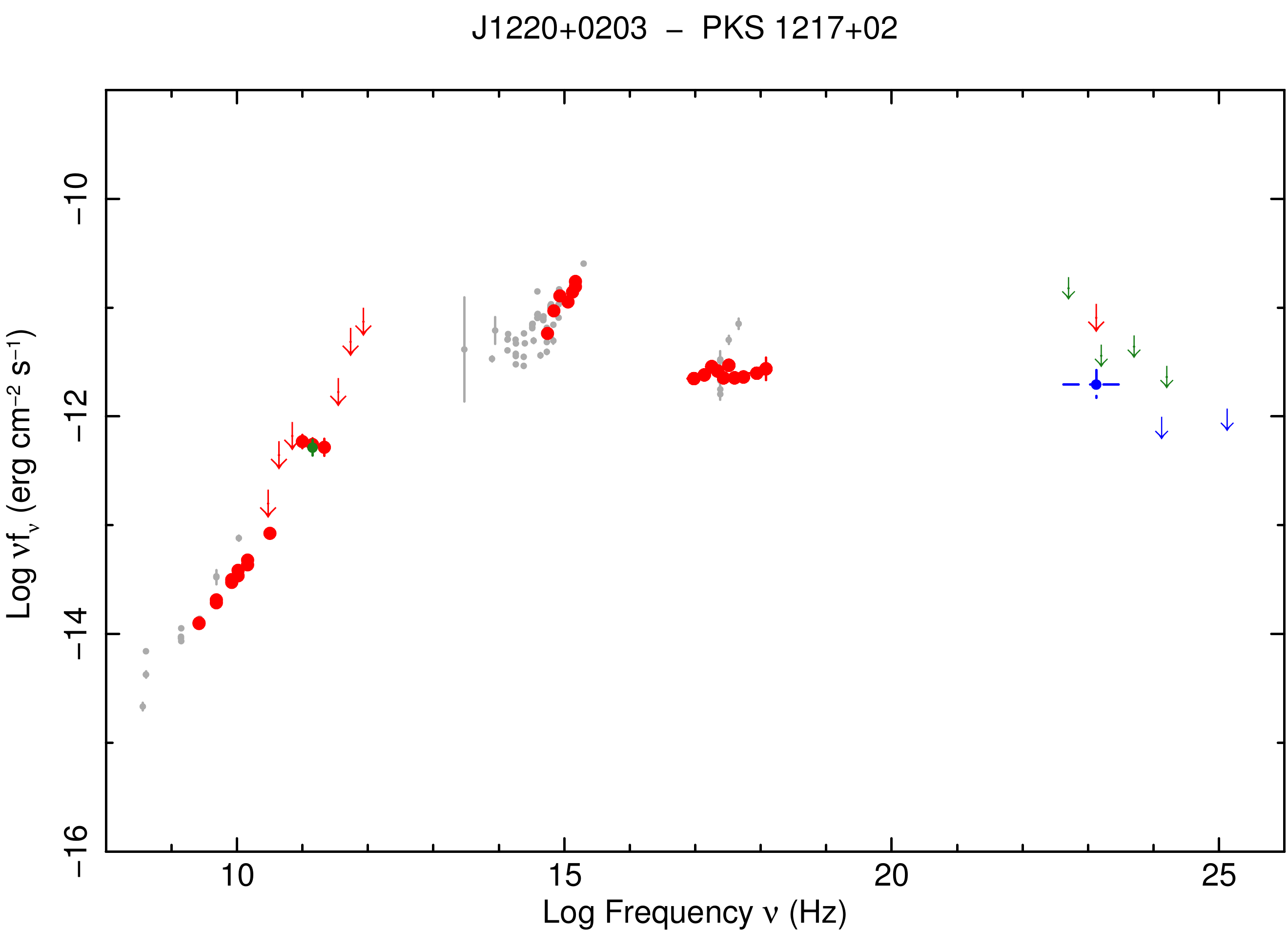}
\includegraphics[width=6.5cm,angle=-90]{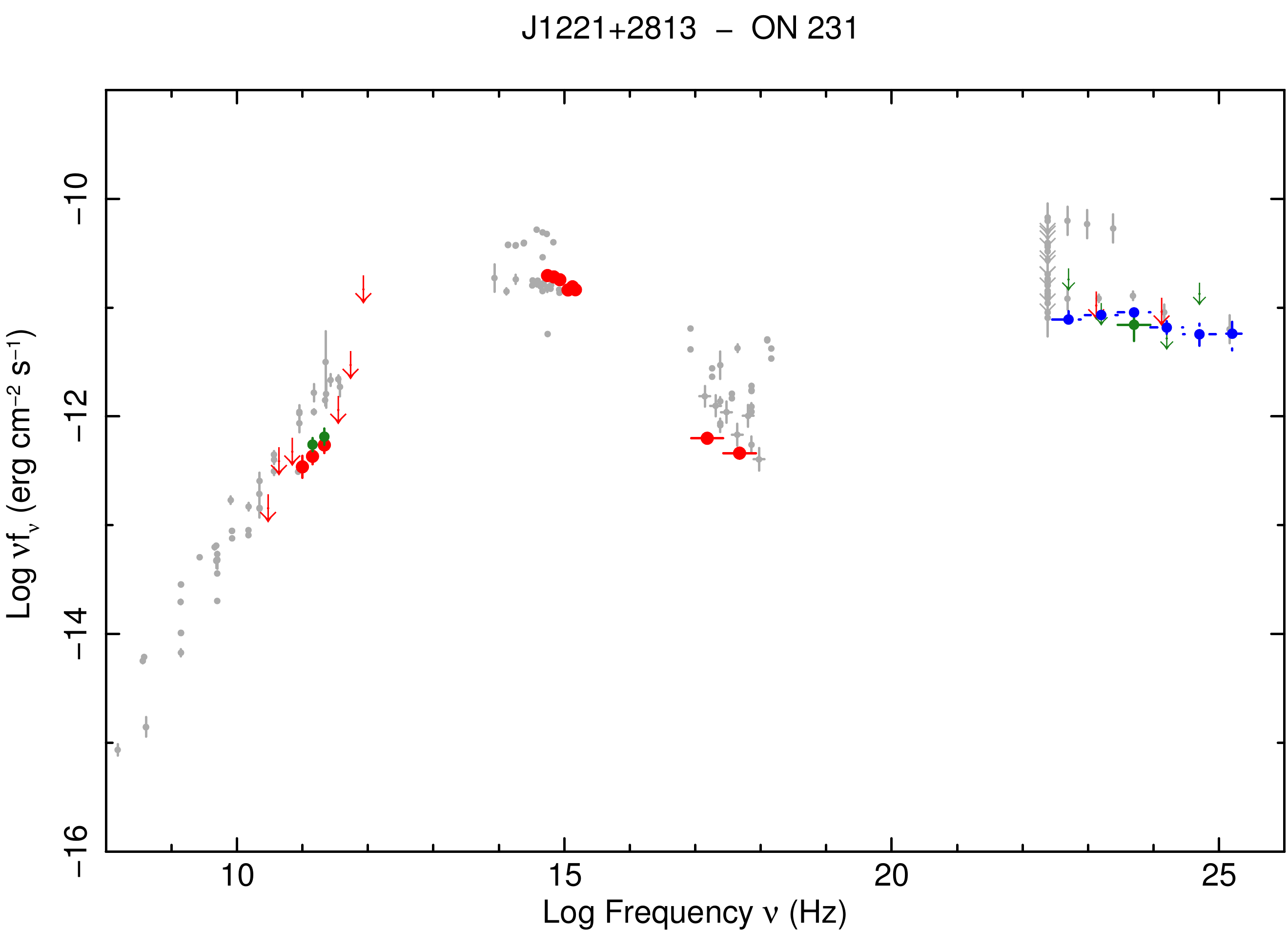}
\includegraphics[width=6.5cm,angle=-90]{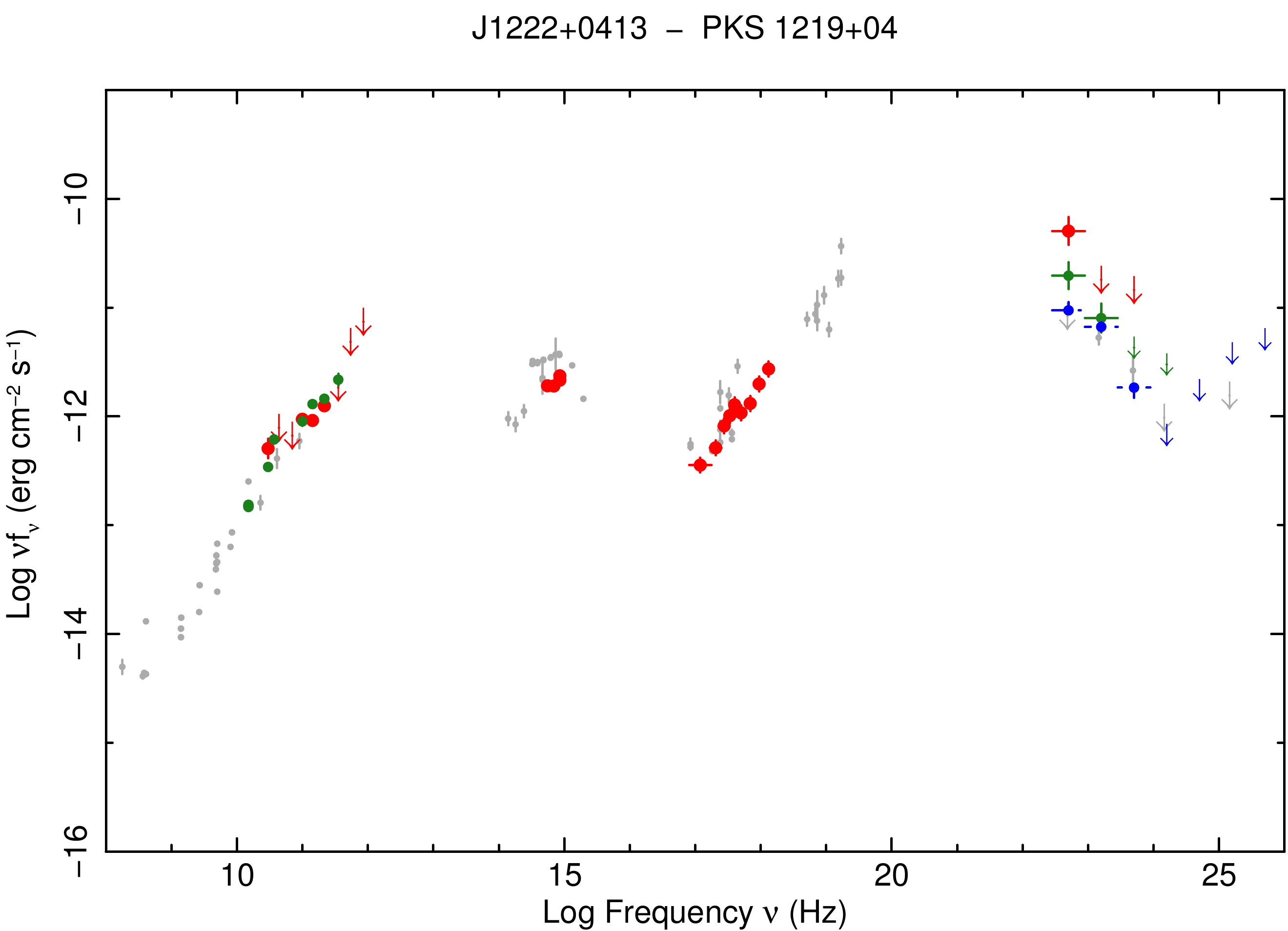}
\includegraphics[width=6.5cm,angle=-90]{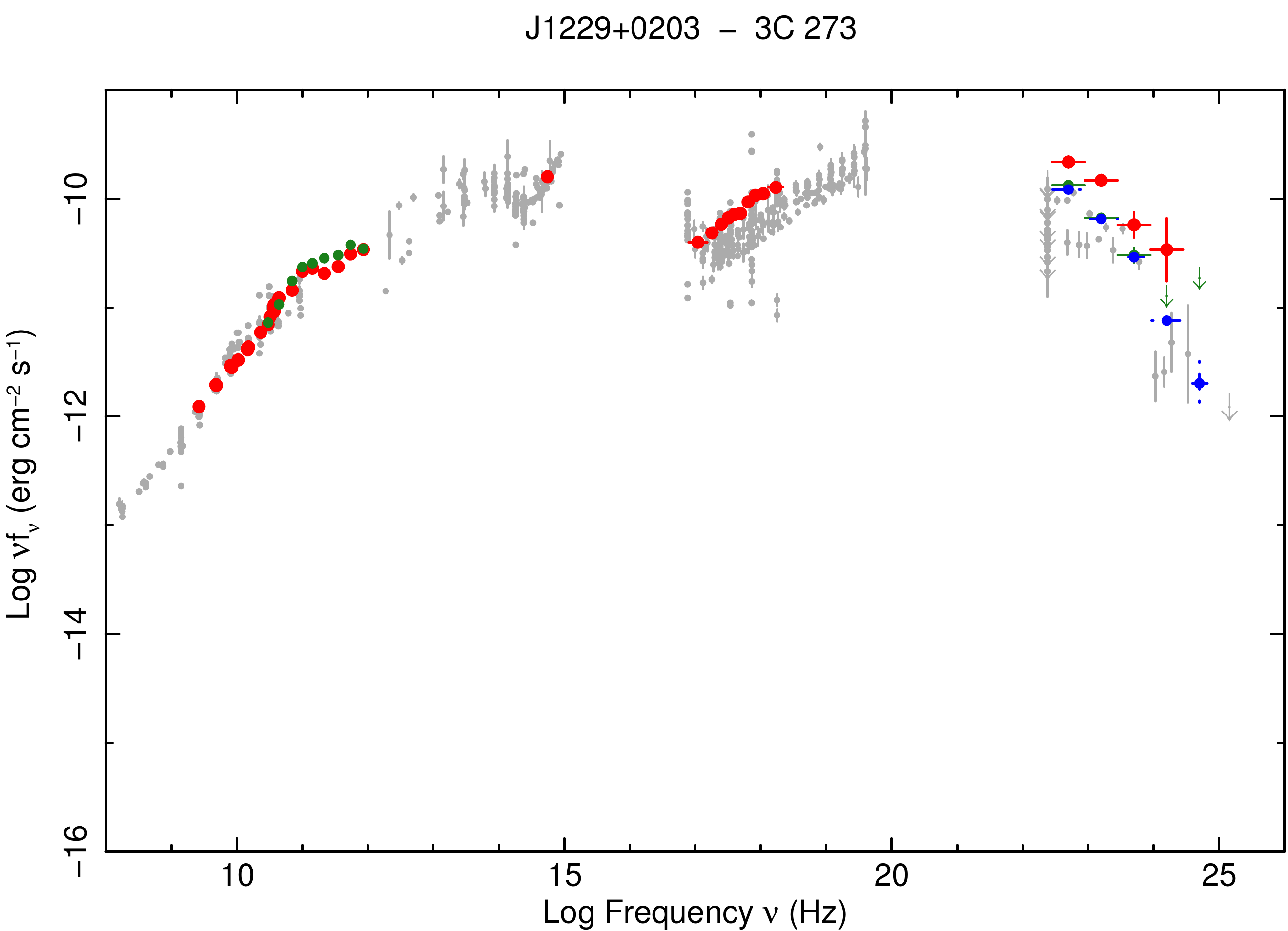}
\caption{The SED of 4C\,29.45 (J1159+2914, top left), ON\,325 (J1217+3007, top right),
PKS\,1217+02 (J1220+0203, middle left), ON\,231 (J1221+2813, middle right),
PKS\,1219+04 (J1222+0413, bottom left), and 3C\,273 (J1229+0203, bottom right). 
Simultaneous data are shown in red; quasi-simultaneous data, i.e. {\it Fermi} data
integrated over 2 months, {\it Planck} ERCSC and non-simultaneous ground based observations
are shown in green; {\it Fermi} data integrated over 27 months are shown in blue;
literature or archival data are shown in light gray.}
\label{fig:sed25}
\end{figure*}

\begin{figure*}
\centering
\includegraphics[width=6.5cm,angle=-90]{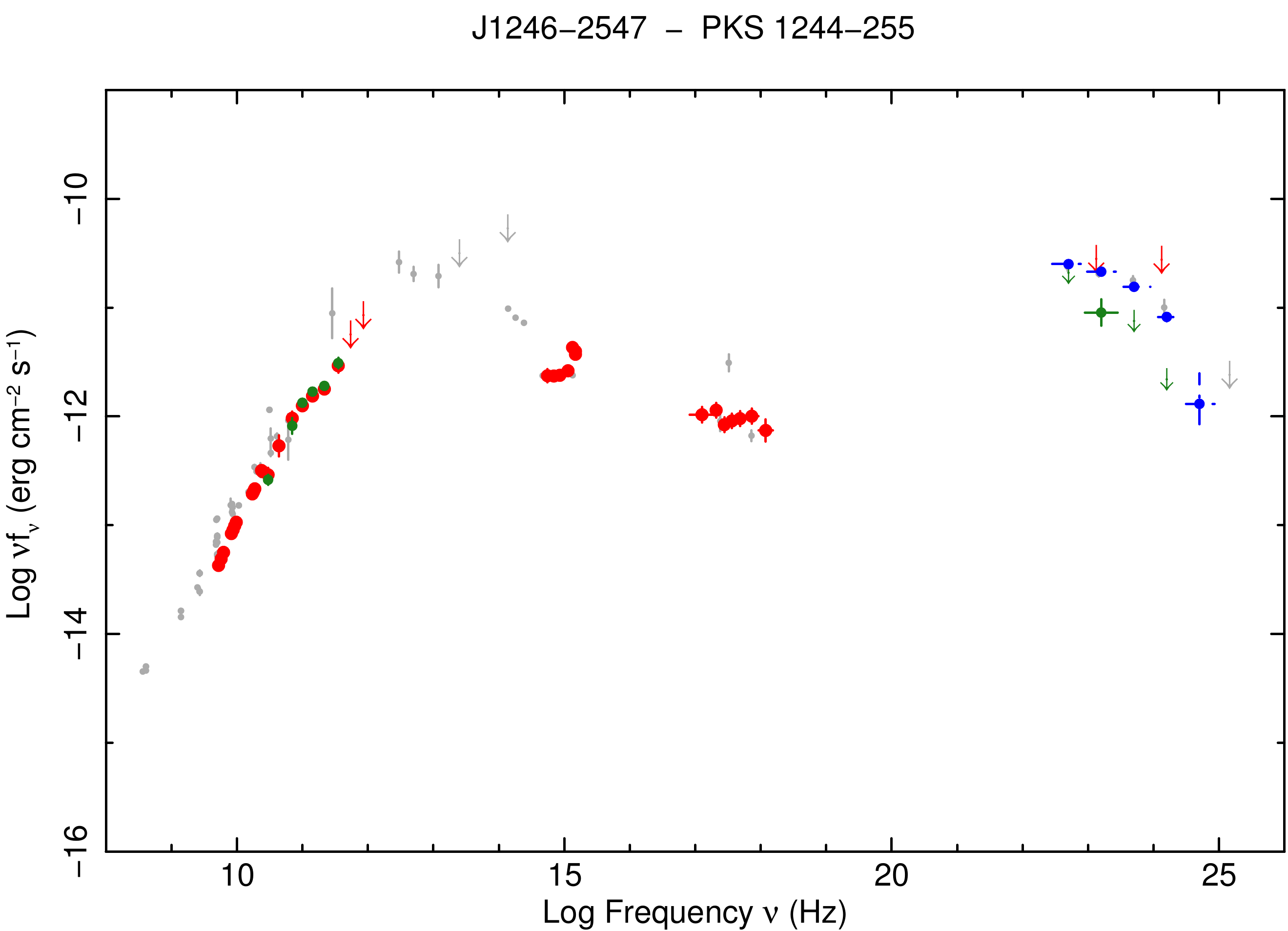}
\includegraphics[width=6.5cm,angle=-90]{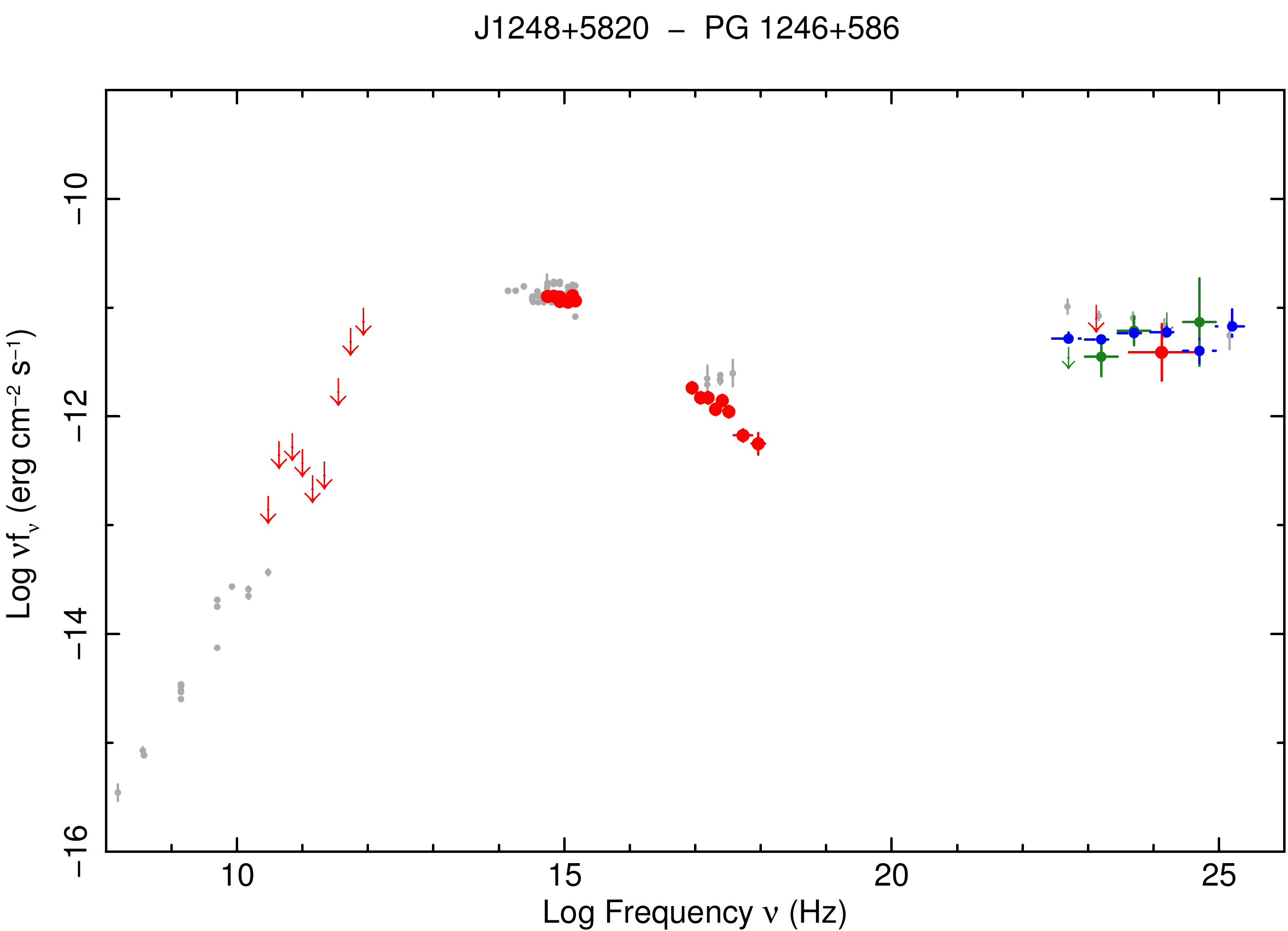}
\includegraphics[width=6.5cm,angle=-90]{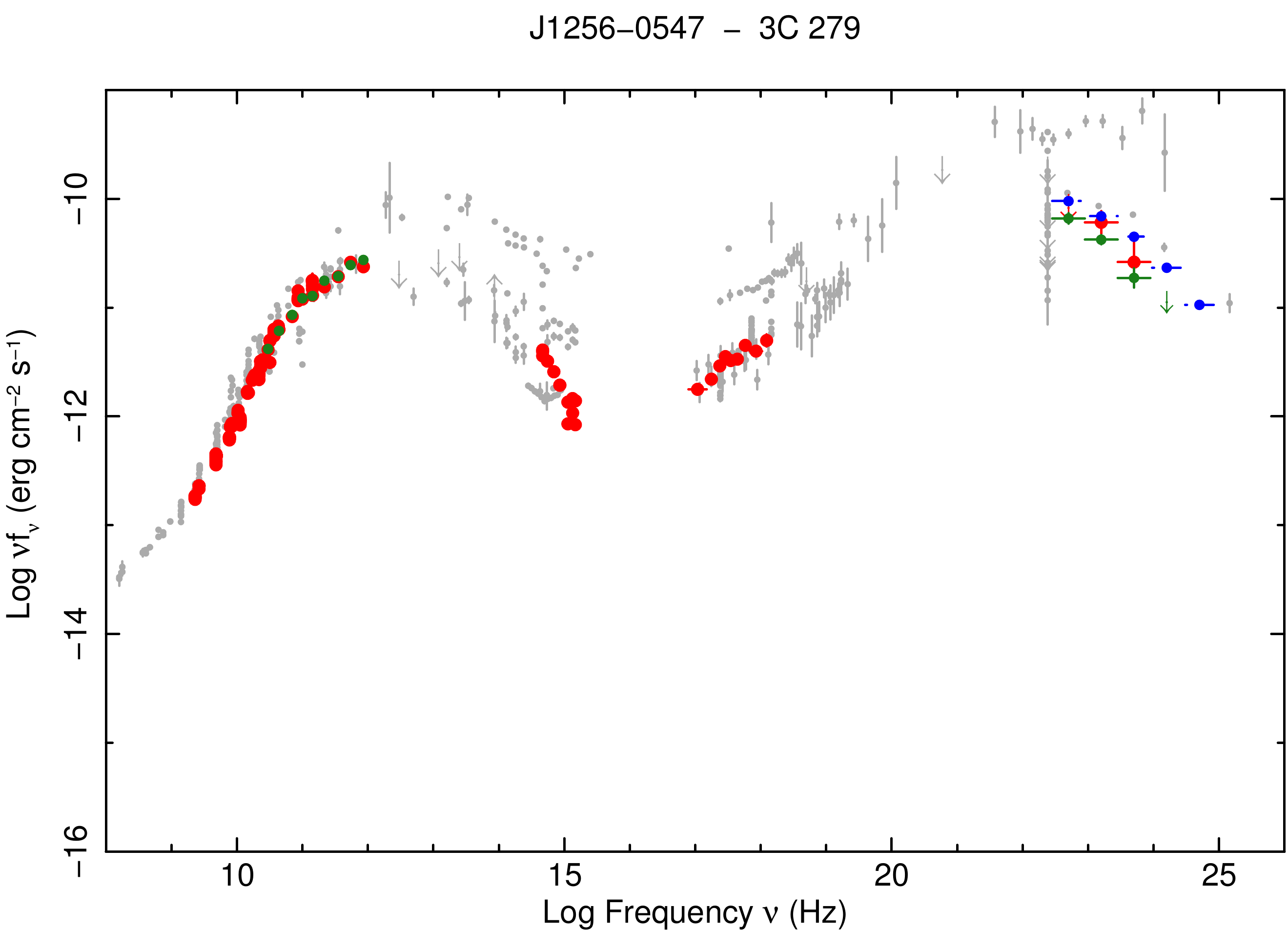}
\includegraphics[width=6.5cm,angle=-90]{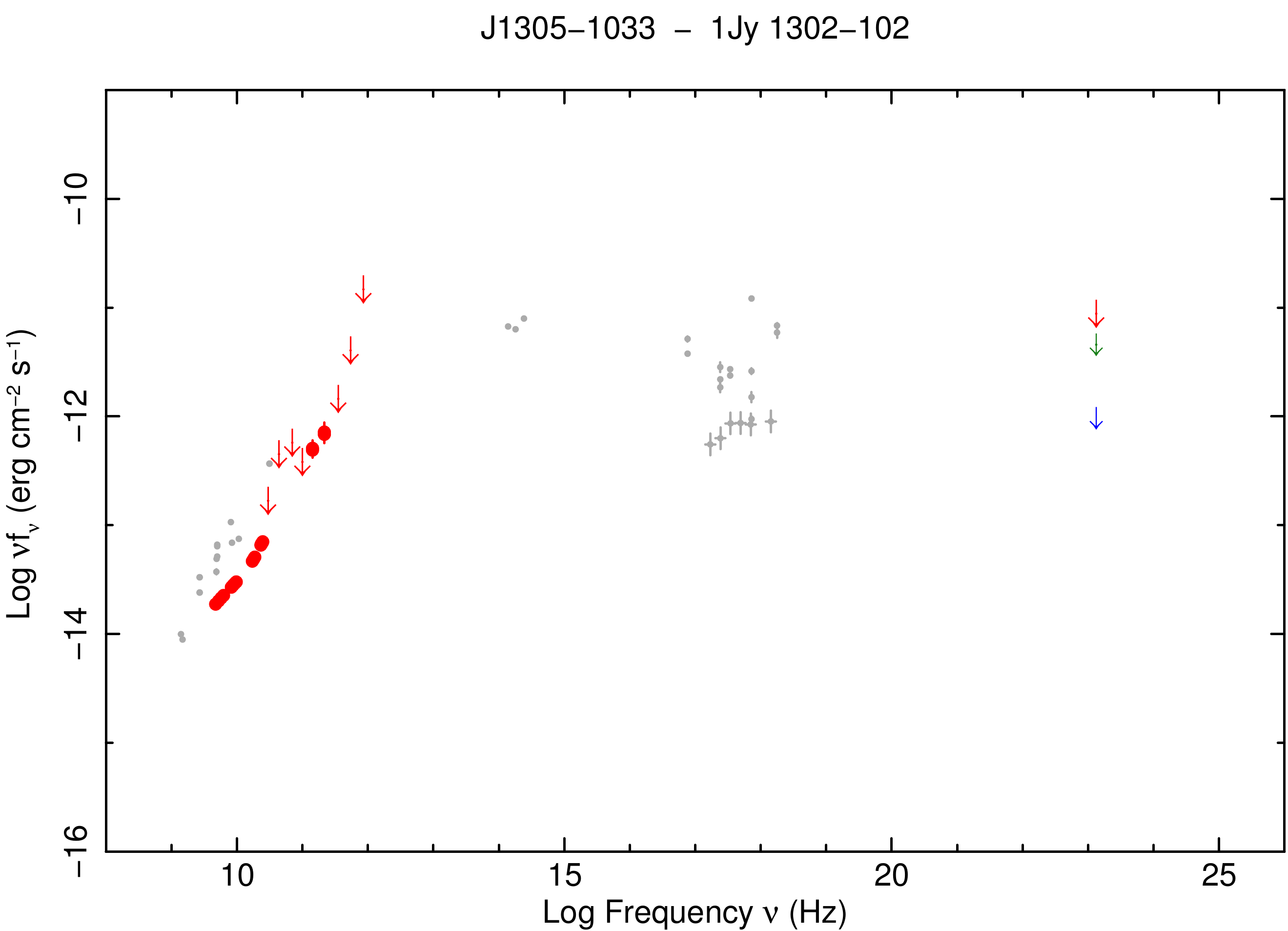}
\includegraphics[width=6.5cm,angle=-90]{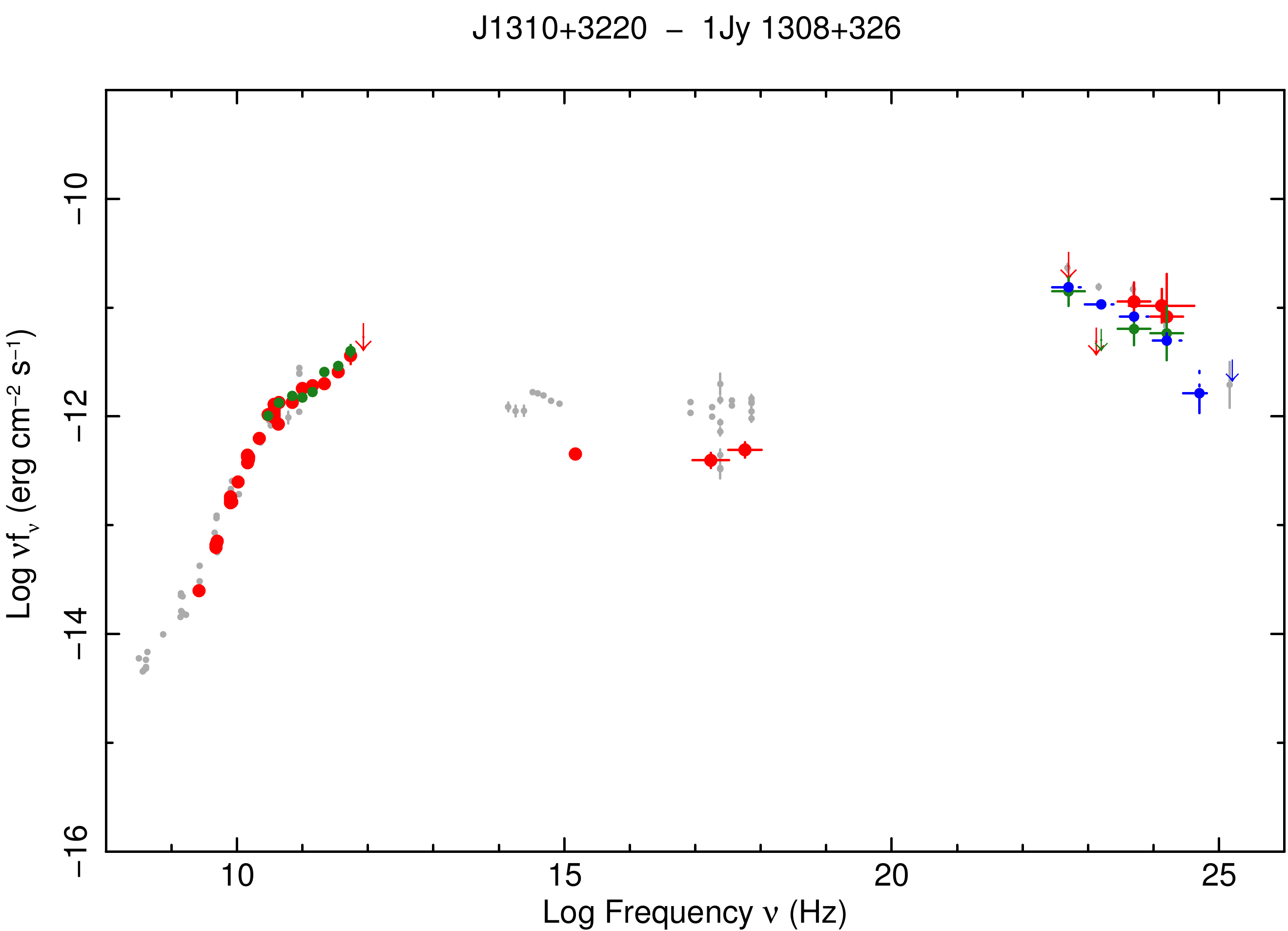}
\includegraphics[width=6.5cm,angle=-90]{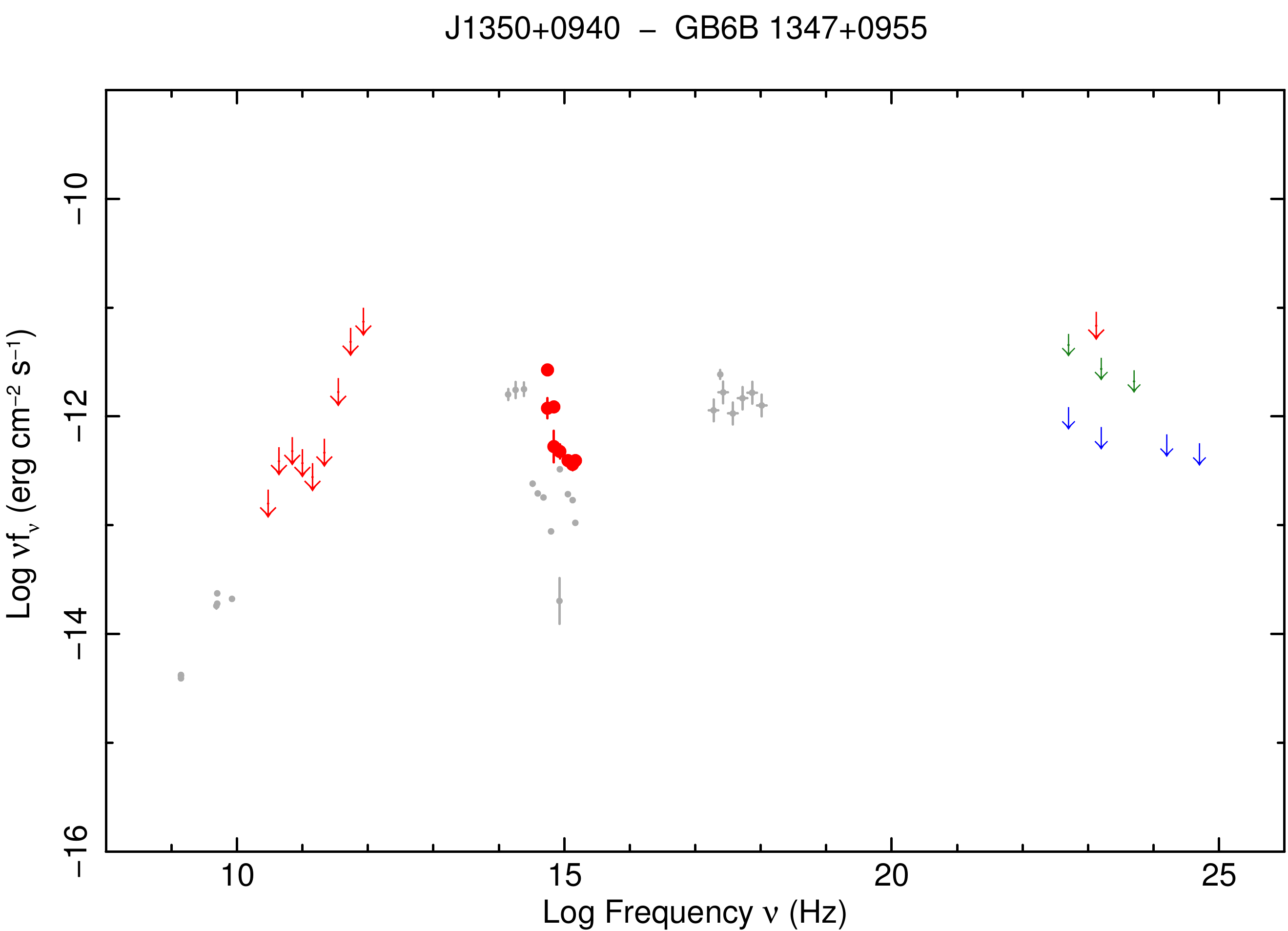}
\caption{The SED of PKS\,1244$-$255 (J1246$-$2547, top left), PG\,1246+586 (J1248+5820, top right). 
 3C\,279 (J1256$-$0547, middle left), 1Jy\,1302$-$102 (J1305$-$1033, middle right),
 1Jy\,1308+326 (J1310+3220, bottom left), and GB6B\,1347+0955 (J1350+0940, bottom right).
 Simultaneous data are shown in red; quasi-simultaneous data, i.e. {\it Fermi} data
integrated over 2 months, {\it Planck} ERCSC and non-simultaneous ground based observations
are shown in green; {\it Fermi} data integrated over 27 months are shown in blue;
literature or archival data are shown in light gray.}
\label{fig:sed28}
\end{figure*}

\begin{figure*}
\centering
\includegraphics[width=6.5cm,angle=-90]{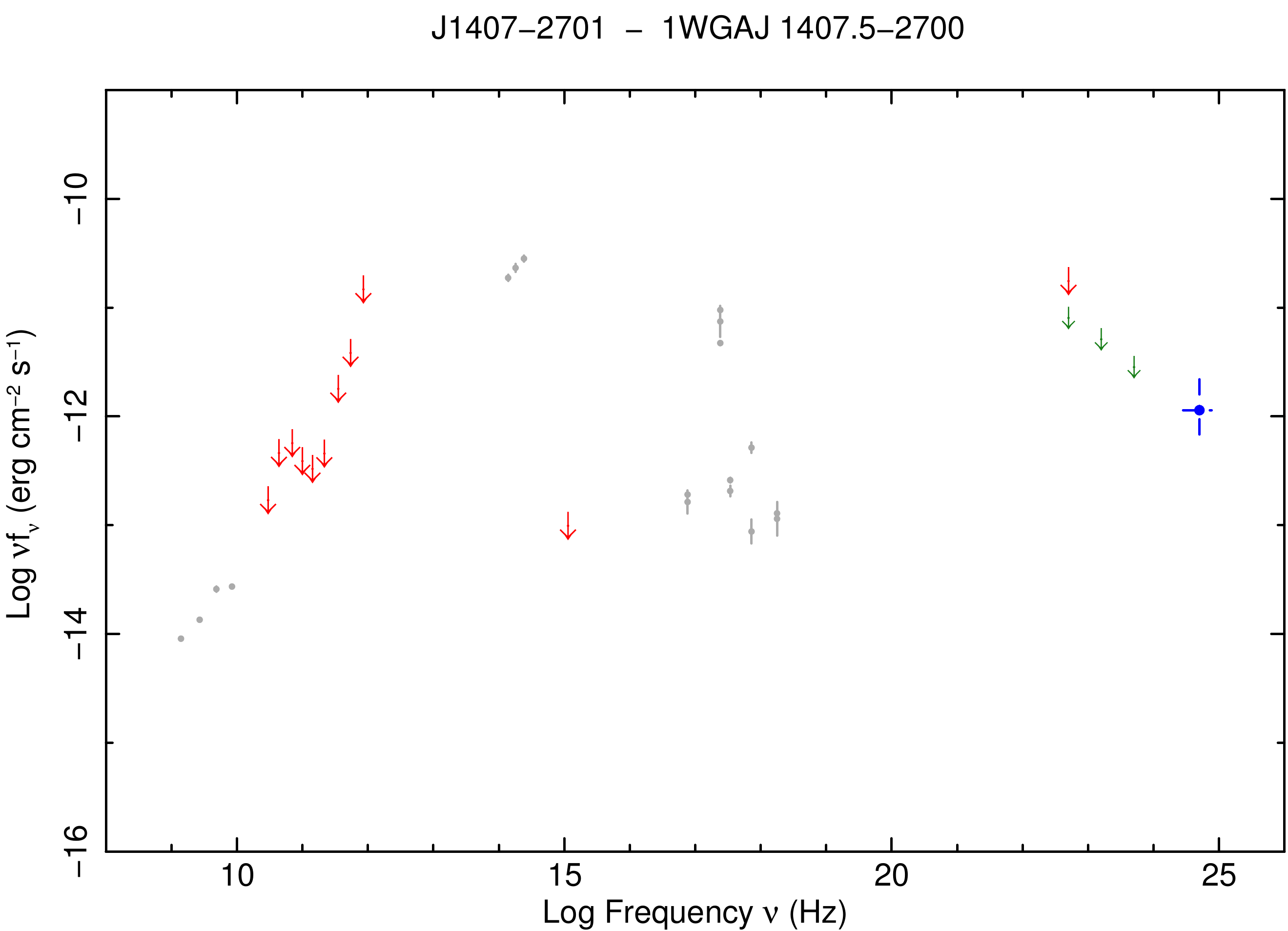}
\includegraphics[width=6.5cm,angle=-90]{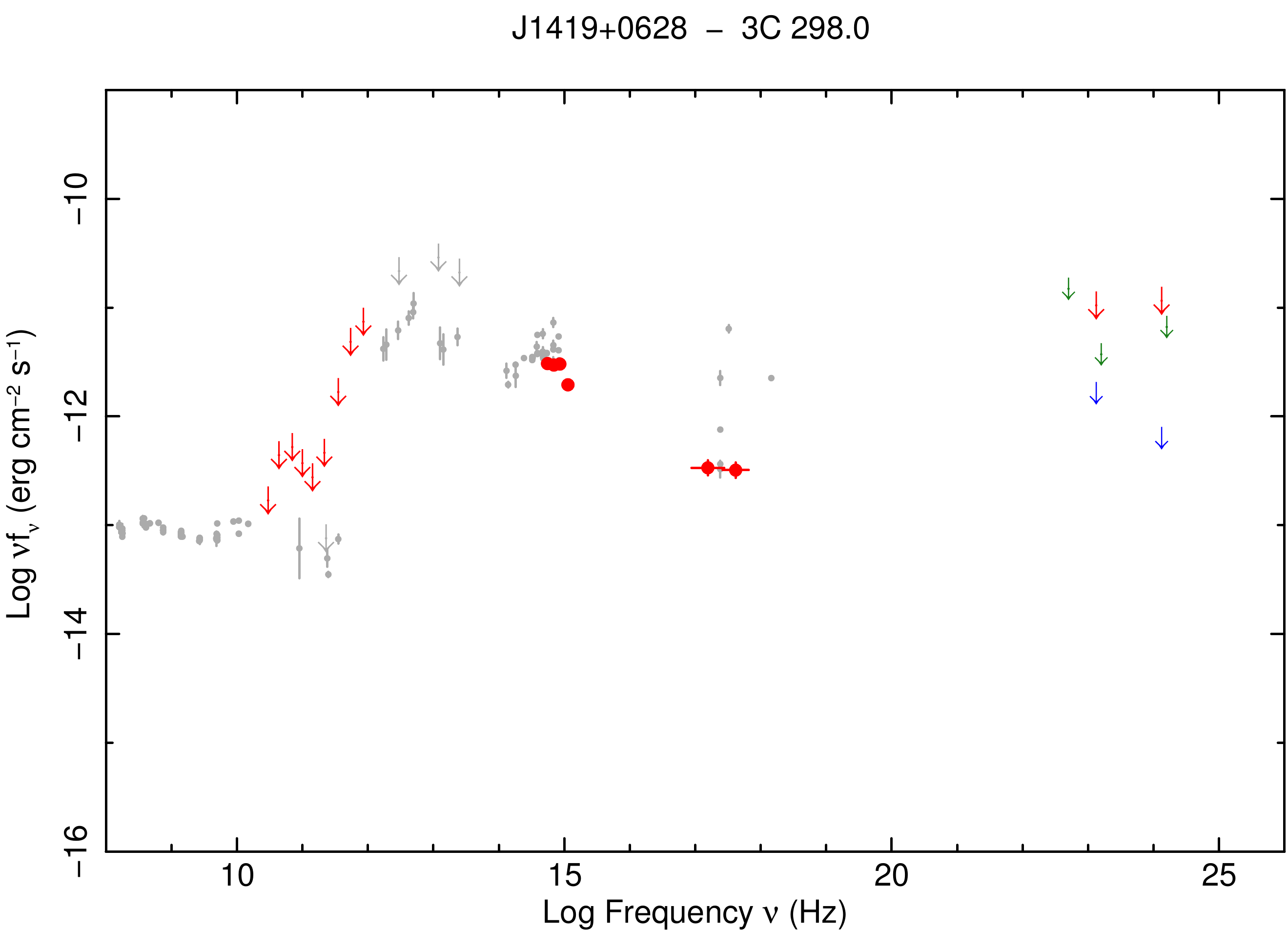}
\includegraphics[width=6.5cm,angle=-90]{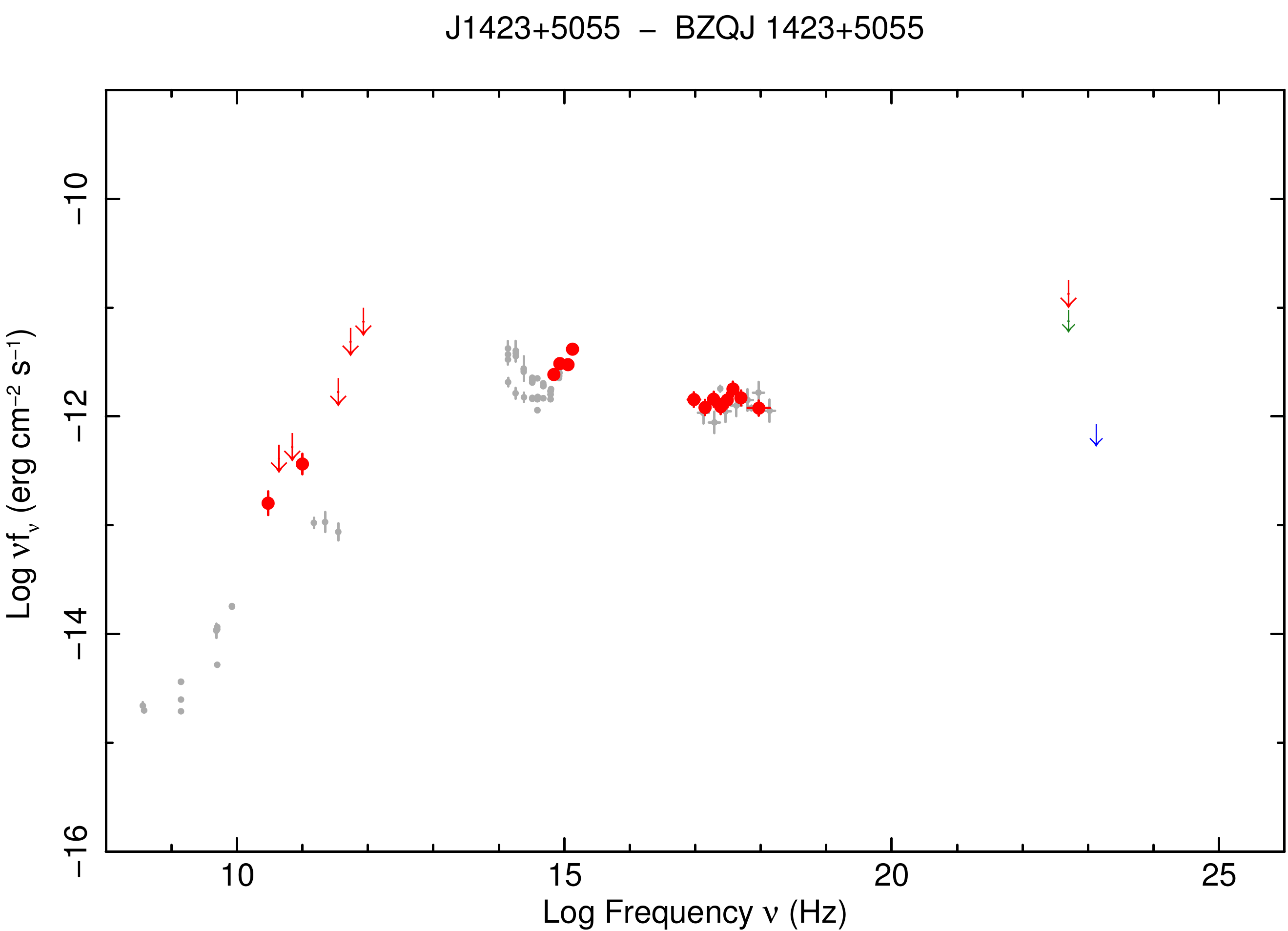}
\includegraphics[width=6.5cm,angle=-90]{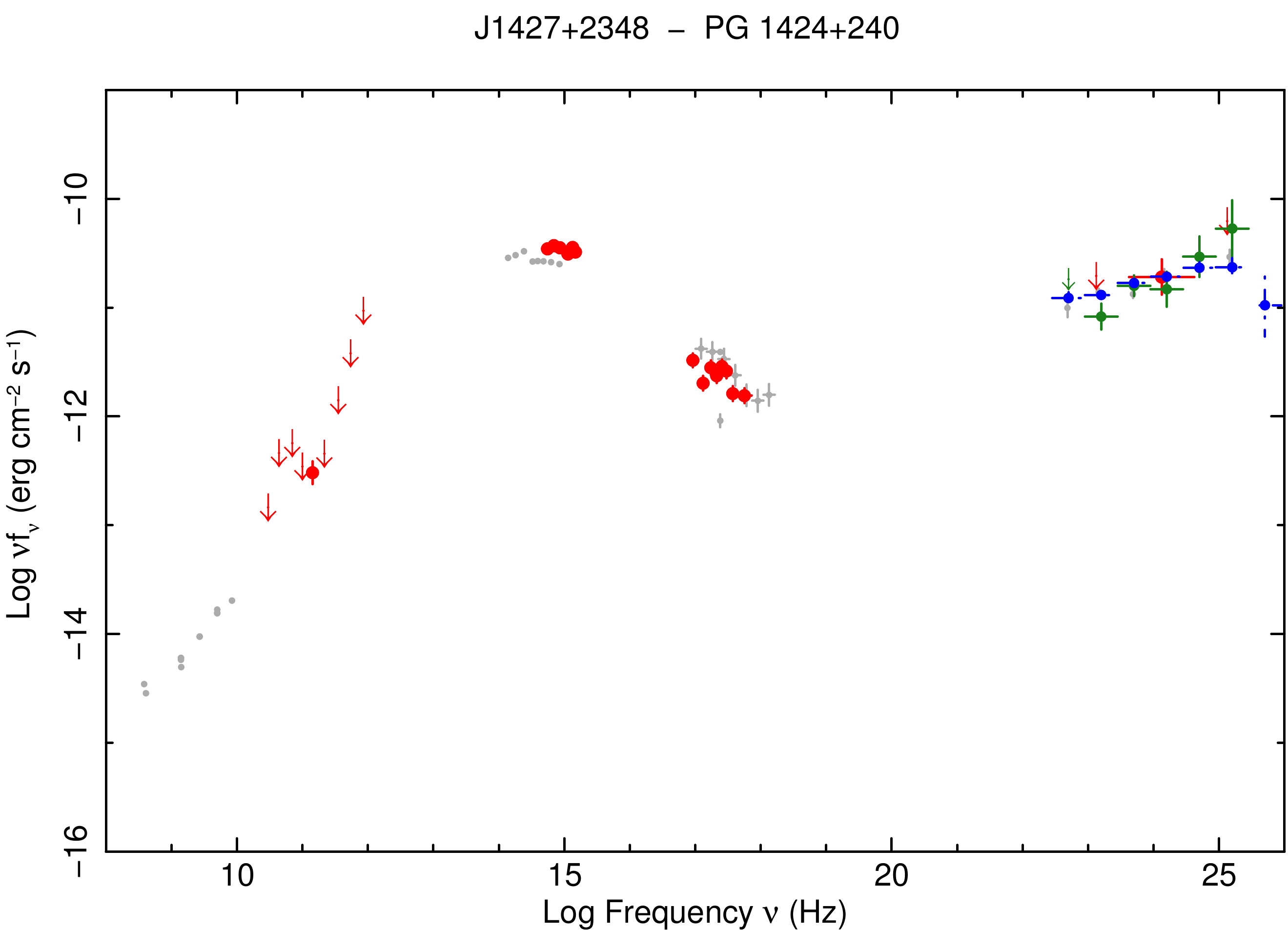}
\includegraphics[width=6.5cm,angle=-90]{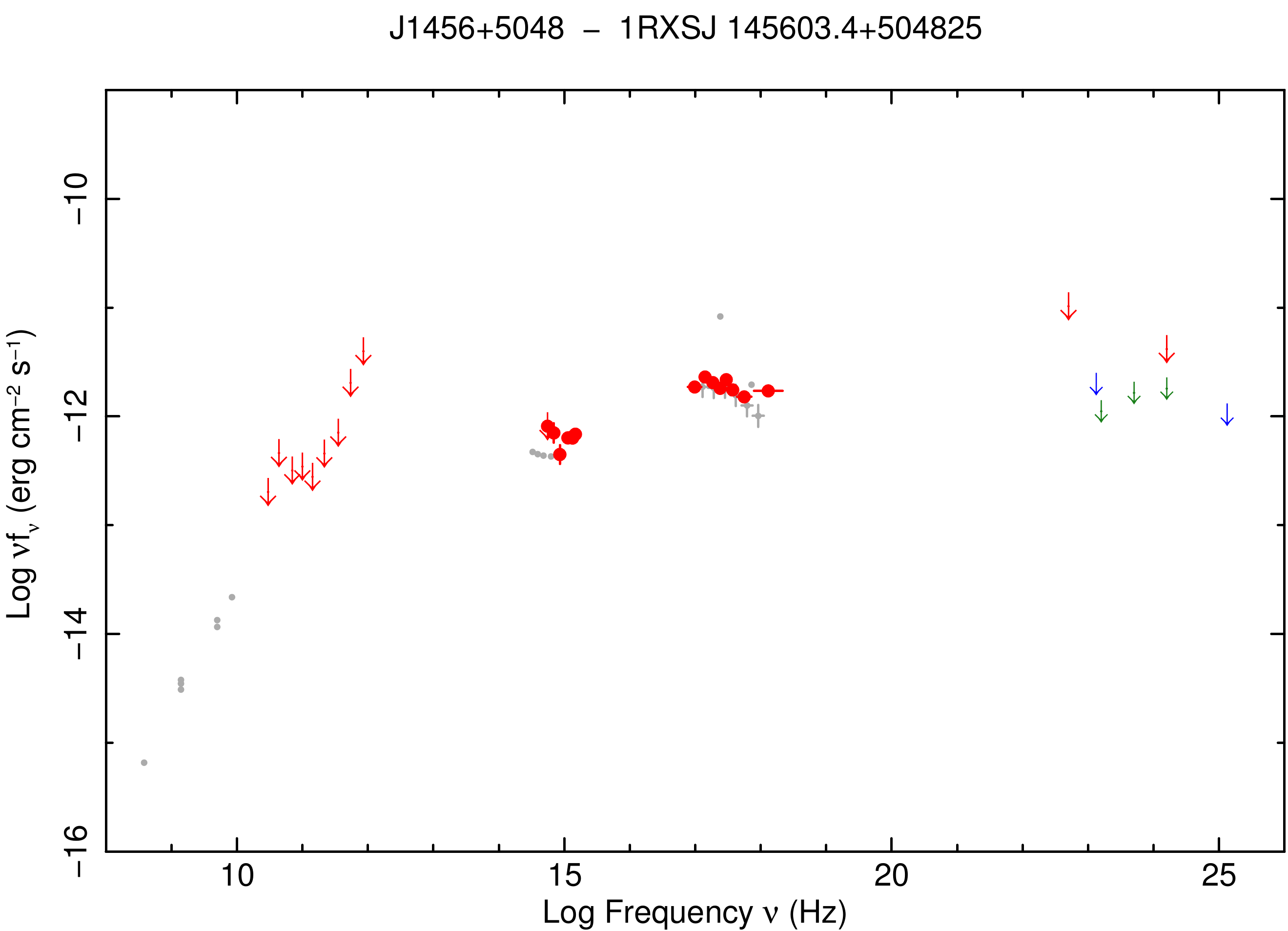}
\includegraphics[width=6.5cm,angle=-90]{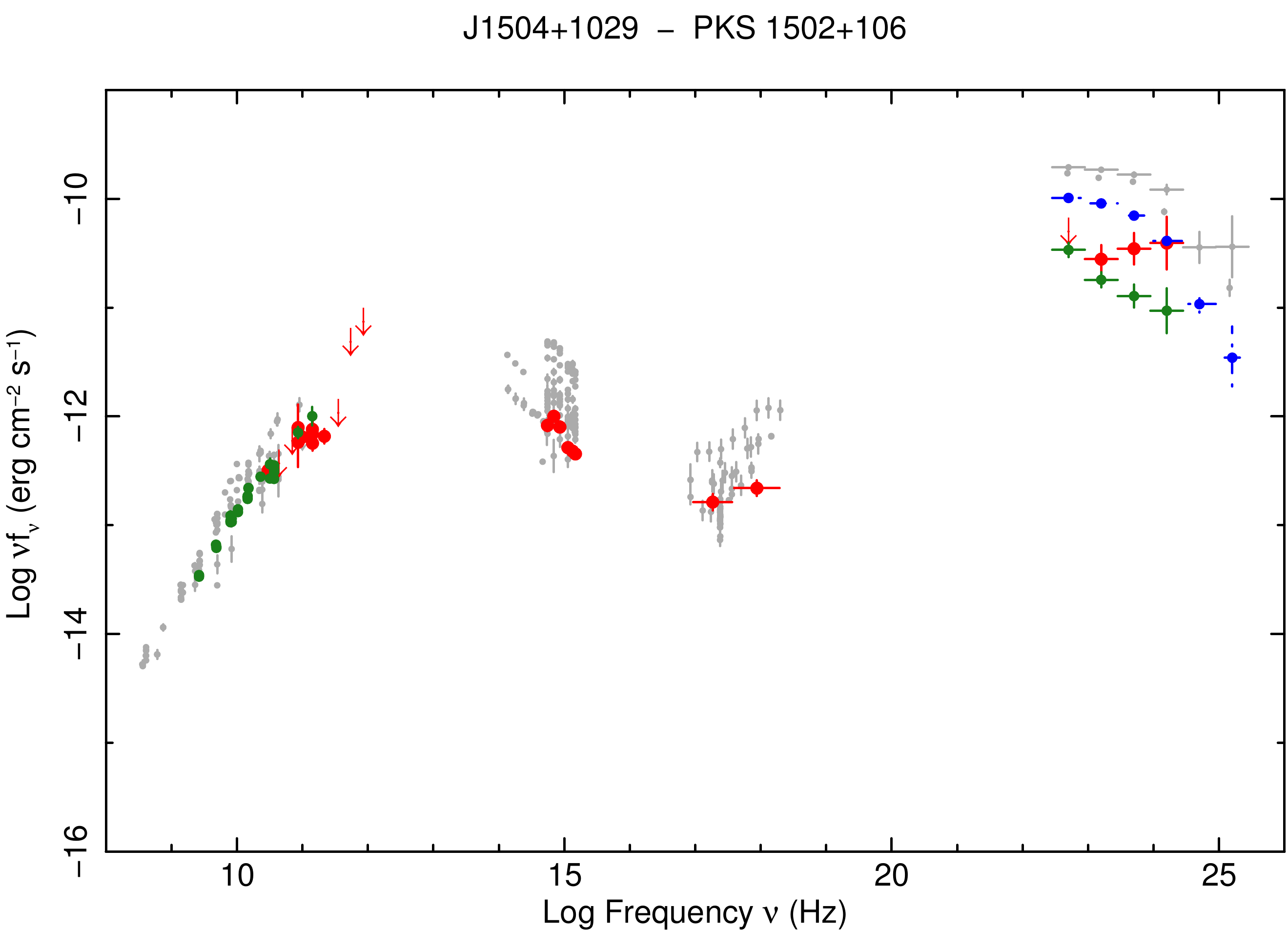}
\caption{The SED of 1WGAJ\,1407.5$-$2700 (J1407$-$2701, top left) and of 3C\,298.0 (J1419+0628, top right),
CSO\,643 (J1423+5055, middle left), of PG\,1424+240 (J1427+2348, middle right),
1RXSJ\,145603.4+504825 (J1456+5048, bottom left), and PKS\,1502+106 (J1504+1029, bottom right). 
Simultaneous data are shown in red; quasi-simultaneous data, i.e. {\it Fermi} data
integrated over 2 months, {\it Planck} ERCSC and non-simultaneous ground based observations
are shown in green; {\it Fermi} data integrated over 27 months are shown in blue;
literature or archival data are shown in light gray.}
\label{fig:sed31}
\end{figure*}

\begin{figure*}
\centering
\includegraphics[width=6.5cm,angle=-90]{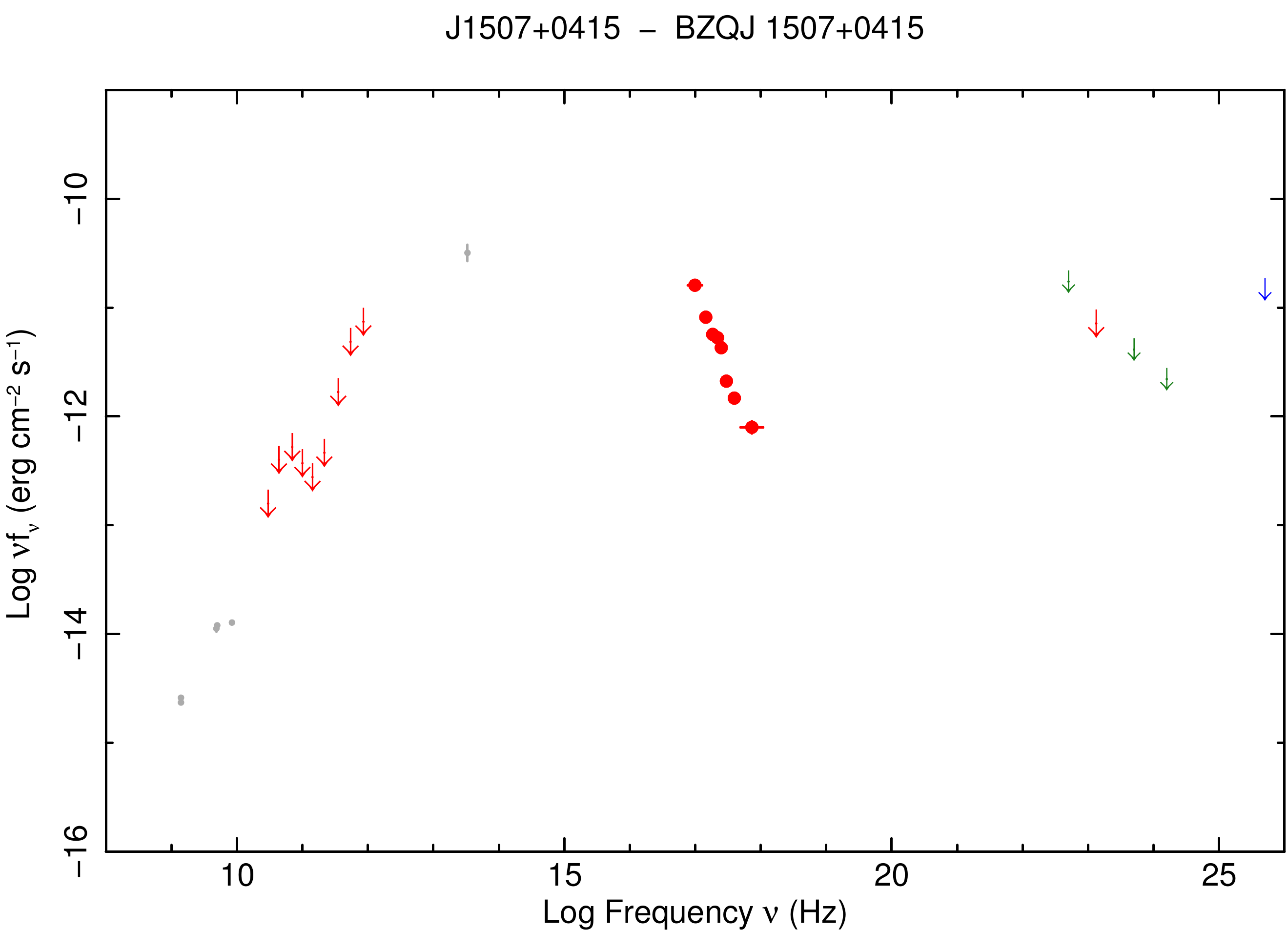}
\includegraphics[width=6.5cm,angle=-90]{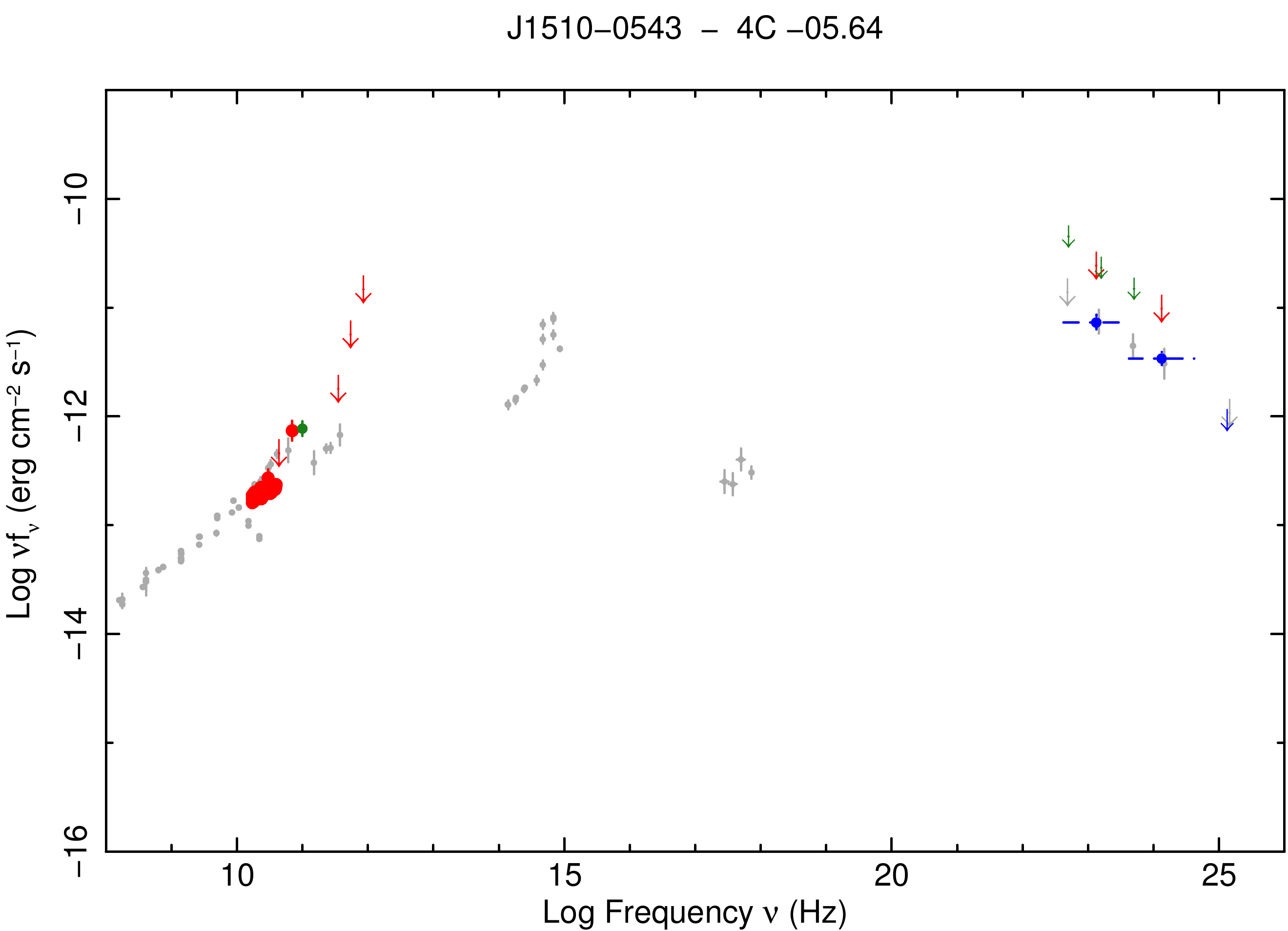}
\includegraphics[width=6.5cm,angle=-90]{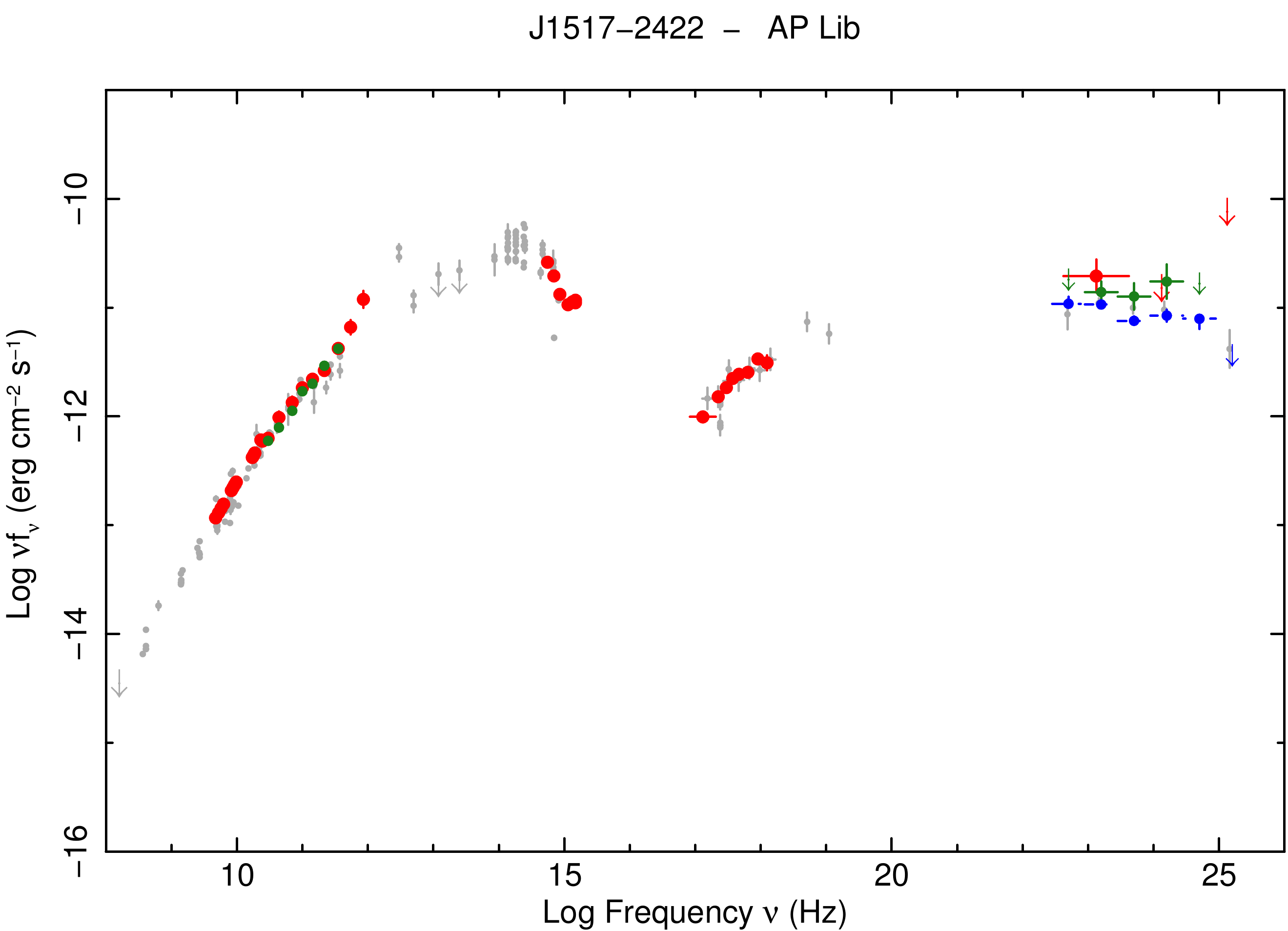}
\includegraphics[width=6.5cm,angle=-90]{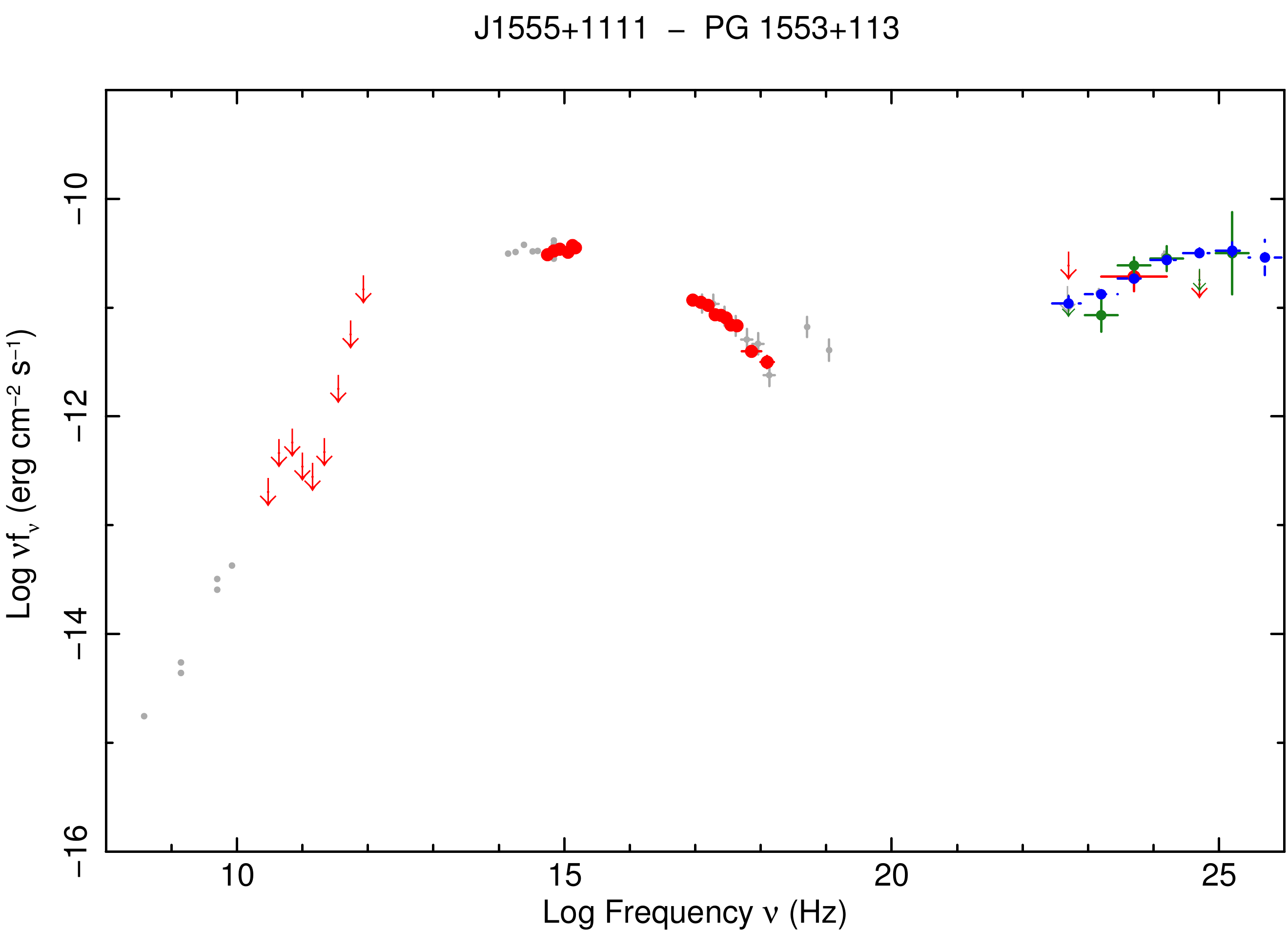}
\includegraphics[width=6.5cm,angle=-90]{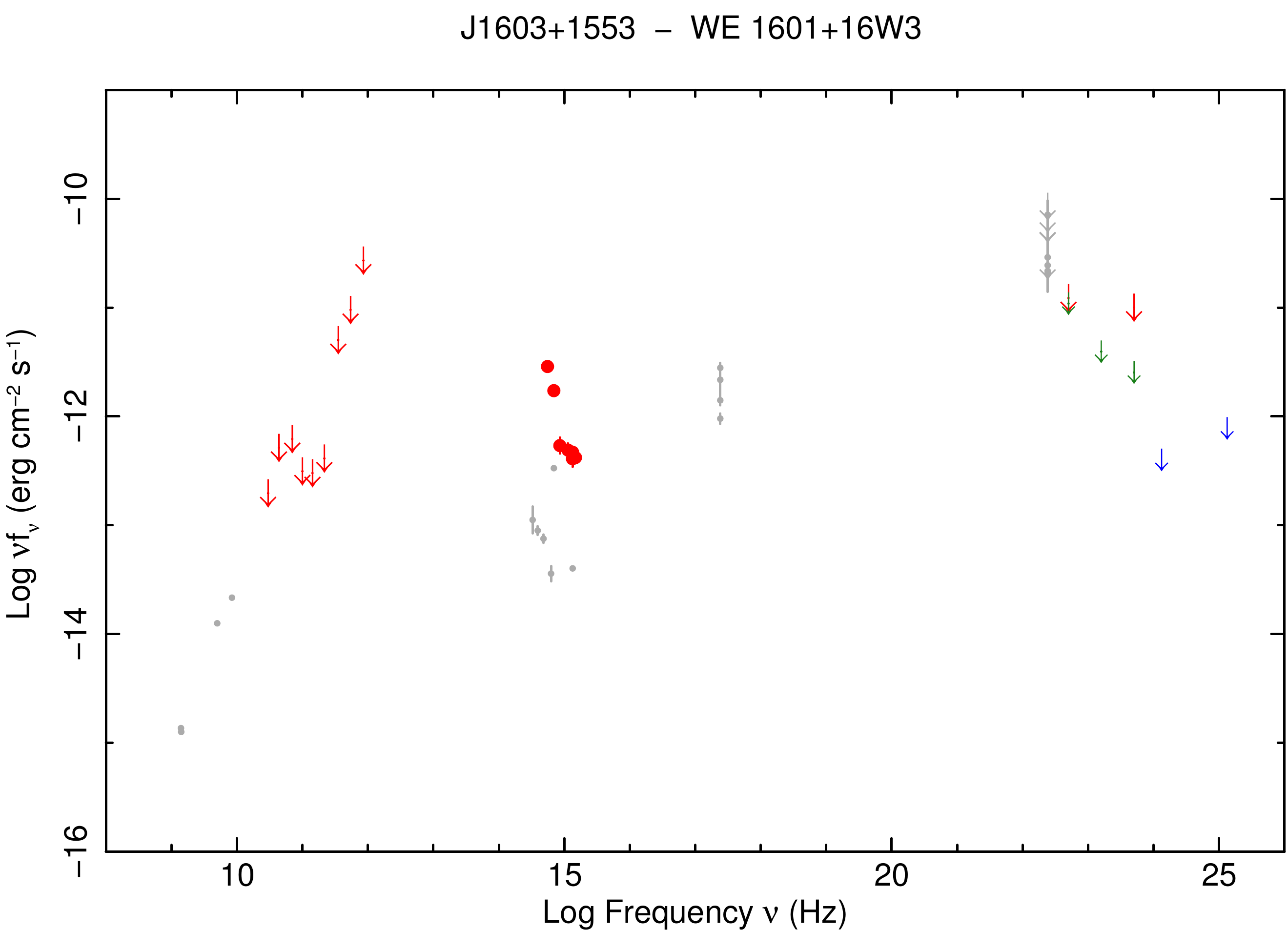}
\includegraphics[width=6.5cm,angle=-90]{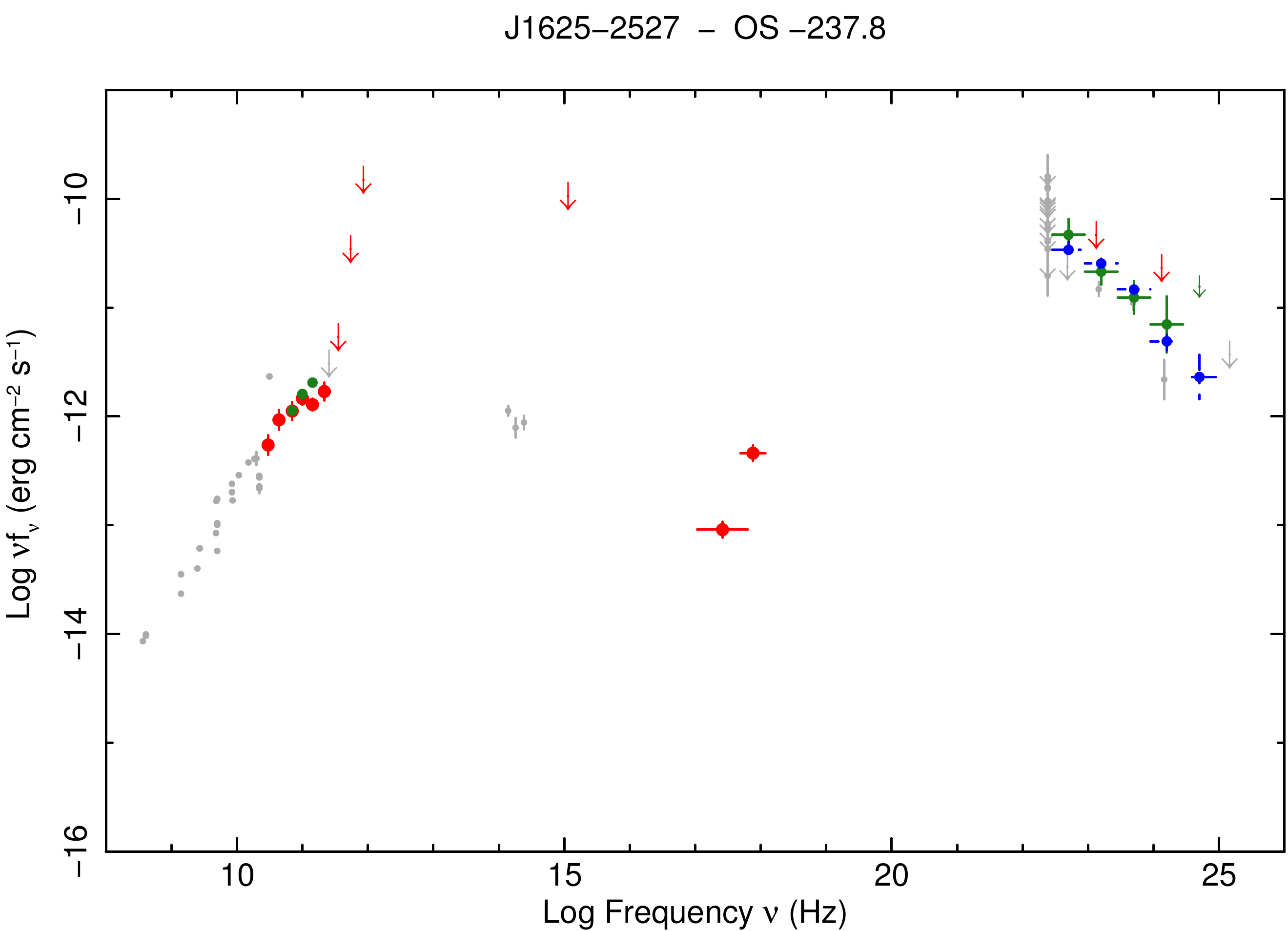}
\caption{The SED of BZQJ\,1507+0415 (J1507+0415, top left), 4C\,$-$05.64 (J1510$-$0543, top right),
AP\,Lib (J1517$-$2422, middle left), PG\,1553+113 (J1555+1111, middle right),
WE\,1601+16W3 (J1603+1553, bottom left), and OS\,$-$237.8 (J1625$-$2527, bottom right). 
Simultaneous data are shown in red; quasi-simultaneous data, i.e. {\it Fermi} data
integrated over 2 months, {\it Planck} ERCSC and non-simultaneous ground based observations
are shown in green; {\it Fermi} data integrated over 27 months are shown in blue;
literature or archival data are shown in light gray.}
\label{fig:sed34}
\end{figure*}

\begin{figure*}
\centering
\includegraphics[width=6.5cm,angle=-90]{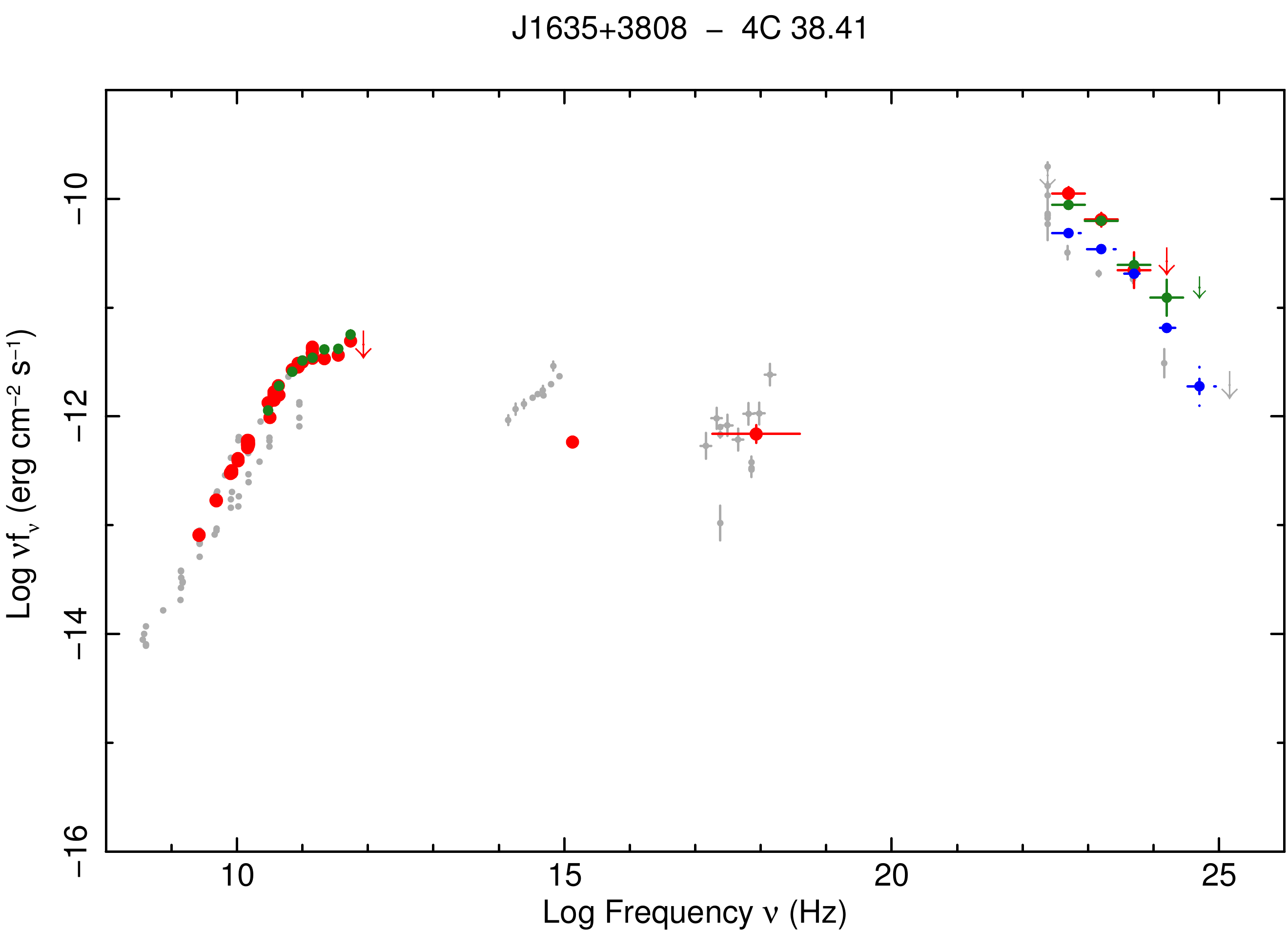}
\includegraphics[width=6.5cm,angle=-90]{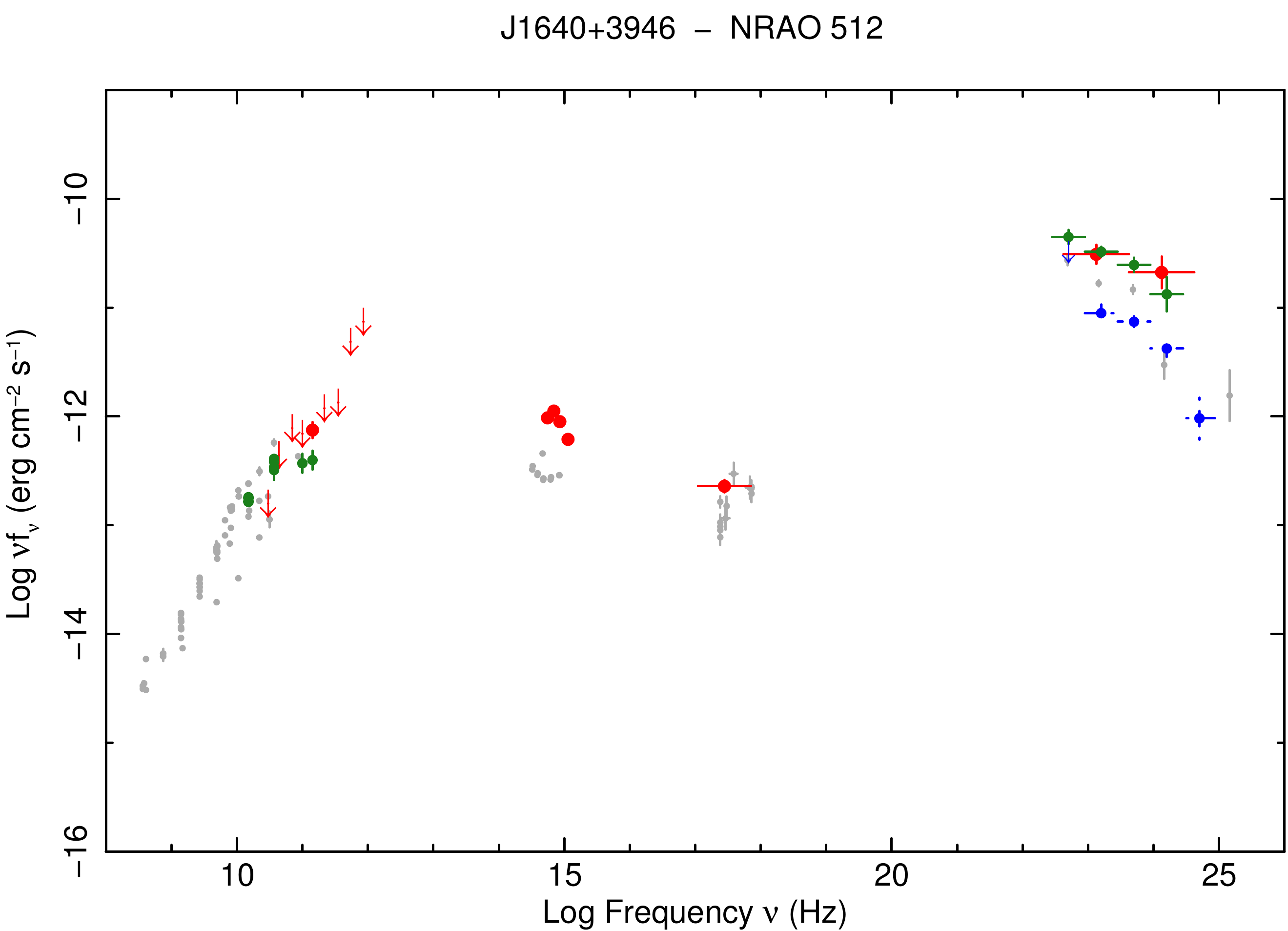}
\includegraphics[width=6.5cm,angle=-90]{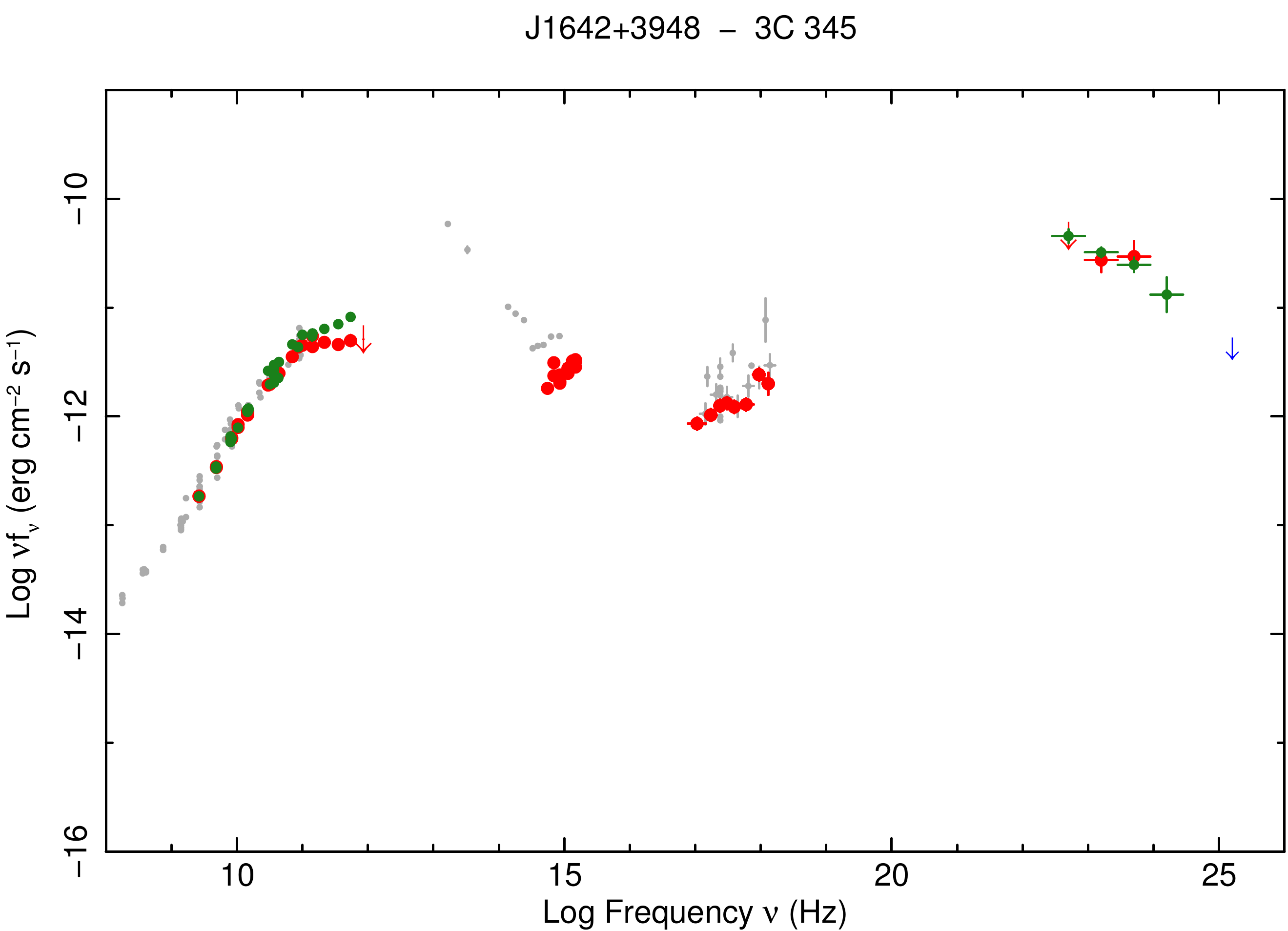}
\includegraphics[width=6.5cm,angle=-90]{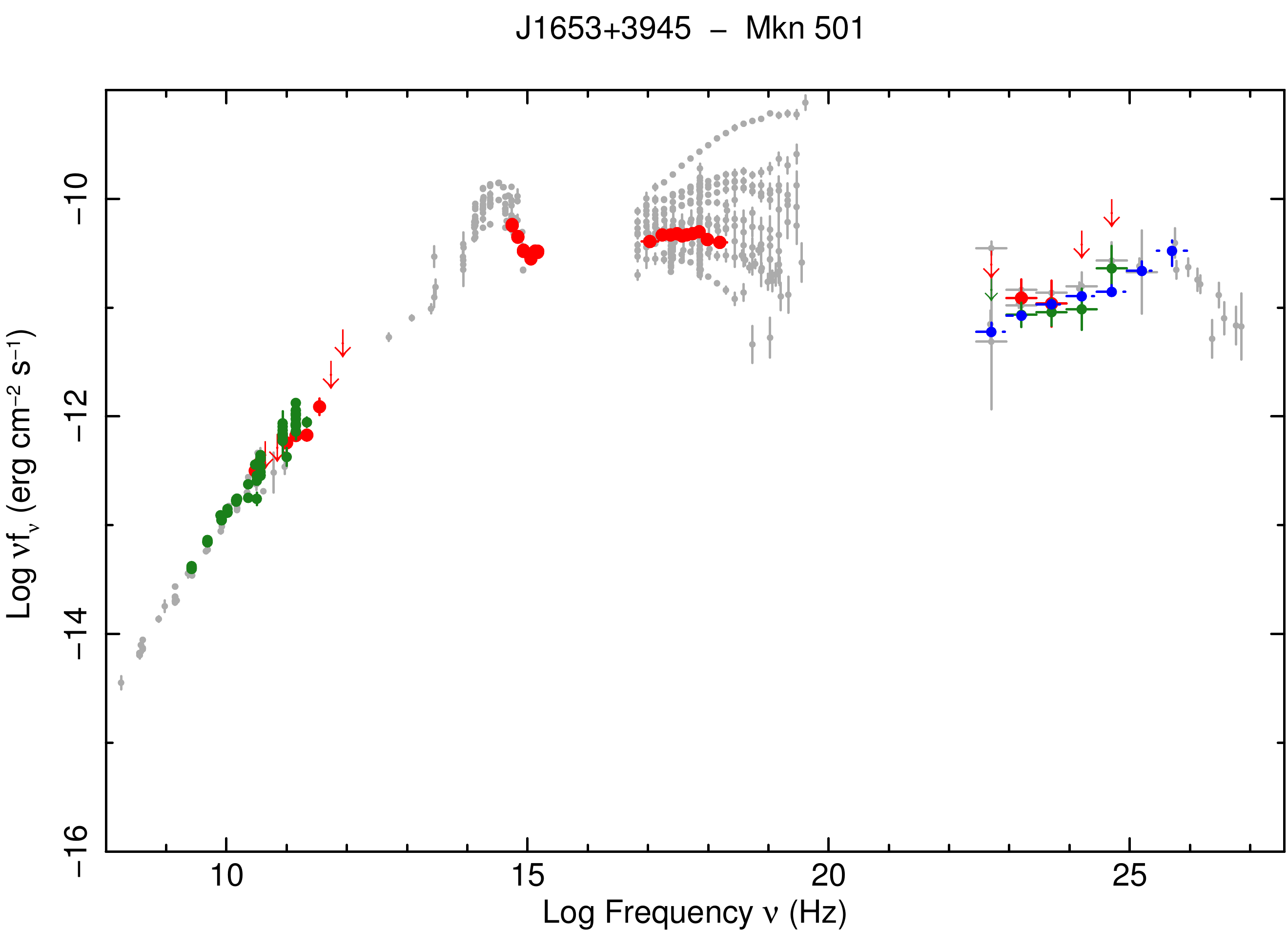}
\includegraphics[width=6.5cm,angle=-90]{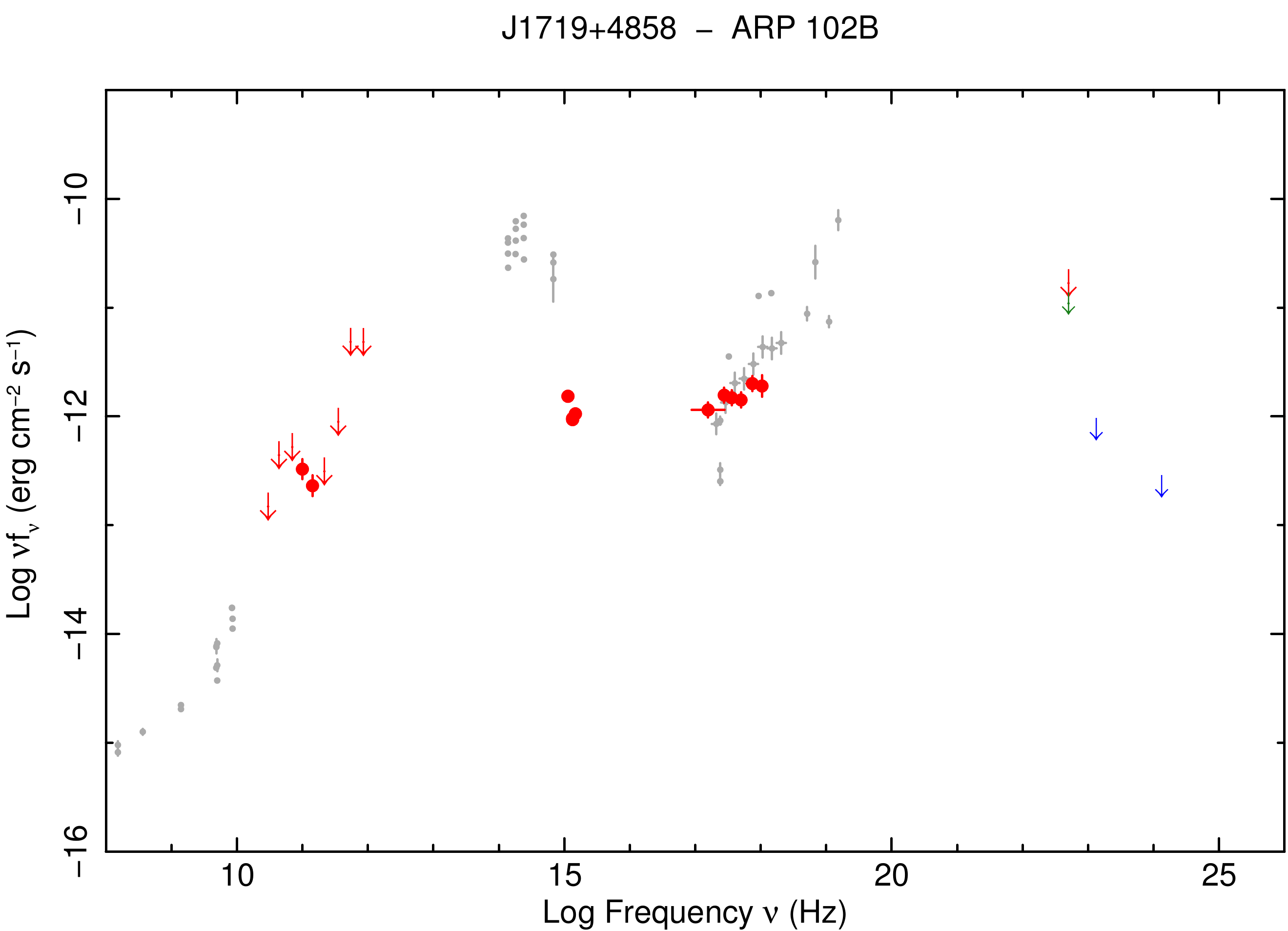}
\includegraphics[width=6.5cm,angle=-90]{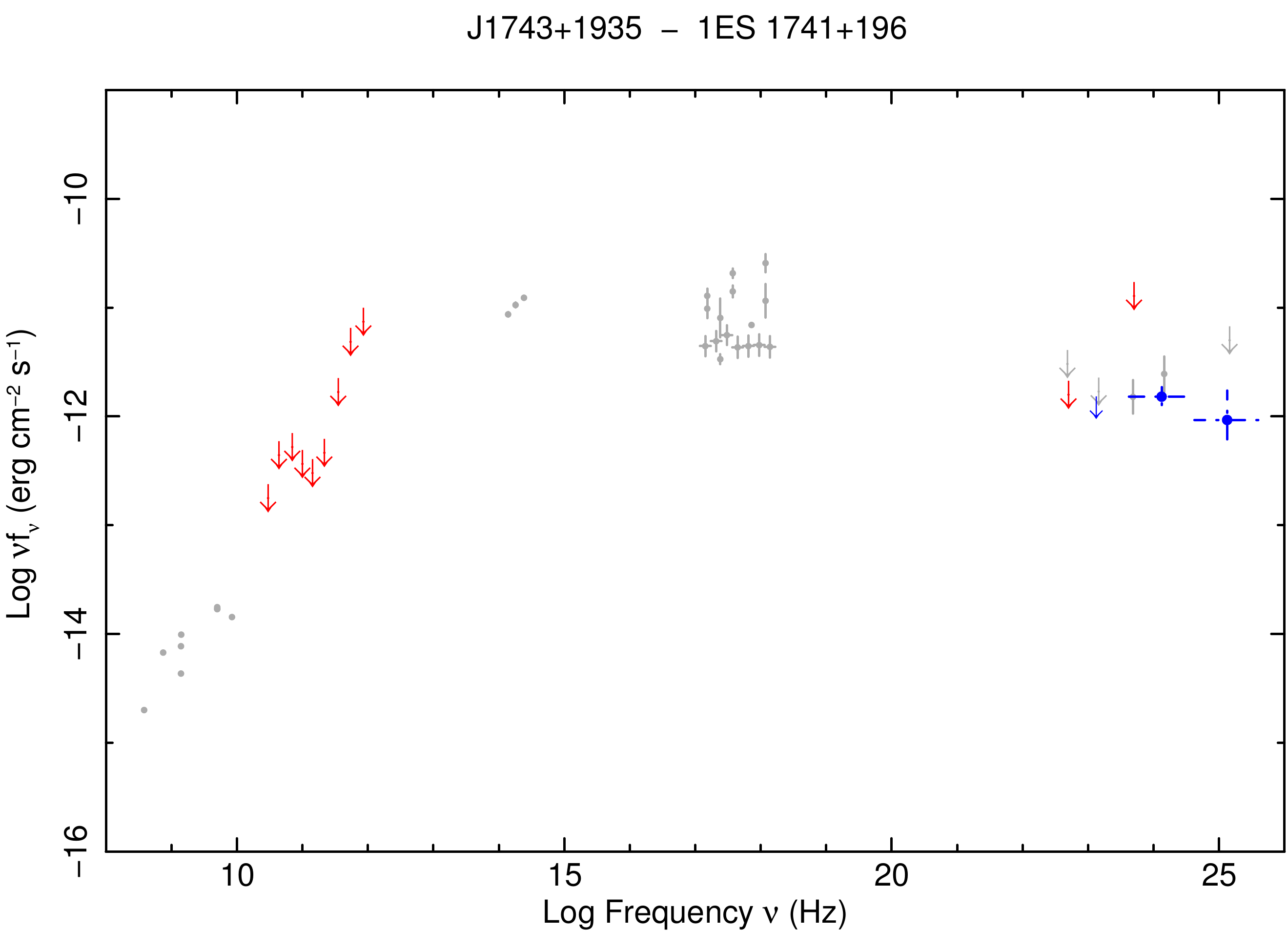}
\caption{The SED of 4C\,38.41 (J1635+3808, top left), NRAO\,512 (J1640+3946, top right),
3C\,345 (J1642+3948, middle left), Mkn\,501 (J1653+3945, middle right),
ARP\,102B (J1719+4858, bottom left), and 1ES\,1741+196 (J1743+1935, bottom right). 
Simultaneous data are shown in red; quasi-simultaneous data, i.e. {\it Fermi} data
integrated over 2 months, {\it Planck} ERCSC and non-simultaneous ground based observations
are shown in green; {\it Fermi} data integrated over 27 months are shown in blue;
literature or archival data are shown in light gray.}
\label{fig:sed37}
\end{figure*}

\begin{figure*}
\centering
\includegraphics[width=6.5cm,angle=-90]{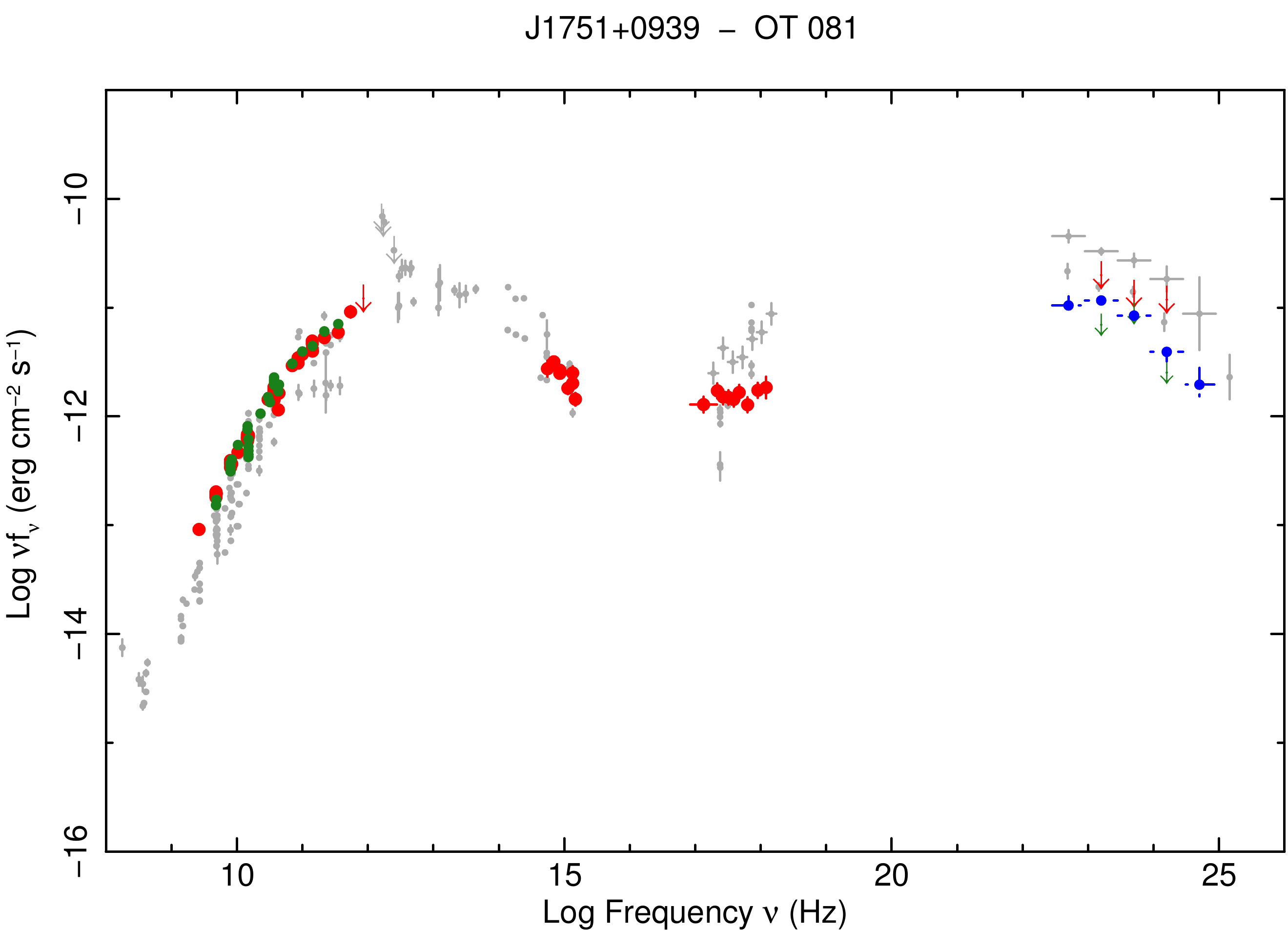}
\includegraphics[width=6.5cm,angle=-90]{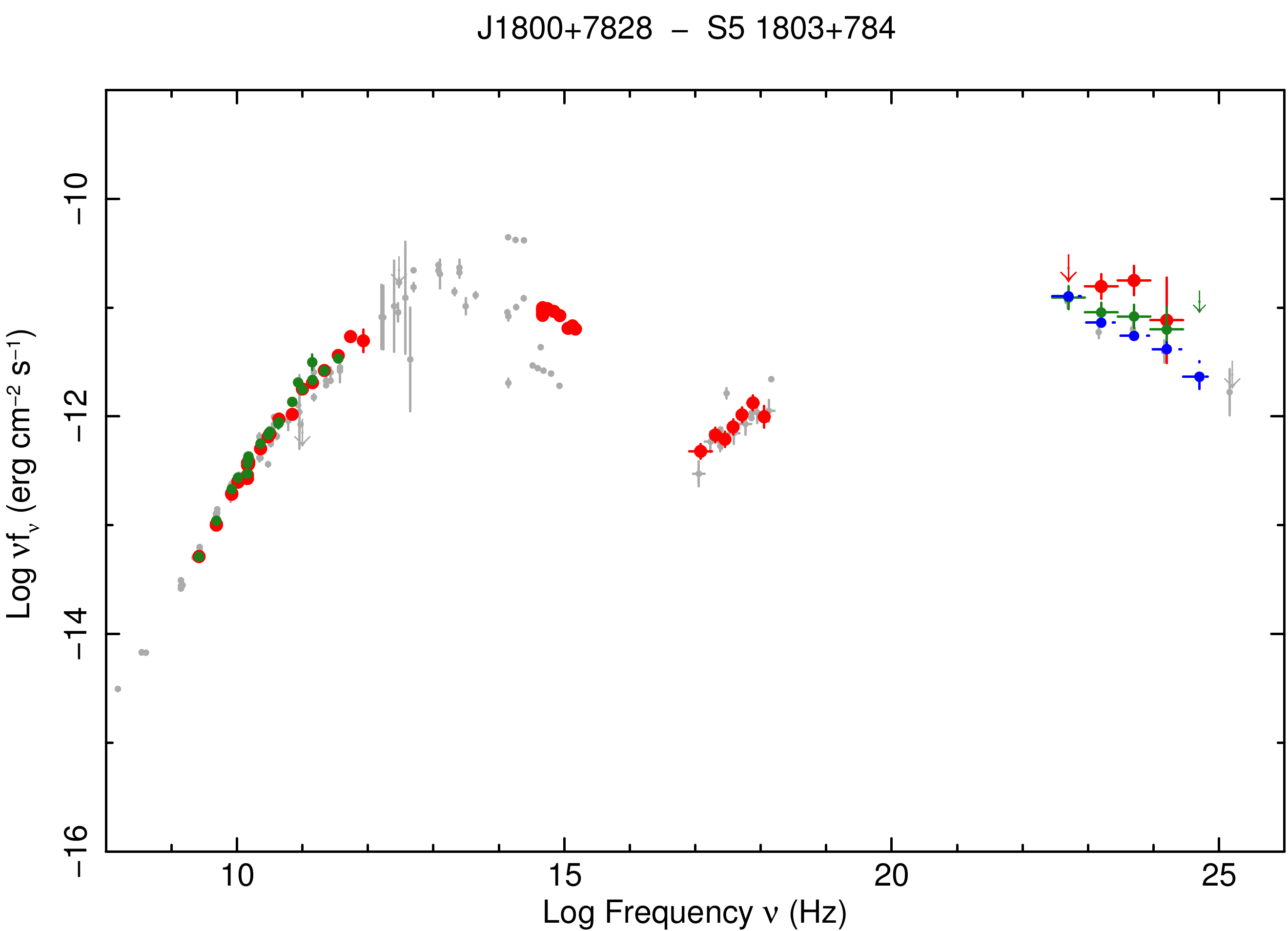}
\includegraphics[width=6.5cm,angle=-90]{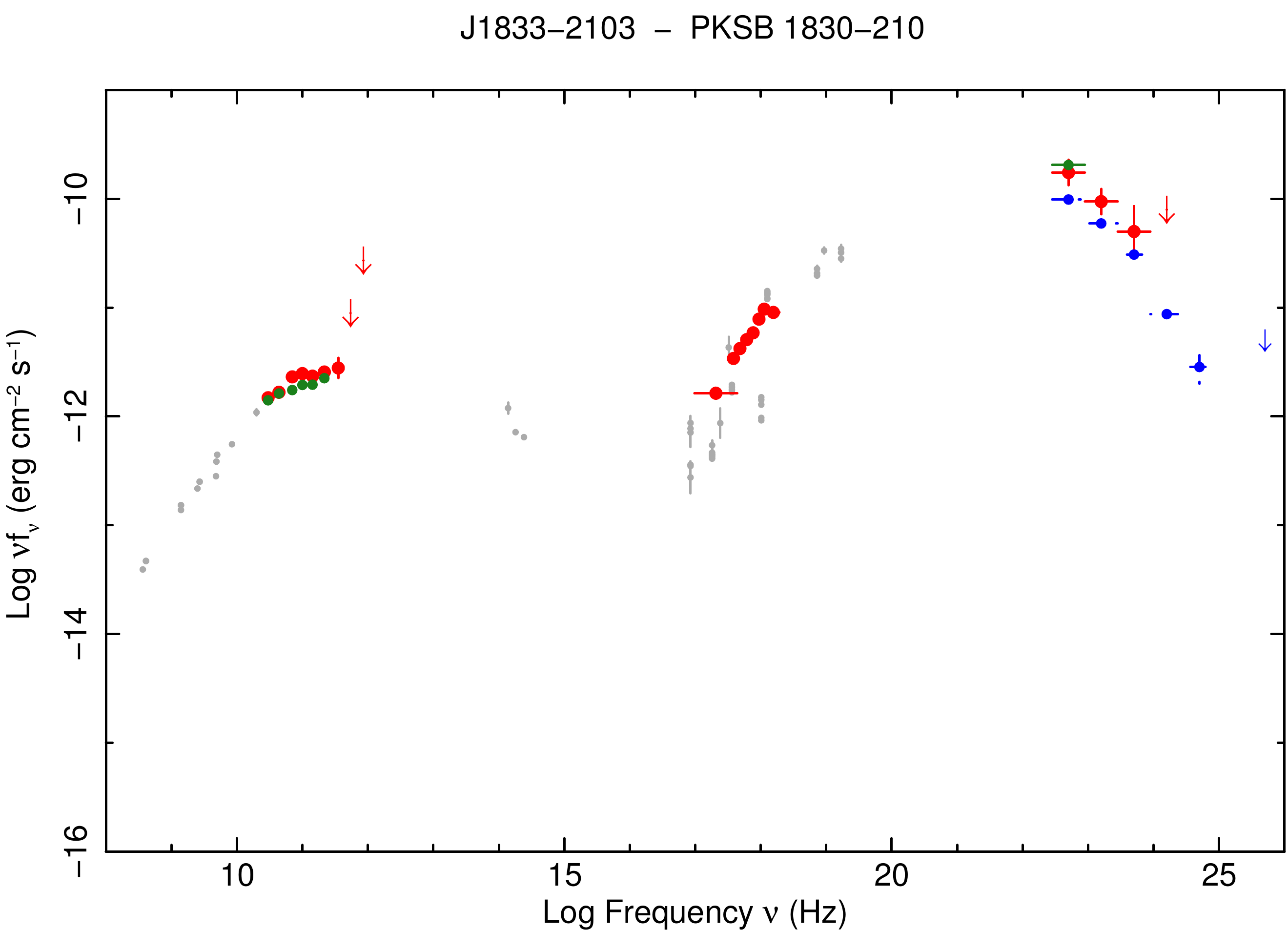}
\includegraphics[width=6.5cm,angle=-90]{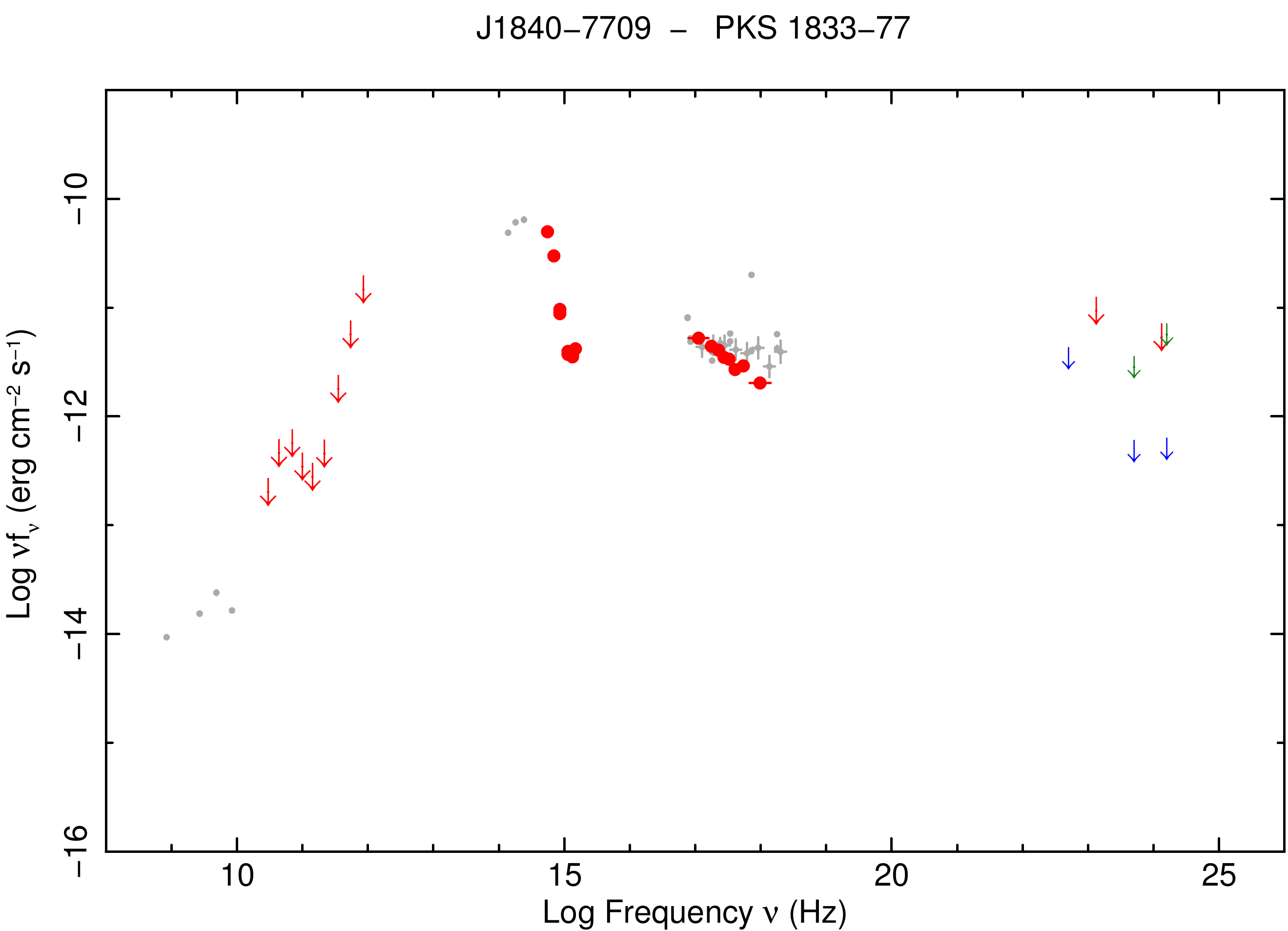}
\includegraphics[width=6.5cm,angle=-90]{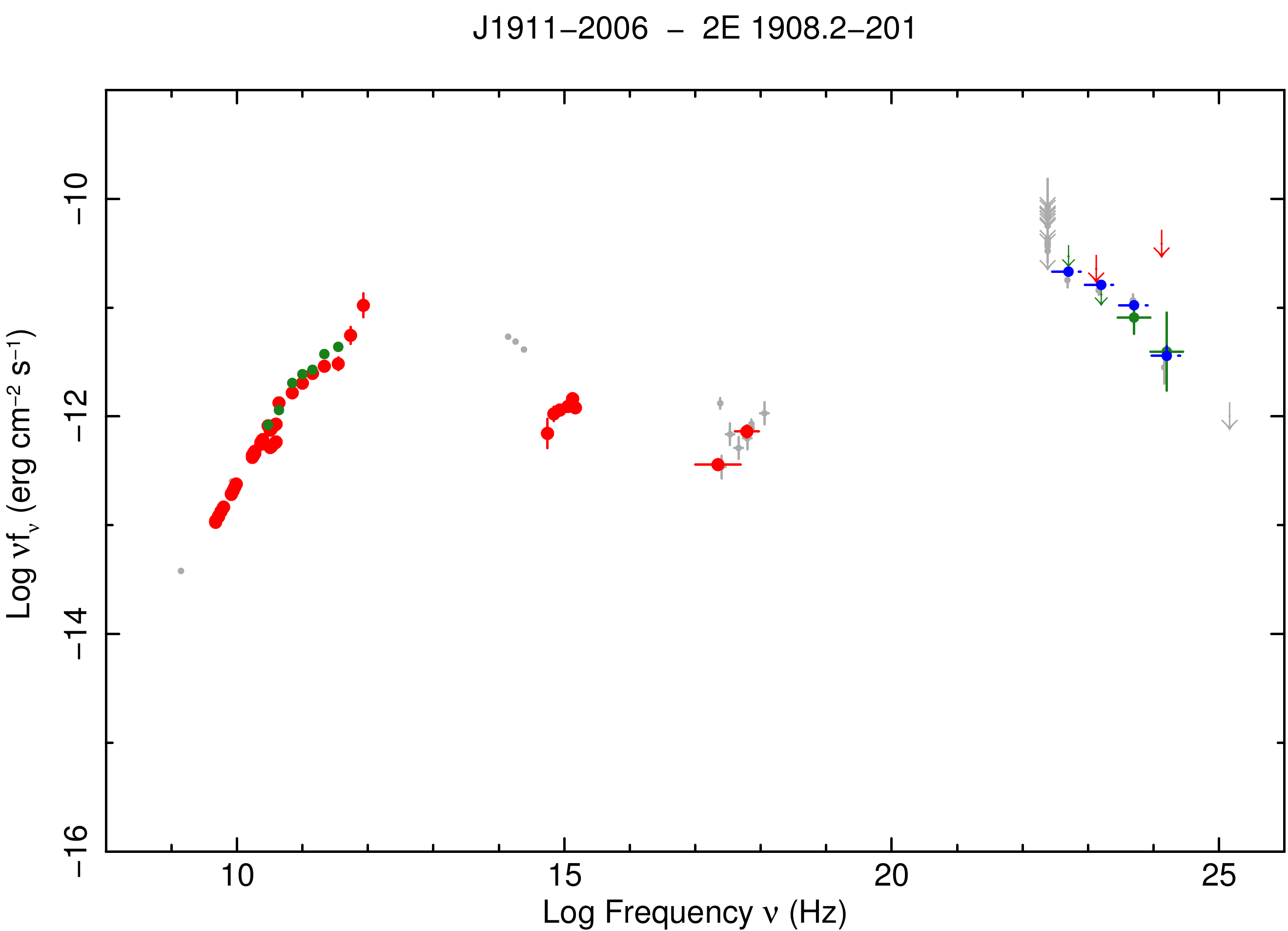}
\includegraphics[width=6.5cm,angle=-90]{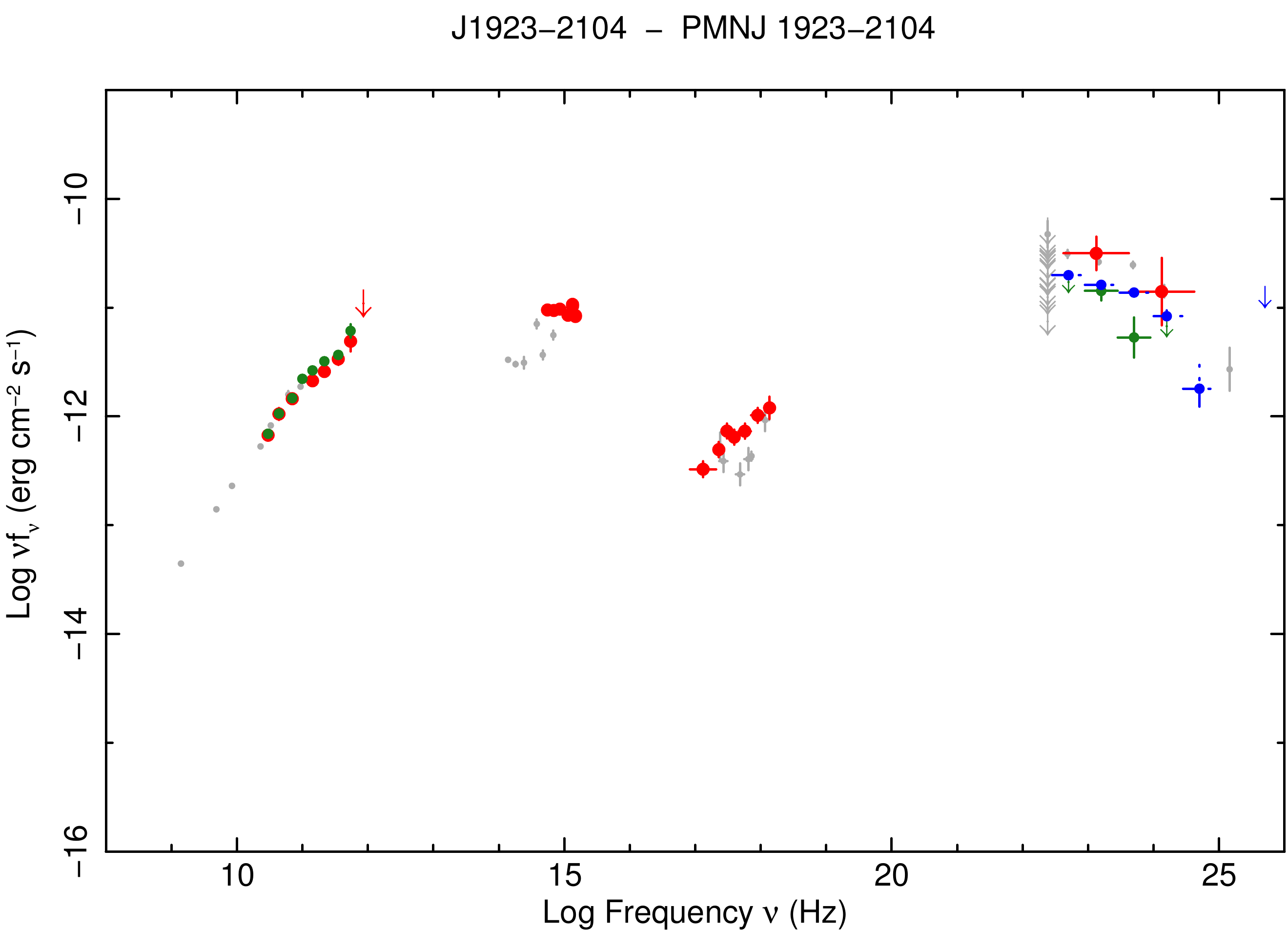}
\caption{The SED of OT\,081 (J1751+0939, top left), S5\,1803+784 (J1800+782, top right),
PKSB\,1830$-$210 (J1833$-$2103, middle left), PKS\,1833$-$77 (J1840$-$7709, middle right),
2E\,1908.2$-$201 (J1911$-$2006, bottom left), and PMNJ\,1923$-$2104 (J1923$-$2104, bottom right). 
Simultaneous data are shown in red; quasi-simultaneous data, i.e. {\it Fermi} data
integrated over 2 months, {\it Planck} ERCSC and non-simultaneous ground based observations
are shown in green; {\it Fermi} data integrated over 27 months are shown in blue;
literature or archival data are shown in light gray.}
\label{fig:sed40}
\end{figure*}

\begin{figure*}
\centering
\includegraphics[width=6.5cm,angle=-90]{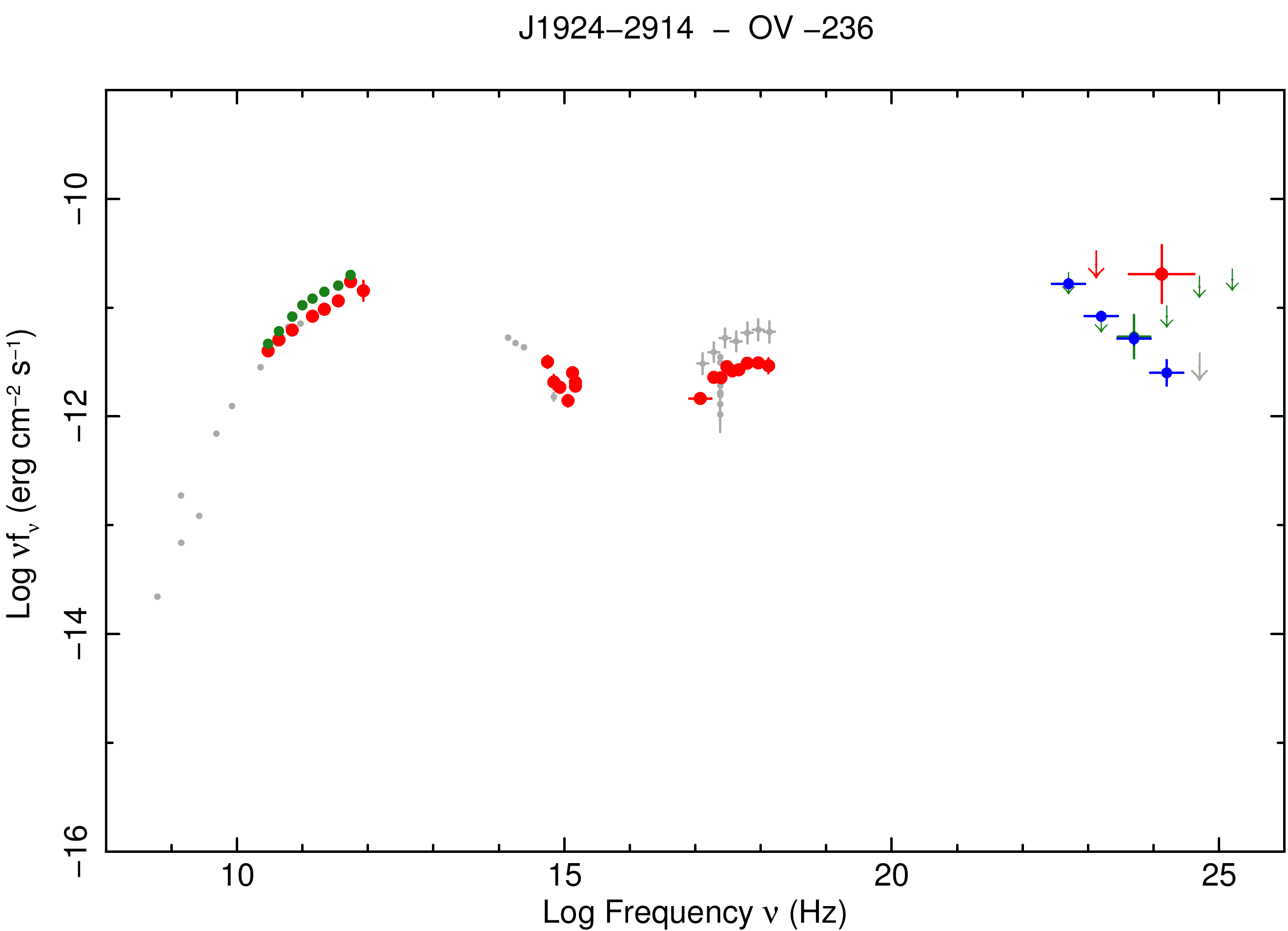}
\includegraphics[width=6.5cm,angle=-90]{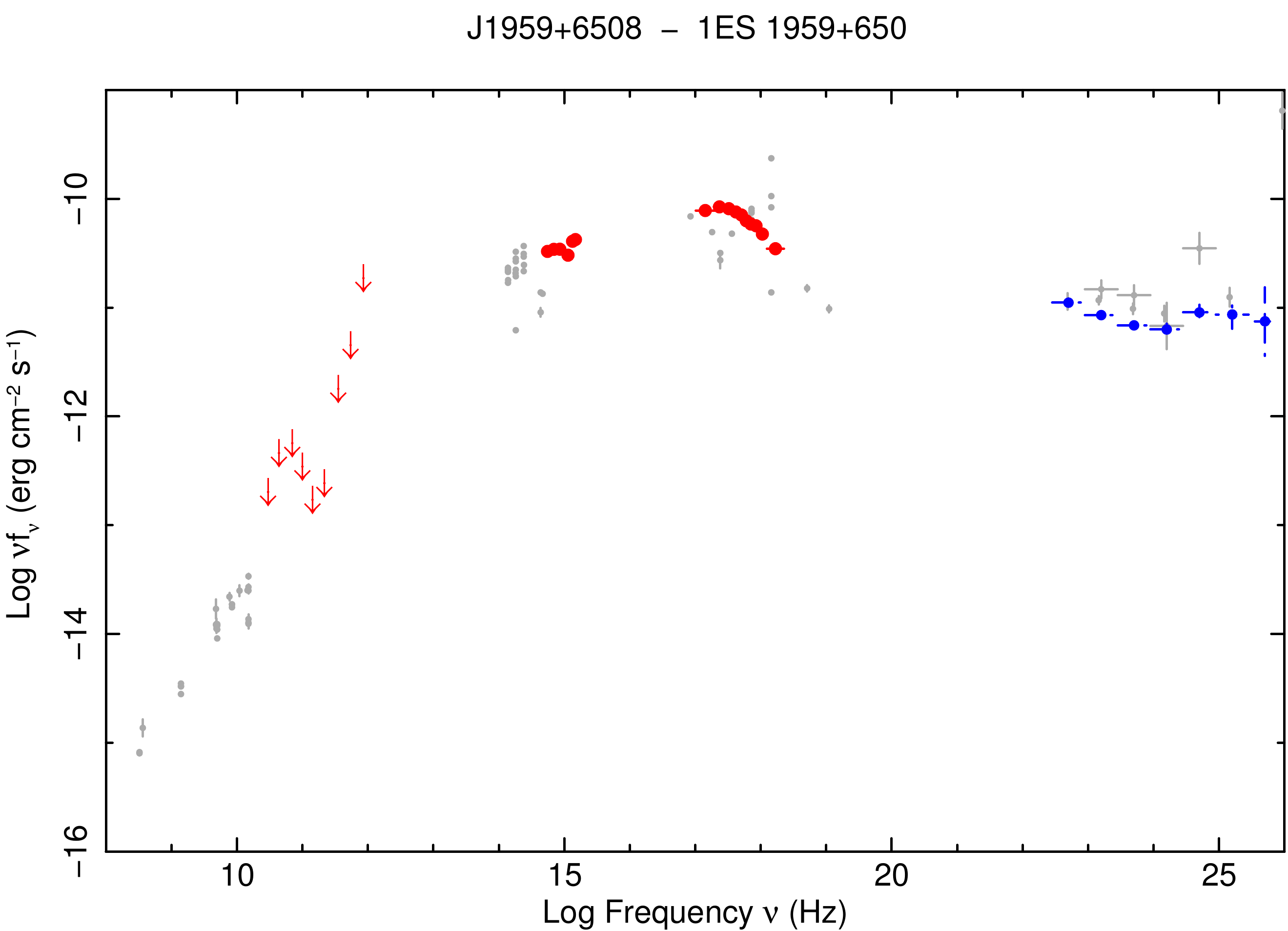}
\includegraphics[width=6.5cm,angle=-90]{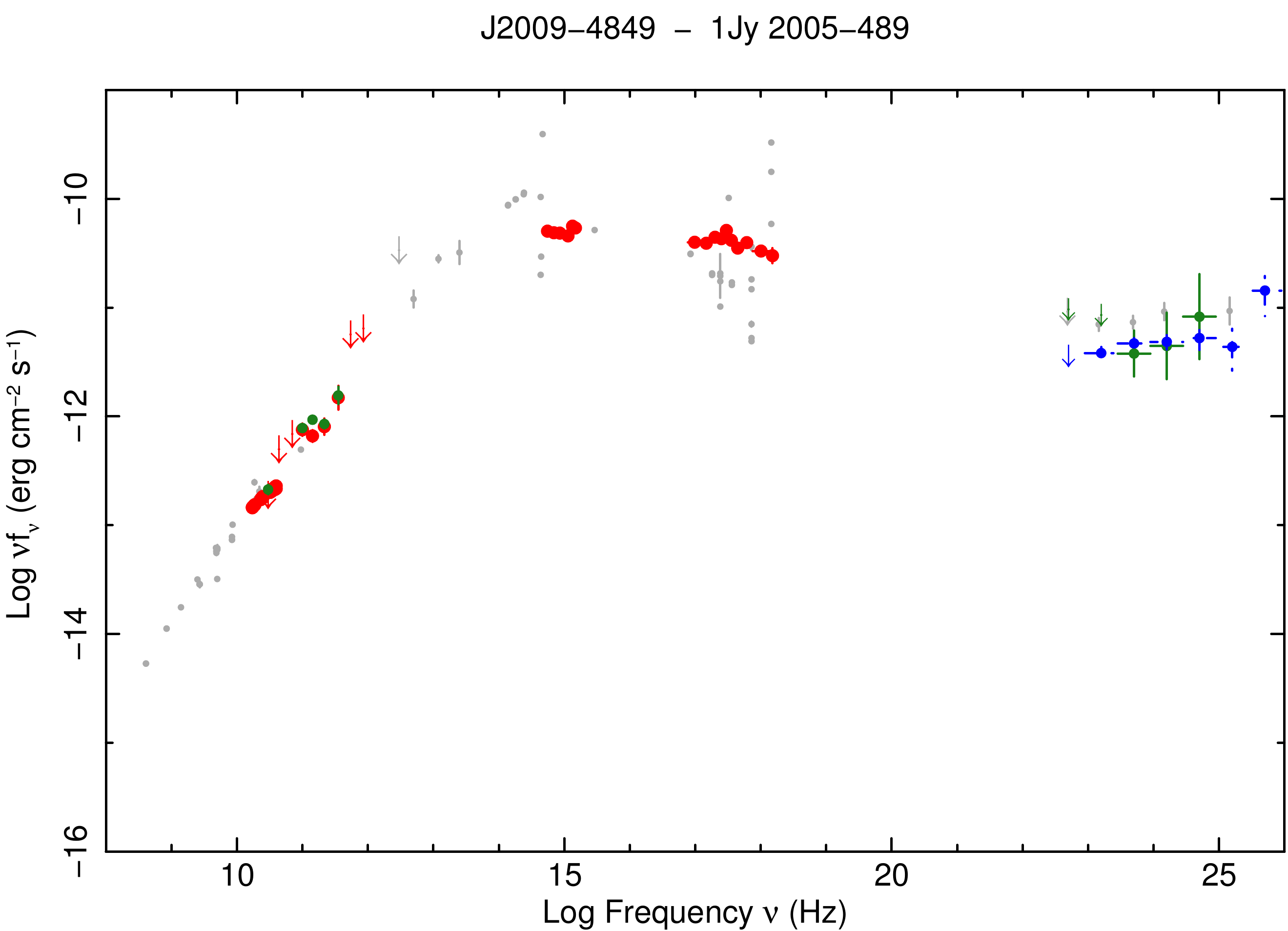}
\includegraphics[width=6.5cm,angle=-90]{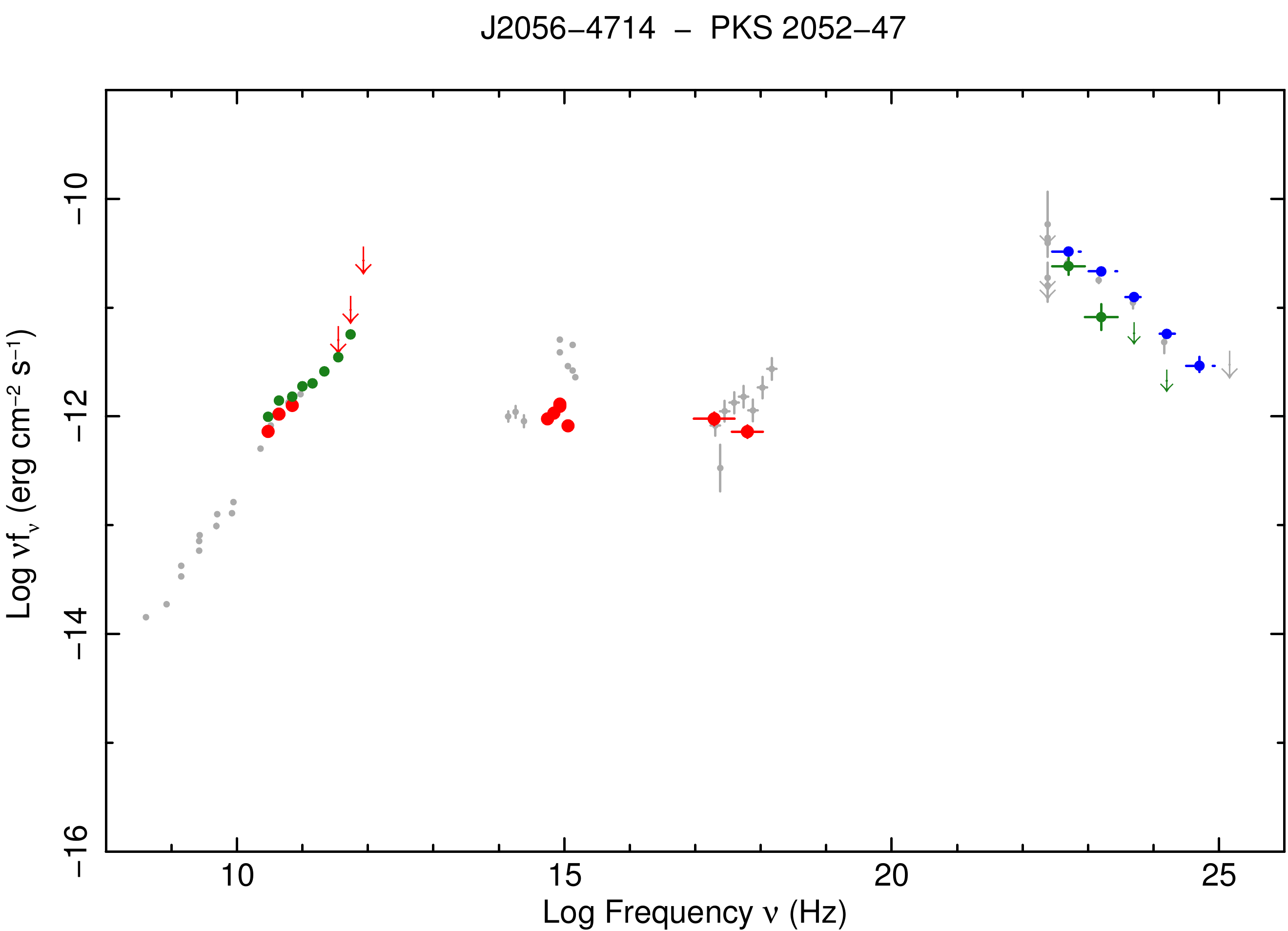}
\includegraphics[width=6.5cm,angle=-90]{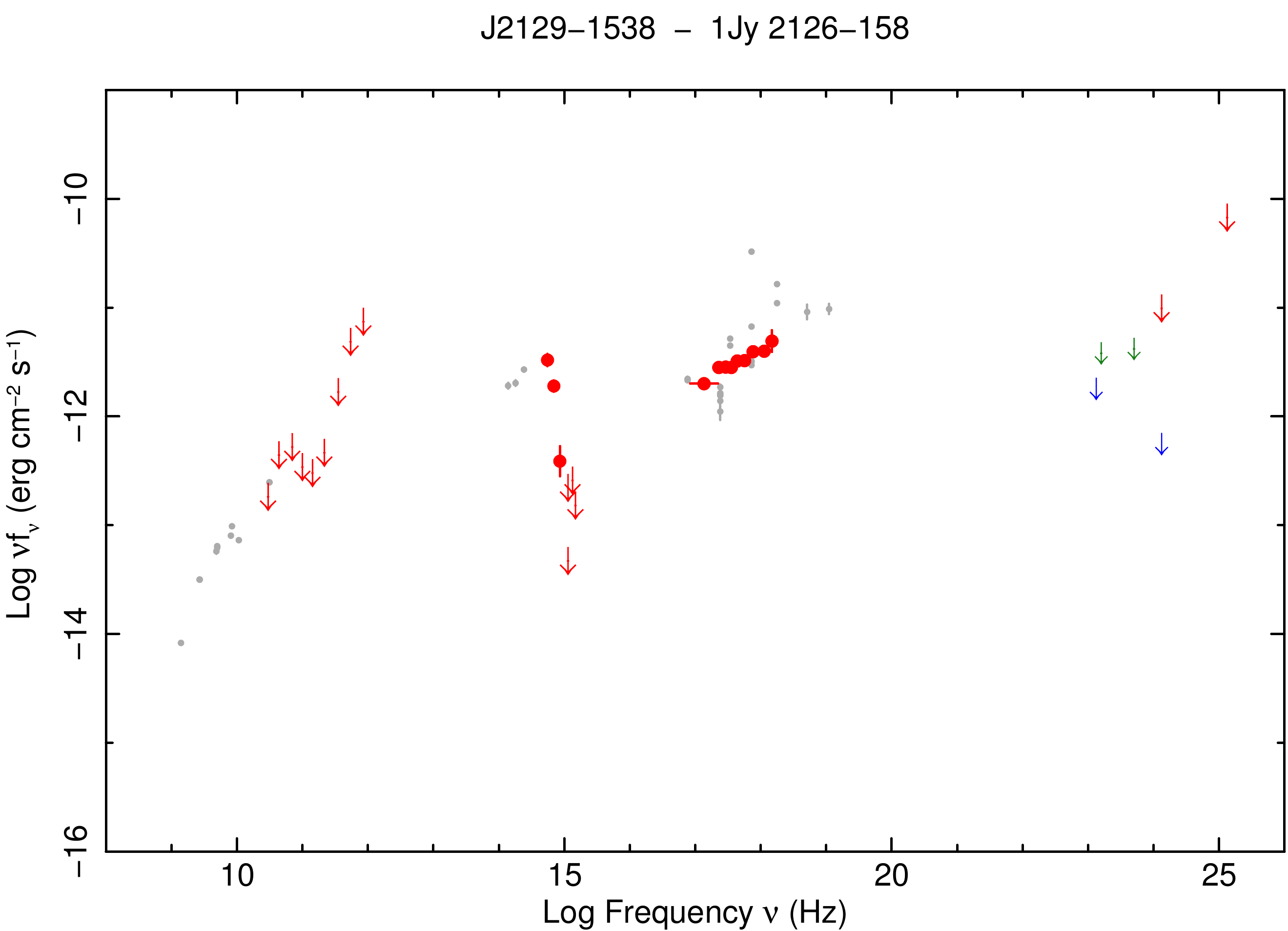}
\includegraphics[width=6.5cm,angle=-90]{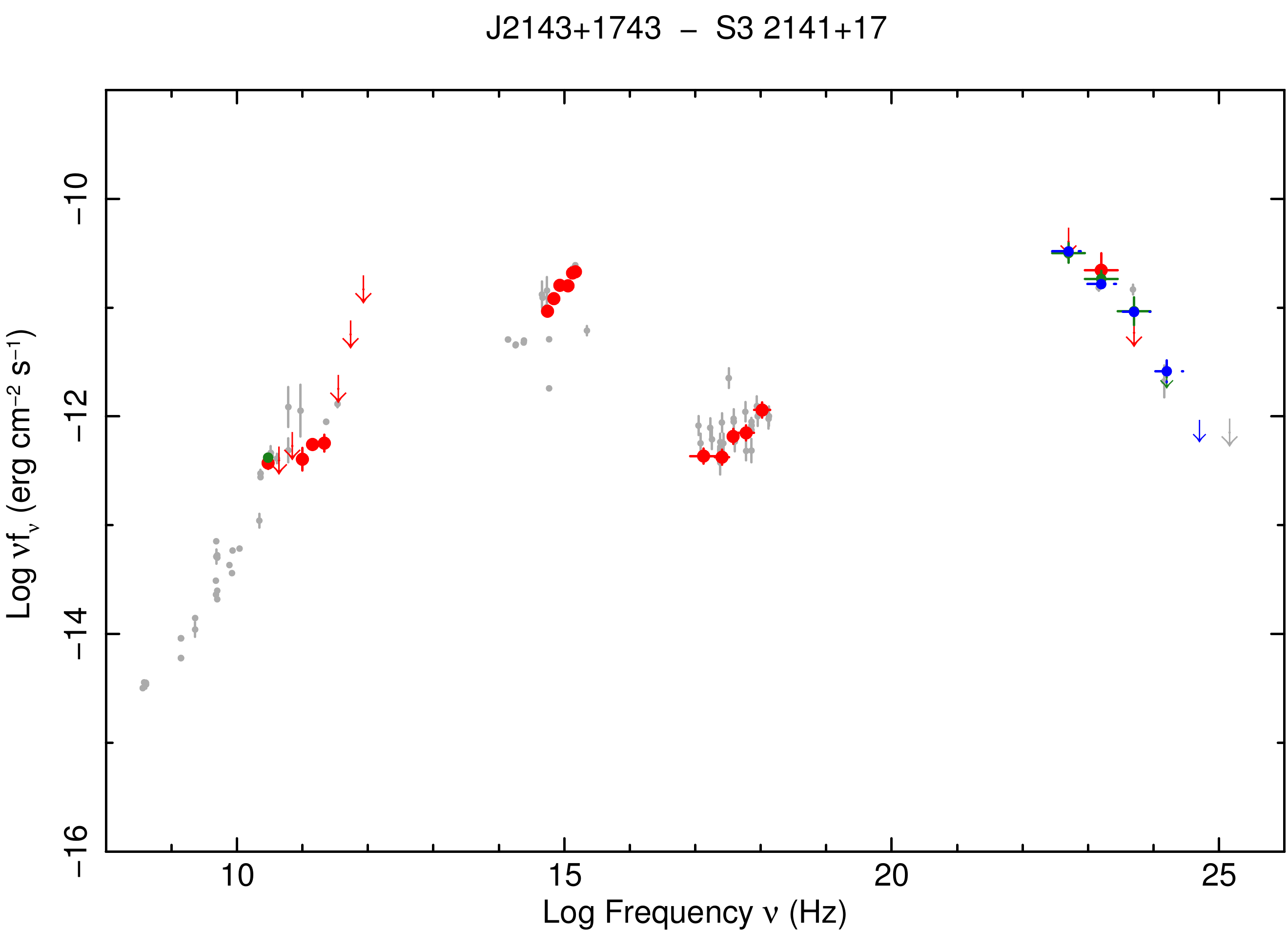}
\caption{The SED of OV$-$236 (J1924$-$2914, top left), 1ES\,1959+650 (J1959+6508, top right),
1Jy\,2005$-$489 (J2009$-$4849, middle left), PKS\,2052$-$47 (J2056$-$4714, middle right),
1Jy\,2126$-$158 (J2129$-$1538, bottom left), and S3\,2141+17 (J2143+1743, bottom right). 
Simultaneous data are shown in red; quasi-simultaneous data, i.e. {\it Fermi} data
integrated over 2 months, {\it Planck} ERCSC and non-simultaneous ground based observations
are shown in green; {\it Fermi} data integrated over 27 months are shown in blue;
literature or archival data are shown in light gray.}
\label{fig:sed43}
\end{figure*}

\begin{figure*}
\centering
\includegraphics[width=6.5cm,angle=-90]{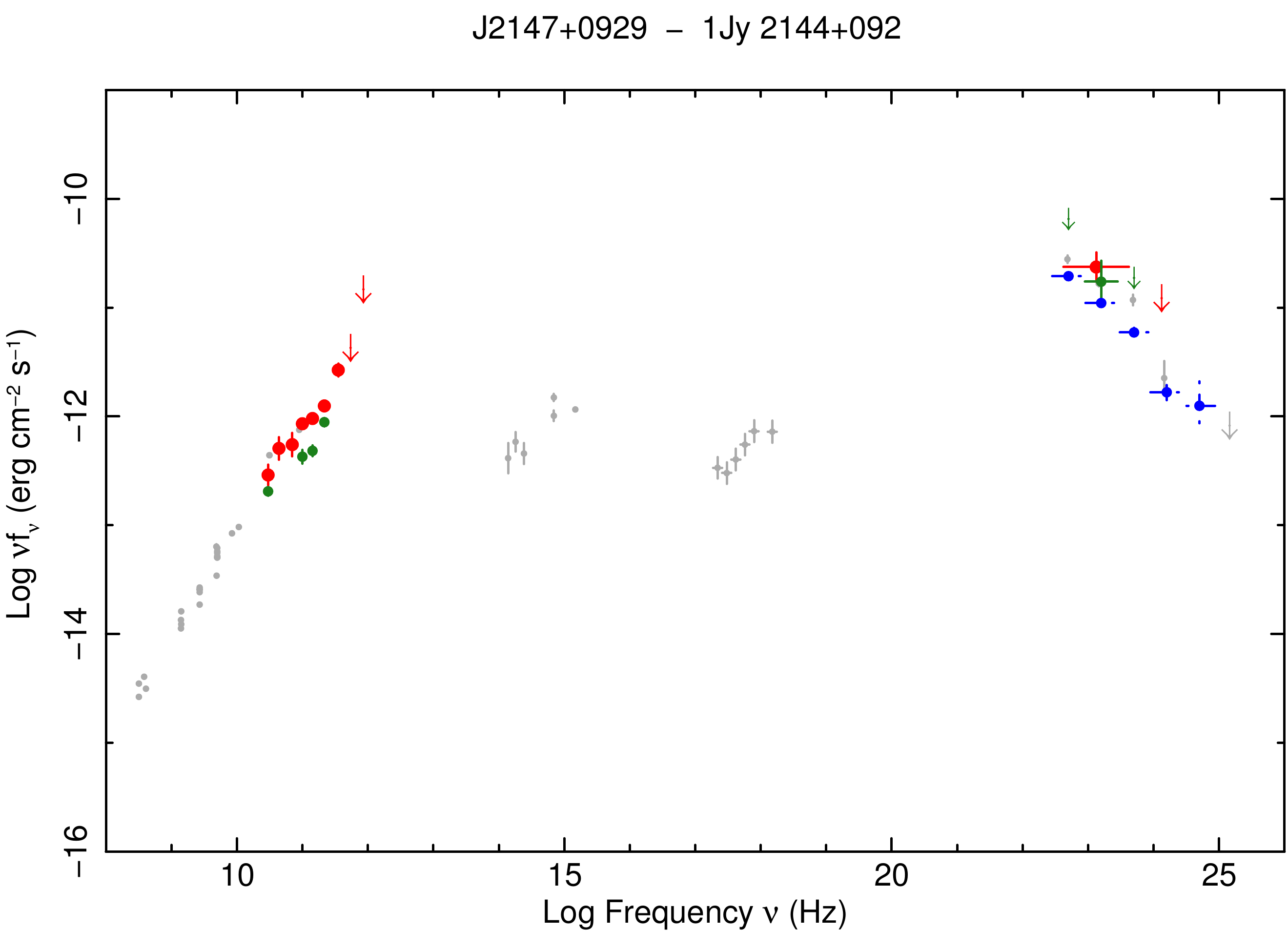}
\includegraphics[width=6.5cm,angle=-90]{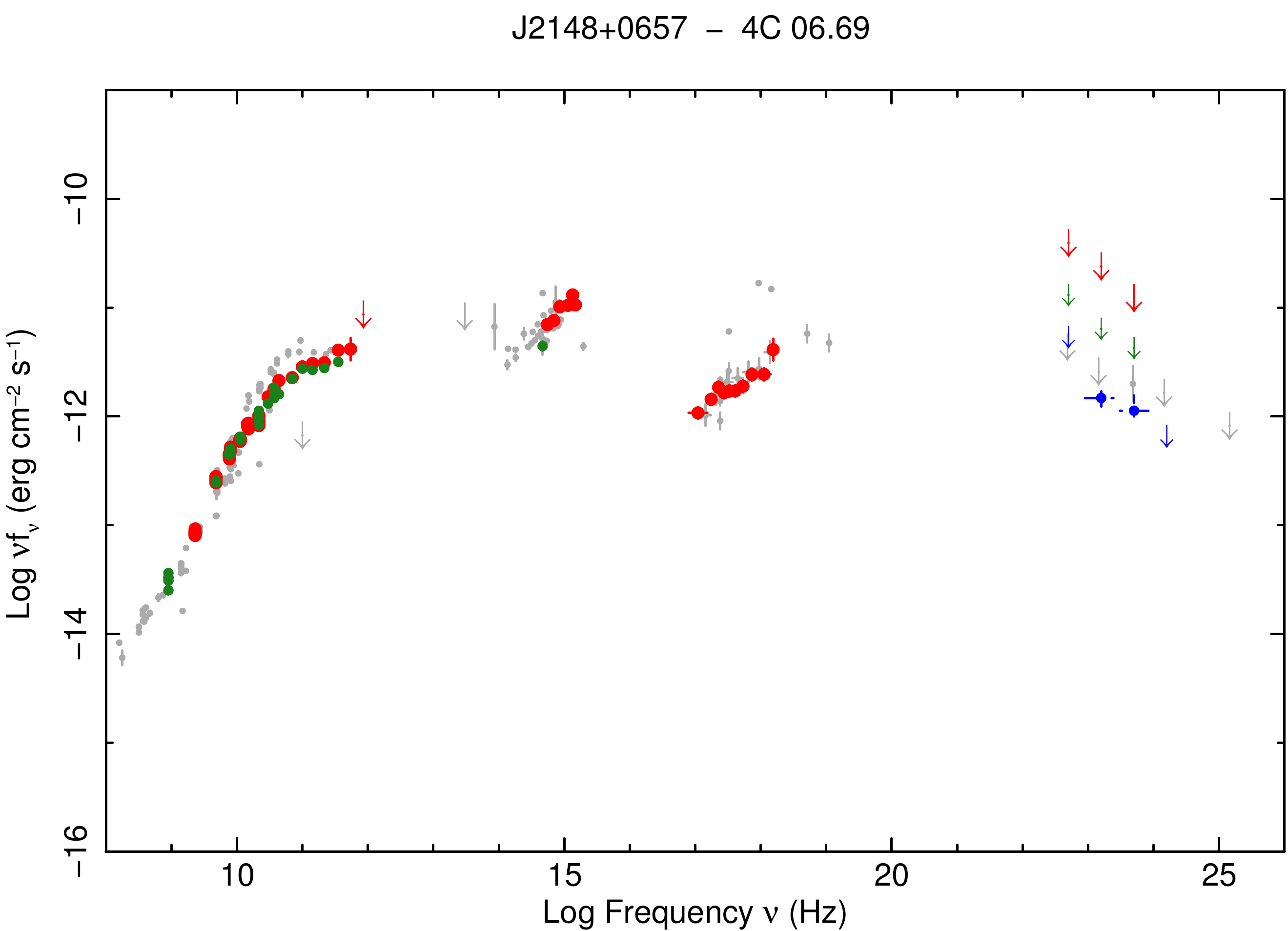}
\includegraphics[width=6.5cm,angle=-90]{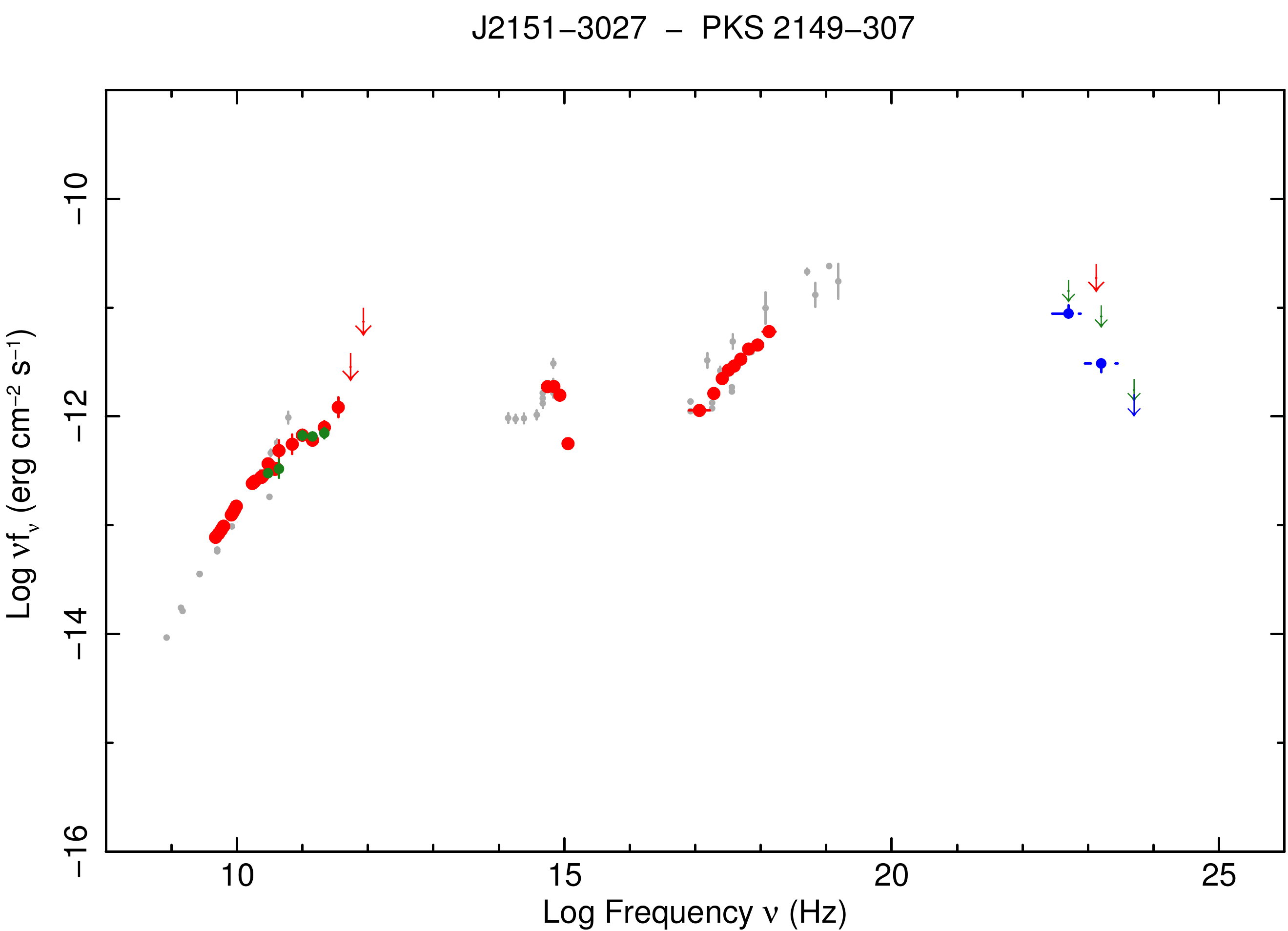}
\includegraphics[width=6.5cm,angle=-90]{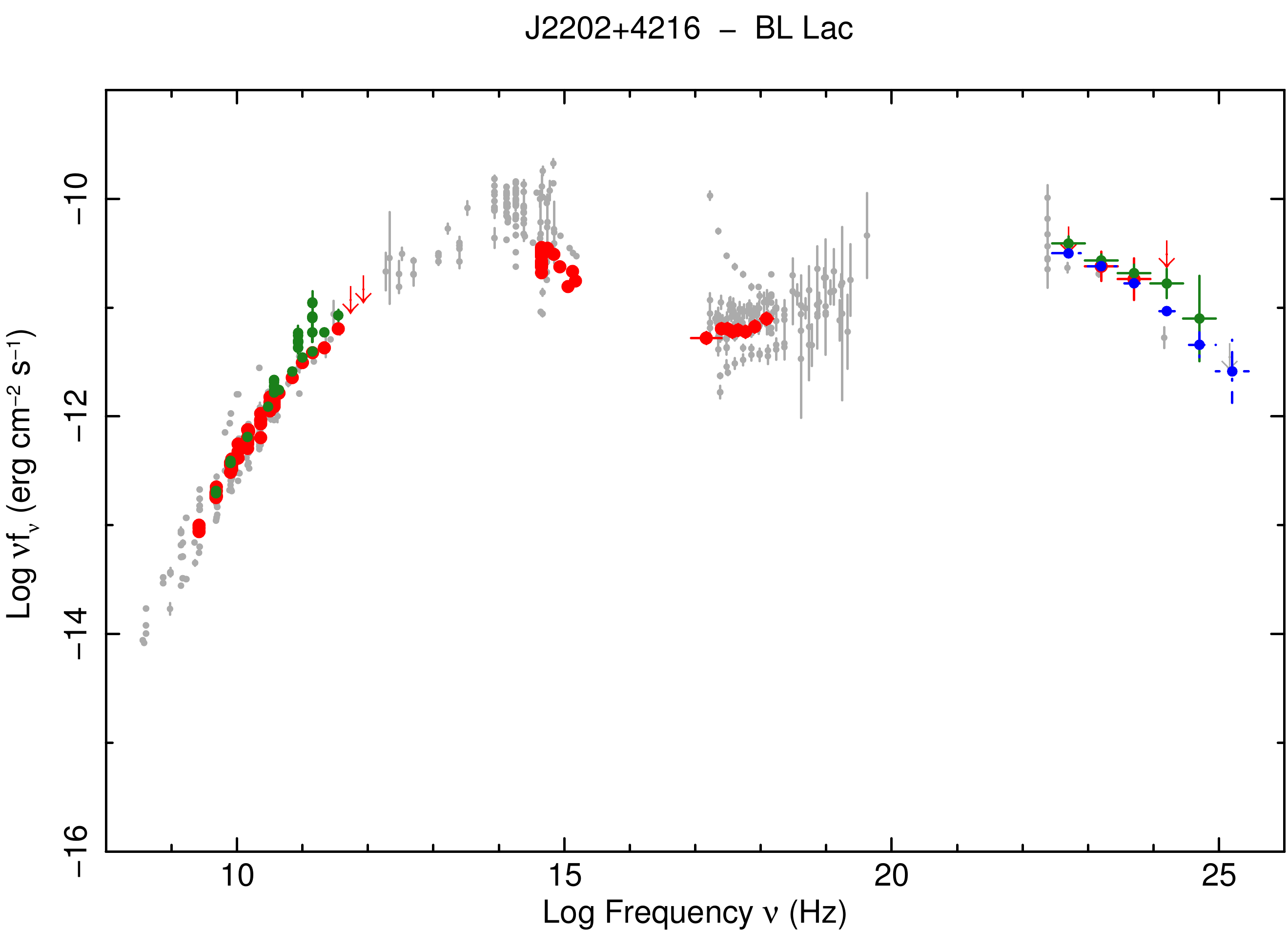}
\includegraphics[width=6.5cm,angle=-90]{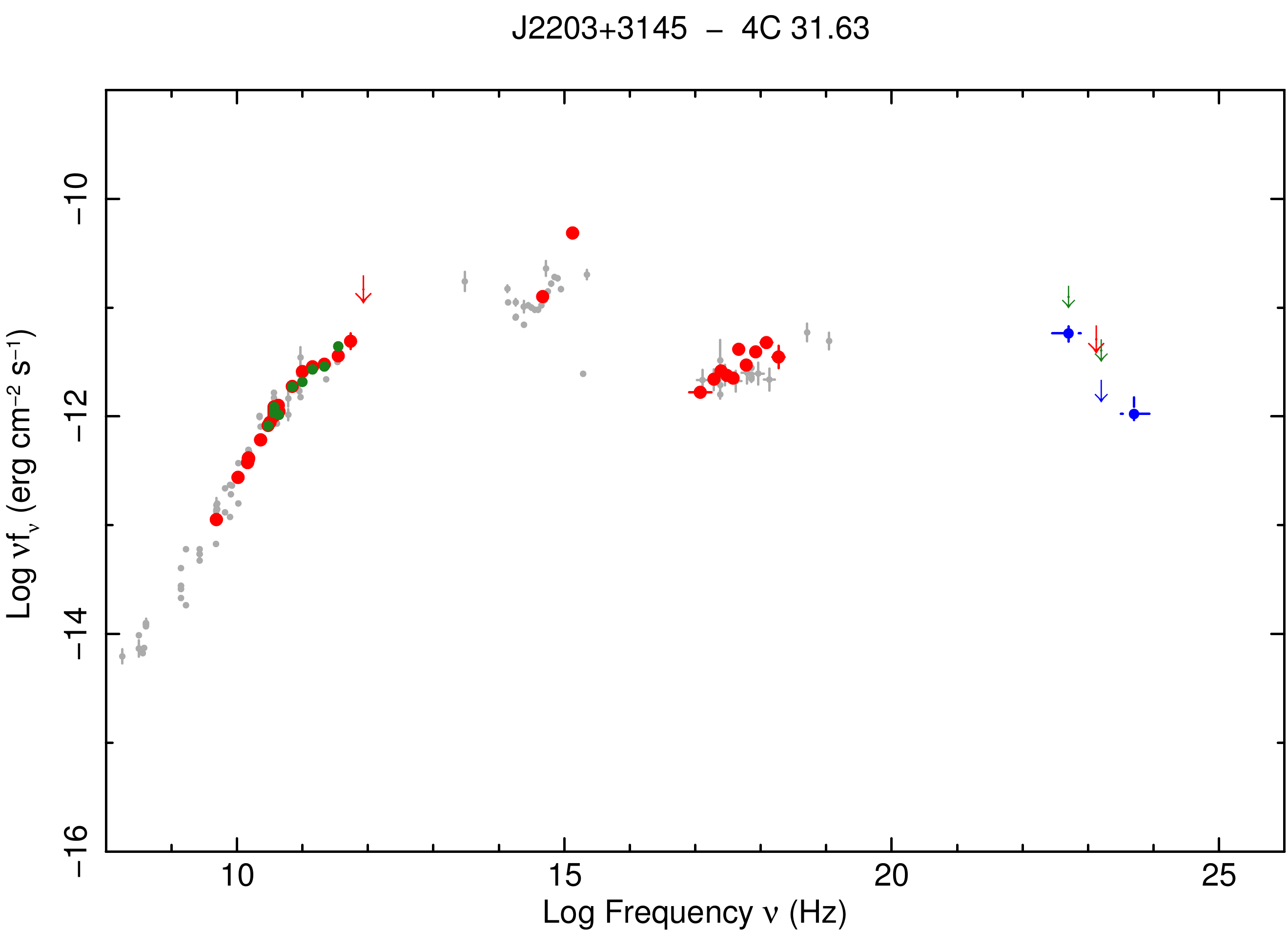}
\includegraphics[width=6.5cm,angle=-90]{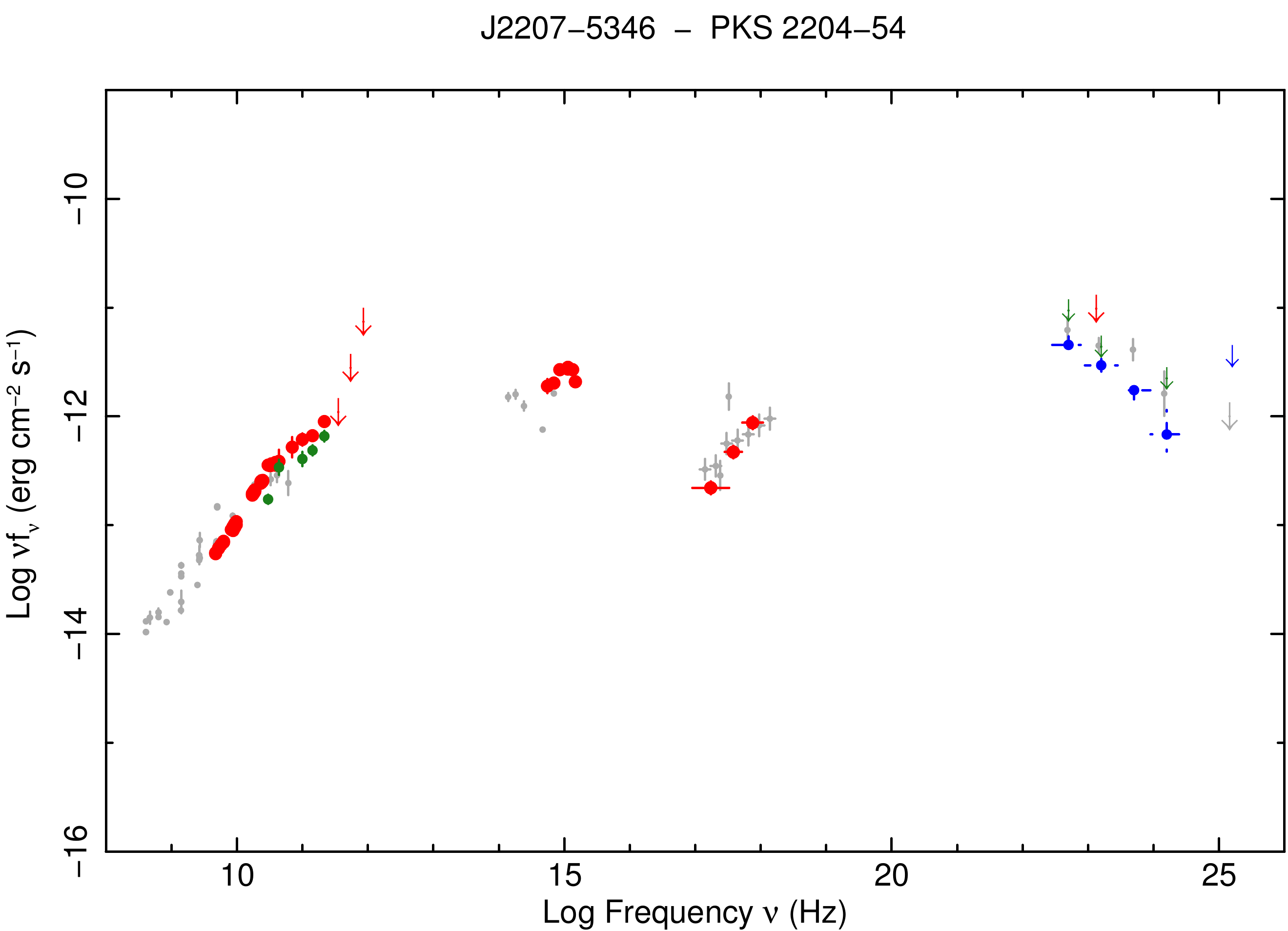}
\caption{The SED of 1Jy\,2144+092 (J2147+0929, top left), 4C\,06.69 (J2148+0657, top right),
PKS\,2149$-$307 (J2151$-$3027, middle left), BL\,Lac (J2202+4216, middle right),
4C\,31.63 (J2203+3145, botton left), and PKS\,2204$-$54 (J2207$-$5346, botton right). 
Simultaneous data are shown in red; quasi-simultaneous data, i.e. {\it Fermi} data
integrated over 2 months, {\it Planck} ERCSC and non-simultaneous ground based observations
are shown in green; {\it Fermi} data integrated over 27 months are shown in blue;
literature or archival data are shown in light gray.}
\label{fig:sed46}
\end{figure*}

\begin{figure*}
\centering
\includegraphics[width=6.5cm,angle=-90]{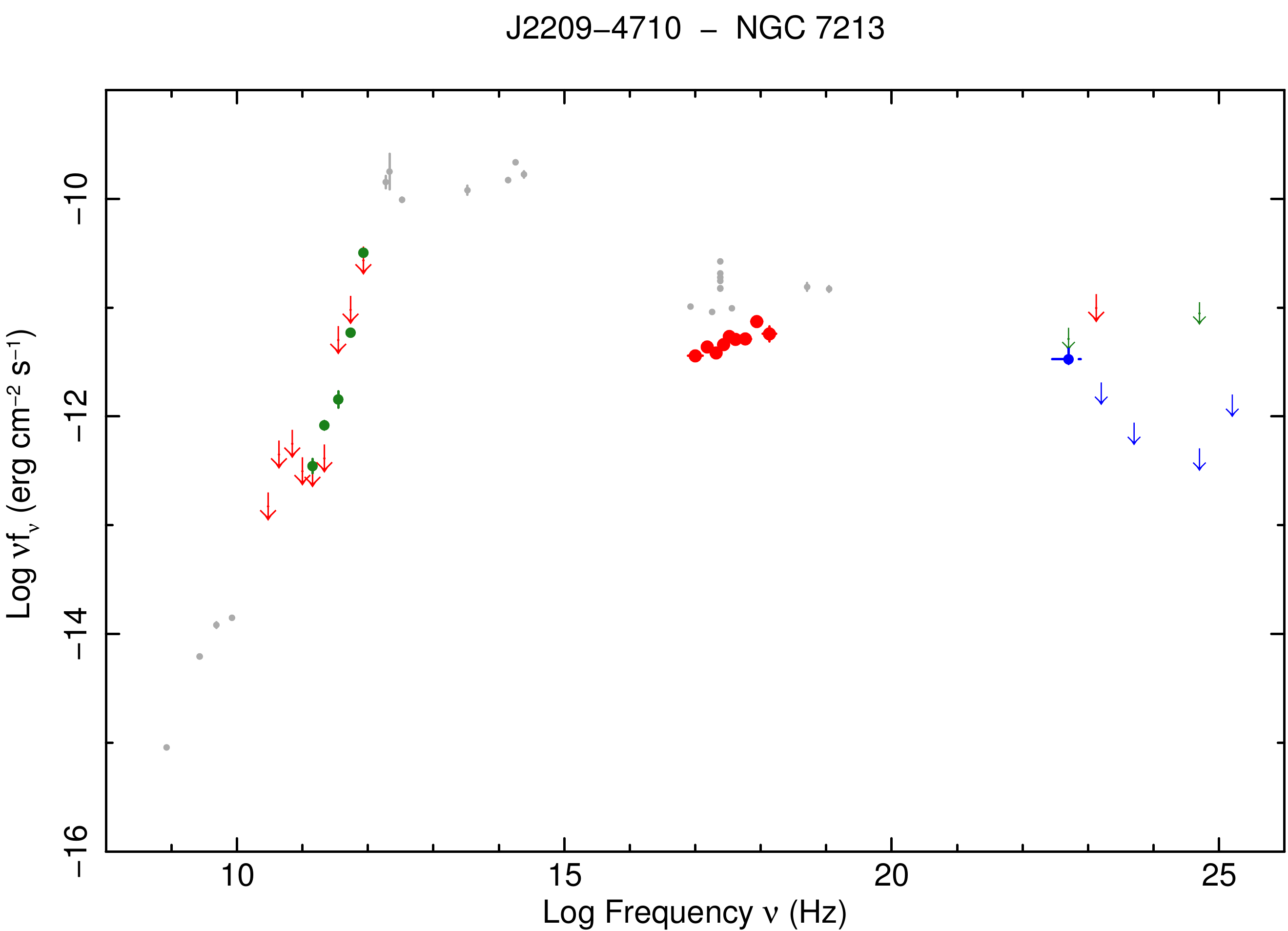}
\includegraphics[width=6.5cm,angle=-90]{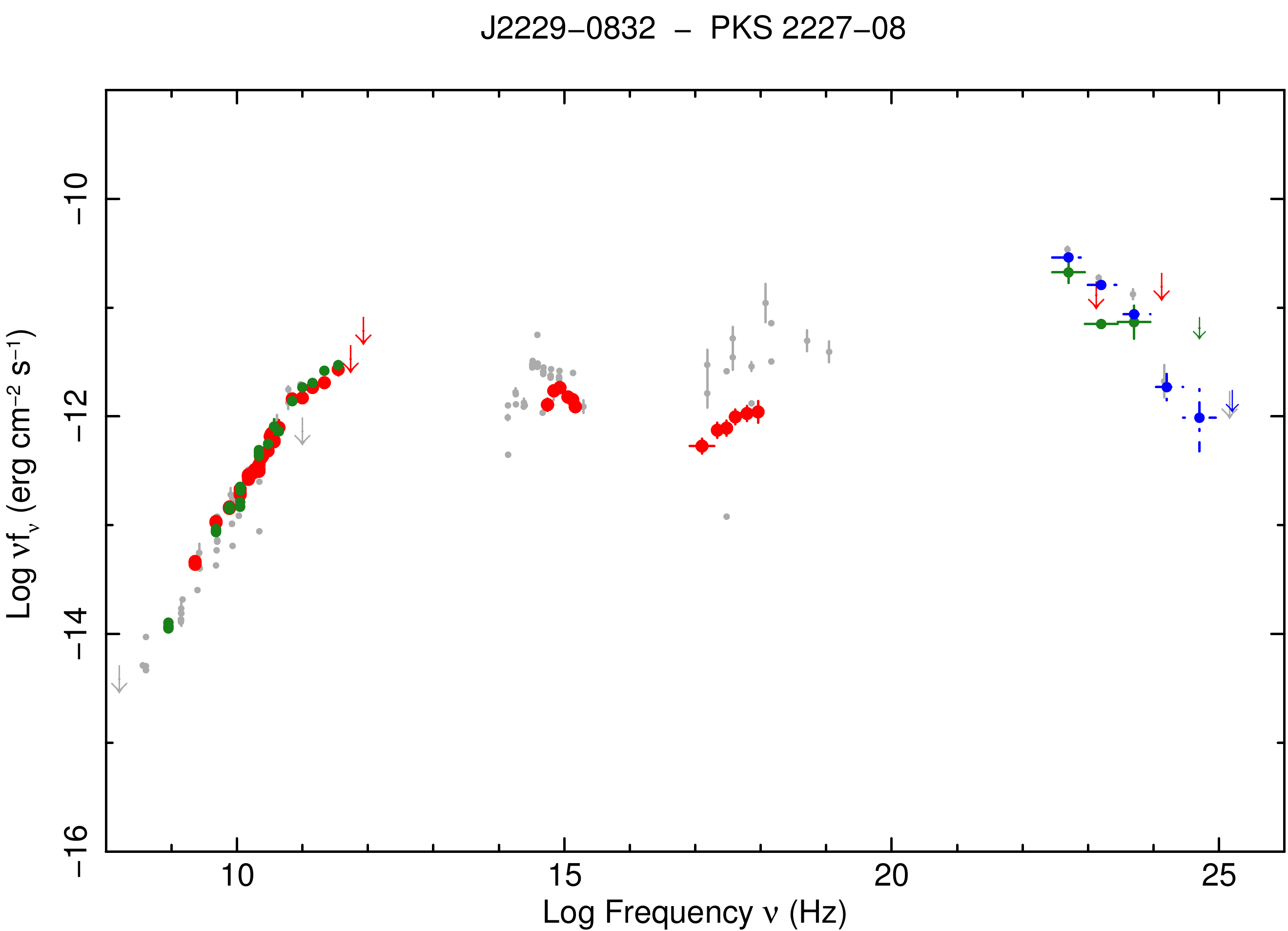}
\includegraphics[width=6.5cm,angle=-90]{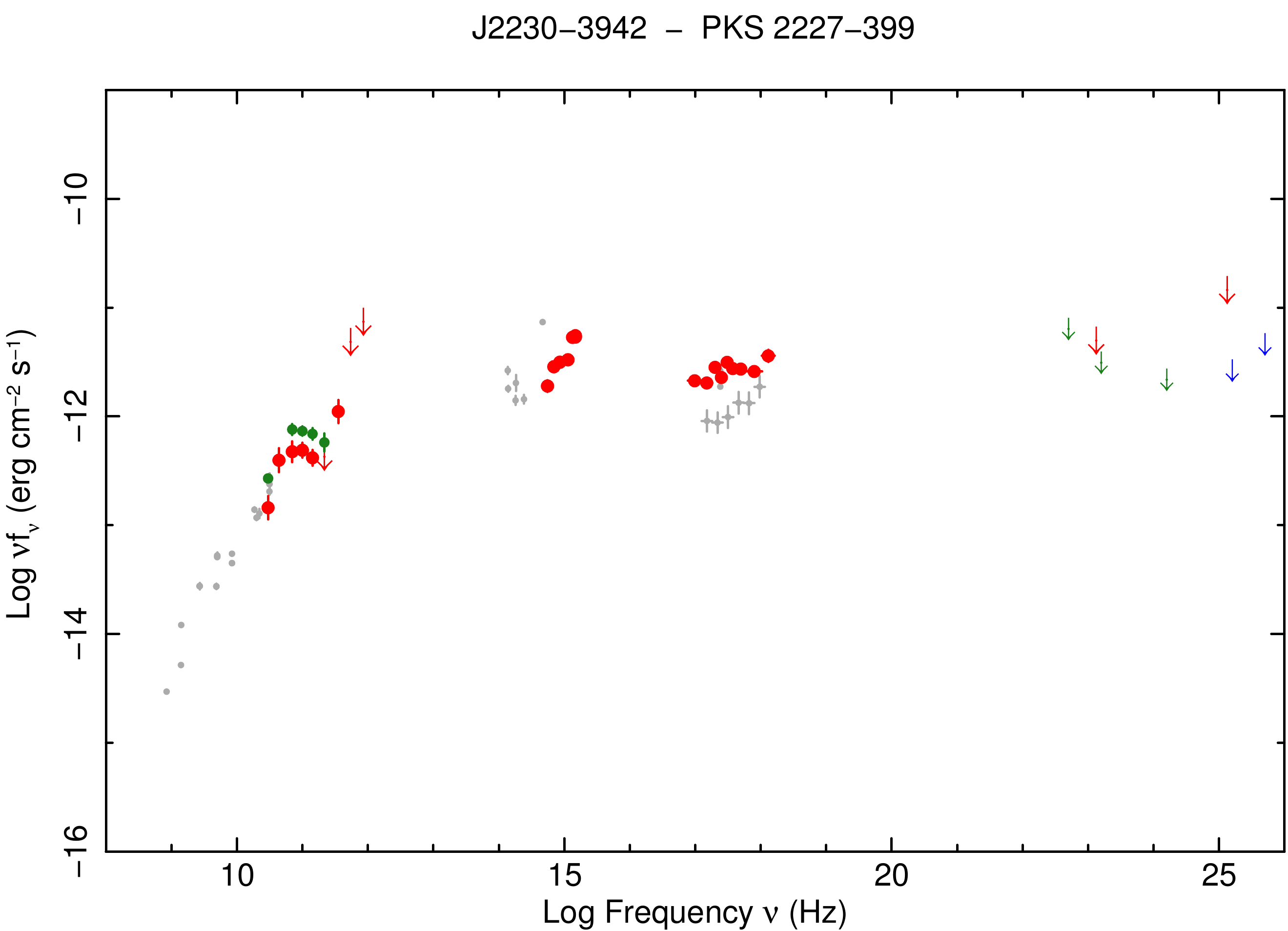}
\includegraphics[width=6.5cm,angle=-90]{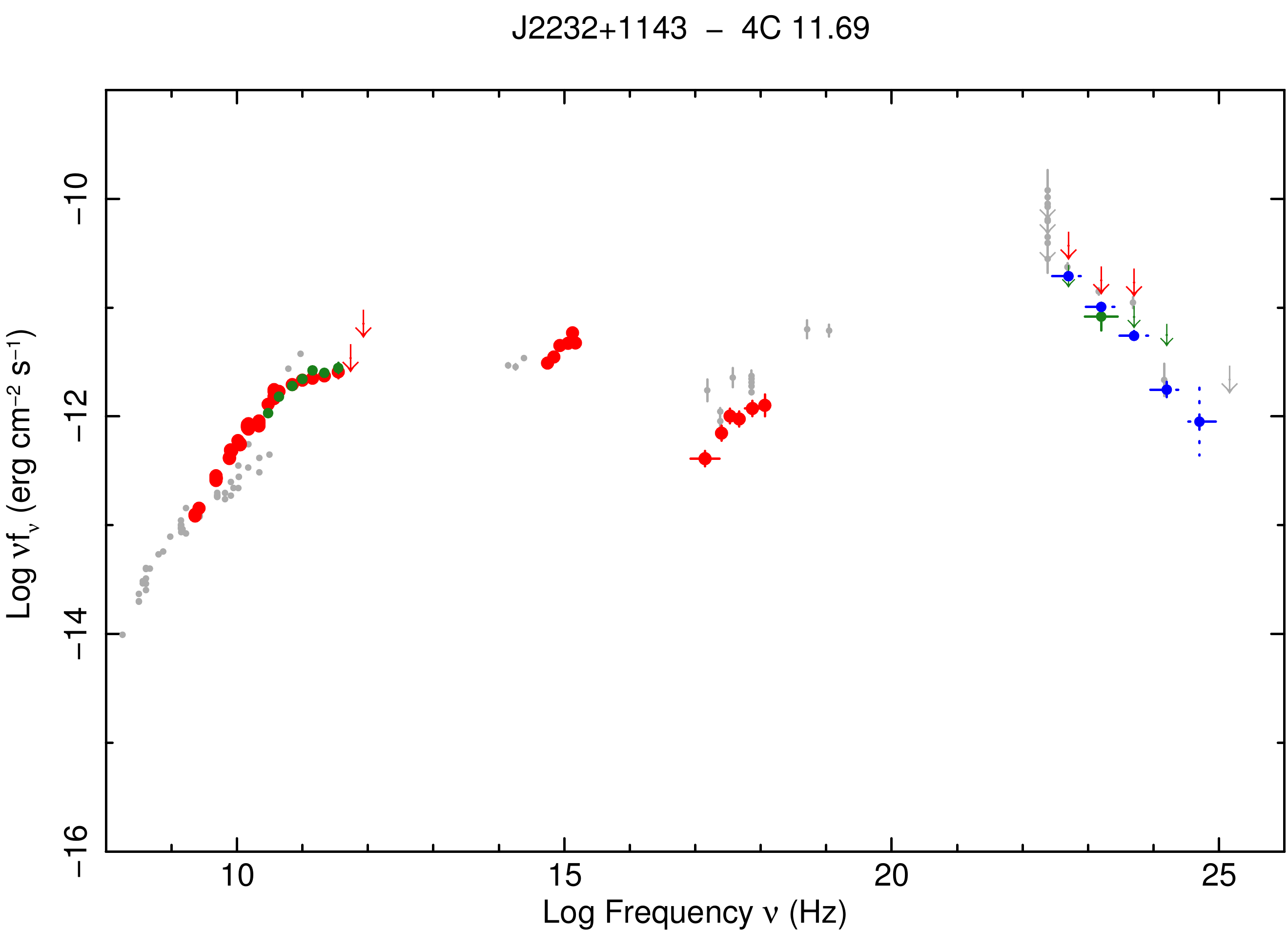}
\includegraphics[width=6.5cm,angle=-90]{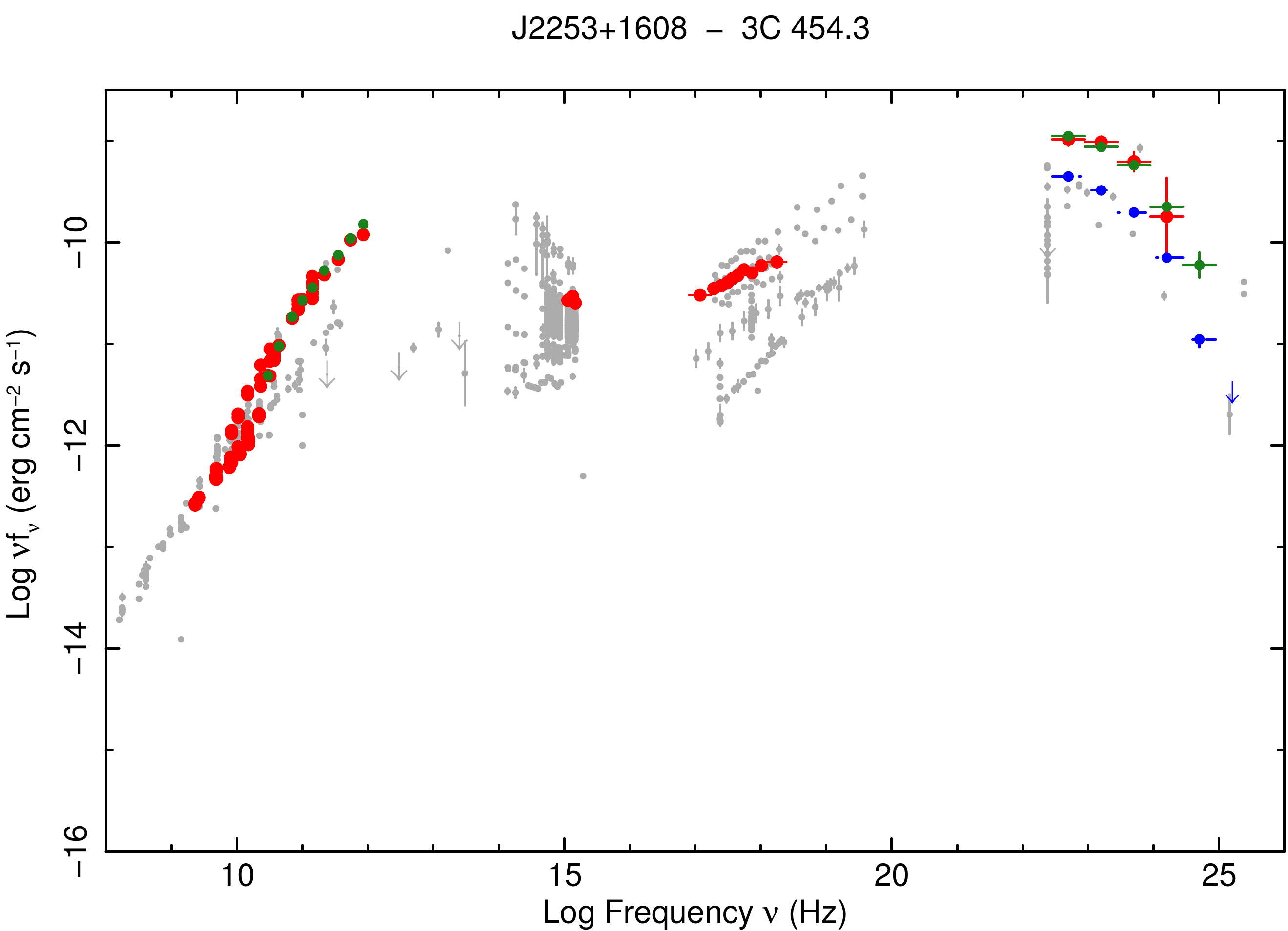}
\includegraphics[width=6.5cm,angle=-90]{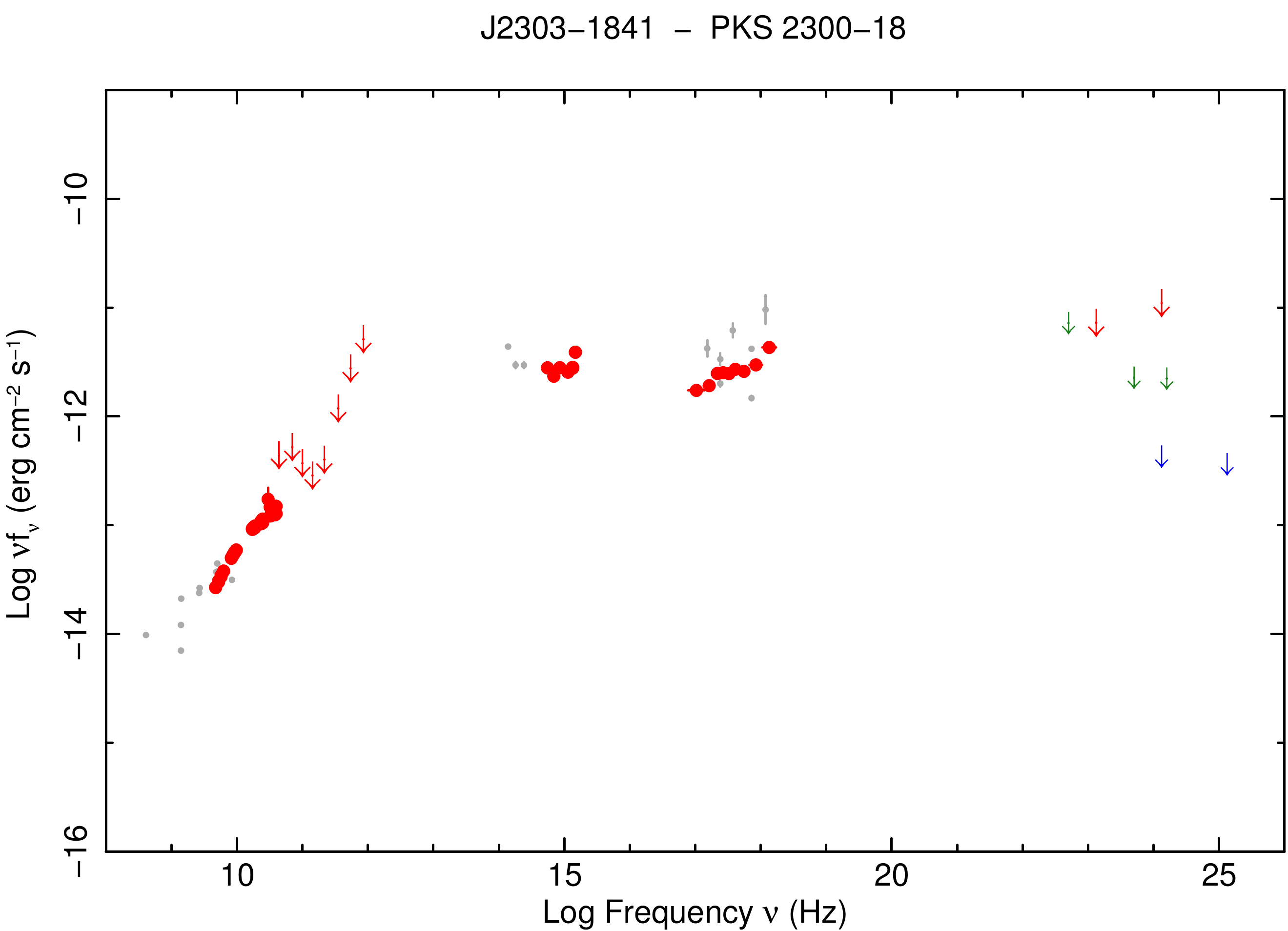}
\caption{The SED of NGC\,7213 (J2209$-$4710, top left), PKS\,2227$-$08 (J2229$-$0832, top right),
PKS\,2227$-$399 (J2230$-$3942, middle left), 4C\,11.69 (J2232+1143, middle right),
3C\,454.3 (J2253+1608, bottom left), and PKS\,2300$-$18 (J2303$-$1841, bottom right). 
Simultaneous data are shown in red; quasi-simultaneous data, i.e. {\it Fermi} data
integrated over 2 months, {\it Planck} ERCSC and non-simultaneous ground based observations
are shown in green; {\it Fermi} data integrated over 27 months are shown in blue;
literature or archival data are shown in light gray.}
\label{fig:sed49}
\end{figure*}

\begin{figure*}
\centering
\includegraphics[width=6.5cm,angle=-90]{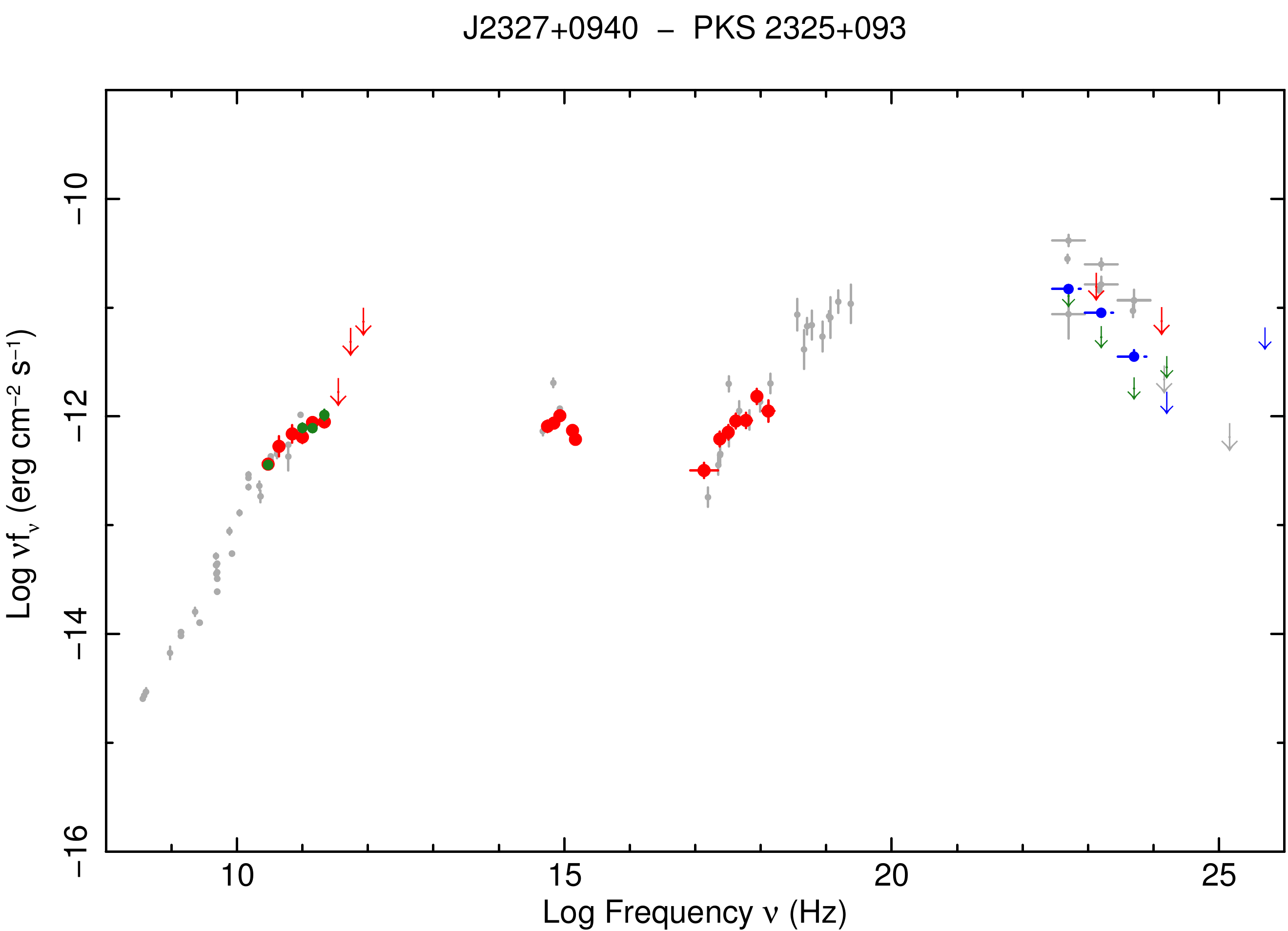}
\includegraphics[width=6.5cm,angle=-90]{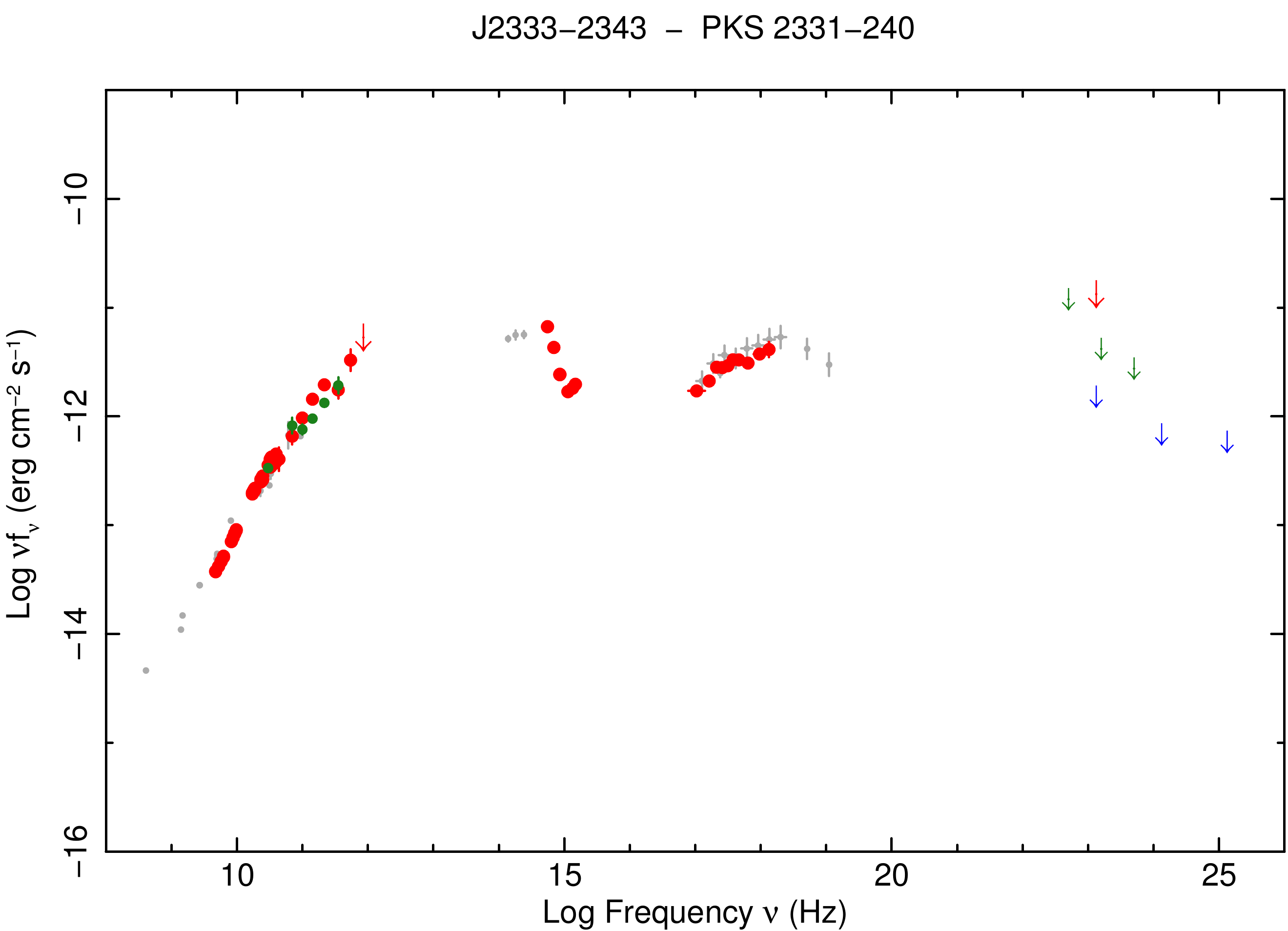}
\includegraphics[width=6.5cm,angle=-90]{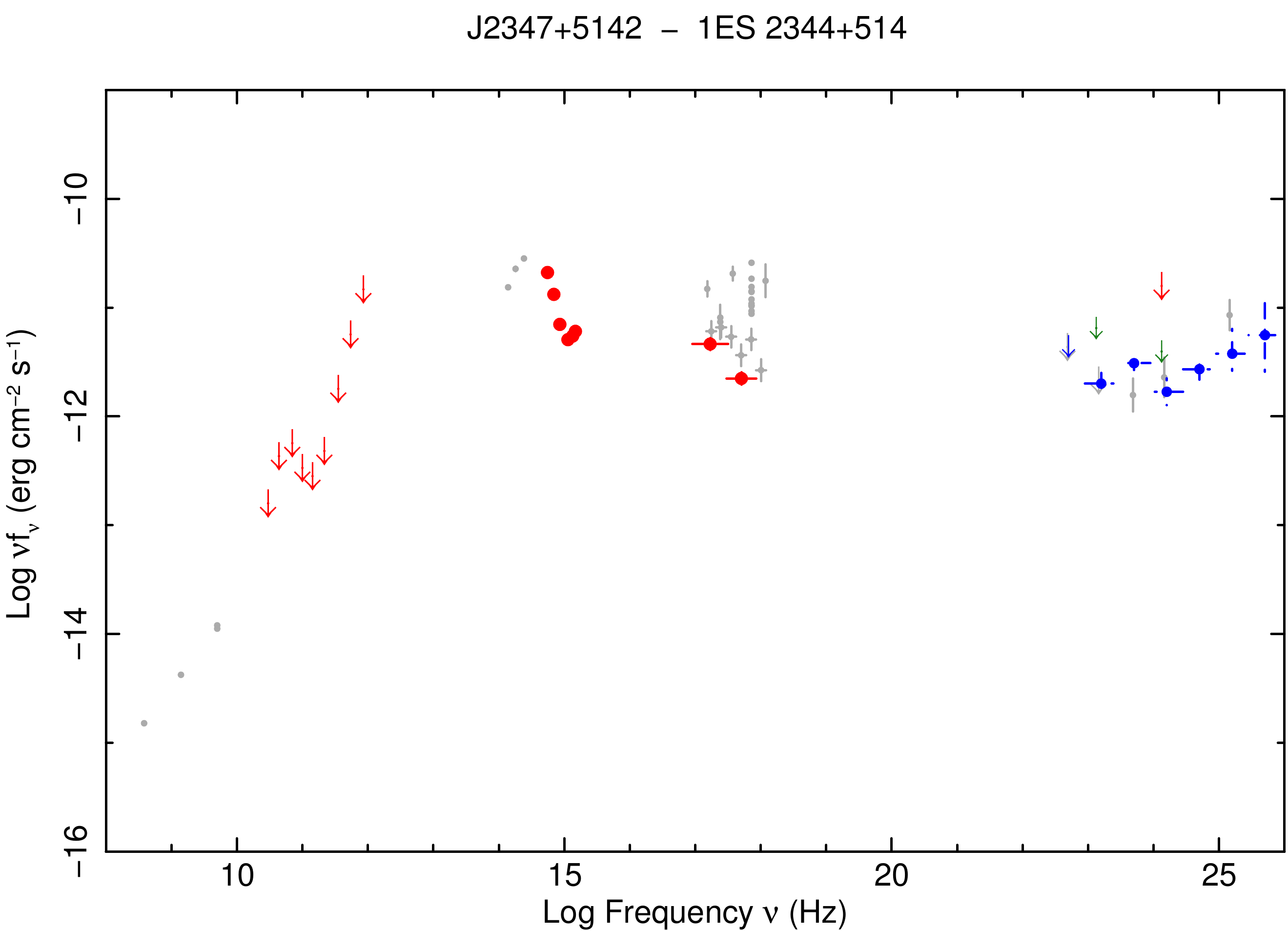}
\caption{The SED of PKS\,2325+093 (J2327+0940, top left), PKS\,2331$-$240 (J2333$-$2343, top right),
and 1ES\,2344+514 (J2347+5142, middle). 
Simultaneous data are shown in red; quasi-simultaneous data, i.e. {\it Fermi} data
integrated over 2 months, {\it Planck} ERCSC and non-simultaneous ground based observations
are shown in green; {\it Fermi} data integrated over 27 months are shown in blue;
literature or archival data are shown in light gray.}
\label{fig:lastSED}
\end{figure*}

The top panel of Fig.~\ref{fig:BlazarSequence} plots the logarithm of the bolometric power, represented by the sum of the synchrotron and inverse Compton peak luminosities [$L_{\rm Bol} \sim \nu^{\rm S}_{\rm peak}L({\nu^{\rm S}_{\rm peak}}) + \nu^{\rm IC}_{\rm peak}L({\nu^{\rm IC}_{\rm peak}})$]
as a function of log(\nupS) for all sources in the four samples considered in this paper and  \cite{planck2011-6.3a} for which an estimate of \nupS\ and the bolometric luminosity was possible.
 We use this plot to test the relationship known as the {\it Blazar Sequence}, which is the strong anti-correlation between bolometric luminosity and \nupS claimed by \cite{fossati98} and \cite{ghisellini98} that remains a subject of lively debate \citep[e.g.,][]{giommi99,padovani03,caccianiga04,nieppola06,padovani07,ghisellinitavecchio08}. 

We stress that to robustly test for the existence of such a relationship it is mandatory to use samples that are {\it unbiased}, that is selected in such a way that  
no particular part of the $L_{\rm Bol}$--\nupS\ diagram is more likely than others to be selected.    
Although our samples are statistically well-defined, they are not unbiased from this viewpoint for the following reasons: 

a) As shown in Figs.~\ref{fig:nupeaks_distr_sy} and~\ref{fig:nupeaks_distr_ic}, the distribution of  \nupS\ strongly depends on the selection method; hence, for 
the various samples, we obtain different samplings of the parameter \nupS. 

b) The area above the dashed line, which represents the luminosity above which the non-thermal emission of a blazar completely dominates the observed optical flux,\footnote{We assume a non-thermal luminosity one order of magnitude higher than that of the host giant-elliptical galaxy of luminosity equal to that found by \cite{scarpa00} and \cite{urryscarpa00} and verified by us to fit our SEDs.} cannot be populated by blazars with no emission lines, such as BL Lacs, as in this case they would appear completely featureless and therefore their redshift could not be measured. 
In this respect, we note that over 40\% of the BL Lacs in the BZCAT catalog, and an even larger fraction of the \fermi-LAT detected BL Lacs, still lack any redshift measurement \citep[][]{bzcat10,shaw09,shaw10}. All the BL Lacs in our samples for which redshifts are unknown are plotted in Fig.~\ref{fig:BlazarSequence} as lower limits estimated assuming that their non-thermal light is ten times brighter than the optical light of the host galaxy.
Some BL Lacs of known luminosity (red points in Fig.~\ref{fig:BlazarSequence}) are above the red line because their redshift was measured from emission lines with rest-frame equivalent widths below the 5\AA~limit; they might be objects with properties in-between those of BL Lacs and FSRQs \cite[see. e.g.,][]{ghisellini11}.

c) The different radio flux-density cuts applied to our samples imply that different subsamples probe different parts of the radio luminosity function. The radio and \gr~selected samples, which are defined using a high radio flux limit ($S\ge 1$\,Jy), probe the high-luminosity end of the radio luminosity function, which, for \nupS = 10$^{13}$\,Hz and the observed redshift and CD distributions, translates into $L_{\rm Bol} \gsim 10^{46} $\,erg\,s$^{-1}$. The soft and hard X-ray selected samples have a radio flux density cut of $S = 0.1$--0.2\,Jy, or about one order of magnitude fainter than that of the radio and \fermi-LAT samples, hence might include significantly less powerful sources, as faint as bolometric luminosities of the order of $10^{44}$\,erg\,s$^{-1}$.

If we remove the \nupS\ dependence of the selection method by considering only FSRQs (which have the same \nupS\ distributions as all samples), we see that  the 
luminosity values span five orders of magnitude from $\sim 10^{44} $ to $\sim 10^{49} $ erg\,s$^{-1}$ and show  no trend with  \nupS, which only ranges  between  $\sim 10^{12.5} $ and $\sim 10^{14}$\,Hz. No obvious correlation is present in each sample separately or in the union  of the four samples.
The L-shaped distribution that is apparent in Fig. \ref{fig:BlazarSequence}, if lower limits are ignored, is similar to that found by \cite{meyer11} who estimated both the
 \nupS\ and peak luminosities of a large sample of blazars using \textit{non-simultaneous} multi-frequency data. These authors, however, instead of considering lower limits to 
 the peak luminosities of blazars with unknown redshifts,  assumed a luminosity corresponding to the redshift that the host galaxy would have for the observed blazar optical magnitude. \cite{meyer11} argued that the strong correlation predicted by the blazar sequence  turns into a \textit{ blazar envelope} when partly misaligned blazars are included in the samples. However, 
\cite{giommisimplified}, by means of detailed Monte Carlo simulations, showed that this envelope, or  L-shaped distribution, is expected when blazars with no redshift measurements are not properly taken into account.

Finally, we consider the subsample of sources that satisfy the {\it same conditions} as in \cite{fossati98}, that is  
$S_{5\,\rm GHz} > 2$\,Jy for FSRQs,  $S_{5\,\rm GHz} > 1$\,Jy for radio-selected BL Lacs, no restrictions for  X-ray selected BL Lacs, and the exclusion of all BL Lacs with 
no redshift information. 
This case is illustrated in Fig.~\ref{fig:BlazarSequence} (bottom panel), which shows a trend very similar to that presented in \cite{fossati98}.

Taking take into account all of the above, we conclude that our data do not show  a correlation of the type predicted by  \cite{fossati98}, owing to  the presence of low luminosity LSP objects and the difficulty in measuring the redshifts of likely high-luminosity HSP sources. That such a correlation becomes evident when the objects are selected by the criteria of \cite{fossati98} supports the hypothesis that the correlation might be the result of a selection effect.

\subsection{Selection effects and sample composition}

Our  decision to select flux-limited samples for widely different parts of the electromagnetic spectrum (radio, soft X-ray, hard X-ray, and \gr) has allowed us to demonstrate 
the strong selection biases that can affect important physical parameters, such as the peak energy of both the synchrotron and  inverse Compton components (see Figs.~\ref{fig:nupeaks_distr_sy} and~\ref{fig:nupeaks_distr_ic}) and the Compton dominance (see Fig.~\ref{fig:cddistr}). Since FSRQs and BL Lacs have significantly different \nupS\ distributions, these selection biases also strongly affect the composition of the samples in terms of the relative abundances of blazar subclasses (FSRQs vs. BL Lacs, LSPs vs. HSPs), as is apparent from Table~\ref{tab:samples}.

Radio-selected samples include sources that are bright in the radio band. If there is no correlation between radio flux/luminosity and other parameters such as \nupS\ and Compton dominance, this is the best selection for measuring
 the distributions of these important physical parameters. If instead there is a strong correlation between radio luminosity and \nupS, then the distribution of \nupS\ should strongly depend on
the radio flux limit.

X-ray selection favors high \nupS\ (and consequently high \nupIC) sources, which are much brighter at X-ray frequencies than low \nupS\ for the same radio flux. X-ray flux-limited 
samples are therefore much richer in high \nupS\ BL Lacs (HBLs or HSP sources) than radio-selected samples. This selection effect has been known since the first soft X-ray surveys became available.

Selection in the \gr\ band favors bright \gr~objects and therefore highly Compton-dominated sources.  \fermi-LAT TS-limited samples contain more sources with flat \gr~spectral slopes or high \nupIC sources. This explains the overabundance of HSP blazars (only BL Lacs) and high CD  blazars (only FSRQs) in \fermi-LAT catalogs.

\begin{acknowledgements}

We thank the entire \swift~team for the help and support and especially the Science Planners and Duty Scientists for their invaluable help and professional support with the
planning and execution of a large number of ToOs.

The \planck\ Collaboration acknowledges the support of: ESA; CNES and CNRS/INSU-IN2P3-INP (France); ASI, CNR, and INAF (Italy); NASA and DoE (USA); STFC and UKSA (UK);
CSIC, MICINN and JA (Spain); Tekes, AoF and CSC (Finland); DLR and MPG (Germany); CSA (Canada); DTU Space (Denmark); SER/SSO (Switzerland); RCN (Norway); 
SFI (Ireland); FCT/MCTES (Portugal); and DEISA (EU).
A full description of the \planck\ Collaboration and a list of its members, indicating which technical or scientific activities they have been involved in, can be found at \url{http://www.rssd.esa.int/Planck}.
We thank the \planck\ team and in particular the members of the Data Processing Centers for their support in the reduction of LFI and HFI data carried out specifically for this work.  

The \fermi-LAT Collaboration acknowledges generous ongoing support
from a number of agencies and institutes that have supported both the
development and the operation of the LAT as well as scientific data analysis.
These include the National Aeronautics and Space Administration and the
Department of Energy in the United States, the Commissariat \`a l'Energie Atomique
and the Centre National de la Recherche Scientifique / Institut National de Physique
Nucl\'eaire et de Physique des Particules in France, the Agenzia Spaziale Italiana (ASI)
and the Istituto Nazionale di Fisica Nucleare (INFN) in Italy, the Ministry of Education,
Culture, Sports, Science and Technology (MEXT), High Energy Accelerator Research
Organization (KEK) and Japan Aerospace Exploration Agency (JAXA) in Japan, and
the K.~A.~Wallenberg Foundation, the Swedish Research Council and the
Swedish National Space Board in Sweden.

Additional support for science analysis during the operations phase from the Istituto Nazionale di Astrofisica in Italy and the Centre National d'\'Etudes Spatiales in France is gratefully
acknowledged.

The Mets\"ahovi team acknowledges the support from the Academy of Finland
to our projects (numbers 212656, 210338, 121148, and others).
This work was also supported by grants 127740 and 122352 of the Academy of Finland. UMRAO is supported by a series of grants from the NSF and NASA, and by the University of Michigan.
This publication is partly based on data acquired with the Atacama
Pathfinder Experiment (APEX). APEX is a collaboration between the
Max-Planck-Institut f\"ur Radioastronomie, the European Southern
Observatory, and the Onsala Space Observatory. This research is
partly based on observations with the 100-m telescope of the MPIfR
(Max-Planck-Institut f\"ur Radioastronomie) at Effelsberg, the IRAM 30-m
telescope, and the Medicina (Noto) telescope operated by INAF ---Istituto di Radioastronomia.
J. Wu and X. Zhou are supported by the Chinese National Natural Science Foundation grants 10633020, 10778714, and 11073032, and by the National Basic Research Program of China (973 Program) No. 2007CB815403.
The OVRO 40-m monitoring program is supported in part by NASA grants
NNX08AW31G and NNG06GG1G and NSF grant AST-0808050.
O.G. King acknowledges the support of a Keck Institute for Space Studies
Fellowship. W. Max-Moerbeck acknowledges support from a Fulbright-CONICYT scholarship. V. Pavlidou acknowledges support provided by NASA through Einstein Postdoctoral Fellowship grant number PF8-90060 awarded by the Chandra X-ray Center, which is operated by the Smithsonian Astrophysical Observatory for NASA under contract NAS8-03060.
The Australia Telescope is funded by the Commonwealth of Australia for
operation as a National Facility managed by CSIRO.

This paper makes use of observations obtained at the Very Large Array (VLA) which is an instrument of the National
Radio Astronomy Observatory (NRAO). The NRAO is a facility of the National Science Foundation operated under cooperative agreement
by Associated Universities, Inc.


We acknowledge the use of data and software facilities from the ASI
Science Data Center (ASDC), managed by the Italian Space Agency (ASI).  Part of this work is based on 
archival data and on bibliographic information obtained from the NASA/IPAC Extragalactic Database (NED) and
from the Astrophysics Data System (ADS).

We thank the anonymous referee for his/her useful and constructive comments.
\end{acknowledgements}

\bibliographystyle{aa}
\bibliography{paper}

\raggedright

\clearpage

\clearpage
 
 \clearpage
 
  \clearpage
  
\clearpage
 
\clearpage

\longtab{6}{
\begin{landscape}
\scriptsize

}

\raggedright
\onecolumn
\end{document}